\documentclass[twoside,12pt,sort,compress]{elsarticle}

\usepackage{epsfig}
\usepackage{graphicx,bm,amsmath,color,scalerel}
\usepackage{xcolor} 
\usepackage{comment}
\usepackage{verbatim}
\usepackage{hyperref}
\hypersetup{
    colorlinks=true,
    linkcolor=blue,
    filecolor=magenta,      
    urlcolor=cyan,
}
\usepackage{latexsym}
\usepackage{acronym}
\usepackage{amssymb}
\usepackage{multirow}
\usepackage{multicol}
\usepackage{gensymb} %adds \degree
\usepackage{ragged2e}

\usepackage[displaymath,mathlines]{lineno}

\newcommand*\linenomathpatchAMS[1]{%
  \expandafter\pretocmd\csname #1\endcsname {\linenomathAMS}{}{}%
  \expandafter\pretocmd\csname #1*\endcsname{\linenomathAMS}{}{}%
  \expandafter\apptocmd\csname end#1\endcsname {\endlinenomath}{}{}%
  \expandafter\apptocmd\csname end#1*\endcsname{\endlinenomath}{}{}%
}

\expandafter\ifx\linenomath\linenomathWithnumbers
  \let\linenomathAMS\linenomathWithnumbers
  \patchcmd\linenomathAMS{\advance\postdisplaypenalty\linenopenalty}{}{}{}
\else
  \let\linenomathAMS\linenomathNonumbers
\fi

\linenomathpatchAMS{gather}
\linenomathpatchAMS{multline}
\linenomathpatchAMS{align}
\linenomathpatchAMS{alignat}
\linenomathpatchAMS{flalign}

\newcommand{\be}{\begin{equation}}
\newcommand{\ee}{\end{equation}}
\newcommand{\bea}{\begin{eqnarray}}
\newcommand{\eea}{\end{eqnarray}}

\begin{document}

\title{Quantum gravity phenomenology at the dawn of the multi-messenger era --- A review}

\author[Sichuan,INFNFrascati]{A.~Addazi}
\author[IGFAE]{J.~Alvarez-Muniz}
\author[UAM/CSIC]{R.~Alves~Batista}
\author[Napoli,INFNNapoli]{G.~Amelino-Camelia}
\author[Milano,INFNMilano]{V.~Antonelli}
\author[Napoli,INFNNapoli]{M.~Arzano}
\author[Zaragoza]{M.~Asorey}
\author[Toulouse]{J.-L.~Atteia}
\author[Tartu,TokyoTech]{S.~Bahamonde}
\author[Napoli]{F.~Bajardi}
% ...

\address[Sichuan]{Center for Theoretical Physics, College of Physics, Sichuan University, 610065 Chengdu, China}
\address[INFNFrascati]{Laboratori Nazionali di Frascati INFN Via Enrico Fermi 54, I-00044 Frascati (Roma), Italy}
\address[IGFAE]{Instituto Galego de Física de Altas Enerxías (IGFAE), Universidade de Santiago de Compostela, 15782 Santiago de Compostela, Spain}
\address[UAM/CSIC]{Instituto de Física Teórica (UAM/CSIC), Universidad Autónoma de Madrid, Cantoblanco, 28049 Madrid, Spain}
\address[Napoli]{Dipartimento di Fisica Ettore Pancini, Università di Napoli ``Fedeico II”, Complesso Univ. Monte S. Angelo, I-80126 Napoli, Italy}
\address[INFNNapoli]{INFN, Sezione di Napoli}
\address[Milano]{Dipartimento di Fisica, Università degli Studi di Milano, via Celoria 16, I-20133 Milano, Italy}
\address[INFNMilano]{INFN, Sezione di Milano, via Celoria 16, I-20133 Milano, Italy }
\address[Zaragoza]{Centro de Astropartículas y Física de Altas Energías (CAPA), Departamento de Física Teórica, Universidad de Zaragoza, C/ Pedro Cerbuna 12, E-50009 Zaragoza, Spain}
\address[Toulouse]{IRAP, Université de Toulouse, CNRS, CNES, UPS, (Toulouse), France}
\address[Tartu]{Laboratory of Theoretical Physics, Institute of Physics, University of Tartu, W. Ostwaldi 1, 50411 Tartu, Estonia}
% ...
\address[IEM/CSIC]{Instituto de Estructura de la Materia, CSIC, Serrano 121, 28006 Madrid, Spain}
% ...
\address[TokyoTech]{Department of Physics, Tokyo Institute of Technology, 1-12-1 Ookayama, Meguro-ku, Tokyo 152-8551, Japan}

\begin{abstract} 
The exploration of the universe has recently entered a new era thanks to the multi-messenger paradigm, characterized by a continuous increase in the quantity and quality of experimental data that is obtained by the detection of the various cosmic messengers (photons, neutrinos, cosmic rays and gravitational waves) from  numerous origins. They give us information about their sources in the universe and the properties of the intergalactic medium. Moreover, multi-messenger astronomy opens up the possibility to search for phenomenological signatures of quantum gravity. On the one hand, the most energetic events allow us to test our physical theories at energy regimes which are not directly accessible in accelerators; on the other hand, tiny effects in the propagation of very high energy particles could be amplified by cosmological distances. After decades of merely theoretical investigations, the possibility of obtaining phenomenological indications of Planck-scale effects is a revolutionary step in the quest for a quantum theory of gravity, but it requires cooperation between different communities of physicists (both theoretical and experimental). This review, prepared within the COST Action CA18108 ``Quantum gravity phenomenology in the multi-messenger approach", is aimed at promoting this cooperation by giving a state-of-the art account of the interdisciplinary expertise that is needed in the effective search of quantum gravity footprints in the production, propagation and detection of cosmic messengers.
\end{abstract}

\maketitle 
\newpage

\tableofcontents

\newpage
\section{Introduction}
\label{section1}

The theory of \ac{QG} still eludes us. To start with, it is not even clear if gravity can be quantized like the rest of the fundamental interactions in Nature. For instance, gravity could be viewed as an emergent force, say entropic, as a result of changes of information associated with the positions of material bodies. If one adopts the former point of view, i.e.\ that gravity has to be quantized, then the important question is to find the extension of Einstein's \ac{GR}, that can accommodate such a quantum nature of the gravitational interactions in a mathematically and physically consistent way. From this point of view, the necessity for going beyond Einstein's \ac{GR} is immediate. The latter is a classical generally covariant, non-linear theory of the gravitational field, in 3+1 dimensional spacetime, with a coupling constant (Newton's (or gravitational) constant) $G$, which carries  dimensions of inverse squared mass. As such, the theory of the quantized gravitational field, viewed as a conventional  \ac{QFT}  extension of \ac{GR}, would be non-renormalizable, in the sense that the higher the order of a perturbative expansion in $G$, the higher the degree of divergence, i.e. at each new order in perturbation theory there would appear new divergent graphs in the \ac{UV} limit of momenta. Thus, it would not be possible to remove the \ac{UV} cutoff in momentum space by absorbing such divergences in a finite number of parameters  (couplings and masses), as is the case of renormalizable theories in flat spacetimes, such as the \ac{SM} of particle physics that describes the electromagnetic, weak and strong interactions in Nature. 

Having discussed the non-renormalizability of \ac{GR}, 
one is tempted to view it as an effective low-energy theory of a more fundamental (renormalizable in some sense) theory, which is valid below some energy scale, characteristic of the larger theory. One can make the analogy with the Fermi theory of the 
$\beta$-decay in nuclear physics, which is a non-renormalizable theory of fermions, involving four-fermion contact interactions, with a dimensionful coupling, the Fermi ``constant" $G_\text{F}$, which also has units of inverse squared mass. The Fermi theory can be viewed as an effective low-energy theory of the \ac{SM}, which can be embedded into a mathematically and physically consistent way, by viewing the four-fermion contact interactions as a low-energy limit of renormalizable \ac{SM} trilinear-vertex interactions, mediated by $SU(2)$ weak gauge bosons in the spontaneously broken phase of the \ac{SM}, in which the latter are massive.
The Fermi theory is obtained at momentum scales of the exchanged gauge boson $ p \ll M_W$.  In this limit, the Fermi constant can be expressed as $G_\text{F} \propto g^2/M_W^2$, where $g$ is the dimensionless $SU(2)$ coupling of the \ac{SM}, and $M_W$ is the mass of the charged-weak-boson ($W^\pm$-bosons), of order $\mathcal O(100\,\rm GeV)$. Thus, the electroweak symmetry breaking scale $\mathcal O(M_W)$
is a characteristic scale, at which corrections to the 
Fermi theory become important, and the theory is no longer adequate to describe physical phenomena.  

A natural question that arises at this point concerns whether there is a corresponding mass (or energy) scale for quantum gravity, $M_{\rm QG}$, 
below which GR can be more or less adequate in describing the natural world, including the gravitational interactions, but at which (and above it), a new consistent quantum theory, renormalizable or even finite appears as the correct natural theory to consider, whose low-energy limit compared to $M_{\rm QG}$ yields \ac{GR}. In fact, surprisingly enough, such a scale existed well before the theory of \ac{GR}, and was due to the German physicist M.\ Planck, who in 1899 suggested that there existed some fundamental natural units for length, mass, time and energy. 
Using dimensional analysis, he postulated the existence of a fundamental length (now called the
Planck length) based only on the speed of light {\it in vacuo}, $c$, Newton's constant $G$ and what is now called the Planck constant $h$. A modern version of the Planck length makes use of the so-called reduced Planck constant $\hbar = h/2\pi$: 
\[ \ell_\text{P} = \sqrt{\frac{\hbar G}{c^3}} \simeq 1.616 255 (18) \times 10^{-35}\,\rm m. \]
From this length unit one then obtains the derived scales of Planck time, Planck mass, and Planck Energy $E_\text{P}$. From a theoretical point of view, the Planck length is the size of a black hole at which its Schwarzschild radius is the same as its Compton wavelength, that is the length (or, equivalently, energy scale) at which gravitational and quantum physics effects are at the same scale. This may be viewed as the shortest possible length scale which can be probed by particle collisions, since probing scales shorter than this, via higher-energy collisions, may result in the formation of black holes.\footnote{It should be stressed, of course, that whether black holes actually form depends on the details of the gravitational dynamics at the Planck scale.}  We may also view the Planck length as the (length) scale, below which the structure of spacetime itself, as perceived by low-energy (compared to Planck energy) observers, no longer exists, and instead a space-time ``foamy'' structure may emerge. This picture can be understood qualitatively, by adopting the picture of Regge (1958) on the effects of the Planck length on a Euclidean geometry. According to Regge, there is an uncertainty in the metric tensor itself, associated with the fact that, as quantum effects make the gravitational field perform zero-point oscillations, the geometry itself oscillates. The metric tensor uncertainty can be argued to be of order $\Delta g \sim \ell_\text{P}^2/\ell^2$, for a region of the Euclidean spacetime with dimensions $\ell$ (although this latter formula can itself depend on the precise \ac{QG} setting). Usually, for scales of order of the atomic or particle ones, such quantum effects are negligible. But for $\ell$ near the Planck scale, the distortion of the geometry due to quantum effects of spacetime becomes significant, which corroborates the common perception that \ac{QG} effects in $(3+1)$-dimensions set in at the Planck scale $\ell_\text{P}$.

Thus the corresponding Planck-energy scale may be taken to be the characteristic scale of \ac{QG}. Following the above reasoning this is probably the highest energy scale, beyond which a low-energy field theory is not well defined, and, in this latter sense, the Planck scale plays the r\^ole of a physical \ac{UV} cutoff. 
Unfortunately, at present \iac{UV} complete theory of \ac{QG} is not known. 
There are several main streams of research in this direction. One is string theory, which although appears as a mathematically  consistent theory that can incorporate \ac{QG} effects, nonetheless has its own shortcomings, the most important one being its space-time background dependence (at least in its present formulation). Also, its landscape problem might be an issue, given the non-uniqueness of the string vacuum. The important feature of string theory is that mathematical consistency seems to require the spacetime to have higher than 3 spatial dimensions. This has, as a consequence, that the higher-dimensional gravitational energy scale, i.e.\ the scale at which quantum gravitational effects become important, could be much lower than Planck. In string theory this is an arbitrary phenomenological scale at present, and it is currently 
restricted by collider experiments to be higher than a few TeV, since string-like or extra dimensional effects have been excluded up to such scales. 

The background independence of \ac{QG} is tackled in other main stream approaches to \ac{QG}, alternative to string theory,
which are 
based on a dynamical
emergence of spacetime from rather abstract fundamental building blocks (``geometry'' quanta), and as such are rather discretized approaches to \ac{QG}:
(i) \ac{LQG} (and its cosmological version, \ac{LQC}), and the spin-foam models; and
(ii) the \ac{GFT} approach to \ac{QG}, consisting of \acp{QFT} of spacetime in which the
base manifold is an appropriate Lie group, describing the dynamics of both the topology and the
geometry of spacetime. All of the above approaches also suffer from inconsistencies at present, for instance the problem of general covariance is not completely resolved, as yet, given that all such models suffer from some unavoidable degree of appropriate ``gauge fixing''. 

The lack of \iac{UV}-complete, mathematically consistent theory of \ac{QG}, which would lead to phenomenological predictions in the low-energy world (a top-down approach), prompted several physicists, to follow a bottom-up phenomenological approach to \ac{QG}, by trying to postulate properties that could characterize a complete theory of \ac{QG}, which however could have effects at much lower scales than the Planck energy scale, and thus have a chance of leading to realistic prospects for phenomenology. In string theory, for instance, the arbitrariness of the extra-dimensional (``bulk'') gravitational scale, leads to, in principle, falsifiable low-scale string models in collider searches (of course if the string scale is close to Planck, the possibility for experimental falsification, or discovery (!), of the pertinent (high-scale) string models is lost for all practical purposes).   

Nonetheless, if one is prepared to sacrifice some of the important symmetries that particle physics phenomenology is based upon, namely Lorentz and/or \ac{CPT} invariance, both of which could be violated at some high-energy scale in the context of \ac{QG}, then one might have effects that could be searched for, in, say, cosmic high-energy messengers. Such effects may characterize models of \ac{QG} explicitly, in the sense of having \ac{LIV} (and perhaps \ac{CPT} Violation) in the \ac{SME}, due to some tensor fields, acquiring vacuum expectation values, which are constant, due to some unknown mechanisms of the \ac{UV} complete theory of \ac{QG}. The phenomenology of such theories of \ac{QG} can be done by writing down appropriate higher-dimension operators
in the context of \acp{EFT}, and identifying the role of various operators on certain physical processes, which can then be constrained by either precision experiments, or high-energy cosmic messenger observations. In this approach, the Planck scale, or in general the scale of \ac{QG}, since the latter is a phenomenological parameter, and thus need not be identified necessarily with the former, is an observer-independent quantity, as is the case of the gravitational constant also in string theory. 

An alternative approach to the study of departures from Lorentz invariance is provided by \ac{DSR} models, in which the Planck length is viewed as an ordinary length, which under 
the ordinary Lorentz transformations of \ac{SR} would undergo the usual length contraction (and the Planck time the corresponding time dilation). However, by postulating appropriate modifications of the Lorentz transformations
one may arrive at mathematically consistent formulations of the pertinent models in which the Planck length (and thus the other derived Planck-value quantities) remain an invariant for all \ac{DSR} observers, just like the speed of light in vacuo in \ac{SR}, whose invariance is also maintained here. While the laws of transformations between inertial frames are modified, the relativistic properties of \ac{DSR} models are not spoiled. This implies a consistent modification of other physical features of these models. 

One of these is the composition law for momenta in the case of multi-particle states. Indeed,  the standard linear addition of momenta no longer applies, and is replaced by a nonlinear law, which is covariant under the modified relativistic symmetry transformations. This modified composition law is the main distinguishing feature of the \ac{DSR} phenomenology with respect to \ac{LIV} models and Lorentz-invariant models.  For this reason it constitutes an important, currently open, active area of research in such theories, aimed at uncovering the possibly observable experimental implications.  

Relativistic consistency with the modified laws of transformations between inertial observers in \ac{DSR} models might also require a modification of the dispersion relations of particle excitations, so that these are the same for all inertial observers. This is an important prediction which can be tested experimentally, at least in principle.
In fact,  the 
\ac{MDR} is such that the effect increases with the energy of the probe, leading, for example, to differences in the time of travel of more energetic photons, compared to lower-energy ones. These differences accumulate over large travel distances, leading to the expectation that \ac{QG} effects of this sort can be best probed by very high energy cosmic messengers, rather than precision laboratory experiments, in contrast to the \ac{SME} models.

Indeed, cosmic messengers, namely gamma rays, neutrinos, cosmic rays, and \acp{GW}, emitted from various astrophysical objects cross cosmological distances before reaching detectors. Moreover, they reach energies which by far exceed the ones attainable in particle accelerators (whether the currently existing ones, or those imaginable in the foreseeable future). Considering that the modifying terms in the \ac{MDR} (if present in the first place) are expected to be minuscule and cumulative, these two aspects of cosmic messengers expose them as excellent probes of \ac{QG} effects. A drawback of employing astrophysical observations for \ac{QG} tests is that the experimental conditions are not controllable. Our present-day knowledge of the emission processes in astrophysical objects is rather poor, making it difficult to resolve propagation effects from the source intrinsic ones. Furthermore, it is possible that messengers propagation is influenced by other phenomena, either expected such as background electromagnetic fields, or possible but not confirmed ones, e.g., mixing with axion-like particles, or some we are not yet aware of. Nevertheless, analysis methods have been developed to tackle these problems, and researchers are constantly working on improving them, as well as improving our understanding of the source emission processes and messenger propagation effects. In this work, we will review different experimental tests of \ac{QG} performed on astrophysical observations, we will briefly present the experiments, discuss the analysis techniques, and compare and critically analyze their results.

Unfortunately, \ac{MDR} characterizes various approaches to \ac{QG}, e.g. \ac{DSR}, non-critical string theory, or effective theories formulated on the so-called Finsler geometries, i.e. geometries in which the background metric depends on phase-space variables (velocities and coordinates of particles), rather than on coordinates alone, as in \ac{GR}. Thus, although an important phenomenological effect, if observed, \ac{MDR} will not point directly to a certain theory model. However, a possible observation of \ac{MDR} would be a spectacular and thrilling result, triggering various follow-up experimental tests. Indeed, numerous additional tests will be required firstly to disentangle \ac{QG} effects from the source intrinsic ones and other propagation effects, and secondly to properly measure and understand the nature and origin of the modification terms. Observing an effect consistent with \ac{MDR} is a necessary first step on this journey.

Nonetheless, such phenomena are of great interest in the so-called phenomenology of \ac{QG}, and are among the focus points of this review. The aim of the latter is to describe a bottom-up phenomenological approach to \ac{QG}, whereby starting from rather generic predictions of \ac{QG} models, such as \ac{MDR}, one could identify physical processes in the high energy universe, such as time delays in the arrival times of more energetic photons from intense cosmic sources, photon decay (and its effects in ultrahigh energy cosmic rays) etc. which could be due to such models. Last, but not least, such modified dispersions, but also the foamy nature of spacetime, that could characterize a \ac{QG} universe at (near) Planck scales, can in principle be probed with accuracy using \ac{GW} detector networks, where such effects will essentially appear as extra fuzzy ``noise" in the interferometers. Disentangling such stochastic effects from other conventional physics effects is a challenge at present, but certainly worthy of further investigations, especially in view of the increased sensitivity of future interferometric devices. Finally, in addition to the aforementioned \ac{MDR}, other important effects of great potential interest to the phenomenology of \ac{QG} are searches for \ac{CPT} violation, which could be either the consequence of \ac{LIV}, as happens in the context of the \ac{SME}, or pure consequence of the foamy nature of spacetime, in which case the \ac{CPT} operator is ill-defined in low-energy effective theories, leading to rather unique effects in the physics of entangled neutral mesons in Laboratory experiments.

This review will focus on the phenomenology of \ac{QG} and the associated tests, with the emphasis placed on the cosmic messengers. Theoretical models of \ac{QG} will only be briefly mentioned, as they serve to motivate the phenomenological and experimental parts of this work. This review is the outcome of the collective efforts of the participants of the \href{https://qg-mm.unizar.es/}{COST Action CA18108 ``Quantum Gravity Phenomenology in the multi-messenger approach''} \cite{QGMMWebsite}, and expresses a very fruitful collaboration between the theorists, experimentalists and observational cosmologists that participate in this endeavor.

We would like to point out that there are several aspects of \ac{QG} phenomenology we will not discuss in this review, including terrestrial precision measurements and cosmological tests. The incorporation of this kind of studies here would require a significant extension of the present discussion beyond the scopes of the current review, which focuses on cosmic messengers. Nonetheless, this should not be interpreted as implying that such tests are not important enough, or exhibit inferior sensitivities as compared to the cosmic probes discussed here. On the contrary, depending on the effect, the sensitivity of terrestrial probes and cosmological observations in some effects could surpass that of astrophysical messengers. Moreover, we envisage that as the quality of data from terrestrial, astrophysical and cosmological tests will improve, it will become more and more important to use all the windows that nature allows us to open on the Planck scale in order to make complementary analyses and reach a thorough understanding of physics in this realm.

The structure of the review is as follows: in the next Sec.~\ref{section2} we discuss briefly the theoretical models, whose predictions will be reviewed in this work. In Sec.~\ref{section3} we discuss the cosmic (high-energy) messengers, and briefly mention their most important properties, which will be of relevance to our subsequent discussion in Sec.~\ref{section4}, where we focus mainly on detection techniques. The latter depend crucially on the type of the messenger. In Sec.~\ref{section5} we discuss the \ac{QG} phenomenology of the various messengers and experimental bounds, which is the main focus of our review. Finally, conclusions and an outlook will be presented in Sec.~\ref{section6}.

This review is complemented by a catalogue, the QG-MM Catalogue \cite{QGMMCatalogue}, containing experimental bounds of \ac{QG} effects on astrophysical data.

%\newpage
\section{Theory}
\label{section2}

In this section we shall first discuss briefly theoretical frameworks and models for \ac{QG}, which are relevant in the subsequent phenomenological and experimental studies of this review. Specifically, in the context of a top-down approach to \ac{QG}, we shall commence our discussion with the so-called 
asymptotically-safe gravity, which is an attempt to quantize the gravitational interaction within the well-known framework of conventional local  \ac{QFT}. 
Next, we move onto the \ac{LQG} framework (and its path integral counterpart, the spin foam model), which is a connection-based canonical quantization approach to \ac{QG} whose configuration variables are holonomies of the connection that resemble the Wilson loops of standard Yang--Mills theories.
We then continue with the group field theory approach to \ac{QG}, a brief mention of deformations of the \ac{HDA}, 
which is the algebra of generators of local diffeomorphism invariance. Afterwards we consider  lattice (discretized) approaches to \ac{QG}, in particular \acp{CDT}, before we briefly discuss aspects of string theory like string cosmology. We then continue with the causality-based theories, in which causal relations between space-time events play a crucial r\^ole. 

In the second part of this section, we follow a bottom-up approach to \ac{QG}, and present phenomenological models, entailing violation and deformations of Lorentz and/or \ac{CPT} symmetry, which could be viewed as low-energy flat space-time limits of some \ac{QG} theories. These models 
include the \ac{SME} \acp{EFT}, the \ac{DSR} models, with their intriguing properties of phenomenological significance, such as \acp{MDR} relative locality and modification in the conservation laws of four momenta. Moreover, we briefly discuss non-commutative space-time geometries, Finsler spacetimes and their Hamilton generalizations. Finally we discuss theoretical models of modified classical gravity, viewed as effective theories of \ac{QG}, as well as the \ac{GUP} in quantum physics, which could be one of the potential consequences of \ac{QG}. Our discussion is focused on phenomenological predictions of the relevant models. 

We close the section with a discussion on some important conceptual issues, including: a discussion on the scale of \ac{QG}, a comparison between models that break Lorentz symmetry versus those that modify it, and the possibility that \acp{EFT} may be inadequate to describe all the effects of \ac{QG}.

\subsection{Fundamental frameworks}
\label{section2.1}

There are numerous approaches to \ac{QG}, ranging from unifying all fundamental forces as in string theory, to frameworks which focus on the quantization of gravity like asymptotic safety, \ac{LQG}, \ac{GFT}, \ac{CDT}, causal sets and causal fermion systems, as well as non-local approaches.

In this section we give a compact non-technical overview over these approaches, before we continue with the phenomenological models of \ac{QG} in the next section.

\subsubsection{Asymptotic safety}
\label{section2.1.1}

\label{asymptotic-safety}

The idea of asymptotically safe gravity is to quantize the gravitational interaction within the well-tested and powerful framework of  \ac{QFT}. In this setting, the fate of the dynamics at very high energies, or conversely small distance scales, is critical. Using perturbation theory, it has been shown that gravity-matter systems are not perturbatively renormalizable \cite{Deser:1974cy,Deser:1974cz}, meaning that the \ac{QFT} description loses predictivity at high energies, as infinitely many different interaction vertices, all coming with their own undetermined coupling parameter, start to play a role. This could either mean that \ac{QFT} is the wrong framework at these energies, or it could mean that an additional symmetry principle is missing. The idea of asymptotic safety is that \emph{quantum scale symmetry} provides the missing principle that restores predictivity \cite{Weinberg:1980gg,Reuter:1996cp}. Scale symmetry is a symmetry that relates the dynamics at different scales and can more intuitively be thought of as a form of self-similarity, as also exhibited by fractals.
Quantum scale symmetry means that such a symmetry is realized in the presence of quantum fluctuations, where competing effects that drive coupling parameters to either larger or smaller values balance out~\cite{Wetterich:2019qzx}. Then, coupling parameters become constant and do no longer change as the energy scale is increased. Therefore, achieving a regime of quantum scale symmetry is only possible when certain relations between the coupling parameters of the dynamics are satisfied.
As a consequence, dimensionful observables, such as, e.g., scattering cross-sections, are expected to exhibit a characteristic energy dependence beyond the energy scale where quantum scale symmetry sets in~\cite{Draper:2020knh}. 
Performing scattering experiments at that characteristic scale is expected to yield an easily visible ``smoking-gun" signature of asymptotic safety. Given that this scale is roughly the Planck scale of about $10^{19}$ GeV, accessing this signature seems prohibitively challenging. Hence, other, more indirect imprints of asymptotic safety in observations are searched for. \\
Intriguingly, there are two independent indications for scale symmetry in nature at high energies: The first is the spectrum of the \ac{CMB}, which is nearly scale-invariant, i.e., exhibits a very particular dependence on the wave-length \cite{Aghanim:2019ame}. The second is the Higgs sector of the \ac{SM}: From a Higgs mass of 125 GeV one can calculate the expected energy dependence of the coupling parameter of the Higgs-particle self-interaction which actually exhibits very nearly scale invariance close to Planckian energies \cite{Shaposhnikov:2009pv,Buttazzo:2013uya}. These promising hints contribute to motivating extensive theoretical studies of the asymptotic-safety paradigm for matter-gravity systems with a set of different \ac{QFT} techniques: functional renormalization group techniques \cite{Percacci:2017fkn,Reuter:2019byg}, lattice techniques \cite{Loll:2019rdj,Laiho:2017htj}, cf.~also Sec.~\ref{section2.1.6} as well as tensor-model techniques \cite{Eichhorn:2018phj}, in addition to perturbative studies \cite{Niedermaier:2009zz}. Compelling evidence for the realization of quantum scale symmetry in gravity-matter systems has been found, see \cite{Eichhorn:2018yfc} for a review and list of references. It must be stressed that this evidence arises within a Euclidean phase of spacetime and the technically more demanding study of Lorentzian signature, i.e., the inclusion of causal structure, is in its infancy \cite{Manrique:2011jc}. All results listed below should be interpreted in light of this caveat; for a more extensive discussion and list of other open problems, see \cite{Bonanno:2020bil}.

\begin{description}

\item {\it Symmetries:} Asymptotically Safe Gravity aims at quantizing gravity maintaining the symmetries of \ac{GR}
 and avoiding the introduction of new degrees of freedom. Thus, modifications to standard dispersion relations both for gravitational as well as matter degrees of freedom are not expected to occur (unless a violation of Lorentz invariance is inserted explicitly at the level of the action \cite{Knorr:2018fdu,Eichhorn:2019ybe}). In other words, the most conservative version of Asymptotically Safe Gravity -- the one based on the symmetries of \ac{GR}
 -- would be falsified by  experiments if a deviation from the standard dispersion relations is detected.\\
In contrast, all interactions compatible with the symmetries of the theory are expected to be non-vanishing at high energies, where they contribute non-trivially to quantum scale-invariance. This also includes non-minimal gravity-matter couplings. While the so far investigated set of induced non-minimal couplings does not introduce new free parameters into the system~\cite{Eichhorn:2017sok,Eichhorn:2018nda}, these couplings, if they persist at sizable values at physically accessible energy scales, might be expected to lead to spin-dependent modifications of the geodesic equation. Based on their higher-order nature, the generic expectation would be for such couplings to be present, but tiny at experimentally accessible scales. Corresponding studies are very much in their infancy, and presently only address the question whether non-minimal couplings between gravity and matter fields are indeed present at trans-Planckian scales~\cite{Narain:2009fy,Percacci:2015wwa,Eichhorn:2017sok,Eichhorn:2018nda}.

\item {\it Standard Model of particle physics as a consistency test:} As a fundamental  \ac{QFT}  of gravity and matter, Asymptotically Safe Gravity naturally offers particle-physics phenomenology as a test of quantum gravity. This includes questions of consistency regarding degrees of freedom and symmetries of the Standard Model, as well as potential indirect signatures.
All studies so far indicate that scale symmetry is compatible with the Standard Model matter content~\cite{Eichhorn:2018yfc}, while there might be constraints on the number of \ac{BSM} matter fields \cite{Dona:2013qba}. This includes the existence of light fermions since quantum fluctuations of spacetime seem to preserve chiral symmetry~\cite{Eichhorn:2011pc}, see also~\cite{Gies:2018jnv,Hamada:2020mug}. Potential indirect signatures might arise because Asymptotically Safe Gravity offers a dynamical unification where scale symmetry (akin to an enhanced grand unified symmetry) could enforce specific relations between some of the Standard Model couplings \cite{Shaposhnikov:2009pv,Harst:2011zx,Eichhorn:2017lry,Eichhorn:2018whv}.

\item {\it Beyond the Standard Model:} In addition to the matter degrees contained in the \ac{SM}, a dark sector could be required to provide dark matter and dark energy. First studies indicate that the predictive power of the asymptotic safety paradigm imposes novel constraints on the parameter space of dark-matter models~\cite{Eichhorn:2017als,Reichert:2019car,Hamada:2020vnf,Eichhorn:2020kca}. Similar constraints might also determine the vacua of scalar potentials and thereby could restore predictive power to grand unified theories \cite{Eichhorn:2019dhg}.

\item {\it Black-hole spacetimes:} The description of spacetime by \ac{GR}
 is expected to break down at Planckian curvature scales, e.g., 
in the very early universe and inside black holes. While robust quantitative predictions from Asymptotically Safe Gravity are currently out of reach, toy models based on scaling arguments suggest that quantum scale symmetry results in a weakening of the effective gravitational strength and can thus resolve singularities at the center of black holes
 \cite{Bonanno:1998ye,Falls:2012nd,Platania:2019kyx,Bosma:2019aiu}. Resulting horizon-scale modifications, in principle observable via black-hole shadows and gravitational waves, typically remain Planck-scale suppressed \cite{Held:2019xde}. 

\item {\it Early- and late-universe cosmology:} \noindent Similarly to the situation close to the center of black holes, a weakening of the gravitational strength is expected at the very beginning of our universe. There, the Big Bang singularity could be replaced by alternative non-singular cosmologies \cite{Kofinas:2016lcz,Bonanno:2017gji}.
Thus, nearly-scale-invariance of the cosmological power spectrum could be explained in terms of the departure from the quantum scale invariance realized at trans-Planckian energies. This possibility has also been explored via asymptotic-safety-inspired models of inflation~\cite{Bonanno:2007wg,Bonanno:2010bt}, aiming at explaining the primordial accelerated expansion of the universe in terms of fluctuations of the quantum geometry. In particular, in the context of asymptotically safe gravity, no ad-hoc inflaton field is required, as higher-derivative operators are expected to play a key role at trans-Planckian energies (see \cite{Bonanno:2017pkg,Platania:2020lqb} for reviews). On the other hand, if an inflaton field is added by hand, asymptotically safe quantum-gravity fluctuations tend to flatten the corresponding inflationary potential \cite{Narain:2009fy,Wetterich:2019qzx}. 

\item {\it Links to other approaches:} The asymptotic-safety program is not only explored with several techniques, bridging the gap between lattice studies and continuum techniques, but potentially also provides points of contact to other approaches at the phenomenological level. For instance, studies in \ac{LQG} suggest the emergence of an effective metric in the early universe \cite{Ashtekar:2011ni}, which is also required in the context of semi-classical scenarios of quantum cosmology \cite{Hartle:1983ai,Vilenkin:1982de, Feldbrugge:2017kzv}. Both lend themselves to a \ac{QFT} approach and a potential connection to an asymptotically safe regime. 
A similar connection has been made more explicit in the context of string theory \cite{deAlwis:2019aud,Basile:2021krk,Basile:2021euh}, potentially connecting string theory in an anti-de Sitter setup with a positive low-energy cosmological constant.
Asymptotic safety induces a tower of higher-derivative interactions and could hence be reinterpreted as a particular type of infinite-derivative-gravity theory \cite{Tomboulis:1980bs, Biswas:2011ar,Modesto:2017sdr, Buoninfante:2018mre} where quantum scale symmetry guarantees the presence of only finitely many free parameters.
\end{description}

%%%%%%%%%%%%%%%%%%%%%%%%%%%%%%%%%%%%%%%%%%%%%%%%%%%%%%%%%%%%%%%
%%%%%%%%%%%%%%%%%%%%%%%%%%%%%%%%%%%%%%%%%%%%%%%%%%%%%%%%%%%%%%%
\subsubsection{Loop quantum gravity}
\label{section2.1.2}

\ac{LQG} is based on a formulation of \ac{GR}
 in terms of connection variables along the lines of standard Yang Mills theories. The quantization adopts techniques from lattice gauge theory but generalized to a background independent context. 
In order to investigate physical consequences and phenomenological implications of \ac{LQG} one usually makes additional assumptions such as symmetry reduction and/or gauge fixings, both operations that in general do not commute with quantization as important ingredients in these models. 
A source of \ac{LQG} phenomenology is the consideration of quantum and classical \emph{test} matter fields, i.e., by neglecting back-reaction. Unequivocal predictions cannot be made due to the fact that solutions to the quantum Einstein's equation corresponding to Minkowski spacetime are not known, among other difficulties. However, under additional simplifications, effects on matter dispersion relations can be made plausible \cite{Gambini:1998it,Sahlmann:2002qj,Sahlmann:2002qk,Ashtekar:2009mb,Agullo:2012sh,Dapor:2012jg,Assaniousssi:2014ota}. Among the simplifications is working with gravity states that are not exact solutions \cite{Ashtekar:1992tm,Thiemann:2000bw,Thiemann:2000ca,Thiemann:2000by,Thiemann:2000bx,Ashtekar:2005dm,Freidel:2010tt}, and making specific choices on quantization ambiguities \cite{Thiemann:1997rt}. Effects involve changes of coefficients,
higher order terms, 
and terms breaking chiral symmetry \cite{Gambini:1998it}. The coefficients in the dispersion relations depend on parameters in the quantum state of the gravitational field which are not well understood up to now. There are relations among the coefficients in one particle species, as well as among those of different species. In these models Lorentz invariance is generically broken, but since the results rest on a number of uncontrolled approximations and assumptions, this cannot be taken as a prediction of \ac{LQG}. 
Using \ac{LQC} for the underlying state, some of the steps can be made more rigorous, \cite{Ashtekar:2009mb,Agullo:2012sh,Assaniousssi:2014ota}, see also in the highlighted paragraphs below. Notably, Lorentz symmetry seems to be preserved in these models \cite{Dapor:2012jg}. 
Modified dispersion relations are also obtained in the context of deformations of the \ac{HDA} which encodes the diffeomorphism invariance at the canonical level of \ac{GR}. 
It has been shown that holonomy and inverse triad corrections, caused by the polymer quantization used in \ac{LQG}, in spherical symmetric and related midispace models of \ac{LQG} \cite{Bojowald:2009ih,kreienbuehl2011model,Kreienbuehl_2012,Bojowald_2014,Brahma_2015,Bojowald_2015a,Bojowald_2015b,Bojowald_2017,Bojowald_2018,Ben_Achour_2017,Ben_Achour_2018,Bojowald:2019fkv} as well as  \ac{LQC} models based on an anomaly-free formulation of cosmological perturbations \cite{Bojowald_2008,Bojowald_2009,Cailleteau_2012,Barrau_2015} lead to such a deformations of the \ac{HDA}.
A deformed Poincar\'e algebra can be obtained from a deformed \ac{HDA} yielding to modified dispersion relations associated with it \cite{Bojowald_2012,Mielczarek:2013rva,Amelino-Camelia:2016gfx,Brahma:2018rrg}. Further physical implications are the possibility of a signature change  \cite{Bojowald_2017,Bojowald_2018} and the realization of a deformation of general covariance and non-classical spacetime structures \cite{Bojowald_2012,Bojowald_2016,Bojowald_2018b}. The small \ac{LQG} effects may be accumulated over large distances in the dispersion relation for photons coming from extragalactic sources, for example \acp{GRB} \cite{AmelinoCamelia:1997gz}. The data provided by gamma-ray astronomy have been used to set limits on the \ac{QG} energy scale \cite{Ellis:2005wr} (for more details, see Ref.~\cite{Horns:2016soz}).
In \ac{LQG}, both the interior and exterior/full spacetime of black holes, particularly the Schwarzschild black hole has been studied. 
The interior corresponds to a minisuperspace model, which is quantized usually using polymer quantization \cite{Ashtekar:2005qt, Bohmer:2007wi, Chiou:2008nm, Alesci:2019pbs, Bodendorfer:2019cyv, Morales-Tecotl:2018ugi, Corichi:2015xia}. Its effective dynamics leads to the resolution of the singularity since the radius of the infalling 2-spheres never reaches zero and bounces back after reaching a finite value. Consequently this leads to a Kretchmann scalar that is everywhere finite, and a black hole-to-white hole transition. An analysis of the Raychaudhuri equation in \ac{LQG} has also shown that the quantum effects become strong close to the singularity and lead to avoidance of infinite focusing of geodesics \cite{Blanchette:2020kkk}. If one considers the exterior/full spacetime, one is then dealing with a midisuperspace. 
In that case, 
the spectrum of the volume operator or the area of spheres of symmetry cannot vanish, and this again leads to the singularity resolution~\cite{Gambini:2013hna,Ashtekar:2018lag,Bojowald:2018xxu}. 

\begin{description}
\item {\it Spin foams:}
Spin foam models \cite{Baez:1997zt} are defined by the assignment of probability amplitudes to cellular complexes labeled by representations (and intertwiners) of a Lie group. They are then used as a covariant encoding of the quantum dynamics of spin networks states in \ac{LQG}, in the spirit of path integrals and lattice gauge theory \cite{Perez:2012wv, Rovelli:2014ssa}. Most modern spin foam models \cite{Barrett:1999qw, Engle:2007wy, Freidel:2007py, Dupuis:2011dh,
Finocchiaro:2018hks}, in fact, result from the quantization of simplicial
geometric structures and can be understood as a rewriting of simplicial
gravity path integrals in group-theoretic variables. Spin foam models arise
also as Feynman amplitudes of \ac{GFT}s
\cite{Krajewski:2012aw, Oriti:2014uga} and are being developed, also in the
direction of phenomenology, within that context. We refer to the \ac{GFT}
Sec.~\ref{section2.2.2}, for these developments.
Like most discrete \ac{QG} formalisms (and \ac{LQG})
they face the difficult challenge of controlling the quantum dynamics of
the fundamental discrete structures well enough to be able to extract an
effective continuum description for spacetime and gravity, and this remains
the biggest obstacle in making solid contact with phenomenology.
Lacking a solution to the problem of the continuum limit/approximation, and
since the complexity of spin foam amplitudes grows quickly with the
combinatorial complexity of the underlying cellular complexes, most
physical applications of spin foam models have been limited to very simple
cellular complexes interpolating very simple boundary spin network states,
encoding discrete geometries with few degrees of freedom.
One direction being pursued aims at capturing cosmological dynamics,
choosing boundary states corresponding to discretizations of cosmological
geometries,  and at making contact with \ac{LQC}
\cite{Bianchi:2010zs, Vidotto:2010kw}.
Another uses similar techniques to provide some backing, with a fundamental
theory, to scenarios      involving black hole-white hole transitions
\cite{Christodoulou:2016vny, Christodoulou:2018ryl}.
While the above results are very preliminary, it is clear that a solid link
between the fundamental spin foam models and cosmological or black hole
scenarios involving \ac{QG} effects would allow, in principle, to
derive predictions directly from the fundamental
formalism. 
A different, less ambitious strategy has been to employ some ingredient
which appears in current spin foam models, 
for example in the derivation of black
hole entropy \cite{Bianchi:2012ui}, or trying to deduce from it
possible physical consequences, 
for example
the existence of a maximal acceleration \cite{Rovelli:2013osa}.

\item {\it Loop quantum cosmology:}
\Ac{LQC} is a symmetry-reduced model of \ac{LQG} which carries out quantization by mimicking the constructions used in the full theory \cite{Ashtekar:2003hd, Ashtekar:2011ni}. 
Quantization of cosmological backgrounds have thus been studied within variety of models in \ac{LQC}  \cite{Bojowald:2001xe,Bojowald:2004bb,Bojowald:2007cd, Ashtekar:2006rx, Ashtekar:2006uz, Ashtekar:2006wn, Ashtekar:2007em,Ashtekar:2006es,Szulc:2006ep,Pawlowski:2011zf,Kaminski:2009pc,Ashtekar:2009vc,MartinBenito:2008wx,MartinBenito:2009qu, Ashtekar:2009um,MartinBenito:2008ej, Garay:2010sk,MartinBenito:2010bh,MartinBenito:2010up, Alesci:2014rra,Dapor:2017rwv,Assanioussi:2018hee,Giesel:2020raf}. 
Their main consequence is the resolution of the big-bang singularity using modifications of the gravitational  as well as  the matter Hamiltonians due to quantum geometry \cite{Ashtekar:2006rx,Bojowald:2004bb,Bojowald:2007cd}. Accordingly, the Big Bang  is replaced by a quantum bounce where a contracting phase should have taken place before the current expanding phase of our universe \cite{Ashtekar:2006rx}. 
The other  important result of \ac{LQC} is the predictions of  perturbations on the cosmological quantum  geometries. Different approaches to treat cosmological perturbations have been developed in \ac{LQC} \cite{Bojowald:2006qu, Ashtekar:2009mb,Agullo:2012fc, Ashtekar:2015dja,Agullo:2013ai, Barrau:2013ula}. 
These studies lead to spectra that are compatible with the current observations \cite{Ashtekar:2015dja,Agullo:2013ai,Girelli:2012ju,Barrau:2013ula,Gomar:2017yww}. 
The other  phenomenological prediction of \ac{LQC} is provided by the  backreactions of quantum matter on the quantum  geometry \cite{Bojowald:2007cd, Dapor:2012jg}.   
It turns out that the backreaction effects can result in the non-classical features of spacetime, for example,  phenomenological concrete deformation on the dispersion relations for  perturbations and \acp{GW}, which subsequently leads to violation of the Lorentz symmetry \cite{Lewandowski:2017cvz, Assaniousssi:2014ota, Dapor:2012jg,Tavakoli:2015fvz}.  This deviation from the standard physics can leave an imprint on the emission spectrum which can be verified by the realistic observations. 

\item {\it Quantum gravity and gravitational waves:} 
One expects that \ac{QG} can modify both the production as well as the propagation of \acp{GW} \cite{AmelinoCamelia:1998ax}. The \ac{LQG} effects on tensor modes of the \acp{GW} have been discussed in Ref.~\cite{Mielczarek:2007zy,Bojowald:2007cd,Calcagni:2008ig,Mielczarek:2009vi,Garcia-Chung:2020zyq},
while the vector mode dynamics in the context of cosmological models based on \ac{LQG} and \ac{LQC} is considered in Ref.~\cite{Bojowald:2007hv} (for more details and implications, see also Refs.~\cite{Bojowald:2008jv,Cailleteau:2012fy,Cailleteau:2013kqa}). One may thus introduce parametrizations for either   the emitted waveform or its propagation. The former could be viewed as representing possible GR modifications in the strong-field regime, and the latter would correspond to weak-field modifications away from the source. Constraints on such parametrizations have been placed using the binary black hole signals from the LIGO-Virgo Catalog GWTC-1~\cite{LIGOScientific:2018mvr}
Moreover, focusing on the propagation, it has been recently shown~\cite{Calcagni:2019ngc,Calcagni:2019kzo} that \ac{QG} modifications can influence the \ac{GW} luminosity distance, the time dependence of the effective Planck mass and the instrumental strain noise of interferometers.  
\ac{LQC} can produce detectable deviations from the standard power spectrum of cosmological perturbations leading to falsifiable predictions~\cite{Calcagni:2012vw,Zhu:2015xsa} (for a review, see Ref.~\cite{Barrau:2013ula}).
\end{description}

%%%%%%%%%%%%%%%%%%%%%%%%%%%%%%%%%%%%%%%%%%%%%%%%%%%%%%%%%%%%%%%
%%%%%%%%%%%%%%%%%%%%%%%%%%%%%%%%%%%%%%%%%%%%%%%%%%%%%%%%%%%%%%%
\subsubsection{Group field theory}
\label{section2.1.3}
\Acp{GFT} encode the idea that spacetime can be built out of discrete building blocks, that
can be glued together to form extended structures. As for any quantum many-body system, the discrete building blocks can form different phases, including ``gas-like'' and ``fluid-like'' phases. The latter would be related to our experience of spacetime as continuous, as described in \ac{GR}. The search for such a phase takes center stage in the \ac{GFT} approach.

More precisely, \acp{GFT} \cite{Oriti:2011jm,Krajewski:2012aw} are field theories for tensorial fields defined on a group manifold, characterized by combinatorially non-local interactions. With fields associated to polyhedra, their non-local interactions generate dynamical processes dual to lattices, summed over in the \ac{GFT} perturbative expansion. These combinatorial aspects are shared by random tensor models \cite{Gurau:2012hl,Gurau:2019qag}. The group-theoretic data allow a rich quantum geometry to be associated to the combinatorial structures, so \ac{GFT}s are  \ac{QFT}s of quantized geometric polyhedra and the Feynman amplitudes associated to the lattices dual to \ac{GFT} Feynman diagrams can be equivalently represented as spin foam models or lattice gravity path integrals \cite{Reisenberger:2000zc,Baratin:2011hp,Finocchiaro:2018hks}.  Also, for appropriate models, \ac{GFT}s offer a second quantized formulation for the quantum states (spin networks organized in a Fock space) and operators of \ac{LQG}\cite{Oriti:2013aqa,Oriti:2017ave}. 

Like in all discrete quantum gravity approaches, a solid contact with phenomenology requires control over the quantum dynamics of highly populated states, and the ability to extract an effective continuum dynamics from the fundamental theory. For this, \ac{GFT}s can take advantage of powerful \ac{QFT} techniques, adapted to the quantum gravity context, for example the renormalization group \cite{Carrozza:2016vsq,Eichhorn:2018phj,Finocchiaro:2020fhl}. 

At the more physical level, the properties of \ac{GFT} condensate states and mean field techniques have been key for extracting effective cosmological dynamics from quantum gravity in \ac{GFT} condensate cosmology \cite{Oriti:2016acw,Gielen:2016dss,Pithis:2019tvp}. The general idea is that continuum gravitational physics should be looked for in the hydrodynamic regime of the fundamental theory, and condensate states allow to extract it rather straightforwardly. The collective condensate wavefunction gives a probability distribution on the minisuperspace of homogeneous universes \cite{Gielen:2014ila}, and its dynamics takes the form of a non-linear extension of quantum cosmology. The resulting cosmological dynamics has been shown to reproduce a flat Friedmann universe in the classical limit, using a variety of techniques \cite{Oriti:2016qtz,Gielen:2013naa,Pithis:2016cxg,Wilson-Ewing:2018mrp,Gielen:2019kae}. In the same hydrodynamic approximation, the Big Bang singularity seems generically replaced by a quantum bounce, with \ac{LQC} appearing as a special case \cite{Oriti:2016qtz,Gielen:2016uft,Gielen:2020fgi,deCesare:2018cts,deCesare:2019pqj}.
The formalism has also the potential to reproduce an emergent universe scenario, going beyond hydrodynamics, with a phase transition to the geometric phase being what truly replaces the cosmological singularity. 
With spacetime emerging from non-spatiotemporal structures \cite{Oriti:2018dsg} and the whole cosmological dynamics emerging from \ac{GFT} hydrodynamics, the potential for quantum gravity effects in cosmology is not confined to singularity resolution or Planck scales. 

The fundamental \ac{GFT} interactions can generate an effective inflationary phase in the early universe (without an inflaton), a cyclic dynamics, and, in some generality, a phantom dark energy-like accelerated expansion at late times (without any phantom field) with asymptotic de Sitter phase  \cite{deCesare:2016rsf,Pithis:2016wzf,Pithis:2016cxg,Oriti:2021rvm}. \ac{GFT} condensate cosmology has been extended to include anisotropies \cite{deCesare:2017ynn,Pithis:2016cxg}, and inhomogeneities \cite{Gielen:2017eco,Gerhardt:2018byq,Gielen:2018xph}, with very promising results, but not going yet beyond consistency with basic observational constraints: dynamical suppression of anisotropies at late times, and an approximately scale invariant power spectrum of inhomogeneities in the early universe, with good classical evolution. More work is needed to arrive a solid quantum gravity predictions, but it is remarkable that this analysis can take place fully within the fundamental \ac{GFT} formalism, beyond toy models).

Recently, the \ac{GFT} formalism has been used to describe the microscopic quantum gravity states of spherically symmetric horizons, and to compute their entropy and holographic properties from first principles, in terms of generalised condensates involving a sum over lattice structures \cite{Oriti:2015rwa,Oriti:2018qty}. 

Another route toward quantum gravity phenomenology in the \ac{GFT} formalism is suggested by a number of works showing how non-commutative field theories of \ac{DSR} type could emerge for perturbations around mean field solutions of the \ac{GFT} quantum dynamics \cite{Girelli:2009yz,Girelli:2010ct}. 

%%%%%%%%%%%%%%%%%%%%%%%%%%%%%%%%%%%%%%%%%%%%%%%%%%%%%%%%%%%%%%%
%%%%%%%%%%%%%%%%%%%%%%%%%%%%%%%%%%%%%%%%%%%%%%%%%%%%%%%%%%%%%%%
\subsubsection{Deformations of the hypersurface deformation algebra}
\label{section2.1.4}

Deformations of the symmetries of  Minkowski space, discussed in Sec.~\ref{section2.2.2},
are one of the leading approaches to the phenomenology of quantum gravity.
The question which naturally arises in this context is whether such deformations 
can be generalised to the fully general relativistic case. Taking up this issue 
requires to pass from the level of the Poincar\'e algebra to the \ac{HDA}, which is the algebra of generators of invariance 
with respect to local diffeomorphisms. While the algebra is known already 
for more than sixty years now \cite{Dirac1}, quantum deformations 
of the algebra were not the subject of investigations until roughly a decade ago. 
 
The first results on deformations of the \ac{HDA}
emerged in the context of \ac{LQG}, discussed in Sec.~\ref{section2.2.1}, thanks to  studies of the effective models with spherical symmetry as well as perturbative cosmological inhomogeneities \cite{Bojowald:2011aa,
Cailleteau:2011kr,Bojowald:2012ux}.
In this case, the deformation is governed by  the square of the speed of 
propagation of inhomogeneities, including \acp{GW}. 

Attempts to link the deformed \ac{HDA} to the 
corresponding deformed Poincar\'e algebra have been made. 
The first approach, based on the mass Casimir operator for the loop-deformed Poincar\'e algebra, was performed in ref. \cite{Mielczarek:2013rva}, where corrections (starting from quadratic orders in momenta) to the dispersion relation were derived. 
Further studies on this subject \cite{Amelino-Camelia:2016gfx} indicated a possible relation between the loop-deformed 
\ac{HDA} and the $\kappa-$Minkowski non-commutative spacetime, described in Sec.~\ref{section2.2.2} . This line of investigations also confirmed leading quadratic energy dependence in the 
speed of light \cite{Brahma:2016tsq}.
Moreover, in Ref. \cite{Mielczarek:2016zfz}, a systematic method of  linking the deformed
\ac{HDA} to the deformed Poincar\'e algebra has been proposed. These studies revealed that 
 quantum effects may also 
produce a maximal speed of propagation for particles, that depends on the particle mass. 
This kind of 
observations open possibility of phenomenological studies in the multi-messenger approach to \ac{QG}, which is the main subject of this review. An interesting question arises in this respect, as to how, if at all, one could distinguish the \ac{HDA}-induced modifications of the dispersion relations from other \ac{QG}-related cases where such modifications arise (see Sec.~\ref{section2.2.2} and Sec.~\ref{section2.1.5}). This issue remains open at present.

%%%%%%%%%%%%%%%%%%%%%%%%%%%%%%%%%%%%%%%%%%%%%%%%%%%%%%%%%%%%%%%
%%%%%%%%%%%%%%%%%%%%%%%%%%%%%%%%%%%%%%%%%%%%%%%%%%%%%%%%%%%%%%%
\subsubsection{String cosmology}
\label{section2.1.5}

The term {\it String cosmology} refers to several approaches to understanding the universe and its evolution within the frameworks of: (i) ``traditional'' (super-)string theory~\cite{Green:1987sp,Green:1987mn} and (ii) its modern extensions, involving branes~\cite{Polchinski:1998rq,Polchinski:1998rr}. 

Assuming  (i), our four-dimensional world is viewed: (a) either as a result of compactification of the six extra dimensions in appropriate configurations, to ensure conformal invariance of the underlying 
world-sheet theory, given that  the critical space-time dimension of superstrings is ten, or (b) a direct four-dimensional construction, with the extra dimensions being associated with appropriate conformal field operators in the first-quantised version of string theory~\cite{Kawai:1986va}, so that the total central charge deficit of the world-sheet theory, vanishes. 
Generalising this framework, {\it super}-critical string theory has been used~\cite{Antoniadis:1988aa,Antoniadis:1988vi}, 
to describe an expanding universe, by identifying the central charge {\it surplus}  of the internal theory with the cosmic time derivative of the dilaton field, whose non-trivial space-time configuration is essential for consistency of the string propagating in four large target-space-time dimensions. 

At a  non-critical level  the underlying (weakly-coupled) string theory 
induces a tree-level dilaton potential that corresponds to dark energy that relaxes to zero at late times. This model can be viewed as an example of a more general non-critical approach, 
in which the time is identified with a time-like Liouville mode of the first-quantised version of string theory~\cite{Ellis:1998dw,Ellis:2005qa}. 
It bears a pertinent 
feature~\cite{Ellis:2006nd,Basilakos:2011kc,Mavromatos:2007zh}, in that it  may alleviate the 
so-called Hubble tension between Planck measurements~\cite{Aghanim:2018eyx} and late-time data~\cite{Riess:2016jrr}.
Moreover, other models with a non-trivial phenomenological dilaton potential have been proposed as an \ac{EFT} description of the early universe: they  may describe a pre Big-Bang history of the universe~\cite{Gasperini:2002bn} that might leave imprints in  the \ac{CMB} spectrum or in primordial \acp{GW}~\cite{Gasperini:2007zz}. 

In framework (ii), one views our universe as a three-brane domain wall propagating in a higher-dimensional bulk spacetime~\cite{Binetruy:1999ut,Randall:1999ee,Langlois:2002pk}. In general, the cosmological evolution on the brane world will bring 
significant  modifications, compared to standard cosmology. 
This is a consequence of gravitational interactions between the brane world and the bulk, which may manifest themselves as dark radiation. In addition, standard model particles are viewed as being ``confined'' on the brane universe, while only fields in the gravitational multiplet of the string, and standard model singlet fields, such as right-handed neutrinos, are allowed to propagate in the bulk. The phenomenology of such brane worlds is rich and diverse, including interesting \ac{GW} physics.

Furthermore, we mention the possibility of consistent \ac{LIV} backgrounds that characterise ground states of string theory~\cite{Mavromatos:2020kzj}, which (a) 
affect the propagation of particles,
in which both photons~\cite{Ellis:2008gg} and \acp{GW}~\cite{Elghozi:2016wzb} may exhibit a quantum-gravity-induced refractive index, with the effective scale of \ac{LIV} being redshift dependent in general~\cite{Ellis:2008gg}, 
(b) lead to consistent growth of structure in brane universes populated with point-like brane defects~\cite{Mavromatos:2012ha,Elghozi:2015jka}, (c) 
allow \ac{CPT} violating leptogenesis~\cite{Mavromatos:2018map}, and (d) share features
with contorted cosmologies~\cite{Hehl:1976kj,Magueijo:2012ug}.

Lastly, let us mention taking string theory 
as a setup to explore quantum cosmology (e.g., minisuperspce formulation). 
Besides~\cite{Gasperini:2002bn}, it has unveiled (albeit phenomenologically) 
several features that determine initial conditions 
associated, e.g., to spatial 
non-commutativity~\cite{Rasouli:2014dba,Rasouli:2018nwi} or
($N=2$) supersymmetric \ac{QM}~\cite{Lidsey:2000rb}. Altogether, 
such aspects imply different vacua whose 
characteristics (non-Bunch--Davies) 
hopefully may be appraised 
observationally in a foreseeable future~\cite{Balasubramanian_2011}.

%%%%%%%%%%%%%%%%%%%%%%%%%%%%%%%%%%%%%%%%%%%%%%%%%%%%%%%%%%%%%%%
%%%%%%%%%%%%%%%%%%%%%%%%%%%%%%%%%%%%%%%%%%%%%%%%%%%%%%%%%%%%%%%
\subsubsection{Causal dynamical triangulations}
\label{section2.1.6}
\Ac{CDT} \cite{Ambjorn:2001cv,Loll:2019rdj} is a lattice field theory that provides a non-perturbative description of quantum gravity by defining its path integral formulation. Since the theory is unphysical at finite lattice spacing, a universal continuum limit -- linked to a suitable higher-order phase transition in the phase diagram, spanned by different couplings of the theory -- has to be found. \ac{CDT}  can thus be seen as a lattice formulation of the asymptotic safety scenario, if this turns out to be realized, to be used as a suitable computational tool like lattice QCD is for continuum QCD. However, \ac{CDT}  provides more general access to gravitational universality classes, i.e., potentially different continuum limits. 

In \ac{CDT}, spacetime geometries are encoded into equilateral simplicial lattices
and the partition function is defined by summing over them using as weight the Regge action for discrete gravity, reduced thus to a function of the combinatorial structure of the lattice. Crucially, discrete random geometries appearing in the path integral are restricted to the ones that admit a global causal structure, suppressing spatial topology change. However, it may be 
sufficient to demand only local causality conditions \cite{Jordan:2013awa}. 

The theory has been solved analytically in 1+1 dimensions (where it coincides with Ho\v{r}ava--Lifshitz gravity \cite{Ambjorn:2013joa}) but in higher dimensions one has to resort to numerical simulations and the nature of spacetime symmetries in the continuum limit is an open question; in particular, full spacetime diffeomorphism symmetry could be very well available in 3+1 dimensions.  \Ac{CDT} in 3+1 dimensions turns out to have a phase diagram containing at least four phases, one of which appears to possess a smooth geometric description in terms of a minisuperspace solution: de Sitter universe (for spherical spatial topology) \cite{Ambjorn:2011ph} or the flat one (for toroidal spatial topology) \cite{Ambjorn:2018qbf}. 

\Ac{CDT} realizes a dynamical dimensional reduction of spacetime in the UV \cite{Ambjorn:2005db,Coumbe:2014noa}. Such an effect may induce deformed dispersion relations for fields on the  
quantum spacetime, with a number of phenomenological implications \cite{Mielczarek:2015cja,Mielczarek:2015xyu}, possibly within the range of the Fermi space telescope data \cite{Mielczarek:2015cja}. On the other hand, the effect of deformed dispersion could be nullified by the scale-dependence of time intervals \cite{Coumbe:2015aev}. 

To bridge the gap towards particle-physics phenomenology, the inclusion of matter fields could be fruitful, and has been explored, e.g., in \cite{Ambjorn:2014pwa,Ambjorn:2015gea}. In the related Euclidean dynamical triangulations approach, gauge fields were studied early on \cite{Ambjorn:1999ix} and
fermionic matter fields have been introduced \cite{Catterall:2018dns}, with first indications that chiral symmetry -- protecting the lightness of fermions compared to the Planck scale -- could remain intact, similarly to the continuum approach to asymptotic safety \cite{Eichhorn:2011pc}. 

%%%%%%%%%%%%%%%%%%%%%%%%%%%%%%%%%%%%%%%%%%%%%%%%%%%%%%%%%%%%%%%
%%%%%%%%%%%%%%%%%%%%%%%%%%%%%%%%%%%%%%%%%%%%%%%%%%%%%%%%%%%%%%%
\subsubsection{Causal fermion systems and causal sets}
\label{section2.1.7}
%short intro
In this section we discuss two distinct approaches to quantum gravity, which are based on causality: \emph{causal fermion systems},  and  \emph{causal sets}. The causal structure and the volume element \cite{Sorkin:2003bx} decouple for both, suggesting non-Riemannian measure theories as an effective description. See \cite{Bensity:2020sfu} for a review on studies in the cosmological setting.

In causal fermion systems \cite{Finster:2016zhe, website}, all spacetime structures and matter therein arise by minimizing the causal action. The causal action is built upon an operator manifold, merging the fundamental mathematical structures of \ac{QM} and \ac{QFT} 
and \ac{GR}. The Lagrangian then incorporates the causal structure between all possible events.
Causal fermion system reproduce the \ac{SM}
 and \ac{GR} in appropriate limits while explaining the relative weakness of gravitation as it appears as a third order effect. For the resulting equations to be well-posed, it requires three generations of fermions. First steps towards cosmological phenomenology can be found in \cite{paganini2020proposal,Finster:2021dqm}. Whether it gives rise to observable effects in cosmology is an open question. 

A causal set \cite{Bombelli:1987aa,Surya:2019ndm} is a discrete, causal network of events; with spacetime points constituting nodes that share a link for events at timelike or null separation. The discreteness does not imply a breaking of Lorentz invariance, because the spacetime points of a causal set do not lie on a regular lattice. Instead, the points of a causal set are embedded into an emergent, effective continuum via a random process, a Poisson-sprinkling.

To extract phenomenological implications, one should firstly understand whether the sum-over-histories (the path integral) over all causal set exists; see \cite{Benincasa:2010ac} for a proposed dynamics and, e.g., \cite{Surya:2011du,Cunningham:2019rob} for Monte-Carlo simulations of the partition function. Secondly, one needs to be able to describe quantum fields of spin 0, 1/2 and 1 and 2 on a causal set to describe \ac{SM} matter as well as \acp{GW}/gravitons. So far, this has only been achieved for the free scalar field \cite{Sorkin:2011pn,Aslanbeigi:2014zva}.
With these caveats in mind, let us describe the state-of-the-art regarding the phenomenology with a focus on aspects relevant to (MM) astrophysics.

Causal set motivates that the observed cosmological “constant” fluctuates in value with a vanishing mean and a magnitude comparable to the critical density at any epoch. This could be tested observationally (see e.g.\cite{zwane2018cosmological} and references therein).
For point particles propagating on a causal set, discreteness is could result in a Lorentz invariant diffusion in momentum space \cite{Dowker:2003hb,Philpott:2008vd} that provides an acceleration mechanism. Further the spacetime discreteness may not result in loss of coherence of light from distant sources  \cite{Dowker:2010pf}.

The scalar-field d'Alembertian \cite{Aslanbeigi:2014zva} features a non-locality scale several orders of magnitude larger than the Planck length (typically assumed to be the discreteness scale). This type of non-locality, if it remains relevant in the full \ac{QFT} and for fermionic fields, could be tested experimentally \cite{Belenchia:2015ake,Belenchia:2016zaa,Belenchia:2019yvw}.

Finally, a change of the effective dimensionally at very small length scales, occurring in several \ac{QG} approaches, is also relevant for causal sets \cite{Eichhorn:2013ova,Carlip:2015mra,Belenchia:2015aia,Abajian:2017qub,Eichhorn:2019uct}, resulting in asymptotic silence \cite{Eichhorn:2017djq}. Whether a direct, observable imprint ensues is an open question.

%%%%%%%%%%%%%%%%%%%%%%%%%%%%%%%%%%%%%%%%%%%%%%%%%%%%%%%%%%%%%%%
%%%%%%%%%%%%%%%%%%%%%%%%%%%%%%%%%%%%%%%%%%%%%%%%%%%%%%%%%%%%%%%
\subsubsection{Non-local quantum and modified theories of gravity}
\label{section2.1.8}
\ac{QM} and \ac{QFT} are non-local theories by construction. On the other hand, Einstein's \ac{GR} is a local theory so it is difficult to reconcile  non-local interactions and probabilistic evolution of phenomena in the framework of gravitational interaction. In view of describing  gravity under the same standard of the  other fundamental interactions and  overcoming  incompatibilities at quantum level,  extensions of the Einstein--Hilbert action containing non-local terms can be taken into account. Such extensions are supposed, at infrared scales,  to account  for phenomenology occurring at cosmological and astrophysical level, as well as to  remove the classical singularities of \ac{GR}, to improve its renormalizability at the quantum level and to preserve unitarity (absence of ghosts and conservation of probability). 

Non-local quantum gravity \cite{Krasnikov:1987yj,Kuzmin:1989sp,Tomboulis:1997gg,Biswas:2005qr,Modesto:2011kw,Biswas:2011ar} is a perturbative  \ac{QFT}  of gravity and matter endowed, at the fundamental classical level, with non-local form factors, which can be described either as operators $\gamma_i(\Box)$, with infinitely many derivatives or as convolutions with an integral kernel \cite{Pais:1950za}. Here $\Box$ is the covariant d'Alembert operator. The appearance of such term in the Lagrangian of a physical field theory can be motivated by the quantum contributions of \ac{QFT}s in curved space, see \cite{Buchbinder:1992rb} or more recently \cite{Modesto:2017sdr,Franchino-Vinas:2018gzr,Benisty:2018fgu,Benisty:2018ufz,CANTATA:2021ktz}. The general action is of the form
\begin{equation}
\mathcal{S}=\frac{c^4}{16\pi G_N}\int\!d^4x \sqrt{-g}\, F(R, R_{\mu\nu}, R_{\mu\nu\rho\sigma}, \gamma_0(\Box)R, \gamma_2(\Box)R_{\mu\nu},\gamma_4(\Box)R_{\mu\nu\rho\sigma}) \, +\mathcal{S}^{(m)}\,,
\end{equation}
where different choices of $F$ fix different models studied in the literature. This form of the action includes fourth-order Stelle gravity and its generalizations \cite{Stelle:1976gc,Asorey:1996hz}, infinite derivative gravity theories (considering analytic transcendental functions of $\Box $ \cite{Modesto:2013ioa}) and integral kernel theories of gravity (considering $\Box^{-1}$ operators \cite{Deser:2007jk} and taking into account IR quantum corrections provided by \ac{QFT} on curved spacetime \cite{Barvinsky:2014lja}). Further extensions of non-local gravity are the application of $\Box$ to the torsion scalar of teleparallel gravity \cite{Aldrovandi:2013wha,Cai:2015emx,Bahamonde:2021gfp}; and a nonperturbative approach to non-local gravity is the revival of Mandelstam's theory of gravity \cite{Gambini:2018adk, Gambini:2020ycb}. The reduced phase space quantization of the latter theory leads to noncommutativity of the Riemann tensors where space-time points are emergent entities, with important consequences, e.g., for physics in the vicinity of a black hole. Important theoretical constraints on the form of $F$ in the Lagrangian come from the perturbative ghost-free conditions constraint around Minkowski spacetime \cite{Biswas:2011ar,Asorey:2018wot} and from experimental constraints~\cite{Belgacem:2018wtb}.

The applications of non-local quantum gravity are numerous. In cosmology and for black holes they are capable to resolve the cosmological~\cite{Biswas:2005qr,Calcagni:2013vra} and the black hole~\cite{Biswas:2011ar,Frolov:2015bia,Frolov:2015usa,Buoninfante:2018xiw,Buoninfante:2018rlq} singularities, as well as the late-time cosmic expansion of the universe without any dark energy, see for example \cite{Bahamonde:2017sdo} in a teleparallel model. In spherical symmetry non-local gravity is capable of fitting the S2 star orbit around Sgr $A^*$, even better than the Keplerian case \cite{Dialektopoulos:2018iph}. Effects on \ac{GW} are usually too small to be detected with current technology \cite{Briscese:2019rii,Koshelev:2020foq,Calcagni:2020}.

\subsection{Phenomenological frameworks}
\label{secction2.2}

After the brief overview on fundamental approaches to \ac{QG}, we present the bottom-up approach by discussing some phenomenological models and their most important predictions, which lead to the phenomenology of \ac{QG}. These phenomenological approaches constitute the focus of the current review. 

%%%%%%%%%%%%%%%%%%%%%%%%%%%%%%%%%%%%%%%%%%%%%%%%%%%%%%%%%%%%%%%
%%%%%%%%%%%%%%%%%%%%%%%%%%%%%%%%%%%%%%%%%%%%%%%%%%%%%%%%%%%%%%%
\subsubsection{Lorentz invariance violation, effective field theory and gravity}
\label{section2.2.1}

The form of the operators in both the \ac{SM} and \ac{GR}
 are heavily constrained by assumed spacetime and internal symmetries.  In contrast, \ac{QG} models often modify these symmetries, either by breaking them directly~\cite{Horava:2009uw, Hinchliffe:2002km}, subsuming them into a larger symmetry group~\cite{Gross:1984dd}, or deforming the action of the underlying transformations~\cite{AmelinoCamelia:2011bm}.  Unfortunately, it is usually quite difficult to directly compute the low-energy effect of any given \ac{QG} model. However, one can still explore the possible quantum \ac{QG} by considering the \ac{SM} and \ac{GR}
 as \ac{EFT}, modifying one or more of the 
constraining symmetries as indicated by whichever \ac{QG}  model one is interested in, writing down the new allowed operators, and testing the resultant physics via propagation, scattering, etc. This approach, which is common in particle physics~\cite{Contino:2016jqw} has been successfully used to study quantized weak gravitational field coupled to gravitational wave detectors~\cite{Coradeschi:2021szx, Guerreiro:2021qgk} and constrain extended gauge symmetries~\cite{Miura:2016krn}, supersymmetry~\cite{Baer:2020kwz}, broken translation invariance~\cite{Dowker:aza,Dowker:2003hb, Kaloper:2006pj}, and, as we shall now focus on, violations of Lorentz invariance~\cite{Colladay:1998fq}.
	
\ac{LIV} have been considered in numerous \ac{QG}  models, including string theory~\cite{Kostelecky:1988zi,Ellis:2008gg,Kiritsis:1999tx, Freidel:2018apz} (see also Sec.~\ref{section2.1.5}), \ac{LQG}~\cite{Bojowald:2004bb} (see also Sec.~\ref{section2.1.2}), Ho{\v r}ava--Lifshitz gravity~\cite{Horava:2009uw}, \ac{CDT}~\cite{Loll:2019rdj} (see also Sec.~\ref{section2.1.6}), non-commutative geometry~\cite{Hinchliffe:2002km} (see also Sec.~\ref{section2.2.2.3}), and emergent gravity. They also occur as a consequence of more general \ac{QG} considerations such as spacetime discreteness or dimensional reduction~\cite{Carlip:2017eud,Carlip:2019onx}.  Whereas \ac{LIV} can be incorporated in an \ac{EFT}, deformations of Lorentz invariance (see Sec.~\ref{section2.2.2}), which occur when spacetime acquires a non-commutative structure or there is a non-trivial geometry in phase space and simply modify the action of the Lorentz group on configuration space variables, are more difficult to handle. For an in-depth discussion of the differences between the Lorentz violating and the Lorentz deforming scenarios and on the effective vs. non-effective field theory approach see Secs.~\ref{section2.3.1} and \ref{section2.3.2} respectively.  Due to the ultraviolet-infrared mixing that often occurs in these scenarios, integrating out high momentum modes to generate a consistent low-energy \ac{EFT} is sometimes problematic and we leave discussion of these issues for Secs.~\ref{section2.3.3}.  

Within the \ac{EFT} framework, \ac{LIV} are incorporated into the \ac{SM} via the introduction of terms constructed by coupling the matter fields and their derivatives with new vector- or tensor-valued background fields. For example, one can consider an additional term coupling a vector field $u^\mu$ and a Dirac field $\psi$, such as $i u_\mu u_\nu \bar{\psi} \gamma^\mu \nabla^\nu \psi$. These tensors can be seen as the vacuum expectation values of some string operators \cite{Kostelecky:1988zi,Kostelecky:1991ak,Kostelecky:1994rn} and explicitly break Lorentz invariance. 	Independent of any other symmetries one can consider the set of all possible Lorentz invariance violating non-dynamical fields and renormalizable operators - in this case one gets the minimal \ac{SME}~\cite{Colladay:1998fq}.  The minimal \ac{SME} is very tightly experimentally constrained, and as a result non-renormalizable operators with \ac{LIV} suppressed by the Planck energy, which also appear more naturally in some quantum gravity models, have also been explored.  For these operators an immediate issue arises: because one is working in an \ac{EFT}, a small (at astrophysical energies) non-renormalizable operator generically induces large renormalizable operators via loop corrections~\cite{Collins:2004bp, Polchinski:2011za}.  (By large, we mean that the dimensionless coefficient in front any operator is of order unity.) To avoid this one must introduce some additional new physics besides \ac{LIV} at a scale $\Lambda_c$, a hierarchy between $\Lambda_c$ and the Lorentz violating scale $\Lambda_{\textrm{LIV}}$, and a custodial mechanism that suppresses generation of renormalizable operators by a power of the ratio of $\Lambda_c/\Lambda_{\textrm{LIV}}$.  Various schemes, including the use of supersymmetry~\cite{GrootNibbelink:2004za} and hierarchies between the \ac{LIV} scale and the Planck scale~\cite{Cortes:2016fye, Pospelov:2010mp}, allow non-renormalizable operators with order unity coefficients to be possible while not violating low-energy constraints.  Even these operators however can often be directly constrained beyond their natural Planck scale suppression by astrophysical tests (c.f. experiments listed in ~\cite{Mattingly:2005re, Liberati:2013xla} or the data tables on \ac{LIV}~\cite{Kostelecky:2008ts}).
	
Besides requiring hierarchies, the \ac{EFT} approach with non-dynamical tensor fields violates general covariance as the fields constitute prior geometry. While the ramifications of this have been explored (c.f.~\cite{Anber:2009qp, Rubakov:2004eb}), the more natural approach from a theoretical perspective is to make the tensor fields dynamical with non-zero vacuum expectation values~\cite{Kostelecky:2003fs,Jacobson:2000xp, ArkaniHamed:2003uy, Blas:2009qj}. In this case, since the fields have associated kinetic terms there is an inevitable coupling to the metric. This generically leads to non-trivial gravitational phenomenology, including modified velocities and polarizations for \ac{GW}~\cite{Jacobson:2004ts, Mewes:2019dhj, Gong:2018cgj} (see also Secs.~\ref{section5.7} and \ref{section7.2.3}), violations of the equivalence principle~\cite{Kostelecky:2010ze}.  Besides the infrared modifications, violating Lorentz invariance in gravity allows one to add higher derivative operators while avoiding the problem of ghosts that plagues standard Lorentz invariant  \ac{QFT}  approaches incorporating higher curvature gravity (c.f.~\cite{Salvio:2018crh} for a discussion).  As a result, one can construct renormalizable \ac{LIV} gravitational models~\cite{Horava:2009uw,Barvinsky:2015kil,Visser:2009ys}, in contrast to perturbatively non-renormalizable \ac{GR}. The net effect of the generic gravitational coupling and loop corrections is that \ac{LIV} in one sector will typically lead to \ac{LIV} in \textit{all} sectors without additional physics and strong hierarchies.
	
The above considerations lead to four classes of phenomena from \ac{LIV} \ac{EFT} that are most relevant for multi-messenger astronomy and astrophysics:  
	
\begin{itemize}
	\item Generically, there will be time-of-flight differences between different matter fields and between matter and gravity that scale with some power of the energy of the excitations~\cite{AmelinoCamelia:1997gz, Mattingly:2005re}.  Time-of-flight differences have been used to constrain electromagnetic, gravitational, and neutrino \ac{LIV}, see Sec.~\ref{section7.1}.
	
	\item The kinematics of particle interactions are modified due to new energy-momentum dispersion relations and anomalous reactions occur, see Sec.~\ref{section5.3}
	
	\item Birefringence occurs for \ac{CPT}-odd operators, see Sec.~\ref{section7.2}.  
	A related effect can be used in the neutrino sector as Lorentz violation will affect oscillations of both astrophysical and solar neutrinos, see Sec.~\ref{section7.5}.
		
	\item New \ac{GW} polarizations are often a consequence of dynamical \ac{LIV} due to mixing of the dynamical tensor field with the usual gravitational excitations.  This can yield both new signals at \ac{GW}
 observatories as well as change the decay rates of inspiraling \ac{GW}
 sources~\cite{Jacobson:2004ts,Hansen:2014ewa, Yagi:2013qpa,Kostelecky:2016kfm,Mewes:2019dhj}, see Sec.~\ref{section5.7}.
\end{itemize}

The plethora of possible signals shows that \ac{LIV} \ac{EFT} is very relevant for astrophysical observations.  At the same time, much of the parameter space has already been explored, with tight constraints on many \ac{LIV} operators up to mass dimension five, or singly suppressed by the Planck scale.  The next frontier is doubly Planck suppressed mass dimension six operators, which due to their high suppression and preservation of other symmetries such as \ac{CPT} invariance, are much more difficult to constrain. Although some progress has been made, constraints are still many orders of magnitude below the Planck scale~\cite{Kostelecky:2008ts}.  

%%%%%%%%%%%%%%%%%%%%%%%%%%%%%%%%%%%%%%%%%%%%%%%%%%%%%%%%%%%%%%%
%%%%%%%%%%%%%%%%%%%%%%%%%%%%%%%%%%%%%%%%%%%%%%%%%%%%%%%%%%%%%%%
\subsubsection{Modified kinematic symmetries}
\label{section2.2.2}

Modified kinematics allows us to accommodate a fundamental length scale, the Planck length, as a relativistic invariant quantity, alongside the speed of light.
Due to the special relativistic effects of length contraction and time dilations, assuming the existence of a fundamental observer-independent length scale leads necessarily to a modification of the relativistic symmetries of particles and observers on spacetime. In this section we give an overview over the most prominent models in the literature  and their phenomenological consequences. The latter will be discussed more quantitatively in Sec.~\ref{section5}

%%%%%%%%%%%%%%%%%%%%%%%%%%%%%%%%%%%%%%%%%%%%%%%%%%%%%%%%%%%%%%%
\paragraph{Doubly (or deformed) special relativity}
\label{section2.2.2.1}
\Acf{DSR} indicates a class of models of relativistic kinematics in which a fundamental observer-independent length scale is present. The main motivating idea behind such models was to circumvent the apparent clash between the contraction of lengths in ordinary \ac{SR} and the widely accepted prediction of \ac{QG} that the Planck length sets a lower limit to distance measurements once quantum and gravitational effects are taken into account\footnote{In more recent times it has been realized that the Planck length does not necessarily play the role of a minimum length (see e.g. \cite{Amelino-Camelia:2008fcv, Lizzi:2019wto}), but it is still regarded as a  fundamental constant in \ac{QG}.} (see \cite{Garay:1994en,Hossenfelder:2012jw} for a review). The general principles of \ac{DSR} theories were first introduced in \cite{AmelinoCamelia:2000mn, AmelinoCamelia:2000ge}. They include the standard principle of \ac{SR} appended by an additional one postulating the presence of an observer independent scale of length (or mass)\footnote{This has to be contrasted with scenarios in which the presence of a fundamental length scale is associated with a violation of Lorentz symmetry (see Sec.~\ref{section2.3.2})}.

The basic ingredients defining a given \ac{DSR}  model for particles kinematics are the laws of transformations between inertial observers, the composition rule of momenta and the dispersion relation. In order to avoid spoiling the relativistic invariance of the model these ingredients must satisfy a number of consistency conditions \cite{AmelinoCamelia:2011yi, Carmona:2012un, Amelino-Camelia:2013sba, Carmona:2016obd}. Preliminary studies indicate that such conditions guarantee the relativistic compatibility of multiple-vertices interactions \cite{AmelinoCamelia:2011nt, Amelino-Camelia:2014qaa, Gubitosi:2019ymi}.

The claim of the early \ac{DSR}  papers, that \ac{DSR}  could be realized using Hopf algebra deformations of Poincar\'e algebra (see below), was shortly confirmed in \cite{KowalskiGlikman:2001gp}. It was also realized \cite{Magueijo:2001cr, Magueijo:2002am} that kinematical models with DSR features can be obtained through non-linear redefinitions of the generators of the Poincar\'e algebra. The algebraic equivalence of these models was shown in \cite{KowalskiGlikman:2002we}, even though this does not necessarily imply equivalence of physical predictions \cite{Amelino-Camelia:2019dfl}. Shortly after that it was shown that this algebraic equivalence can be understood geometrically. Namely, \ac{DSR}  theories could be thought of as theories with curved momentum spaces of constant curvatures, and the observer-independent mass scale is associated with the curvature of momentum manifold \cite{KowalskiGlikman:2003we,Lobo:2016blj,Carmona:2019fwf}.
This observation was, in turn, the starting point of the relative locality principle \cite{AmelinoCamelia:2011bm}, see Sec.~\ref{section2.2.2.2}. Working at the level of momentum space, one can generate \ac{DSR}  models that have no correspondence with Hopf algebras \cite{Amelino-Camelia:2013sba, Arzano:2014jua,Carmona:2019fwf}.

A commonly accepted view is that \ac{DSR}  models offer an effective description of some flat-space-time limit of \ac{QG} i.e., heuristically, the limit in which Newton and Planck's constant go to zero, $G\to 0$ and $\hbar\to 0$, but their ratio remains constant $\hbar/G\neq 0$ \cite{AmelinoCamelia:2011bm,AmelinoCamelia:2011pe}. In this limit the Planck length goes to zero, which smooths out the small-scale structure of spacetime, and the minimal length is replaced by a cutoff in the energy or the momentum; the quantum fluctuations of spacetime effectively vanish, but the momentum space curvature stays. Attempts to generalize this picture to curved spacetimes with curved momentum space are discussed in Sec.~\ref{section2.2.2.4}.

By far the most studied model with \ac{DSR}-like features is the $\kappa$-deformed kinematics based on a Hopf algebra deformation of relativistic symmetries known as the $\kappa$-Poincar\'e algebra. Historically it was introduced as a contraction of a quantum deformation of the anti-de Sitter algebra $SO_q(3,2)$ \cite{Lukierski:1991pn,Lukierski:1992dt}. At the space-time level, this deformation of the Poincar\'e algebra is reflected in the emergence of non-commuting coordinates, the so-called $\kappa$-Minkowski non-commutative spacetime \cite{Lukierski:1993wx,Majid:1994cy} (see Sec.~\ref{section2.2.2.3}). Four-momenta, on the other side, become elements of a non-abelian Lie group given, as a manifold, by half of de Sitter space \cite{KowalskiGlikman:2003we,KowalskiGlikman:2002ft, Gubitosi:2011hgc}. The curved manifold structure of four-momentum space reflects in non-linear commutators between translation generators $P_a$ and generators of boosts $N_a$ 
\begin{align}\label{kalg}
[P_0, N_a] & = -i P_a\,, \nonumber\\
[P_a, N_b] & = -i\delta_{ab} \left(\frac{\kappa}{2} \left(1 - e^{-2 P_0/\kappa}\right) + \frac{1}{2\kappa} P_a P^a\right) + \frac{i}{\kappa} P_a P_b
\end{align}
while all other commutators are unmodified. The non-linearities appearing in the $\kappa$-deformed commutator lead to finite boost transformations in which $\kappa$ plays the role of the invariant fundamental energy scale. Indeed boosting a four-momentum with an infinite boost parameter will lead to infinite energy but finite linear momentum with Planckian magnitude \cite{KowalskiGlikman:2001gp,Bruno:2001mw}. Obviously this non-linear action is incompatible with an ordinary quadratic relativistic energy-momentum dispersion relation and indeed the mass Casimir of the  $\kappa$-Poincar\'e algebra is itself a non-linear function of energy and linear momentum
\begin{align}\label{}
\mathcal{C}= 4\kappa^2 \sinh^2\left(\dfrac{P_0}{2\kappa}\right) + e^{P_0/\kappa} P_aP^a\,.
\end{align}

The most distinctive feature of this Hopf algebra based \ac{DSR}  model is the non-abelian nature of four-momentum space. Indeed we can express the product of two group elements $g = e^{-i k^a X_a} e^{i k_0 X_0}$ and $h = e^{-i l^a X_a} e^{i l_0 X_0}$ as
\begin{equation}\label{}
g h = e^{-i (k^a \oplus l^a) X_a} e^{i (k_0 \oplus l_0) X_0}\,,
\end{equation}
with a non-abelian composition law $(k_0 \oplus l_0, k_a \oplus l_a)$ given by
\begin{equation}\label{copr1}
k^a \oplus l^a = k^a + e^{-k_0/\kappa} \, l^a\,,\qquad k^0\oplus l^0 = k^0 + l^0\,.
\end{equation}
In fact, it is possible to find some coordinates on momentum space (called classical basis) such that the Poincar\'e algebra of the \ac{DSR}  model is the standard one and the whole deformation is contained in the modified energy-momentum composition rules: this shows that the key ingredient of \ac{DSR}  is a non-symmetric deformed composition of energy and momentum, which should be contrasted with the starting point used in the proposal of \ac{DSR}  as based on a modified energy-momentum relation for a free particle. 

The Lie group structure of momentum space provides a solid link with gravity at least in the much simplified case of $2+1$ space-time dimensions. Indeed in this case \ac{GR}
 is a topological theory with no local degrees of freedom and particles are coupled to the theory as space-time defects. It turns out that their three-momentum is no longer given by a three vector but by an element of the local isometry group of spacetime i.e.~the Lie group ${\rm SL}(2, \mathbb{R})$ \cite{Matschull:1997du,Schroers:2007ey}. It turns out that the symmetries governing the kinematics of the particle are given again by a Hopf algebra which can be seen as a deformation of the three dimensional Poincar\'e algebra \cite{Joung:2008mr} and its kinematics shows the distinctive \ac{DSR}  feature as one would expect \cite{Arzano:2014ppa}. The connection between (quantum) gravity and \ac{DSR}  models in $3+1$ is far less clear even though there are some preliminary indications that Hopf algebra relativistic symmetries might be connected to Planck scale kinematics also in this more realistic case \cite{AmelinoCamelia:2003xp,Girelli:2004md}. 

%%%%%%%%%%%%%%%%%%%%%%%%%%%%%%%%%%%%%%%%%%%%%%%%%%%%%%%%%%%%%%%
\paragraph{Relative locality and Born geometry}
\label{section2.2.2.2}

The proposal of a ``principle of relative locality'' \cite{AmelinoCamelia:2011bm, AmelinoCamelia:2011pe} resulted from a deepening in the understanding of the fate of locality in \ac{DSR}  theories. After ten years of developments in \ac{DSR}  research, a series of papers \cite{AmelinoCamelia:2010qv, Hossenfelder:2010tm, Smolin:2010xa} put the accent on the fact that theories of \ac{DSR}  type generally fail to have a property of absolute locality. In particular, in  \cite{AmelinoCamelia:2010qv} it was realised that within \ac{DSR}  models events observed as coincident by nearby observers may be described as non-coincident by some distant observers. Locality of physical processes can still be enforced, as long as one specifies that it holds only for observers local to the process itself. 
For distant observers, the amount of non-locality crucially depends on momenta carried by the probes used to reveal the spacetime structure (like photons used to synchronize clocks in \ac{SR}). The transition from absolute locality in \ac{SR} to relative locality in \ac{DSR}  is analogous to the transition from  the absolute time of Galilean physics to the relativity of simultaneity of \ac{SR}.

The formalization of relative locality, first proposed in \cite{AmelinoCamelia:2011bm, AmelinoCamelia:2011pe}, is based on the nontrivial geometry of momentum space implied by \ac{DSR}  type theories\footnote{Note that, while \ac{DSR}  symmetries imply a nontrivial geometry of momentum space, only some geometries of momentum space are compatible with (deformed) relativistic invariance \cite{Amelino-Camelia:2013sba}.}, see Sec.~\ref{section2.2.2.1}. In this framework, the metric on momentum space is related to the form of the energy-momentum dispersion relation. Specifically,  the mass $m$ of a particle is given by the geodesic distance of the point in momentum space corresponding to the particle’s momentum $p$  from the origin (see \cite{Amelino-Camelia:2013sba} for more details):
\begin{equation}
    m^2=d^2(p,0)\,.
\end{equation}
This equation gives the particle's dispersion relation. The affine connection on momentum space is determined by the law of composition of momenta, see \cite{Amelino-Camelia:2013sba,Carmona:2019fwf} for a discussion of the various proposals to formalize this relation. 

In the relative locality framework spacetime emerges as one considers the dynamics of particles. This is obtained from a
variational principle based on the action
\begin{equation}
    S=\sum_J S^J_{\text{free}} + S_{\text{int}},
\end{equation}
where $S_{\text{free}}$ is the free part of the action, of the form
\begin{equation}
    S_{\text{free}}=\int d\lambda \left(-x^\mu \dot p_\mu+\mathcal{N}\cdot \left(d^2(p)-m^2\right)\right),
\end{equation}
($\mathcal{N}$ being a Lagrange multiplier enforcing the dispersion relation), which is summed over particles entering and leaving the interaction vertex, and $S_{\text{int}}=\mathcal{K}_\mu(0)z^\mu$, where $\mathcal{K}_\mu(0)$ is the conservation law, which depends on the law of composition of momenta at the interaction point ($\lambda=0$), and the $z^\mu$ are Lagrange multipliers that implement the energy-momentum conservation in the interaction. The variational principle applied to this action then gives the coordinates of the $J$-th particle, with momentum $k_\mu^J$, at the interaction point:
\begin{equation}
    x_J^\mu(0)=z^\nu\frac{\partial \mathcal{K}_\nu}{\partial k_\mu^J}(0).
\end{equation}
One then sees that the coordinates of the particles at the interaction point do not coincide, unless $\mathcal{K}$ is the usual conservation law of \ac{SR} (additive composition law), or the $z^\mu$ are zero, which then gives $x_J^\mu(0)=0$ for all $J$ (the observer is local to the interaction).
So this formalism developed to describe interaction vertices provides a  description of spacetime coordinates consistent with the relativity of locality of interactions (see \cite{Gubitosi:2019ymi, Amelino-Camelia:2014qaa} for an in-depth discussion).

\noindent \textit{Born geometry.} In order to reconcile the idea of minimal length/time with relativity we are led to a new development in
thinking about \ac{QG} presented in~\cite{Freidel:2015uug, Freidel:2016pls, Freidel:2013zga, Freidel:2014qna, Freidel:2015pka,  Freidel:2017xsi,  Freidel:2018apz,   Freidel:2019jor, Minic:2020oho,  Freidel:2017wst, Freidel:2017nhg}.
This new approach follows the line of reasoning well known from Einstein's special theory of relativity, albeit
in the quantum context.
In order to reconcile the minimal length with relativity one is led to covariant relative locality
\cite{AmelinoCamelia:2011bm, AmelinoCamelia:2011pe}, which in this setting means ``observer dependent spacetimes''.
Such observer dependent spacetimes
appear as ``sections" of a new and larger concept  - quantum spacetime.
(This is again in analogy with what happens to observer dependent space and observer dependent time
in special theory of relativity - they appear as ``sections'' of a larger concept of spacetime.)
The new concept of quantum (or modular) spacetime \cite{Freidel:2015uug} is implied by a novel formulation of quantum physics
\cite{Freidel:2016pls} in terms of
unitary variables which explicitly contain fundamental length and time (such observables are directly related to modular variables).
The concept of modular spacetime also illuminates the problem of general quantum measurements.

It  turns out that non-locality (as defined by covariant relative locality) together with causality
essentially determines the known structure of quantum theory \cite{Freidel:2016pls}.
The geometry of such causal non-locality, that is, the geometry of quantum spacetime, is
called Born geometry \cite{Freidel:2013zga, Freidel:2014qna} and it unifies the symplectic, orthogonal and double metric geometries.
Once again, in analogy with Einstein's general theory of relativity,
a dynamical quantum spacetime (a dynamical modular spacetime) appears as the natural
arena for \ac{QG} (which ``gravitizes the quantum'' \cite{Freidel:2014qna}). This structure is precisely
realized in a T-duality covariant and intrinsically non-commutative formulation of string theory
(called ``metastring theory'' \cite{Freidel:2015pka, Freidel:2017xsi}), and it can
be shown to be unitary and causal in the metaparticle limit \cite{Freidel:2018apz}. In this theoretical setting quantum gravitational effects
appear both at short distance (\ac{UV}) and at long distance (\ac{IR}) \cite{Freidel:2019jor, Minic:2020oho}.

%%%%%%%%%%%%%%%%%%%%%%%%%%%%%%%%%%%%%%%%%%%%%%%%%%%%%%%%%%%%%%%
\paragraph{Non-commutative spacetimes}
\label{section2.2.2.3}

Generically, a non-commutative spacetime is non-commutative (and associative) algebra generated by a set of spacetime operators $\hat x^\mu$ whose non-commutativity is governed by a parameter which is assumed to be related to the Planck scale.  Non-commutative spacetimes are often called `quantum spacetimes' and provide mathematical realizations of the idea that spacetime at the Planck scale cannot be described by a differentiable manifold. If spacetime operators $\hat x^\mu$ are assumed to play the role of observables whose eigenvalues provide the spacetime localization of a particle coordinate in a certain \ac{QG} regime, their non-commutativity implies the existence of non-vanishing uncertainty relations between them, thus precluding full localization for simultaneous measurements of pairs of spacetime observables. The Poisson version of non-commutative spacetimes is also frequently used in the literature, since the interpretation of \ac{DSR} as a limit $\hbar\to 0, G\to 0, \hbar/G\neq 0$ (see Sec.~\ref{section2.2.2.1}) leads to the formalism with the non-commutativity of spacetime at the classical level, in which commutators are replaced with Poisson brackets. 

Different non-commutative spacetimes arise in many approaches to \ac{QG}, since they provide a consistent framework to introduce a minimal length scale, which is a generic ingredient of these theories \cite{Garay:1994en,Hossenfelder:2012jw}. In particular, non-commutative spacetimes are obtained in a natural way when the quantization of a point particle in (2+1) gravity is considered~\cite{Matschull:1997du}. Moreover, there exists a duality between spacetime non-commutativity and non-vanishing curvature of the associated momentum space \cite{Snyder:1946qz}, and, by adding compatible momenta operators $\hat p^\mu$ to the non-commutative spacetime $\hat x^\mu$, non-commutative deformed phase space algebra arises from which the associated \ac{GUP} can be derived. Therefore, phenomenological consequences described in Sec.~\ref{section2.2.4} are also applicable for the \ac{GUP} obtained for non-commutative spacetimes. Experimental or observational data can be used to provide constraints on the range of allowed values of the non-commutativity parameter. For instance, in~\cite{Kobakhidze:2016cqh} the \ac{GW} signal GW150914 is analysed.

The seminal notion of non-commutative spacetime is due to Heisenberg (see~\cite{Hossenfelder:2012jw}), and the first specific proposal, aimed to introduce a UV cut-off in momentum space, was the Snyder non-commutative spacetime~\cite{Snyder:1946qz} 
\begin{equation}
[\hat x^0, \hat x^k]= i\,{a^2}\, M^{0k}\,,\qquad [\hat x^k, \hat x^j]= i\,a^2\,M^{kj}\,,
\end{equation}
where $M^{0k}$ and $M^{kj}$ are, respectively, boosts and rotations generators of the Lorentz algebra, and $a$ is the non-commutativity parameter. By construction, this non-commutative spacetime is Lorentz invariant, the spectrum of the spatial coordinates operators $\hat x^i$ is discrete and its momentum space is a de Sitter manifold with a non-vanishing constant curvature $a^2$. Relative locality phenomenology of the Snyder model
has been studied in~\cite{Mignemi:2018tdg}, including  time delay and dual curvature lensing, and the composition law of momenta is shown to be deformed by terms quadratic in the inverse Planck energy. An \ac{EFT} of a self-interacting scalar $\phi^4$ on 
Snyder space is given in
\cite{Franchino-Vinas:2019nqy}. Its 
one-loop contributions show that a change of sign in the curvature parameter may happen as a consequence of the renormalization flow.

Another outstanding non-commutative spacetime is the so-called canonical or $\theta$-non-commutative spacetime~\cite{Doplicher:1994zv}
\begin{equation}
[\hat x^\mu, \hat x^\nu]= i\,\theta^{\mu\nu}\,,
\end{equation}
where $\theta$ is a skew-symmetric tensor whose entries are assumed to commute with the spacetime operators $\hat x^\mu$, and are thus often considered simply as constants. The Euclidean version of this non-commutative spacetime is known as the Moyal non-commutative spacetime, whose field theory has been thouroughly studied (see~\cite{Szabo:2001kg,Douglas:2001ba}) including non-commutative versions of gravitation (see, for instance,~\cite{Szabo:2006wx}). In \cite{Bhattacharyya:2018pnt},  \ac{GW}
 data are used as an effective probe of non-commutative structure of spacetime by considering the effect of Moyal non-commutativity on resonant bar detectors. The main prediction is a modification in the resonant frequencies of the detector, modeled as a harmonic oscillator, and phenomenological restrictions on $\theta$ are reviewed.

A covariant yet non-commutative spacetime is the so-called modular spacetime, predicted by
a generic non-commutative formulation of string theory, called metastring theory.
This model has an implementation in the
context of ``metaparticle physics'', the zero modes of the metastring.
Modular spacetime implements covariant relative, or observer-dependent,
locality, and it is tied to the concept of Born geometry (see Sec.~\ref{section2.2.2.2}).
The generic predictions of this non-commutative spacetime, are the
existence of dual matter and dual gravity sectors, as well as very particular 
kinematical features of the ``metaparticle'' degrees of freedom,
which result in an infrared deformation of the dispersion relation~\cite{Freidel:2018apz,   Freidel:2019jor, Minic:2020oho,  Freidel:2017wst, Freidel:2017nhg}.

In non-commutative spectral geometry, gravity and the Standard Model fields are put together into matter and geometry on non-commutative space made from the product of a
four-dimensional Riemannian manifold and an internal zero-dimensional discrete space~\cite{Connes:1994yd,1414300,vanSuijlekom:2015iaa}. 
For the bosonic spectral action, the existence of some energy cut-off $\Lambda$ leads to an effective action that depends on the spectrum of the Dirac operator truncated at the cut-off scale, and the asymptotic expansion of the leading term in $\Lambda$ is computed.
To address issues related with the cut-off bosonic spectral action, a zeta function regularization has been later proposed ~\cite{Kurkov:2014twa}. 
Most of the Standard Model phenomenology obtained through this approach is based on considering either Dirac or Majorana massive neutrinos and employing the see-saw mechanism ~\cite{Chamseddine:2006ep}.
Following however a consistent treatment of both the fermionic and the bosonic parts of the action ~\cite{Sakellariadou:2019ihz}
one obtains nontrivial corrections leading to nonminimal fermion couplings and a potentially rich phenomenology.
Non-commutative spectral geometry leads to an extended gravitational theory, where the gravitational sector includes additional terms beyond the ones of the Einstein--Hilbert action ~\cite{Sakellariadou:2013ve}. Their cosmological consequences and constraints from observational data have been discussed in~\cite{Nelson:2008uy,Nelson:2010ru,Lambiase:2013dai,Buck:2010sv}.

In \ac{DSR} models the Lorentz invariance condition is replaced by the invariance under a deformed symmetry (see Sec.~\ref{section2.3.2}) given by a so-called quantum Poincar\'e group~\cite{Lukierski:1991pn,Lukierski:1992dt}. The non-commutative spacetimes which are covariant under the action of quantum groups are called quantum spaces. The main example is given by the $\kappa$-Poincar\'e algebra described in Sec.~\ref{section2.2.2.1}, whose associated $\kappa$-Minkowski non-commutative spacetime is~\cite{Majid:1994cy}
\begin{equation}
[\hat x^i, \hat x^0]=\frac{ i }{\kappa}\,\hat x^i \, ,
\qquad [\hat x^i, \hat x^j]=0 \, .
\label{eq262-kappaMinkowski}
\end{equation}

\ac{DSR} (see Sec.~\ref{section2.2.2.1}) contains modified composition laws of energy and momentum. If one accepts the conjecture that a consequence of a theory of \ac{QG} is a nontrivial composition of  energy and momentum (the state of a multiparticle system is  determined not only by the state of each particle) then one can derive the non-commutativity of spacetime from  different perspectives. In the algebraic approach based on quantum groups, where one considers a quantum Poincar\'e  (Hopf) algebra~\cite{Majid:1988we} as a deformation of the Poincar\'e algebra of \ac{SR}, the composition of energy and momentum stems from the energy-momentum co-algebra, and the new spacetime commutator algebra is derived through the so-called pairing procedure~\cite{KowalskiGlikman:2002we,KowalskiGlikman:2002jr}. 
If one identifies the composition of energy and momentum as a translation in a curved momentum space of constant curvature, then $\kappa$-non-commutative spacetime coordinates are generators of translations in momentum space~\cite{Lobo:2016blj,Carmona:2019fwf}, and  $\hat x^\mu$  in (\ref{eq262-kappaMinkowski}) are just the generators $X_\mu$ of the Lie group whose group law is the composition law for momenta given in Sec.~\ref{section2.2.2.1}.
Alternatively, the nonlocality (in canonical spacetime) of interactions defined by a nontrivial composition of energy and momentum, leads to identify a non-commutative spacetime where the interactions are local~\cite{Carmona:2019vsh}. 
This point of view has several phenomenological consequences. First, the kinematics of processes have to include modified energy-momentum composition laws, which modify the standard analysis (e.g.~computation of thresholds) done in the \ac{LIV} framework (see Sec.~\ref{section2.3.2}). Second, even if the analysis of an observable does not in principle involve the kinematics of a process (e.g. time delays) the non-commutativity of spacetime may play a role in the appropriate definition of that observable \cite{Aschieri:2017ost,Aschieri:2020yft} (see Secs.~\ref{section2.3.2}, \ref{section2.3.4} and \ref{section7.1}).

%%%%%%%%%%%%%%%%%%%%%%%%%%%%%%%%%%%%%%%%%%%%%%%%%%%%%%%%%%%%%%%
\paragraph{Geometry of curved phase space: Hamilton, Finsler and beyond}
\label{section2.2.2.4}

Effectively the propagation of particles interacting with the quantum nature of gravity can be described by \acp{MDR}, emerging from \ac{DSR} or non-commutative models, see Secs.~\ref{section2.2.2.1} and \ref{section2.2.2.3}, leading to a non trivial phase space geometry, i.e.\ an intertwined curved spacetime and curved momentum space geometry.

In \ac{SR}, the symmetries of Minkowski spacetime, given by the Poincar\'e group, and the compatibility of fundamental physical field theories, such as spinors and gauge potentials, with these symmetries, imply that the 4-momentum of a physical particle must satisfy  $ -E^2 + \vec p^2 = \eta^{\mu\nu}p_\mu p_\nu = - m^2$. When gravity is included, i.e.\ passing from global to local Lorentz symmetry, the dispersion relation becomes $g^{\mu\nu}(x)p_\mu p_\nu = - m^2$, where $g^{\mu\nu}(x)$ defines a Lorentzian metric on a curved spacetime manifold.

Early approaches to formalize the idea of a non-trivial phase space geometry introduced an energy dependent spacetime metric, which yields a dispersion relation $g^{\mu\nu}(p_0)p_{\mu}p_{\nu} = -m^2$, where $p_0=E$ is the energy an observer associates to particles probing the spacetime, see \cite{Magueijo:2002xx} and references therein. A main drawback of this approach is the observer dependence of the energy dependent spacetime geometry.

The fist step towards a covariant curved spacetime formulation of the geometry of phase space, is to realize that it is possible to introduce the dispersion as the level sets of a covariant Hamilton function on phase space $H(x,p)=\text{constant}$, and, that one can introduce a phase space metric $g^{\mu\nu}(x,p)$ with a covariant transformation behavior that is compatible with coordinate transformations on the base manifold.

From these insights it turned out that three geometric frameworks are promising candidates to understand the curved geometry of spacetime and momentum space:
\begin{description}
	\item \emph{Finsler geometry}: To reach Finsler geometry from a dispersion relation on phase space one employs a Legendre transform to map the Hamilton function $H(x,p)$ to a homogeneous point particle Lagrangian  $L(x,\dot x)$ on position and velocity space, technically the tangent bundle. Finsler geometry then yields a curved intertwined velocity dependent geometry of position and velocity space that is derived from $L$. Point particles follow Finslerian geodesics and the geometry can be determined dynamically by Finslerian generalizations of the Einstein equations \cite{Girelli:2006fw,Gibbons:2007iu,Pfeifer:2011xi,Amelino-Camelia:2014rga,Letizia:2016lew,Lobo:2016xzq,Fuster_2016,Hohmann:2018rpp,Fuster_2018,Lammerzahl:2018lhw,Pfeifer:2019wus,Hohmann:2019sni,Elbistan_2020,Hohmann:2020mgs,Heefer:2020hra}. A position and velocity dependent metric, the so called Finsler metric, is a secondary object, derived from the Finsler Lagrangian.
	\item \emph{Hamilton geometry}: Instead of performing the above mentioned Legendre transformation, one can derive the geometry of phase space, technically the cotangent bundle, directly from the dispersion relation defining Hamilton function  $H(x,p)$ \cite{Barcaroli:2015xda}. The resulting curved position and momentum space geometry is the dual formulation of a corresponding Finsler geometry on position and velocity space \cite{Lobo:2016lxm,Torri:2019gud}. In this approach point particles follow solutions of the Hamilton equations of motion. A position and momentum dependent metric, the so called Hamilton metric, is a secondary object, derived from the Hamilton function.
	\item \emph{Generalized Hamilton spaces}: The step from Hamilton geometry to generalized Hamilton spaces is to consider a phase space metric as fundamental and the dispersion relation defining Hamilton function as a derived object \cite{Relancio:2020zok}. The geometry of phase space is derived from the metric in a way such that it is consistent with a split of phase space into position and momentum space. Point particles follow particular geodesics of the phase space metric in this picture \cite{Pfeifer:2021tas}.
\end{description}

From the point of view of phenomenology, the most relevant feature of these geometric frameworks resides in the link to \iac{MDR}, which would in turn have implications for the propagation of particles, in particular for those of astrophysical origin.

For the different approaches mentioned above, this link can go both ways. In the case of Finsler/Hamilton geometry, one takes a dispersion relation for granted, derives the phase space geometry and works out related phenomenological implications. In the generalized Hamilton phase space framework, one assumes a metric geometry on the cotangent bundle and derives the dispersion relation and other relevant/observable quantities.

Either way, the robustness of the link between a given dispersion relation and the associated geometry on the phase space is still matter of debate in the literature. Other issues concern whether the link between a geometry and a dispersion relation is unique, and whether the geometry (and thus the geodesics followed by particles) is mass-dependent -- an issue especially relevant for multi-messenger observations.

%%%%%%%%%%%%%%%%%%%%%%%%%%%%%%%%%%%%%%%%%%%%%%%%%%%%%%%%%%%%%%%
\paragraph{Phenomenological implications}
\label{section2.2.2.5}

The phenomenological consequences of all the frameworks discussed in the previous sections reflect the fact that the kinematics of particles and fields propagating on quantum spacetime are affected through their interaction with quantum spacetime. We now list the most prominent effects:

\begin{description}
    \item {\textit{Time delays and distance changes}} (see also Sec.~\ref{section7.1}) \cite{AmelinoCamelia:1997gz,AmelinoCamelia:2011nt,Amelino-Camelia:2016ohi,AmelinoCamelia:2011gy,Relancio:2020zok,Barcaroli:2016yrl,Pfeifer:2018pty,Relancio:2020mpa,Barcaroli:2015eqe,Laanemets:2022rmn}:
    The time of arrival of elementary particles such as photons and neutrinos is derived from the modified momentum space geometry. On the basis of Finsler and Hamilton geometry, where one simply solves the Hamiltonian, respectively geodesic, point particle equations of motion, one obtains a non-vanishing time delay, depending on the particles mass and energy. Taking into account that distances measured by particles probing spacetime may also depend on their momentum, the time delay may by absorbed into the energy dependent notion of distance.
    
    \item {\textit{Redshift}} \cite{Relancio:2020zok,Barcaroli:2016yrl,Barcaroli:2017gvg,Hasse:2019zqi,Relancio:2020mpa,Barcaroli:2015eqe}:
    The gravitational redshift is derived from the trajectories of photons (light rays), solving the Hamiltonian, respectively geodesic, point particle equations of motion, and an observer/clock model, which implements how an observer measures the frequency of a photon. Depending on the observer model employed different magnitudes of a redshift are derived. On cosmological scales a deviation from the general relativistic redshift is obtained to first order in the Planck length, while the redshift in spherically symmetric spacetimes is only of second order.
    
    \item {\textit{Photon orbits (black hole shadows) and gravitational lensing}} \cite{Barcaroli:2017gvg,Pfeifer:2019zhc,Glicenstein:2019rzj,Lammerzahl:2012kw,Laanemets:2022rmn}: 
    The photon sphere, respectively the photon regions around a black hole characterize its shadow, which is nowadays directly detectable, thanks to the Event-Horizon telescope \cite{EventHorizonTelescope:2019dse}. A non-trivial momentum space geometry makes these regions frequency- dependent, which leads to the prediction that photons of different frequency propagate in different photon regions, and thus eventually one obtains a frequency-dependent black hole shadow, literally a rainbow effect. Looking beyond the photon regions around black hole, the photon trajectories of a  non-trivial momentum space geometry lead to a frequency-dependent gravitational lensing.
    
    \item{\textit{\Acp{UHECR} and the \ac{GZK} cut-off}} \cite{Torri:2018qka,Torri:2019gud,Torri:2020fao,Torri:2021hpj}:
    \Acp{UHECR} can interact with the \ac{CMB}, dissipate energy during their propagation and are attenuated in a way that depends on their energy and nature (see Sec.~\ref{sec:UHECRprop}); therefore, the universe is opaque to their propagation (the \ac{GZK} cut-off). The \ac{GZK} opacity sphere can be enlarged by modifications of the momentum space available for \ac{CMB} interaction processes, introduced via modifications of the kinematics related to a Hamilton/Finsler geometry, whose metric depends on the species and energy of the considered particles (see also Sec.~\ref{hadronic}).
    
    \item {\textit{Neutrino oscillations}} \cite{Antonelli:2018fbv,Torri:2020dec}:
    The flavor oscillation phenomenon is governed by a phase that depends on the ratio of the propagation length divided by the particle energy. Modifications of the kinematics can again be introduced via a Hamilton/Finsler geometry in a covariant scenario. The phase governing the oscillation is amended by the introduction of a term given by the product of the particle energy and the perturbation term related to the geometry.
    
    \item {\textit{Synchrotron radiation}} \cite {Relancio:2020mpa}:
    One can study the synchrotron radiation with a space-time metric depending on momenta in such a way that electrons are always subluminal. The modified metric is crucial in order to obtain the result, making a momentum-dependence appear between coordinate and proper time and spatial coordinate and physical distance probed by a particle. One finds in this scheme that there is not a maximum frequency of the emitted photons, in contrast with the \ac{LIV} case.

    \item {\textit{Laboratory experiments}} \cite{Lammerzahl:2008di,Itin:2014uia}:
    Finsler geometry implies a violation of Lorentz invariance. Accordingly, tests of the  Michelson--Morley type and atom spectroscopy can by used to constrain Finslerian parameters. In the theoretical description the whole experiment, that is, the electromagnetic fields as well as the interferometer arms and the atomic matter have to be treated in the Finslerian framework. Further spectroscopic observables which can be used to constrain Lorentz symmetry violations and deformations are the spectrum of a hydrogen atom, which has been been derived in these frameworks, as well as cold atom systems which open a window to test \acp{MDR} at the scale of order $10^{-7} \mathrm{eV}^2$.
    
    \item {\textit{Non-localizabillity of events}} \cite{Lizzi:2018qaf,Ballesteros:2019vws,Ballesteros:2021bhh}:
    Limits on the localizability of events and properties of reference frames for the $\kappa$-Minkowski spacetime, as well as its non-commutative space of worldlines have been studied.
    
    \item {\textit{Emergence of a cosmological constant}} \cite{Mignemi:2008ek,Ballesteros:2019hbw,Ballesteros:2016bml,Ballesteros:2017zrx,Ballesteros:2017pdw}:
    \ac{DSR} models including spacetime curvature at cosmological distances through a nonvanishing cosmological constant have been considered: $\kappa$-(A)dS curved momentum spaces have been constructed, as well as the non-commutative $\kappa$-(A)dS spacetime, which is  covariant under the $\kappa$-(A)dS quantum group, in which the cosmological constant appears as a deformation parameter. 
    
    \item {\textit{Threshold anomalies} (see also Sec.~\ref{section2.3.2})} \cite{AmelinoCamelia:2010pd}:
    The deformed energy-momentum dispersion relation and four-momentum conservation laws which characterize \ac{DSR}  models could affect the kinematics and energy threshold conditions for certain particle production processes, like the production of electron-positron pairs in photon scattering $\gamma \gamma \rightarrow e^+ e^-$.
    
    \item {\textit{Muon lifetimes}} \cite{Arzano:2019toz,Arzano:2020rzu,Lobo:2020qoa}:
    Muon physics turns out to be sensitive to deformations of Lorentz symmetry. Deformations of \ac{CPT} symmetry and time dilations between different rest frames can be constrained by precision measurements of the muon's lifetime.
    
    \item {\textit{Decoherence}} (see also Sec.~\ref{section5.3}) \cite{Arzano:2014cya}: 
    Time evolution in a \ac{DSR}  setting, as described by time translation generators obeying deformed composition and inversion rules, might cause a fundamental decoherence.
    
    \item {\textit{Dark Matter and axion-like backgrounds}} \cite{Minic:2020oho,Berglund:2019ctg, Berglund:2019yjq,Ho:2010ca, Edmonds:2017zhg,Freidel:2017wst,Freidel:2017nhg,Benisty:2017eqh,Benisty:2018qed,Benisty:2018oyy}:
    In Born geometry metaparticles are endowed with a non-trivial propagator and an IR sensitive dispersion relation, that could be investigated in various experimental settings. Such fields are inherently bi-local and non-commutative, and each observed quantum field has a ``dual'', which could be associated with dark matter, and, in the gravitational sector, dark energy, as well as axion degrees of freedom.
\end{description}

%%%%%%%%%%%%%%%%%%%%%%%%%%%%%%%%%%%%%%%%%%%%%%%%%%%%%%%%%%%%%%%
%%%%%%%%%%%%%%%%%%%%%%%%%%%%%%%%%%%%%%%%%%%%%%%%%%%%%%%%%%%%%%%
\subsubsection{Classical modified gravity as an effective description of quantum gravity}
\label{section2.2.3}

Alternative theories of gravity are important tools in the effort to bridge the gap between \ac{QG} and observations. 
New physics that resolves the conflict between \ac{GR} and \ac{QFT} should actually appear at some (curvature) scale {\em above} the Planck length. Theoretical considerations, such as the conundrums that arise in \ac{QFT} near black hole horizons and the cosmological constant problem, suggest that one needs to keep an open mind about how large that lengthscale will be. Moreover, dark matter and dark energy are striking examples of low-energy new physics that appeared in observations unexpectedly.

\Ac{QG} usually follows a bottom-up approach: starting from fundamental assumptions about the quantum nature of gravity, one aims to construct a self-consistent and complete \ac{QG} model and eventually develop it enough to obtain observational signatures that can be used to test it. Modified gravity can follow the inverse route: it can use observations to quantify in which ways and how much  one can deviate from \ac{GR} classically, and then feed into \ac{QG} model building. One of its key strengths is that an alternative theory of gravity should describe gravitational phenomena in a very wide range of systems, from the universe to compact objects. Hence one can combine different observations in order to improve bounds on parameters. This is particularly relevant in the era of multi-messenger astronomy. It is also highly complementary to other strands of \ac{QG} phenomenology, where one obtains constrains using theory-independent, but system-specific, parametrizations.

\paragraph*{Uniqueness of GR and modified gravity} Lovelock's theorem \cite{Lovelock:1971yv} is one of many ways to show that GR is unique among gravity theories. It assumes diffeomorphism invariance, 4 spacetime dimensions, that the metric is the only independent field and that it satisfies second order differential equations. Violating any of these assumptions will thus lead to a way to deviate from GR: higher-dimensional models, e.g.~the DGP model \cite{Dvali:2000hr}, Lorentz-violating theories, such as Einstein-aether \cite{Jacobson:2000xp,Jacobson:2008aj} and Ho\v rava gravity \cite{Horava:2009uw,Sotiriou:2009gy,Sotiriou:2009bx,Blas:2009qj,Sotiriou:2010wn}, higher-order theories. e.g.~4th-order gravity \cite{Stelle:1976gc} (see also Sec.~\ref{section2.1.8}) or $f(R)$ gravity \cite{Sotiriou:2008rp,DeFelice:2010aj}, and a wide variety of models with extra fields \cite{Clifton:2011jh,Capozziello:2011et,Benisty:2021sul,Benisty:2021cmq,Benisty:2021cin}. The latter class includes theories with independent connections, e.g.~\cite{Hehl:1976kj,Sotiriou:2006qn,Olmo:2011uz,Aldrovandi:2013wha} and massive gravity models \cite{Fierz:1939ix,deRham:2010kj,deRham:2014zqa}\footnote{We do not consider here nonlocal theories, which are discussed in Sec.~\ref{section2.1.8}.}. 

In the context of \ac{QG}, where the symmetries and the choice of variables being quantized are crucial, different ways to violate Lovelock's theorem are expected to lead to distinct approaches. This can help link alternative theories to a \ac{QG} candidate. However, at a purely classical level, which is usually sufficient for phenomenology and confrontation with observations, classifications based on how one circumvents the uniqueness theorem are far less rigid \cite{Sotiriou:2015lxa}. Alternative theories are seen as \acp{EFT}. Hence, higher-dimensional models can be compactified to yield a 4-dimensional model with new fields, describing the extra dimensions, diffeomorphism invariance can be restored by introducing new (Stueckelberg) fields \cite{Ruegg:2003ps}, and field redefinitions can be used to re-write higher-order theories as second-order theories with a larger field content. 

This perspective reduces significantly the theory space and suggests a way to navigate it: define a field content and construct a 4-dimensional, diffeo-invariant, second-order theory, adhering to the principles of \ac{EFT}.   
One can then further restrict focus on the subclass of models that have interesting phenomenology for a given system. As a characterstic  example, fix the field content to one extra scalar that respects shift-symmetry (and hence can be massless). Members of the shift-symmetric Horndeski class \cite{Horndeski:1974wa,Deffayet:2009mn} can then be reasonable \acp{EFT}. This is a large class but if one further restricts attention to black holes, then there is a unique term that gives rise to hair \cite{Sotiriou:2013qea,Sotiriou:2014pfa} and gives the leading-order contribution to deviations in \ac{GW}  emission from black hole binaries \cite{Witek:2018dmd}.

\paragraph*{Outlook} Testing gravity with multimessenger observations requires modeling the corresponding system beyond GR \cite{Berti:2015itd,Barack:2018yly,Sathyaprakash:2019yqt,Barausse:2020rsu}. Modified gravity theories offer the framework for doing that in the nonlinear regime and this makes them an indispensable tool for linking fundamental physics and observations. In the dawn of \ac{GW} astronomy, modeling waveforms and \ac{GW} propagation is the new frontier in alternative theories of gravity. Some related theoretical challenges include: (i) identifying theories that have interesting strong-field phenomenology, such as black hole hair \cite{Sotiriou:2015pka,Herdeiro:2015waa} and strong field phase transitions \cite{Damour:1993hw,Silva:2017uqg,Doneva:2017bvd}; (ii) addressing the open problem of well-posedness and nonlinear evolution beyond \ac{GR} \cite{Benkel:2016rlz,Benkel:2016kcq,Okounkova:2017yby,Cayuso:2017iqc,Witek:2018dmd,Okounkova:2019dfo,Okounkova:2019zjf,Cayuso:2020lca}; (iii) linking \ac{QG} candidates to low-energy \ac{EFT}s, as a way to access observations.

%%%%%%%%%%%%%%%%%%%%%%%%%%%%%%%%%%%%%%%%%%%%%%%%%%%%%%%%%%%%%%%
%%%%%%%%%%%%%%%%%%%%%%%%%%%%%%%%%%%%%%%%%%%%%%%%%%%%%%%%%%%%%%%
\subsubsection{Generalized uncertainty principle}
\label{section2.2.4}

\Ac{GUP} models were originally derived in the context of string theory \cite{Amati:1988tn, Konishi:1989wk}, which is characterized by a minimal length given by the inverse of the square-root of the Regge slope \cite{Green:1987mn, Green:1987sp, Polchinski:1998rq, Polchinski:1998rr}. \ac{GUP} models were later generalised to cases beyond strings \cite{Maggiore:1993rv, Garay:1994en, Kempf:1994su}, as providers  of phenomenological deformations  of \ac{QM} that could accommodate the presence of a  minimal length, a feature emerging in several approaches to \ac{QG} \cite{Garay:1994en, Hossenfelder:2012jw, Chang:2011jj}.
Since the presence of a minimal length is not compatible with  Heisenberg's uncertainty principle, heuristic arguments led to propose \ac{GUP} of the form $\Delta x \Delta p \sim \frac{\hbar}{2} (1 + \beta p^2 + \mathcal{O}(p^4))$ with the \ac{GUP} parameter $\beta$ related to the square of the minimal length. Such modified \ac{GUP} is usually encoded \cite{Kempf:1994su,Das:2008kaa,Ali:2010yn,Todorinov:2018arx,Bosso:2020aqm} into deformed Heisenberg commutators between space coordinates and spatial momenta, $[x,p]=i \hbar (1+\beta p^2+ \mathcal{O}(p^4))$, even though some phenomenological studies focus directly on the modification of the Heisenberg uncertainty relation \cite{Scardigli:2014qka,Lambiase:2017adh,Scardigli:2018jlm,Ong:2018zqn,Casadio:2020rsj}.
Covariant extensions of \ac{GUP} models were developed in \cite{Quesne:2006fs}, and were found to be related to one of the oldest non-commutative spacetime models, proposed by H.\ Snyder \cite{Snyder:1946qz, Quesne:2006fs, Ballesteros:2019mxi} (see also Sec.~\ref{section2.2.2.3})

\paragraph*{Phenomenological implications}

Given the above-mentioned relation of (at least some of) the \ac{GUP} models with spacetime non-commutativity, a consistent part of the \ac{GUP} phenomenology can be traced back to deformed kinematics (see e.g. \cite{Mignemi:2013aua, Pramanik:2013zy, Majhi:2013koa, Bosso:2018uus}).  Here we focus on  works that make explicit reference to \ac{GUP} and have direct relevance for multi-messenger astrophysics, even though in some cases we notice that the effects mentioned  are generic predictions of models with modified kinematics, discussed in more detail in  Sec.~\ref{section2.2.2.5} above.

\begin{description}
    \item {\textit{Implications for \ac{GW} detection}}: Explicit probes of \ac{GUP} can be found in opto-mechanical devices for the \ac{GW}  detection \cite{Chen:2013sgs,Girdhar:2020kfl}. In a different direction, an explicit probe of \ac{GUP} corrections at the black hole horizon with \ac{GW}s was investigated in Ref.~\cite{Addazi:2018uhd}. Another well-known proposal is to employ the \ac{GW}  bar detectors such as the \textit{Antenna Ultracriogenica Risonante per l'Indagine Gravitazionale Astronomica} (AURIGA) that can set an upper limit to possible modifications of the uncertainty principle \cite{Marin:2013pga,Marin:2014wja}. Using the modified kinematics implied by \ac{GUP}, in Refs.~\cite{Feng:2016tyt, Das:2021lrb}  the dimensionless \ac{GUP} parameters  were constrained using the  propagation of the observed \ac{GW}  signal. 

    \item {\textit{Implications for neutrino physics}}: One of the first suggestions regarding the experimental detection of the minimal length involves neutrino physics \cite{Mead:1964zz}. This suggestion has been revisited in Ref. \cite{Raghavan:2012sy}  in the context of time-energy \ac{GUP}. Neutrino oscillations have also been suggested as probes of the minimal length \cite{Sprenger:2010dg}.

    \item {\textit{Implications for cosmology}}: \ac{GUP} effects might influence the Planck era of the universe  \cite{Basilakos:2010vs} and  the post-inflation preheating \cite{Chemissany:2011nq}.  Ref. \cite{Djordjevic:2018mzs} gives a prediction for the lower limit of the density of matter at the beginning of the GUT phase and an approximate value for the \ac{GUP} parameter. In the limit of high temperatures, \ac{GUP} leads to a drastic reduction in the degrees of freedom, to varying speed of light as a function of energy, and to MDR \cite{KalyanaRama:2001xd}. Recently, it was investigated whether the \ac{GUP} scenarios are compatible with cosmological data and new cosmologically-motivated constraints on the \ac{GUP} parameter have been derived \cite{Giardino:2020myz}.

    \item {\textit{Implications for black holes}}: A fully general-relativistic description of \ac{GUP} is still not available but one can study the effective consequences of \ac{GUP} to the theory up to a certain order. A first such study has been pursued in the context of canonical \ac{QG} \cite{Bosso:2019ljf,Bosso:2020ztk,Blanchette:2020aca}. At odds with astrophysical observations, a  positive \ac{GUP} parameter permits the white dwarfs not to become arbitrarily large and, thus, their gravitational collapse into a black hole might not occur. This problem is solved by requiring a negative \ac{GUP} parameter  and simultaneously by maintaining a finite temperature at the end of Hawking evaporation \cite{Ong:2018zqn}. Thus, the small white dwarfs observations  support the choice  of negative \ac{GUP} parameter and suggest that physics becomes classical again at Planck scale (see also Refs.~\cite{Jizba:2009qf,Carr:2015nqa}).\ac{GUP} also induces quantum corrections to the Hawking temperature and to the Bekenstein entropy as well as influence the flux of Hawking radiation \cite{Scardigli:2016pjs,Alonso-Serrano:2018ycq}. Due to these modifications, \ac{GUP} critically affects the black hole evaporation  \cite{Adler:2001vs,Casadio:2013aua}, especially at the final stage causing the black hole to leave a remnant of Planck size. The latter provides a hint towards a possible resolution of the black hole information paradox \cite{Gim:2017rmn,Vagenas:2018pez}. 

\end{description}

\subsection{Conceptual issues}
\label{section2.3X}

In this subsection we discuss some important issues, which may help the reader appreciate further the difficulties that characterize the various approaches to \ac{QG}, not only at a formal level, but also at a conceptual one.

\subsubsection{Effective vs.\ non-effective field theory}
\label{section2.3.1}

The \ac{EFT} framework (see also Sec.~\ref{section2.2.1}) allows one to study the low-energy expansion of any relativistic  \ac{QFT} ~\cite{Weinberg:1995mt}. Although we still do not know which are the basic features of a quantum theory of gravity, there are different arguments leading to the conclusion that it will require to go beyond the framework of relativistic \ac{QFT} and possibly also that of \ac{EFT} (\cite{Oriti:2009zz,Crowther:2016hhh}; see also below).

A common ingredient to many theoretical approaches to quantum gravity is a minimal length~\cite{Hossenfelder:2012jw}, implemented in many cases through a non-commutativity of spacetime. The formulations of  \ac{QFT}  in a non-commutative spacetime~\cite{Szabo:2001kg} introduce, in general, a mixing of the infrared and ultraviolet regimes~\cite{Minwalla:1999px}, violating the decoupling which is a characteristic of any local  \ac{QFT} ~\cite{Appelquist:1974tg,Weinberg:1980wa,Ovrut:1979pk}. This is an argument pointing to the necessity to go beyond the \ac{EFT} framework.

The previous arguments about a minimal length, as well those about the creation and evaporation of virtual black holes as an effect of a quantum theory of gravity~\cite{Kallosh:1995hi}, point towards possible modifications of the space-time symmetries that could affect the kinematics of processes at high energies as one of the possible traces of gravitational quantum effects.

Along this line, a first possibility is the presence of a small \ac{LIV}, whose inclusion in an \ac{EFT}~\cite{Colladay:1998fq} presents difficulties~\cite{Collins:2004bp,Polchinski:2011za,Cortes:2016fye}  to make  the effects of quantum corrections compatible with the very stringent constraints on this violation from different experiments~\cite{Kostelecky:2008ts}. This is another hint that it could be necessary to go beyond the \ac{EFT} framework to study the low-energy limit of a quantum gravity theory if it includes a modification of special relativity.

A second possibility is that Lorentz invariance is modified, but not broken.  
\ac{DSR} models emerged as a deformation of special relativity kinematics where the existence of a minimal length can be made compatible with relativistic invariance (see Sec.~\ref{section2.2.2}). The deformation is indeed characterized by the presence of an energy scale (which would correspond to the inverse of such minimum length in an appropriate quantum model) which modifies the standard energy-momentum composition laws to a non-linear composition dependent on this energy scale as a necessary ingredient to make the deformation compatible with the relativity principle. The transformations generated by the new non-linear total momentum of the system $\mathcal{P} = \bigoplus_I p_I$ correspond then to momentum-dependent translations, $x_I\to x_I+f({p})$, so that a local interaction for one observer is seen as non-local for a \ac{DSR}-translated observer, and absolute locality is lost~\cite{AmelinoCamelia:2011bm}. These two features (modified energy-momentum composition laws and a notion of locality which is observer-dependent) are two features of quantum gravity models which cannot be captured by the \ac{EFT} framework.

As a byproduct of the necessity to go beyond the \ac{EFT} framework to include quantum gravity effects, it is interesting to point out that, if the energy scale that limits the domain of validity of local  \ac{QFT}  is many orders of magnitude lower than the Planck energy scale (see Sec.~\ref{section2.3.4}), one has a new perspective of different fine-tuning problems (separation of Planck and Fermi scales, smallness of the cosmological constant) related to the assumption that the \ac{EFT} framework can be extended up to the Planck scale.

\subsubsection{Deforming versus breaking Lorentz symmetry}
\label{section2.3.2}

The requirements of \ac{DSR} (see Sec.~\ref{section2.2.2.1}) are that the laws of physics involve both a fundamental velocity scale $c$ and a fundamental inverse-momentum scale $\ell_\text{DSR}$ (or a corresponding length scale) as relativistic invariant, so that each inertial observer can establish the same measurement procedure to determine the value of $\ell_\text{DSR}$ (besides the invariant measurement procedure to establish the value of the speed of light). 
Conversely, in \ac{LIV} theories a fundamental scale $\ell_\text{LIV}$ is introduced as a scale at which the covariance under Lorentz transformations of the laws of physics involving $\ell_\text{LIV}$ is lost. 
From an algebraic perspective (see Secs.~\ref{section2.2.2.1} and Sec.~\ref{section2.2.2.2}), a breaking of the Lorentz symmetry corresponds to a loss of invariance of the theory under the Poincaré group, while in the case of a deformation, one still has an invariance under translations and Lorentz transformations if the Poincaré algebra is replaced by a nontrivial Poincaré Hopf algebra.

In the \ac{LIV} framework the physical definition of the fundamental scale changes under (Lorentz) boost transformations. Observers connected by (Lorentz) boost transformations would describe different laws of physics, which implies the existence of a sort of ``quantum gravity aether'', i.e.\ a preferred reference frame in which the laws of physics take a specific form.
As an illustrative example, let us consider the case in which the scale of deformation/breaking $\ell$ 
(shorthand for $\ell_\text{DSR}$ or $\ell_\text{LIV}$ depending on the framework)
modifies the form of the dispersion relation as (let us work at leading order in $\ell$)
\begin{equation}
E^{2}-\vec{p\/}^{2}c^{2}+\ell E\vec{p\/}^{2}c-m^{2}c^{4} =0\ .
\label{DSRDispersion}
\end{equation}
Each inertial observer can establish the value of $\ell$ (the same
value for all inertial observers) by determining the dispersion relation
(\ref{DSRDispersion}) in its reference frame.
It is evident that under a Lorentz transformation Eq.~(\ref{DSRDispersion}) changes its form. 
Suppose that Eq.~(\ref{DSRDispersion}) represents the particle's dispersion relation for an observer Alice, so that in her coordinatization she describes the dispersion relation \[
 E_\text{A}^{2} - \vec{p\/}_\text{A}^{2}c^{2} + \ell E_\text{A}\vec{p\/}_\text{A}^{2}c - m^{2}c^{4}=0.
\] For an observer Bob, connected by infinitesimal (undeformed) Lorentz boost with rapidity $\vec{\xi}$, the dispersion law takes the form 
\[
E_\text{B}^{2} - \vec{p\/}_\text{B}^{2}c^2 + \ell E_\text{B}\vec{p\/}_\text{B}^{2}c - \ell\vec{\xi}\cdot\vec{p}_\text{B}\vec{p\/}_\text{B}^{2}c^2 -  2\ell\vec{\xi}\cdot\vec{p}E_\text{B}^{2} - m^{2}c^4 = 0.
\] If Bob adopted the same operative procedure used by Alice to measure the value of the fundamental scale $\ell$, he would 
obtain a 
different value of the scale (different from $\ell$). 
This is the perspective of \ac{LIV} approach: the law~(\ref{DSRDispersion}) holds only for a preferred (in this case Alice's) reference frame.

As already manifest from this simple example, the laws of physics,
in presence of a fundamental inverse-momentum scale $\ell$, cannot
be invariant under ordinary Lorentz transformations. 
The assumption of \ac{DSR} theories is that the principle of relativity
is not violated, even in presence of an observer-invariant inverse-momentum
scale $\ell_\text{DSR}$, so that for example the dispersion relation
(\ref{DSRDispersion})
must have the same form for all inertial observers. It is clear then from
the above discussion that in \ac{DSR} theories the laws of transformation
between inertial observers must be modified respect to the special
relativistic ones. 
This is achieved through a $\ell_\text{DSR}$-deformation of special relativity transformation laws and composition of momenta, and thus of the energy-momenta conservation law (see Sec.~\ref{section2.2.2}), which makes a radical difference between the two scenarios at the kinematic level.

The distinct conceptual approach of the two scenarios implies in some cases significant phenomenological differences~\cite{AmelinoCamelia:2008qg,Mattingly:2005re}.
A crucial distinction is due to the full relativistic framework of \ac{DSR}: effects of departure from Lorentz symmetry which are not compatible with a relativistic scenario, are indeed forbidden in \ac{DSR} theories. On the contrary, this kind of effects are allowed in \ac{LIV} scenarios, and tend to produce much more virulent phenomenological consequences than the effects allowed in both \ac{DSR} and \ac{LIV} approaches.

For example, particle processes which are forbidden by kinematics in ordinary special relativity, can be allowed, above a certain threshold energy, in a preferred frame scenario (see Sec. \ref{section7.5}), and lead to observable effects~\cite{Kifune:1999ex,Aloisio:2000cm,AmelinoCamelia:2000zs,AmelinoCamelia:2000qz,Protheroe:2000hp}. 
The situation is different for \ac{DSR} theories. 
If a process is forbidden in standard special relativity, it must be forbidden also for low-energy particles in \ac{DSR}: since in a relativistic theory a particle which is low-energy for one (local) observer Alice has different higher energy for another relatively-boosted observer Bob (also local to the particle) it must then be the case that the process is forbidden for particles of any energy, given that, evidently, the laws that establish whether or not a process can happen must be observer independent.
We are therefore assured that there cannot be any anomalous thresholds
in \ac{DSR} theories, since they do not admit preferred frames~\cite{AmelinoCamelia:2002dx,Heyman:2003hs,AmelinoCamelia:2008qg,AmelinoCamelia:2012zza}. 

Let us exhibit as an example the case, forbidden in standard
special relativity (and thus also in \ac{DSR}), of a photon decay into a positron and an electron, $\gamma\rightarrow e^{+}+e^{-}$, particularly useful \cite{Jacobson:2001tu,AmelinoCamelia:2001qf,Bertolami:2003yi,Konopka:2002tt} in setting limits on some schemes for departures from Lorentz symmetry.
We assume for both the \ac{LIV} and \ac{DSR} scenarios the dispersion relation~(\ref{DSRDispersion}), $\ell$ standing for $\ell_\text{LIV}$ or $\ell_\text{DSR}$ accordingly. The
conservation laws of momenta are respectively
\begin{align}
\begin{aligned}
    E_{\gamma} &= E_{+}+E_{-}, \\ \vec{p}_{\gamma} &= \vec{p}_{+}+\vec{p}_{-}
\end{aligned} &\quad \text{(LIV),}
&
\begin{aligned}
E_{\gamma} &= E_{+}+E_{-}, \\
\vec{p}_{\gamma} & =\vec{p}_{+} + \vec{p}_{-} - \ell_\text{DSR} E_{+}\vec{p}_{-}
\end{aligned}
&\quad \text{(DSR).}
\label{PhenConservation}
\end{align}
Combining Eq.~(\ref{PhenConservation}) with (\ref{DSRDispersion}) one obtains, for the \ac{LIV} case (denoting $\theta$ the angle between $\vec{p}_{+}$ and $\vec{p}_{-}$)
\begin{equation}
\cos\theta\simeq\frac{m^{2}+E_{+}E_{-}-\ell_\text{LIV}\left(E_{+}E_{-}^{2}+E_{-}E_{+}^{2
}\right)}{\sqrt{\left(E_{+}^{2}-m^{2}\right)\left(E_{-}^{2}-m^{2}\right)}}\ .
\label{eq:PhenDecayLSB}
\end{equation}
For $\ell_\text{LIV}\neq0$, there would be some high values of $E_{+},E_{-}$
(and correspondingly $E_{\gamma}$) for which a real $\theta$ could
solve (\ref{eq:PhenDecayLSB}). 
It must be noticed that the energies needed for this are ultra-high but below-Planckian: photon decay starts to be allowed already at scales roughly of order $(m{}^{2} \ell_\text{LIV}^{-1})^{1/3}$
(which indeed, for $m$ the electron mass and $\ell_\text{LIV}^{-1}$
roughly of order the Planck scale, is $\ll \ell_\text{LIV}^{-1}$
). 
If one performs the same analysis for the \ac{DSR} case, it turns out that the deformed conservation law~(\ref{PhenConservation}) combines with (\ref{DSRDispersion}) so that the formula for $\theta(E_+,E_-,m)$ is the same as in standard special relativity, so that photon decay is forbidden by kinematics (photon stability).

Some effects of departure from Lorentz symmetry can be contemplated in both \ac{LIV} and \ac{DSR} scenarios, like for instance in-vacuo dispersion (see Sec.~\ref{section7.1}) and spacetime fuzziness (see Sec.~\ref{section7.4}). The very different picture of spacetime emerging in the two scenarios must be carefully considered when comparing the effects~(see for instance \cite{Rosati:2015pga}).
Moreover, due to the milder modification of Lorentz symmetries characteristic of \ac{DSR} theories, the quantitative predictions derived within \iac{LIV} scenario tend in some cases to be larger (and thus more constraining).

\subsubsection{IR versus UV effects}
\label{section2.3.3}

In approaches such as \ac{EFT} \cite{Georgi:1993mps}, the \ac{UV}/\ac{IR} divide is clearly spelled out by a cut-off scale. However, non-perturbative effects can mix \ac{UV}  and \ac{IR}  modes in such a way that it is not possible, or at least useful, to disentangle them meaningfully.
Additionally, the \ac{UV}/\ac{IR}  divide might not proceed in accordance with canonical scaling dimensions of couplings in  \ac{QFT}.  Therefore, couplings which appear as free, so-called relevant parameters of the \ac{IR}  theory (e.g., the Standard Model) can actually be fixed uniquely by the underlying \ac{UV}  physics. Examples of this can be found, e.g., in \cite{Shaposhnikov:2009pv} and \cite{Eichhorn:2017ylw}, where is has been suggested that the values of the Higgs mass and the top quark mass are ultimately consequences of particular \ac{UV}  physics, in this case asymptotically safe quantum gravity.

An example of a phenomenon whose nature is not clearly \ac{UV}  or \ac{IR}  is dark energy. Although it affects \ac{IR}  (cosmological) scales, its origin might as well be in the \ac{UV} . In supergravity, dark energy is interpreted as a cosmological constant given by the vacuum energy of matter fields in the quantum theory. This is an instance where the cosmological constant problem is regarded as due to some unknown physics above a certain energy scale (in this case, the scale $M_{\rm SUSY}$ of supersymmetry breaking). However, there is also a possibility that the root of the problem essentially lies in the \ac{IR}. For example, if we do not want to abandon local perturbative  \ac{QFT} , we can revisit the untested assumption that it is valid at arbitrary large scales and introduce an \ac{IR}  cut-off of order of the Hubble horizon, in analogy with the notion that, in the presence of a black hole, ordinary  \ac{QFT}  overcounts the number of degrees of freedom within a given volume and ceases to be valid at large scales \cite{tHooft:1993dmi}. Similarly, in the presence of a cosmological horizon it turns out that \iac{UV}  cut-off must be related to an \ac{IR}  cutoff in such a way as to avoid fine tuning \cite{Cohen:1998zx}. However, on the one hand the \ac{IR}  cut-off is not sufficient to explain the observed value of $\Lambda$ and, on the other hand, some \ac{UV}  modification of physics (quantum gravity?) may still be needed to explain the fundamentals of the overcounting of degrees of freedom. In this and many other cases, it is therefore difficult, or perhaps artificial, to reduce solutions of the cosmological constant problem to purely \ac{UV}  or purely \ac{IR}  effects.
A fundamental quantum gravity origin (thus ``\ac{UV}'' from the point of view of \ac{EFT}) of large scale cosmological dynamics (thus ``\ac{IR}'' from the same perspective) is a general feature of emergent spacetime scenarios and of quantum gravity approaches where spacetime itself emerges as a collective notion. One example is group field theory condensate cosmology \cite{Gielen:2016dss,Oriti:2016acw,deCesare:2016rsf}, where an effective cosmological dynamics, including possible accelerated periods, emerges from the hydrodynamics of the underlying quantum gravity dynamics.

In some models of quantum gravity, the cosmological constant arises dynamically from a condensate of fermionic degrees of freedom \cite{Alexander:2008zf}. Although these degrees of freedom are UV, they combine into Cooper pairs with a long-range interaction, so that the overall effect is an \ac{IR}  phenomenon.

The \ac{UV}/\ac{IR}  dichotomy may not be obvious in scenarios when spacetime is endowed with certain symmetries. For example, a spacetime with discrete scale invariance would possess a self-similar (fractal) structure replicating across all scales, from the \ac{UV} up to the \ac{IR} , with observable consequences \cite{Calcagni:2017via}.

In string theory, due to the separation of the scales, i.e., string scale and the Planck scale, there is a connection between \ac{UV} and \ac{IR}  effects, in particular one can see this in non-commutative field theory \cite{Minwalla:1999px,Matusis:2000jf}, non-local field theory \cite{Buoninfante:2018gce}. Both, non-local field theories and non-commutative field theory have one common root; string field theory \cite{Witten:1985cc,Witten:1992qy,Sen:2004nf,deLacroix:2017lif}. Non-commutative field theory arises as a low energy limit of open string field theory discussed in the seminal papers by \cite{Witten:1985cc,Witten:1992qy,Seiberg:1999vs}.
Typically, in string field theory the interactions have non-local interaction vertices which tend to softened the UV interactions. Non-local field theories, such as $p$-adic strings \cite{Freund:1987kt,Brekke:1987ptq}, and its gravitational counterpart as infinite derivative theories of gravity \cite{Biswas:2005qr,Biswas:2011ar} have similar features, in relating the scale of non-locality to be emerged in the infrared as we increase the number of interaction vertices \cite{Buoninfante:2018gce}.
In fact, a blurred dichotomy between \ac{UV} and \ac{IR} physics is a rather general feature of both non-local and non-commutative field theory, regardless of any connection to string theory, and appears in models derived from or inspired by other quantum gravity formalisms~\cite{Sorkin:2007qi}. 

\subsubsection{The scale of quantum gravity effects}
\label{section2.3.4}

It is standard lore that \ac{QG}  effects should manifest themselves at the Planck scale. Nonetheless, \ac{QG} scenarios have recently been suggested where this does not need to be the only relevant scale for phenomenology.

Typical examples include string/brane theory (and its extension to cosmology -- see Sec.~\ref{section2.1.5}), with \ac{QG}  effects that may be manifest already at scales lower than the Planck (mass) scale $ M_\text{P}$ and emergent \ac{QG} scenarios (e.g.~\ac{GFT} or causal sets, see Sec.~\ref{section2.1.3} and~\ref{section2.1.7} respectively), where spacetime emerges as a collective phenomenon from non-spatiotemporal entities and fundamental constants and scales, including the Planck scale, turn out to be functions of more elementary parameters \cite{Oriti:2018tym, Singh:2021ake}.
Below we will briefly discuss the scales at which we expect \ac{QG}  effects, and comment on present and future
constraints on them.

\begin{description}
\item {\it String Theory:} The string length scale $l_{\rm s}$~\cite{Polchinski:1998rq,Polchinski:1998rr}, which plays the role of minimal length in spacetime, is experimentally constrained by the LHC center-of-mass energy. Cosmological bounds are typically not very stringent, because observables usually depend on a number of free parameters, such as expectation values of Calabi--Yau moduli.
Standard Model physics also constrains the mass scale of non-local infinite derivative theory of gravity~\cite{Biswas:2005qr,Biswas:2011ar,Biswas:2014yia}, some versions of which are close to string field theory.
These and other theories, however, can be constrained with astrophysical and cosmological measurements (e.g.\ \acp{GRB}, \ac{GW}
 speed
of propagation and luminosity distance) when they predict \iac{MDR}.
A notable example of such \acp{MDR} in string theory is the so-called {\it non-critical string theory}
framework~\cite{Martinec:2003ka} and its cosmological extension, where \ac{QG}  effects can appear at scales much lower than the Planck scale~\cite{Ellis:1998dw,Ellis:2005qa} (see Sec.~\ref{section2.1.5}).

\item {\it Large extra dimensions:} 
The scale of \ac{QG}  can be much lower than $ M_\text{P}$ if gravity is embedded in a more fundamental theory that includes $n$ (compact)
extra dimensions and/or D(irichlet)-brane worlds~\cite{ArkaniHamed:1998rs,Antoniadis:1998ig,Shiu:1998pa,Kakushadze:1998wp,Randall:1999ee}. In such models, gravity is 4-dimensional in the  \ac{IR} up to a (mass) scale which is proportional either to the inverse of the typical size $R$ of the extra dimensions or to the possible curvature $k$ along, say, one of the dimensions~\cite{Randall:1999ee,Giudice:2004mg,Illana:2020jpi}.
The large number of Kaluza--Klein excitations of the graviton above this mass scale implies then that gravity becomes strong at the fundamental (bulk) scale $\bar M_D\ll  M_\text{P}$.
In these models, the scale $M_D$ of the quantum gravity effects may take any value between TeV and $ M_\text{P}$. Above $ M_\text{P}$, we would enter in a trans-Planckian regime where all interactions are dominated by gravity \cite{Amati:1987wq,tHooft:1987vrq}.
In models with flat extra dimensions, the scale $R^{-1}$ of the first KK excitation and the scale $ M_D$ where gravity becomes strong are related by the volume $V_n$ of the compact space, $ M_\text{P}^2 = M_D^{2+n}\,  V_n$. As a consequence, low values of $M_D$ will require large volumes that, in turn, will bring very light KK gravitons. Constraints from astrophysics \cite{Hannestad:2003yd} and cosmology \cite{Hannestad:2001nq} on these gravitons imply then indirect bounds around $M_D > 100\,\mathrm{TeV}$. A nonzero curvature $k$, however, breaks the correlation between the two scales: it provides larger KK masses and couplings with matter, and for $k>50$ MeV these constraints are avoided. In that case, values $ M_D > 10$ TeV would be consistent with the direct bounds obtained from colliders \cite{Giudice:2004mg,Franceschini:2011wr}.

\item {\it Other QG scenarios with low-energy scale-predictions:} The back-of-the-envelope conclusion that \ac{QG} theories cannot leave signatures in late-time cosmology or \ac{GW} physics is based on the idea that perturbative
corrections are quadratic in the curvature and strongly subdominant at energy or curvature scales well above the Planck length $\ell_{\rm P}$.
However, such conclusions are evaded in \ac{QG} theories, like Ho\v rava gravity~\cite{Horava:2009uw}, where the scale of new physics in the gravitational sector does need to be lower (less than $10^{16}$\,GeV) and hence comparable with the scale of inflation. Another way to get observable effects at large scale is through nonperturbative mechanisms,
either through the generation of one or more effective intermediate scales $L$, as in \ac{LQC} (see Sec.~\ref{section2.1.2}) with anomaly cancellation~\cite{Bojowald:2011hd}, or via a long-range modification of spacetime geometry and dimensionality that cumulatively affects the propagation of \ac{GW}s on cosmological distances \cite{Calcagni:2019kzo}.
In both cases, what one constrains is not so much the scale of the effects, which is $O(H^{-1})$ by default, but, rather, the presence of certain features in cosmological observables (large-scale enhancement of the power spectrum in the first case, and deviations from the standard luminosity distance at redshifts $z\sim O(1)$ in the second case).
Experiments leading to constraints for the aforementioned \ac{QG}-induced non-local \acp{EFT} are quite different.
In particular, a constraint $\ell_\text{nl}<10^{-19}$\,m was found using LHC data at 8\,TeV \cite{Biswas:2014yia}. Constraints of order $\ell_\text{nl}<10^{-26}$\,m are expected in the forthcoming high precision optomechanical experiments~\cite{Belenchia:2015ake, Belenchia:2016zaa}. Possible extra lengths scales associated with the \ac{GUP} (see Sec.~\ref{section2.2.4}) are presently constrained  
from resonant gravitational bars~\cite{Marin:2013pga, Marin:2014wja} and 
from astrophysical observations~\cite{Giardino:2020myz}.
Non-local effects from a quantum gravity theory, e.g.\ as in the relative locality scenario 
(see Secs.~\ref{section2.2.2.1} and~\ref{section2.2.2.2}), may also require to go beyond the \ac{EFT} framework (see Sec.~\ref{section2.3.1}).
In particular, the relative locality scenario is necessarily associated with a deformation of the kinematics of special relativity based on modified energy-momentum composition laws, which on the one hand guarantees the compatibility
with the relativity principle, and on the other hand
involves a new energy scale, different in general from the Planck scale.
Such a scale can be constrained in future high energy particle accelerators and in the new windows to high energy multi-messenger astrophysics~\cite{Acciari:2020kpi,Carmona:2019oph,Carmona:2020whi,Aloisio:2000cm}, e.g.\ with the analysis of effects in resonances or the physics of the universe transparency to high-energy gamma rays, which put constraints on this new scale of O(TeV)~\cite{Albalate:2018kcf,Carmona:2020whi} (see also Secs.~\ref{section2.2.2.2},~\ref{section2.3.2} and ~\ref{section7.2}).

\item {\it Minimal length QG Models:} 
Apart from first-quantized string theory, discussed above,
there are many other \ac{QG} scenarios involving a minimal length, which could be larger than the Planck length,
for instance, Lorentz-symmetry breaking \ac{QG} scenarios, like Ho\v rava gravity~\cite{Horava:2009uw},
the low energy limit of string field theory~\cite{Eliezer:1989cr}, causal set theory~\cite{Sorkin:2007qi,Belenchia:2014fda},
\ac{GUP} scenarios with linear and quadratic terms in momentum \cite{Das:2008kaa,Das:2009hs,Ali:2009zq,Das:2010sj, Das:2010zf,Basilakos:2010vs},
and axiomatic schemes, based on the geometry of phase space~\cite{Hossenfelder:2012jw}
(for extensive reviews see e.g.~\cite{Garay:1994en,Hossenfelder:2012jw}).
The reconciliation of  the idea of minimal length/time
leads to new developments in our thinking about \ac{QG}~\cite{Freidel:2015uug, Freidel:2016pls, Freidel:2013zga, Freidel:2014qna, Freidel:2015pka,  Freidel:2017xsi,  Freidel:2018apz,   Freidel:2019jor, Minic:2020oho,  Freidel:2017wst, Freidel:2017nhg}, such as
covariant relative locality \cite{AmelinoCamelia:2011bm, AmelinoCamelia:2011pe},
which in this setting means ``observer dependent spacetimes''. The latter  appear
as sections of a new and larger concept  ---quantum spacetime with
Born geometry~\cite{Freidel:2013zga, Freidel:2014qna},
realized in a T-duality covariant, unitary and intrinsically non-commutative formulation of string theory (``metastring theory'', \cite{Freidel:2015pka, Freidel:2017xsi,Freidel:2018apz,Freidel:2019jor, Minic:2020oho}).
This approach also sheds new light on
the \ac{GUP}, and
leads to observable hallmarks, such as:
(i) the ``metaparticles'' (bi-local, non-commutative field quanta in the modular representation), endowed with a non-trivial propagator and an IR sensitive dispersion relation, that could be investigated via different multi-messenger probes,
(ii) their ``duals'',
associated with dark matter degrees of freedom \cite{Minic:2020oho}, and axion-like backgrounds, \cite{Freidel:2017wst, Freidel:2017nhg}, and
(iii) a ``dual'' geometry,
leading to dark energy in the observed spacetime, modelled as a positive cosmological constant \cite{Berglund:2019ctg, Berglund:2019yjq} and
consistent with observations \cite{Ho:2010ca, Edmonds:2017zhg}.
\end{description}

%\newpage
\section{Cosmic messengers}
\label{section3} 

Cosmic messengers are particles and waves emitted by astrophysical objects, carrying information about their sources and about the intergalactic and interstellar space from their sources to Earth.  They include gamma rays, neutrinos, cosmic rays (i.e.\ charged particles, mostly protons and other atomic nuclei) and \acp{GW}.  Each type of cosmic messengers has certain advantages for performing tests of \ac{QG}.

\subsection{Types of cosmic messengers}

\subsubsection{Gamma rays}
Gamma rays are the most energetic form of electromagnetic radiation. By convention, every photon with energy above 100\,keV ($\nu \approx 10^{19}$\,Hz) is considered to be a gamma ray. This significant chunk of the electromagnetic spectrum can be further subdivided into \ac{LE}, \ac{HE}, \ac{VHE}, \ac{UHE} and \ac{EHE} as:
\begin{align*}
    &\text{\ac{LE}},  & 100\,\mathrm{keV} < E &< 100\,\mathrm{MeV}, \\
    &\text{\ac{HE}},  & 100\,\mathrm{MeV} < E &< 100\,\mathrm{GeV}, \\
    &\text{\ac{VHE}}, & 100\,\mathrm{GeV} < E &< 100\,\mathrm{TeV}, \\
    &\text{\ac{UHE}\footnotemark}, & 100\,\mathrm{TeV} < E &< 100\,\mathrm{PeV},\ \text{and} \\
    &\text{\ac{EHE}}, &                     E &> 100\,\mathrm{PeV}. 
\end{align*}\footnotetext{{Note that the phrase ``ultra-high energy'' has two different meanings for gamma rays ($100\,\mathrm{TeV} < E < 100\,\mathrm{PeV}$) and for cosmic rays ($E \ge 1\,\mathrm{EeV}$).}}
Gamma rays provide a valuable probe of the largest energy transfers throughout much of the universe. 

They propagate on straight lines and their sources are relatively easily determined.
Moreover, sources of gamma rays are abundant, with thousands of sources visible in \acs{HE} and hundreds in \acs{VHE} bands. On the other hand, gamma rays interact with the so-called extragalactic background light (\ac{EBL}, see Sec.~\ref{section3.2}), the electromagnetic radiation in infrared (\ac{IR}), optical and ultraviolet (\ac{UV}) bands which permeates the universe. Through these interactions \acs{VHE} gamma rays are converted into electron-positron pairs, meaning that they have finite energy-dependent mean free paths. This results in a cosmic gamma-ray horizon at a redshift of about $z = 1$ at $E\sim 100\,\mathrm{GeV}$. Though this could be seen as a fallback for gamma rays when compared to messenger for which the universe is more transparent, such as neutrinos or \acp{GW}, it does provide in fact an additional powerful handle to test the phenomenology described in this review since it predicts possible changes in the opacity of the universe.
Moreover, gamma rays trace the emission of the charged particles in the cosmic sources, no matter if they are leptonic or hadronic in origin.

A distinctive feature of cosmic gamma rays with respect to the rest of cosmic messengers discussed in this report, is that their experimental detection is rather easier. Their large interaction cross sections and their rates facilitate their detection in the \ac{HE} regime by using moderate-sized particle detectors space-born missions, in the \ac{VHE} regime using arrays of Cherenkov Telescopes deployed on ground at moderate altitudes, and in the \ac{UHE} regime using high-altitude extended detector arrays, all of them much more compact and cheap than the detectors needed for the rest of messengers.

This has allowed the study of the extreme universe targeted in this report, to progress much faster in gamma rays that in the rest of messengers, enabling the detection, and the detailed study, of a few thousands of sources in the \ac{HE} regime, a few hundreds in the \ac{VHE} regime and a few tens in the \ac{UHE} regime already. This new window complements the observations in the rest of the electromagnetic spectrum (``Multiwavelength Astronomy") and has implied a crucial step forward for the understanding of the astrophysics of the extreme universe sources at the highest energies, that shall be complemented now with the one coming from the other high-energy messengers discussed in this review (``Multi-messenger Astronomy''), see Fig.~\ref{fig:fluxsummary}.

Furthermore, the current detection techniques for cosmic gamma rays have proven to be really efficient and easy to scale, and a next generation of instruments aiming at tenfold improvements in sensitivity and energy coverage is already in construction.

\subsubsection{Neutrinos}
Neutrinos are privileged cosmic messengers, offering a link between astrophysics and particle physics. Being electrically neutral and weakly interacting, they can reach us from cosmological distances unaffected by background fields and point to their source. Therefore, they provide a way of looking for signals of new physics, including those due to \ac{QG} corrections, in a complementary approach to high-energy gamma-rays and cosmic rays studies and reaching energies not accessible to human-made accelerators. 

\begin{figure}[t]
    \centering
    \includegraphics[width=0.8\textwidth]{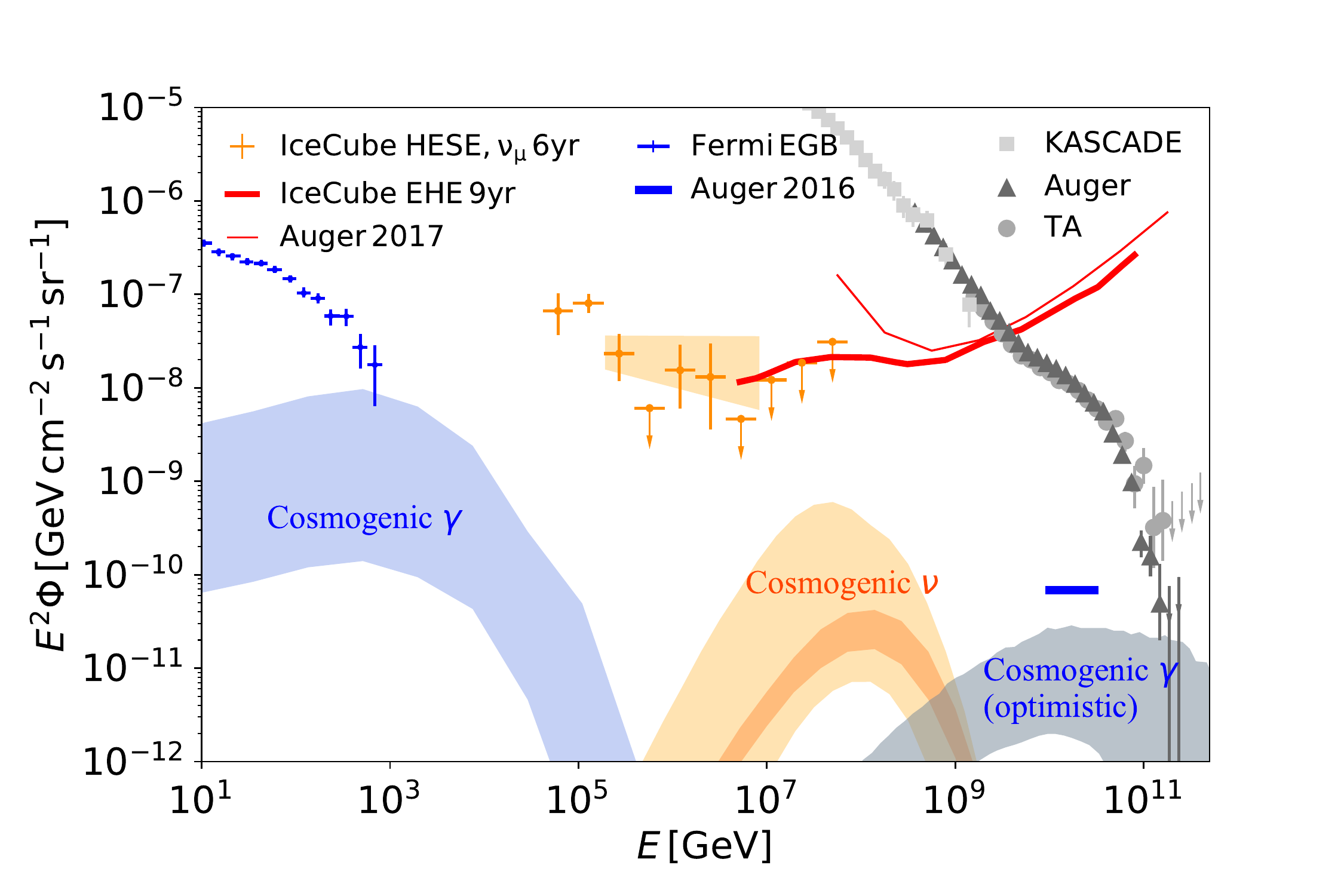}
    \caption{Compilation of measurements of the flux of various cosmic messengers. The diffuse extragalactic gamma-ray background measured by Fermi-LAT is represented by blue markers~\cite{Fermi-LAT:2014ryh, Fermi-LAT:2015otn}. High-energy starting events (HESE) detected by IceCube~\cite{Kopper:2017zzm} are shown as orange squares, and the muon-neutrino events~\cite{Aartsen:2016xlq} are indicated by the overlayed orange band. The grey triangles, squares, and circles correspond to cosmic-ray measurements by the Pierre Auger Observatory~\cite{Aab:2015bza}, KASCADE-Grande~\cite{Apel:2013uni}, and the Telescope Array~\cite{Abbasi:2018ygn}, respectively. The red lines represent the 90\% C.L. upper limits on the extreme-high-energy (\ac{EHE}) neutrino flux by IceCube~\cite{IceCube:2018fhm} (thick line) and Auger~\cite{Bellido:2017cgf} (thin line). The blue line corresponds to the differential flux limit by Auger for photons with energies between 10 and 30~EeV~\cite{Aab:2016agp}. The shaded bands represent the cosmogenic fluxes (i.e., those produced through interactions of UHECRs with cosmological photon fields) of photons (blue band) and neutrinos (orange bands) compatible with measurements by the Pierre Auger Observatory at a 90\% (lighter region) and 99\% (darker region)~\cite{AlvesBatista:2018zui}. The shaded grey band at the highest energies represent the flux of cosmogenic gamma rays expected from optimistic scenarios with a high fraction of UHE protons~\cite{Decerprit:2011qe}. Figure from~\cite{AlvesBatista:2019tlv}.}
    \label{fig:fluxsummary}
\end{figure}

In the \ac{SM}, neutrinos are included as massless left-handed fermionic fields that interact only through the weak force.  But the discovery of the flavor oscillation phenomenon and, consequently, the fact that they are massive particles, provided the first clear indication of the need to extend the \ac{SM}.  
The ``standard" paradigm, that assumes that the flavor eigenstates (in which neutrinos are created and detected) are coherent superpositions of massive neutrino states and the flavor is not conserved during propagation, is quite well settled. 
However, there are still important points to clarify in neutrino physics, like the neutrino mass ordering (normal  or inverted hierarchy), the absolute mass scale, the exact values of the oscillation parameters, the confirmation of leptonic \ac{CP} violation and the determination of the \ac{CP} phase. Moreover, there are even more fundamental questions to answer, like the discovery of the exact neutrino mass generation mechanism and the determination of the real neutrino nature (Dirac or Majorana fermion). The experiments and phenomenological analysis trying to solve these puzzles represent in most cases also an ideal playground to look for \ac{QG} effects, since these effects are typically proportional to the neutrino pathlength and its energy in the form of $L \times E$, instead to $L/E$ as in the standard oscillation scenario. An ideal context to search for \ac{QG} effects are therefore astrophysical neutrinos, since they involve large (cosmological) pathlengths and high energies. The breakthrough in this sector occurred in 2013, when the IceCube collaboration first announced the detection of a flux of high-energy neutrinos compatible with an astrophysical origin~\cite{Aartsen:2013bka}. This first indication of an astrophysical neutrino flux, two events in two years of lifetime with an energy of about 1\,PeV and incompatible with the atmospheric neutrino background at a significance of 2.8$\sigma$, was later unambiguously established with the detection of more than 100 additional events in about ten years of livetime using different data sets and analysis techniques~\cite{Abbasi:2020jmh,Aartsen:2016xlq,Stettner:2019tok}. Figure~\ref{fig:fluxsummary} shows the high-energy neutrino flux measured by IceCube from about 100\,TeV to a few PeV\@. Atmospheric neutrinos, not shown in the figure, dominate below a few tens of TeV, while the flux of astrophysical neutrinos takes over above $100\,\mathrm{TeV}$. The low statistics collected so far above a few hundred TeV does not allow for a precise characterization of the astrophysical flux. Whether the flux shows one component or more, or a cut-off, is still an open question that, for example has a bearing on the possibility to set bounds on the energy scale of \ac{LIV}.

The same weak interaction that makes neutrinos good cosmic messengers, makes them notoriously difficult to detect, needing huge volume detectors which are challenging to build (see Sec.~\ref{section4.2.2}). Neutrino telescopes have an intrinsically limited angular and energy resolution, which can induce relatively large uncertainties on establishing a source location, unlike the case with gamma rays. This drawback can be ameliorated by using multimessenger observations of transient sources, where signal timing and location is established with data from different telescopes.

\subsubsection{Ultra-high-energy cosmic rays}
\label{sec:uhecr}
Cosmic rays are high-energy charged particles (mostly protons and other atomic nuclei) of extraterrestrial origin.
Cosmic rays have been detected with energies up to a few hundred EeV ($1\,\mathrm{EeV} = 10^{18}\,\mathrm{eV} \approx 0.16\,\mathrm{J}$), see Fig.~\ref{fig:fluxsummary}.
Their energy spectrum roughly follows a power law~$dN/dE \propto E^{-\gamma}$ with~$\gamma\sim 3$ over ten orders of magnitude (from a few GeV to a few tens of EeV), after which it starts decreasing with energy much more steeply.
Cosmic rays with energies above~$1\,\mathrm{EeV}$ are known as \acp{UHECR}.  The flux of cosmic rays with energies above a certain threshold,~$E_{\min}$, is of the order of $10^2(E_{\min}/\mathrm{EeV})^{-2}$~particles per square kilometer per year, hence, unlike lower-energy cosmic rays ($E \lesssim 10^{15}\,\mathrm{eV}$), they are too rare to be directly detected by single calorimeters on balloons or spacecraft.  On the other hand, \acp{UHECR} entering the atmosphere generate
\acp{EAS} of particles scattered over large areas (Sec.~\ref{section4.1.2}),
which \ac{UHECR} experiments can detect using sparse arrays of particle detectors on the ground covering vast areas (Sec.~\ref{uhecr-detectors}).

Measurements of the \ac{UHECR} energy spectrum \cite{Aab:2020gxe,Aab:2020rhr,Ivanov:2020rqn} show several slight deviations from a simple power law, with the spectral index decreasing from~$\gamma \approx 3.2$ to $ \gamma \approx 2.6$ at an energy~$E \approx 10^{18.7}$\,eV (the so-called ``ankle''), increasing back to~$\gamma \approx 3.0$ at~$E \approx 10^{19.2}$\,eV (the so-called ``instep''), and finally further increasing to~$\gamma \approx 5$ at~$E \approx 10^{19.8}$\,eV (the ``cutoff''). 

Current estimates of \acp{UHECR} masses \cite{PierreAuger:2014gko,Bellido:2017cgf,Hanlon:2019onl,Yushkov:2019hoh} suggest a composition around the ankle dominated by protons and helium nuclei with a few tens of percent of heavier elements.  At higher energies, the composition appears to gradually become heavier and less mixed, though with progressively larger statistical uncertainties.  Only preliminary estimates exist in the cutoff region \cite{ToderoPeixoto:2020rta}, which appear to show that the increase in mass might be slowing down at the highest energies.  
These estimates generally require assumptions about properties of hadronic interactions in kinematic regimes where they are poorly known (see Sec.~\ref{section4.1.1}) and are affected by severe systematic uncertainties. A nearly model-independent estimated of whether the composition is pure or mixed, obtained from the correlation between two different observables \cite{Aab:2016htd,Yushkov:2020nhr} excludes any pure element and any proton--helium mixture around the ankle, requiring a mixture of both light and heavier elements, regardless of the hadronic interaction model. The results become consistent with a less mixed composition at higher energies.

Since \acp{UHECR} are electrically charged, their trajectories are deflected by the \ac{IGMF} as well as \ac{GMF}, so it is not straightforward to reconstruct the positions of their sources from their arrival directions (``cosmic-ray astronomy''), unlike for photons, neutrinos, and \acp{GW}.
On the other hand, the deflections of higher-energy cosmic rays are expected to be relatively small (typically a few tens of degrees for $E/Z=10$\,EeV, Sec.~\ref{magnetic}), so that the distribution of the arrival directions of \ac{UHECR} may retain some information about the location of their sources.
The observed distribution of \ac{UHECR} arrival directions is consistent with being isotropic at energies below 4\,EeV \cite{Abbasi:2016kgr,PierreAuger:2020fbi} and at higher energies it exhibits a large-scale modulation of amplitude~$d \approx 5(E/10\,\mathrm{EeV})\%$ \cite{Aab:2017tyv,Aab:2018mmi} towards a direction $\approx 120^\circ$ from the Galactic Center, whose statistical significance in the $[8~\mathrm{EeV}, +\infty)$~range has now reached~$6\sigma$ \cite{Aab:2020xgf}.
This suggests that the origin of \acp{UHECR} is predominantly extragalactic: a Galactic population of protons is predicted to result in much larger anisotropies \cite{Abreu:2012ybu,Tinyakov:2015qfz,Abbasi:2016kgr}, and all experiments agree that a large fraction of \acp{UHECR} are protons at least below the ankle energy.

At higher energies, anisotropies are expected to be stronger due to both the reduced magnetic deflections and the smaller number of contributing sources because of larger attenuation. However, the steeply decreasing number of events makes it statistically much harder to detect a given level of anisotropy.  
So far, no medium- or small-scale anisotropies have been conclusively~($\ge 5\sigma$) detected, but several indications have been reported (see Refs.~\cite{diMatteo:2020dlo,diMatteo:2021b6} and references therein).

The interpretation of \ac{UHECR} data is complicated by propagation effects (see Sec.~\ref{section3.3}), involving various poorly known quantities, as well as by major systematic uncertainties in the measurements.
Scenarios in which the observed cutoff is almost entirely due to propagation effects (the \ac{GZK} limit) are now disfavored \cite{PierreAuger:2016use}, but it is still unclear whether it is due almost entirely to the maximum injection rigidity of the source or both effects contribute.

\subsubsection{Gravitational waves}
\Acfp{GW} are transverse, propagating oscillations of the spacetime curvature, emitted, for example, by accelerated masses \cite{2009LRR....12....2S,2014LRR....17....2B}.  In four dimensions and in \acs{GR}, \acp{GW} only have two independent states of polarization, $+$ and $\times$, with corresponding amplitudes $h_+$ and $h_\times$.  In a plane normal to their direction of propagation, \acp{GW} cause a tidal deformation of a circular ring of test masses into an elliptical ring with the same area, with the deformations aligning along different axes for the two  polarizations. This effect is exploited in laser interferometer detectors, where suspended mirrors at the ends of orthogonal arms serve as test masses. For a realistic source, the polarization of the emitted \acp{GW}  depends on the orientation of the dynamics inside the source relative to the observer.

\acp{GW} can be of different types. Those produced by a non-spherically symmetric, single massive spinning object are called continuous \acp{GW}, as the signals are mainly sinusoidal with only small intrinsic variations in the amplitude and frequency over long times. Such signals are expected from clouds of ultralight bosons around black holes, or mountains on neutron stars, among other sources. A second type of signals are those from compact binary systems, when two massive compact objects revolve around each other and, in time, lose orbital energy and angular momentum through the emission of \acp{GW}, causing them to spiral together and eventually merge. Typical astrophysical binary system sources of \acp{GW} involve white dwarfs, neutron stars, and black holes with masses ranging from stellar scales (or smaller in the case of potential primordial black holes) all the way to supermassive black holes, whose masses are of the order of millions of solar masses or larger. Various possibilities for exotic compact objects also exist, for instance from dark matter or other \ac{BSM} fields in the early universe that condense over cosmic time, in addition to numerous proposed scenarios in which black holes differ from the standard classical \acs{GR} description. A third type of \acp{GW} come as bursts from sudden energetic events such as for example supernov\ae\, or black hole formations from highly eccentric binaries or cosmic strings. A fourth type of \acs{GW} signal is the \ac{SGWB}, the incoherent superposition of \acp{GW} from different directions. Its origin may be astrophysical due to unresolved individual sources such as binary systems contributing to the background, or cosmological from possible \acp{GW} generated in the early universe, for instance, during phase transition or inflation. To date, only \acp{GW} of compact binaries have been observed. 

The \acp{GW} of the \ac{SGWB} are characterized by a continuous spectrum on a wide range of frequencies. On the other hand, \acp{GW} from astrophysical individual sources are characterized by the waveform, which encodes detailed information about the source physics in the time-evolution of the phase and amplitude of the \acs{GW}. The phase evolution during the inspiral primarily encapsulates the intrinsic properties of the compact objects and their dynamics. Another interesting observable for these systems is luminosity distance, which can be determined either from the amplitude (\acs{GW} luminosity distance) or from the electromagnetic counterpart of the \acs{GW} event (photon luminosity distance). Sources of both \acp{GW} and light are called standard sirens, while those without light are referred to as dark sirens. The \acs{GW} amplitude also depends on a number of other parameters, for example the inclination angle between the orbital angular-momentum vector and the line of sight \cite{Sesana:2013dma}, and the antenna pattern of the detectors. Since the amplitudes of the two polarization modes have a different dependence on the inclination, information on polarization can remove degeneracies in our knowledge of the amplitude.

\Acp{GW} are the only type of messengers which are not detected as single particles. In this case, what is registered is their amplitude and phase, providing ample information about the distance of the event and the type of source. Even more scientific insights can be gained with \acp{GW} in a multimessenger context, as was the case with the detection of the wide range of electromagnetic counterparts to the GW170817 \ac{GW} event~\cite{Smartt:2017fuw}.

In principle, \acs{QG} effects can modify the production \cite{Yunes:2016jcc,Kobakhidze:2016cqh,Berti:2018cxi,Tahura:2018zuq,Maselli:2018fay,Giddings:2019ujs,Agullo:2020hxe} and the propagation \cite{Yunes:2016jcc,Ellis:2016rrr,Calcagni:2016zqv,Arzano:2016twc,Wang:2017igw,Calcagni:2019kzo,LISACosmologyWorkingGroup:2019mwx,Calcagni:2019ngc,Wang:2020pgu,Garcia-Chung:2020zyq} of \acp{GW}. The advantage of using \acp{GW} as a probe of \acs{QG} is that \acp{GW} are direct messengers of gravitational processes. The \acp{GW} from the inspirals, mergers, and ringdowns or postmergers of binary systems with different constituents and mass ratios explore a wide range of the nonlinear, strong-field dynamics of gravity, are sensitive to the detailed nature of the compact objects, and can involve phenomena such as resonances that may greatly enhance otherwise small effects. Furthermore, the stochastic \acs{GW} background is also sensitive to modifications of gravity~\cite{Mottola:2016mpl} and contains unique information on the very early universe, since the CMB is transparent to \acp{GW}. Among the challenges are that \acs{QG} can barely leave an imprint on the luminosity distance of late-time individual \acs{GW} sources, unless the theory predicts cumulative effects at large distances \cite{Calcagni:2019ngc}, and not many models of \acs{QG} predict a blue-tilted tensor spectrum at the frequencies of \acs{GW} interferometers \cite{Calcagni:2020tvw}. 

\subsection{Astrophysical origins}
\label{section3.2}

This brief description of astrophysical sources is restricted to sources interesting for tests of \ac{QG}.
These are \acp{AGN}, pulsars, \acp{GRB}, black hole mergers, etc. 
The features which makes an astrophysical source appealing for tests of \ac{QG} are:
\begin{itemize}
    \item brightness (easy to detect, with small uncertainties on fluxes);
    \item broadband emission (extended towards the highest energies);
    \item fast flux variability (especially useful for the time-of-flight method);
    \item well constrained emission scenario;
    \item large distance (amplifies the effect of time-of-flight differences)
\end{itemize}
It is not trivial to find a source in which all these characteristics are combined, but any of the mentioned features is significant for \ac{QG} phenomenology investigations. 

The mechanism by which so highly energetic particles can be produced is a subject of an ongoing debate. Two logical possibilities --- new physics phenomena at very high energy scale (e.g., decay of topological defects or superheavy particles) and acceleration of known particles to high energies --- are usually referred to as {\textit{top-down}} and {\textit{bottom-up}} scenarios. Many top-down mechanisms have been proposed (see Ref.~\cite{Bhattacharjee:1998qc} for a review), 
but they generally predict significant fluxes of high-energy photons and neutrinos, on which there are stringent limits as mentioned above, disfavoring such mechanisms as the dominant source of \acp{UHECR} except possibly at the highest energies~$\gtrsim 100~\mathrm{EeV}$.

\begin{figure}
    \centering%
    \includegraphics[width=0.45\textwidth]{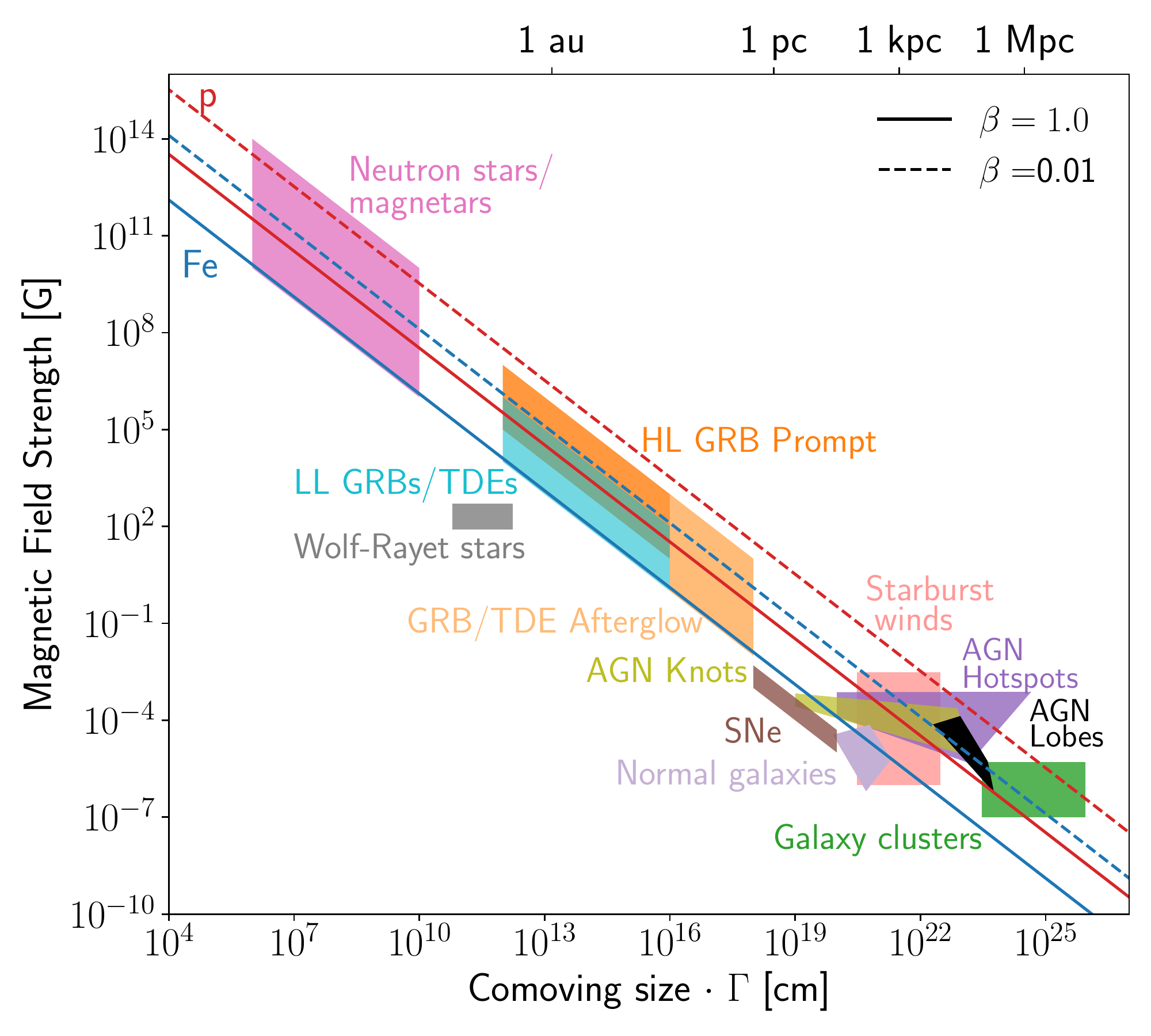}%
    \hspace{0.05\textwidth}%
    \includegraphics[width=0.45\textwidth]{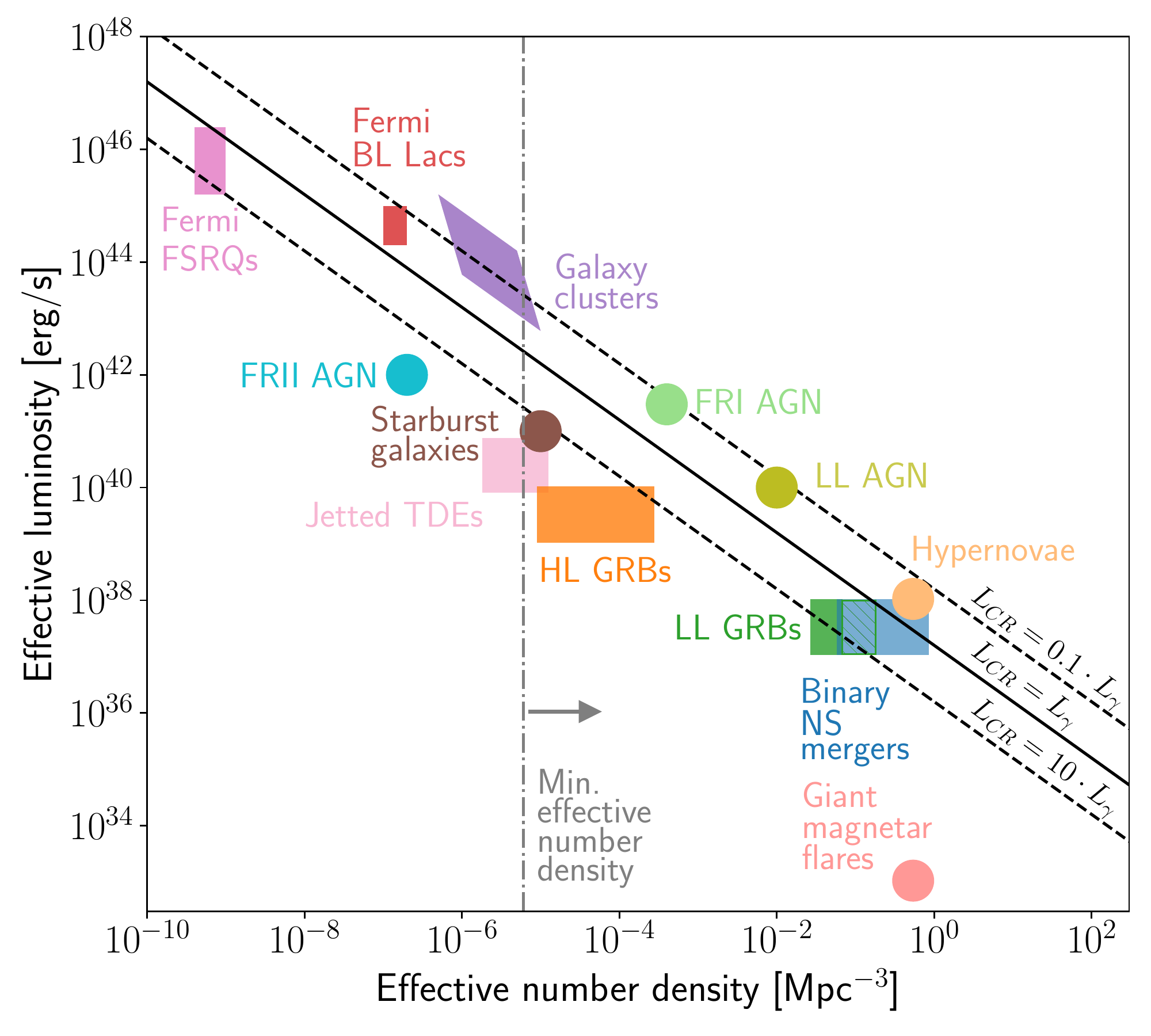}%
        \caption{Left: Hillas plot showing candidate \acp{UHECR} acceleration sites as colored areas on the size (horizontal axis) versus magnetic field (vertical axis) plane. The acceleration to $10^{20}$~eV is not possible below diagonal lines which are red for proton, blue for iron, solid for maximum shock velocity $\beta=1$ and dashed for $\beta = 0.01$. $\Gamma$ is the Lorentz factor of the acceleration region with respect to observer. Right:  luminosity of the candidate sources versus their space number density (steady sources), or effective luminosity versus effective number density (transient sources, assumed lifetime $3\times 10^5$~yr). The black solid line shows the \acp{UHECR} energy production rate of $5\times 10^{44}$~erg~Mpc$^{-3}$~yr$^{-1}$. The grey vertical line gives the lower limit on the \acp{UHECR} source number density.  Plots adopted from Ref.~\cite{AlvesBatista:2019tlv}}
    \label{figure5-sources}
\end{figure}

The other possibilities are the bottom-up scenarios, where particles are gradually accelerated to ultra-high energies. Only stable charged particles --- protons and heavier nuclei --- are relevant for \acp{UHECR}; being too light, electrons cannot reach ultra-high energies because of large synchrotron losses proportional to $1/m_e^4$. Most popular are stochastic acceleration mechanisms which are powered by shocks in magnetized media. Depending on the environment different types of acceleration can take place (see Ref.~\cite{Blandford:1987pw} for a review).
The maximum energy that can be achieved in these mechanisms is generically limited by a simple condition known as the ``Hillas condition'', which is essentially a requirement that the Larmor radius of an accelerated particle in the magnetic field $B$ fits inside the acceleration site. Accounting for a possible Lorentz factor $\Gamma$ of the accelerator with respect to the observer, this condition is written as~$E \lesssim e ZRB \Gamma$, where $R$ is the size of the acceleration site, $e$ electron charge, and $Z$ the atomic number of the nucleus. A modern version of the Hillas diagram from \cite{AlvesBatista:2019tlv} showing potential acceleration sites on the $R$--$B$ plane is presented in \autoref{figure5-sources}. 
Note that the Hillas criterion is a necessary but by no means sufficient condition. Other constraints, notably those resulting from energy losses, usually further limit the maximum achievable energy. Several candidate acceleration sites have been proposed:

\subsubsection{Active galactic nuclei}
\label{section3.1.1}
\Acp{AGN} represent a large population of objects characterized with luminous electromagnetic radiation produced in compact volumes. They are believed to be galaxies containing a massive accreting black hole in their center~\cite{Antonucci:1993sg}. 
There are several classes of active galaxies with substantially different characteristics~\cite{Urry:1995mg,Padovani:1997qt}. The \acs{AGN} classification schemes assume that the differences are basically due to the strongly anisotropic radiation patterns, meaning that the pointing directions are more important than the differences in intrinsic physical properties. \\
About 15\%~\cite{Padovani:2017zpf} of \acp{AGN} have relativistic jets, whose emission is strongly boosted. When pointing at us, these jetted sources are called blazars. Blazars are currently the largest class of gamma-ray emitters. The dramatically enhanced fluxes of the Doppler-boosted radiation, coupled with the fortuitous orientation of the jets towards the observer make these objects perfect laboratories to study the underlying physics of relativistic outflows through multi-wavelength and multi-messenger observations of temporal and spectral characteristics of emission. 
Blazars are further divided in two categories: BL Lacs, with weak or absent broad emission lines, and flat-spectrum radio quasars, with strong broad emission lines. 
The difference between the two flavors of blazars concerns the importance of the broad emission lines with respect to the underlying continuum, once the blazar nature is confirmed (strong radio emission with respect to the optical, strong X-ray emission, possibly strong gamma-ray emission). If the (rest frame, when the redshift is known) equivalent width of the lines is larger than~$5\,\text{\AA}$ the source is a flat-spectrum radio quasar, while for equivalent width less than~$5\,\text{\AA}$ the source is classified as a BL Lac~\cite{Urry:1995mg}.
Cosmic-ray acceleration could take place at various locations, for instance close to the black hole or within jets \cite{Rieger:2009kf}.
The broadband electromagnetic \ac{SED} produced by the jet of blazars presents two broad bumps, peaking in the \ac{IR} -- X-ray band and in the MeV-TeV band. Often fluxes in different bands vary in a coordinated way, suggesting that most of the \ac{SED} is produced by the same electrons in a specific zone of the jet.
The emission region must be compact to account for the observed very fast variability. The radio emission from this region is strongly self-absorbed, at all but the shortest radio wavelengths (sub-mm), therefore must be produced in other larger regions.
Electromagnetic radiation can be emitted by leptons (e.g.\ electrons and positrons), hadrons (mainly protons), or combination of both. 
If \acp{UHECR} are accelerated in the jets (protons as well as heavier nuclei, as for instance reported in \cite{Murase:2014foa}), the neutrinos produced in their interactions are expected to point directly to the source, and might help as a secondary messenger to trace the distribution of \ac{UHECR} sources \cite{Aartsen:2016lir,IceCube:2021pgw}. 

\begin{figure}
    \centering
    \includegraphics[width=12cm]{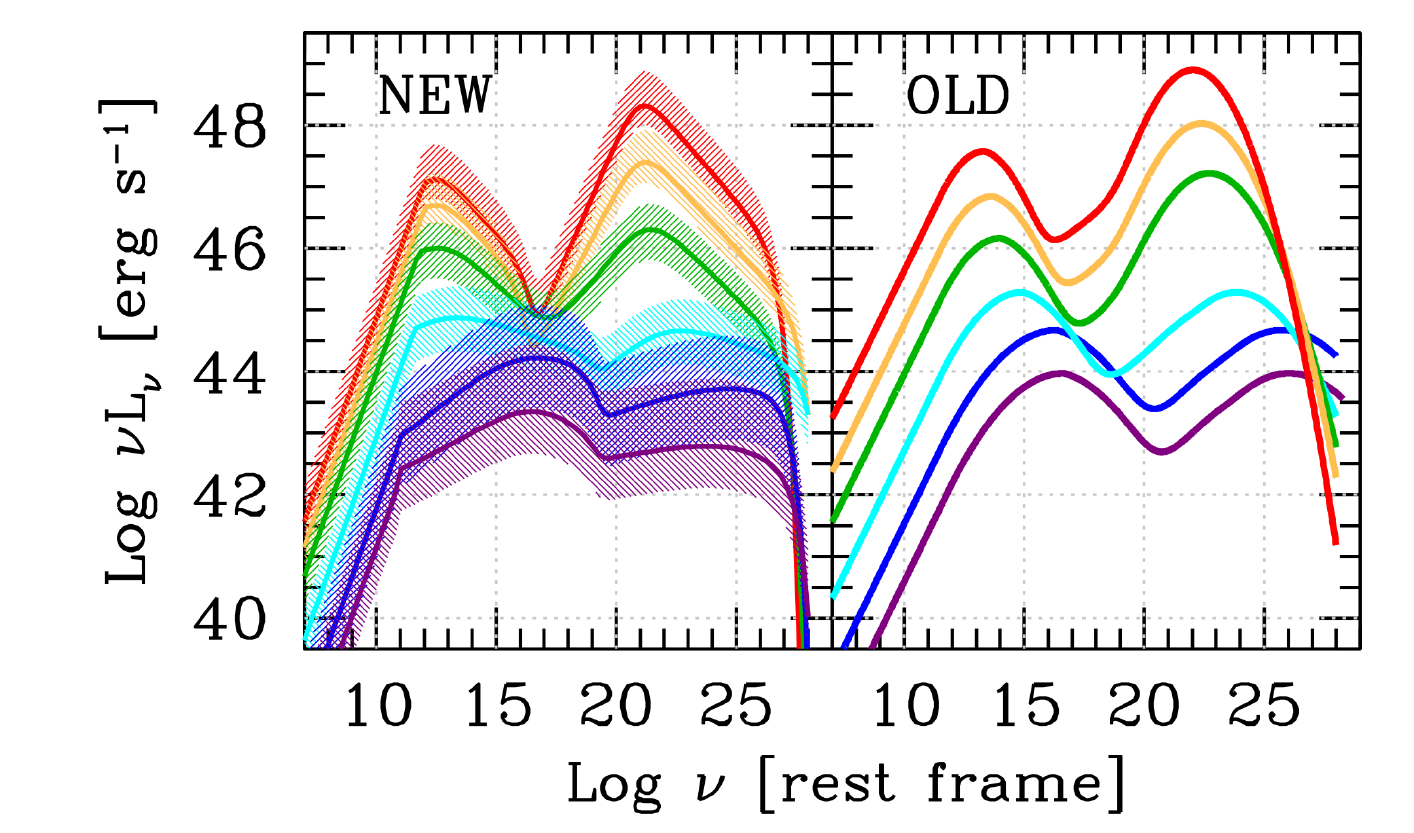}
    \caption{Comparison between the new and the original blazar sequence for all blazars. The original blazar sequence considered five radio luminosity bins, while the new one considers bins in the gamma-ray band. Reprinted with permission from~\cite{Ghisellini:2017ico}.
    }
    \label{figure3.1-blazar_sequence_ghisellini}
\end{figure}

The historical classification of blazars follows the so-called blazar sequence~\cite{Fossati:1998zn,Donato:2001ge}. 
In~\cite{Fossati:1998zn}, 126 objects were considered, belonging to different complete (flux limited) samples: one was X-ray selected and two were radio selected. 33 of those blazars (detected in the gamma-ray band by the EGRET instrument onboard the Compton Gamma Ray Observatory) were the brightest gamma-ray blazars at that time. After dividing the objects on the basis of their 5\,GHz radio luminosities, their fluxes were averaged~\cite{Fossati:1998zn} at selected frequencies, to construct the average \ac{SED} for blazars belonging to 5 radio luminosity bins. Average \ac{SED}s were improved in~\cite{Donato:2001ge} by adding average slopes of the X-ray emission for the same objects. The result is shown in the right panel of Fig.~\ref{figure3.1-blazar_sequence_ghisellini}. 

With the introduction of new telescopes, either in space or ground based, new blazars have been discovered as gamma-ray emitters. The blazar sequence has been then revisited  in the search of a more appropriate description of the blazar class according to the most recent data~\cite{Ghisellini:2017ico} (left panel of Fig.~\ref{figure3.1-blazar_sequence_ghisellini}). 
\acp{AGN} are very interesting for \acs{LIV} studies especially when detected in the gamma-ray band during enhanced states or flares. On those occasions they can produce \acs{VHE} gamma rays with harder spectra and possibly show fast flux variability, making them a very good tool for the study of \acs{LIV} effects with time-of-flight measurements. 

\subsubsection{Starburst galaxies}
\label{section3.1.2}
Starburst galaxies present a very high level of star formation activity \cite{Gao:2003qm}, mainly located in their cores. The rapid star forming activity, corresponding to an enhanced far-infrared luminosity, implies a high supernova rate. This might suggest that cosmic rays are efficiently produced in these sources. The winds of starburst galaxies have been also proposed as re-acceleration sites: particles escaping from the nuclear region with energies of $10^{15}$\,eV could be re-accelerated at the terminal shock of the galactic superwind generated by the starburst. This phenomenon could also ensure the survival of nuclei heavier than hydrogen, the wind being less dense than the starburst galaxy nucleus \cite{Anchordoqui:1999cu}. Analyses of correlations among the positions of extragalactic gamma-ray populations (as for instance \acp{AGN} and starburst galaxies) and the arrival directions of the \acp{UHECR} \cite{Aab:2018chp} can significantly improve our knowledge of \ac{UHECR} sources.

\subsubsection{Tidal disruption events}
\label{sec:tde}
Tidal disruption events have been proposed to be able to accelerate \acp{UHECR} \cite{Farrar:2008ex,AlvesBatista:2017shr,Zhang:2017hom,Guepin:2017abw,Biehl:2017hnb}. These phenomena take place when a star is torn apart by the strong gravitational force of a nearby supermassive black hole, implying some level of accretion.  Tidal disruption events with a high level of accretion should generate a relativistic jet \cite{Rees:1988bf}, which could be able to accelerate protons or nuclei to ultra-high energies. High-energy neutrinos could provide also in this case a window onto the acceleration mechanisms in these candidate sources \cite{Stein:2020xhk}.

\subsubsection{Galaxy clusters}
\label{section3.1.3a}
Possible acceleration sites also include clusters of galaxies \cite{Colafrancesco:1998us}, where the intergalactic medium is heated, or accretion shocks around galaxy clusters \cite{Kang:1995xw}. Clusters of galaxies are the largest gravitationally bound objects in the universe \cite{Voit:2004ah}. They can be considered as ``closed boxes'' that retain all their gaseous matter. Most of the cosmic rays produced in clusters of galaxies remain confined within the cluster and produce high-energy gamma rays and neutrinos~\cite{2016ApJ...828...37F, 2021MNRAS.507.1762H} by interactions with the intracluster baryonic gas. A multi-messenger approach would provide insights to probe if galaxy clusters are sources of \acp{UHECR}. 

\subsubsection{Gamma-ray bursts}
\label{section3.1.4}
\Acfp{GRB} are intense and short pulses of gamma rays lasting from a fraction of a second to several hundred seconds. They are followed by an afterglow --- lower-energy, long-lasting emission in the X-ray, optical and radio \cite{Piran:2004ba},
up to gamma rays~\cite{MAGIC:2019lau, MAGIC:2019irs, HESS:2021dbz}. 
They are transients events occurring at an average rate of a few per day throughout the universe~\cite{DAvanzo:2015kdp}, whose radiation, for a period of seconds, completely flood with their radiation an otherwise almost dark gamma-ray sky~\cite{Paciesas:1999tp}. 

The first \acs{GRB} was detected in 1967 by the Vela satellites. For the next 30 years, \acsp{GRB} were undetected at any wavelengths other than gamma-rays, which provided poor directional information and hence there were no direct clues about their site of origin.
Despite the mystery on their origin, the first observations suggested the existence of two classes of \acsp{GRB}, short and long, based on the duration of the prompt phase, with a boundary around 2\,s~\cite{Kouveliotou:1993yx}.
They are thought to be caused by unusually powerful supernovae (collapsars, long \acsp{GRB}) or as an effect of mergers of two neutron stars or a neutron star and a black hole (short \acsp{GRB}~\cite{Eichler:1989ve,Dai:2017dzh}).
In early 1997 the BeppoSAX satellite succeeded in detecting \acsp{GRB} in X-rays~\cite{Cota:1997cg},
which after a delay of some hours yielded sufficiently accurate positioning for large ground-based telescope follow-up observations and discovery of fading optical afterglows~\cite{vanParadijs:1997wr}.
These proved that they were at cosmological distances, comparable to those of the most distant galaxies and quasars known in the universe~\cite{Metzger:1997wp,Kulkarni:1998nd}.
Even at these extreme distances (up to several Gpc) they outshine galaxies and quasars by a very large factor, albeit briefly.

The bursts might be produced by the dissipation of the kinetic energy of a relativistic expanding fireball. The \ac{GRB} itself could be produced by internal dissipation within the flow while the afterglow via external shocks with the circumburst medium. The physical conditions in the dissipation region \cite{Waxman:1995vg} imply that protons may be Fermi-accelerated in this region to energies $> 10^{20}$\,eV \cite{Baerwald:2014zga}. Acceleration of nuclei is also studied as well as their survival probability \cite{Murase:2008mr,Biehl:2017zlw}, within several shock models. The associated neutrino flux compared to recent IceCube data might help in constraining the \ac{GRB}--\ac{UHECR} paradigm \cite{Aartsen:2017wea}. However, so far such association was not found putting strong upper limits on this model (current limit suggest that unless GRBs put 10 times more energy in UHECRs than they do in gamma-rays they cannot be the sources of UHECRs). Low-luminosity \acp{GRB} are also studied as possible \ac{UHECR} sources \cite{Murase:2006mm,Boncioli:2018lrv}.

The field of \acs{GRB} observations has recently seen significant breakthroughs with the detection of emission with \acp{IACT}, namely from GRB\,180720B~\cite{Arakawa:2019cfc} and  GRB\,190829A~\cite{2019ATel13052....1D} by the \ac{H.E.S.S.}, and GRB\,190114C~\cite{Acciari:2019dxz,Acciari:2019dbx} and GRB\,201216C~\cite{2020GCN.29075....1B} by the \ac{MAGIC}.

\subsubsection{Pulsars}
\label{section3.1.5}
Pulsars are rapidly rotating neutron stars, originally seen in the radio band~\cite{Hewish:1968bj}. They possess strong magnetic, electric and gravitational fields. Particles accelerated to high energies in the magnetospheres of pulsars can interact near the surface of the neutron star to produce gamma rays through curvature radiation, synchrotron radiation or inverse Compton scattering. 
The first pulsar detected in gamma rays, using a gamma-ray telescope on a balloon, was the Crab pulsar~\cite{1971NPhS..232...99B}. Vela was the second one detected~\cite{Thompson:1975dq} by the Second Small Astronomy Satellite (SAS-2). EGRET quickly expanded the number of gamma-ray pulsars~\cite{Thompson:2003tj}, currently counting around 200. Most of them have been discovered thanks to the \ac{LAT} instrument onboard the \textit{Fermi} satellite~\cite{Smith:2017ynv}.
Pulsars have been observed from radio to \acs{VHE} gamma rays so far. Due to their low spin periods (milliseconds to seconds), pulsars have highly variable light curves. Besides, their strong magnetic field strength (up to $10^{12}$\,G) can accelerate particles up to TeV energies.\par
Pulsed emissions from four pulsars have been detected by Cherenkov telescopes at the \acs{VHE} regime so far. The Crab Pulsar was the first detected one above 25\,GeV~\cite{Aliu:2008hc} by the \acs{MAGIC} telescopes, and the detection of the pulsed emission reports followed by \ac{VERITAS} above 100\,GeV~\cite{Aliu:2011zi}, 
and \acs{MAGIC} up to 1.5\,TeV~\cite{Ahnen:2015fda}. 
The second detected \acs{VHE} gamma-ray pulsar is Vela, whose pulsed emission up to 100~GeV was reported by \ac{H.E.S.S.}~\cite{Abdalla:2018vik}. 
In contrast to young Crab and Vela,
the third \acs{VHE} gamma-ray pulsar Geminga is middle-aged. The pulsed emission was reported by \acs{MAGIC}~\cite{Acciari:2020zvo}. The last detected  \acs{VHE} gamma-ray pulsar is PSR~B1706-44, with pulsed emission up to 100~GeV, was detected by \ac{H.E.S.S.}~\cite{Spir-Jacob:2020jth}. \par
Although pulsars are close-by sources, compared to \acp{AGN} and \acp{GRB}, their pulsed emission with short periods make them another good candidate for time-of-flight studies. Accumulation of more statistics in their lightcurves by increasing the observation time improves the sensitivity to \acs{LIV} time delays. The \acs{LIV} limits set by \acs{VERITAS} and \acs{MAGIC} using Crab pulsar data are comparable to the ones obtained using \acsp{AGN} and \acsp{GRB} data, especially for the quadratic case. Deeper observations and possible detection of pulsed emission above TeV energies would further improve the sensitivity. 
Further advancements could be made through discoveries of additional pulsars in \acs{VHE}  gamma rays, specifically of older and fast rotating neutron stars and millisecond pulsars, mainly because of their short spin periods and relatively large distances. In addition, unlike the young pulsars, emission from older ones is not affected by the gamma-ray background from their supernova remnants.

\subsubsection{Binaries of black holes and other compact objects}
\label{section3.1.6}
\paragraph{Black holes and their origin} While \ac{GR}
 reproduces Newtonian results in the weak-field limit, in the strong-field regime it predicts an enhanced gravity, as illustrated for example by the Oppenheimer--Volkoff equation of hydrostatic equilibrium in stars \cite{1939PhRv...55..374O}. An immediate consequence is the occurrence of compact objects, resulted from the uncontrolled gravitational collapse of massive stars. The simple and idealized model of pressureless fluid collapsing under its own gravitational pull under spherical symmetry leads to the Oppenheimer--Snyder collapse into a black hole, a curvature singularity surrounded by an event horizon \cite{PhysRev.56.455}. Such black hole solutions of the Einstein equation have been worked out under various symmetry assumptions and for vacuum or electrovacuum there are important uniqueness theorems leaving only a handful of them \cite{2012LRR....15....7C}.

Despite the rich physics involved in stellar evolution encompassing extreme
gravity, plasma physics, hydrodynamics, electromagnetic fields and
radiation, the ``no-hair theorem'' of \ac{GR}
 predicts that the final black hole product is always characterized by only a
few parameters: the mass, electric charge, and the angular momentum.
Gravitational radiation is responsible for carrying away all additional
information, hence in principle its detection could shed light to some of
the lost physics through the plethora of parameters characterizing a
gravitational waveform. 

The characteristics of black holes as end products of the stellar evolution
sensibly depend on the stellar evolution itself. In the early universe gas
clouds being larger and gravity stronger, more massive stars were born,
evolving at the end of their thermonuclear cycle into gargantuan objects, the
so-called super massive black holes, with masses from one hundred thousand solar masses to billion solar masses. They reside at the center of
galaxies and when during cosmological evolution such galaxies merged,
eventually so did their super massive black holes, after overcoming the infamous last parsec in
their separation \cite{2013degn.book.....M}. Their merger is among the primary targets of the
forthcoming space-born \ac{LISA} \ac{GW} observatory.

By contrast, the much lighter second generation stars evolved through
hypernova explosions or \acp{GRB}s into stellar-mass black holes, with
masses ranging between 5 and a few dozens solar masses. There is no generically agreed 
mechanism yet on how stellar evolution could have resulted in
the formation of black holes with masses ranging from
$10^2$ up to $10^5$ solar masses. They were also shown to exist,
both from analyzing the data of the Chandra COSMOS-Legacy survey \cite{2018MNRAS.478.2576M} and in a recent \ac{GW} detection \cite{LIGOScientific:2020iuh}. The prospective Einstein
Telescope and Cosmic Explorer laser interferometer systems will target \acp{GW} from
such intermediate mass black hole binaries.

Nevertheless there are two more mechanisms leading to the growth of black hole mass. The first one is accretion: whenever the black hole environment is rich in surrounding matter, accretion leads to a sensible mass increase. The second mass-increasing process is the merger of black holes. The alternating sequence of these events, known as a merger-tree could lead to the observed distribution of black hole masses in the universe \cite{2015ApJ...799..178K,2020ApJ...895...95P}. Both processes also spin up the resultant black hole, due to the orbital angular momentum transport into the final product. Typical black holes should spin, hence in \ac{GR}
 they ought to be Kerr black holes. By contrast, in certain modified gravity theories they would also have additional hair (simple examples include a scalar charge or a tidal charge). It is a challenge to understand therefore why the first 50 detections of \acp{GW} identified mostly nonspinning black hole mergers.

\paragraph{Black hole mergers} When two black holes orbit each other, a post-Newtonian (PN) expansion can be used to approximate their orbital evolution during the early inspiral phase at large separation. The orbit gradually shrinks due to gravitational radiation
escaping the system. General relativistic corrections affect the orbit at
the 1PN level (this is where the Schwarzschild perihelion shift appears).
The rotation of the black holes contributes starting from 1.5PN orders
through the spin-orbit coupling (fractional orders appear as the PN
parameter scales with the square of the orbital velocity). When the spins
and Newtonian orbital angular momentum are misaligned, precessional effects
occur \cite{PhysRevD.2.1428,PhysRevD.12.329,1979GReGr..11..149B}. From the
analysis of VLBI radio data of a binary spanning over 18 years the spin
precession of the dominant supermassive black hole was identified 
\cite{2014MNRAS.445.1370K}. Additional general relativistic and spin-spin corrections appear
at 2PN, nevertheless both the energy and the total angular momentum are
conserved up to this order  \cite{1993PhRvD..47.4183K,1995PhRvD..52..821K,2005PhRvD..71h4027W,2010PhRvD..81h4025G,2010PhRvD..82j4031G}. The PN approximation breaks down in the nonperturbative regimes close to the merger, which are explored with numerical relativity simulations. The final epoch of the merger is the ringdown signal from the final remnant black hole, when it radiates away the perturbations through quasi-normal-modes. 

Gravitational radiation dissipating away energy, momentum and angular
momentum only kicks in at the 2.5PN order \cite{PhysRev.136.B1224,2014LRR....17....2B}. This can drastically change the
direction of the dominant spin in binaries where the components have
significantly different masses \cite{2009ApJ...697.1621G}. This spin-flip may explain
the formation of X-shape radio galaxies \cite{2012RAA....12..127G}. Several higher-order
contributions to the gravitational radiation were also computed.

\paragraph{Neutron stars} Not all end products of stellar evolution are black holes. At lower masses
the contraction stops in a dense state of matter comparable with a giant
atom, a neutron star. Frozen-in magnetic fields generate a polar
magnetic structure, at the poles of which electromagnetic radiation can
escape. Whenever the magnetic symmetry axis and the rotation axis are
misaligned, this radiation swepts a cone periodically, appearing as a pulsar
signal. The imprint in the signal of the orbital decay due to gravitational
radiation leaving the binary system in the Hulse-Taylor pulsar (PSR B1913+16) has
been observed for 40 years, spectacularly confirming the prediction of
\ac{GR}
 \cite{2016ApJ...829...55W}. The network of pulsar signals
monitored nowadays is like a system of extremely well-tuned precision
clocks. A \ac{GW} sweeping through space would modify the
distances, hence affect the pulsar timing array, which hence has the
potential to detect \acp{GW} \cite{2010CQGra..27h4013H}.   

Neutron stars have a much richer inner structure than black holes, involving not only strong-field gravity but also extreme states of matter. This manifests in several of their global properties. For example, a characteristic dimensionless rotational mass quadrupolar parameter, which is 
$1$ for black holes, but lies in the range $w_{i}\in $ $\left( 2,14\right) $
for neutron stars \cite{1999ApJ...512..282L}, \cite{2013MNRAS.433.1903U}. More exotic compact objects
were also proposed, like gravastars with $w_{i}\in \left( -0.8,1\right) $ 
\cite{1997PhRvD..55.6081R} or boson stars with $w_{i}\in \left( 10,150\right) $ 
\cite{2016PhRvD..94f4015U}. The quadrupolar parameter affects the binary dynamics at 2PN orders
through the mass quadrupole - mass monopole (QM) contribution 
\cite{1998PhRvD..57.5287P,2003PhRvD..67b4020G,2008PhRvD..78d4021R,2010PhRvD..81l3016H,2011PhRvD..83h4051V,2013PhRvD..87d4009B}, essentially affecting the stability of spin configurations of the binaries with respective components \cite{Keresztes:2021mdl}. Likewise, a number of tidal effects associated with the excitation of isolated quasi-normal-modes dependent on the internal structure of the NS impact the dynamics and \acp{GW}. The most dramatic signatures of an object's interior structure occur at the merger, which differs depending on the progenitor properties.       

\subsubsection{Primordial black hole binaries}
\label{section3.1.7}
Several mechanisms could have led to the formation of primordial black holes in the early universe: the gravitational collapse of large inhomogeneities \cite{Zeldovich:1967lct,Hawking:1971ei,Carr:1974nx,1975Natur.253..251C,Polnarev:1986bi} produced during inflation~\cite{Dolgov:1992pu,Ivanov:1994pa} or phase transitions~\cite{Bianchi:2018ula}, scalar field fragmentation~\cite{Cotner:2016cvr,Cotner:2017tir,Cotner:2018vug}, vacuum bubbles~\cite{Deng:2017uwc,Kusenko:2020pcg} or the collapse of topological defects like cosmic strings~\cite{Polnarev:1988dh, Jenkins:2020ctp}.  They can constitute from a negligible fraction to the totality of dark matter.  Various astrophysical limits exist and are reviewed in~\cite{Carr:2016drx,Carr:2009jm,Carr:2020gox,Carr:2020xqk}.  Given their uncertainties, certain mass windows are  still allowed, including the stellar-mass range, and some of these limits may not apply to black holes with modified-GR thermodynamics~\cite{Capela:2012uk, Sebastiani:2013fsa, Hendi:2017fxp}. The first \ac{GW} detections have rekindled the interest for primordial black holes in this range~\cite{Bird:2016dcv,Clesse:2016vqa,Sasaki:2016jop,2016ApJ...823L..25K,Blinnikov:2016bxu}, for which the \ac{QCD} transition should introduce universal features in the primordial black hole mass function~\cite{Niemeyer:1997mt,Jedamzik:1996mr,Byrnes:2018clq,Carr:2019kxo}, which could be used to distinguish the origin of black holes~\cite{Carr:2019kxo}.   Primordial black holes  may be also linked to the baryogenesis~\cite{Garcia-Bellido:2019vlf,Dolgov:2008wu}, be the seeds of supermassive black holes~\cite{Clesse:2015wea} and galaxies~\cite{Carr:2018rid}, and would provide new ways to test the existence of particles like  WIMPs~\cite{Adamek:2019gns}.  
Observing them would thus have groundbreaking implications for fundamental physics.  \Acp{GW} and multi-messenger astronomy offer multiple ways to probe and constrain the existence of primordial black holes, as well as their formation scenarios:  

\begin{description}
\item{\textit{Subsolar black holes:}}  Detecting a black hole of mass below the Chandrasekhar mass would almost unambiguously point towards a primordial origin.  Subsolar searches have been carried out by LIGO/Virgo~\cite{Abbott:2018oah,Authors:2019qbw}. 
\item{\textit{Black holes in the neutron star mass range and low mass gap:}}  LIGO/Virgo observations have revealed the existence of compact objects in the mass-gap, between the highest mass of known neutron stars and the lowest mass of astrophysical black holes \cite{Abbott:2020uma,Abbott:2020khf}.  Black holes in the mass gap also form when neutron stars merge~\cite{Gupta:2019nwj}, contaminating a hypothetical primordial black hole population.  Multi-messenger astronomy will probe the origin of these objects, eventually revealing their primordial origin if no EM counterparts are detected~\cite{Unal:2020mts}.    
\item{\textit{Intermediate-mass black holes:}}  Above $60 M_\odot$, pair-instability should prevent black holes from forming from single stellar explosions.  Primordial black holes are not sensitive to this limit and may lead to mergers in this range, like GW190521~\cite{LIGOScientific:2020iuh,Abbott:2020mjq,Clesse:2020ghq,DeLuca:2020sae}.  Spin measurements could be used to distinguish them from secondary mergers in dense environments~\cite{Farmer:2019jed}.  Heavier primordial black hole binaries could also be detected with future ground-based and space-based detectors and be the seeds of the super massive black holes at the center of galaxies ~\cite{Clesse:2015wea,Carr:2019kxo}.    
\item{\textit{Black hole mergers at high redshift:}}  The third generation of \ac{GW} detectors will have an astrophysical reach of~$20 < 1+z < 100$, before star formation. Any black hole merger detection would thus point to a primordial origin.  
\item{\textit{Distinguishing primordial black holes vs stellar black holes with statistical methods:}}  Bayesian statistical methods and model selection~\cite{LIGOScientific:2018jsj} applied to  the rate, mass, spin and redshift distributions will help to distinguish PBHs from stellar scenarios~\cite{Kocsis:2017yty,Ali-Haimoud:2017rtz,Clesse:2017bsw,Fernandez:2019kyb,DeLuca:2019buf,Carr:2019kxo,Gow:2019pok,Jedamzik:2020omx,Jedamzik:2020ypm,Bhagwat:2020bzh,DeLuca:2020qqa,DeLuca:2020fpg,DeLuca:2020bjf,Dolghov:2020hjk,Dolgov:2020xzo,Belotsky:2014kca}. 
\item{\textit{Stochastic backgrounds}:}  If primordial black holes contribute to a non-negligible fraction of dark matter, their binaries generate a detectable \ac{SGWB}~\cite{Mandic:2016lcn,Clesse:2016ajp,Wang:2016ana,Wang:2019kaf}, whose spectral shape depends on the primordial black hole mass distribution and binary formation channel.   
\item{\textit{Continuous waves (CWs) from planetary-mass binaries}:}  Such primordial black holes would form binaries emitting CWs waves in the frequency range of detectors, years before they merge.  Projected constraints for a wide range of equal-mass primordial black hole binaries have been obtained for future gravitational-wave detectors, including Einstein Telescope, that show that searches for these long lived sources will be able to obtain meaningful limits on the fraction of dark matter that PBHs could compose, \cite{Miller:2020kmv}. Furthermore, some constraints using upper limits from the first half of LIGO/Virgo's third observing run have been placed, and are very close to becoming physically relevant for asymmetric mass ratio binary systems \cite{Miller:2021knj}. Present and future detectors will detect or set new limits on primordial black holes in the mass range $[10^{-8}-10^{-3}]M_\odot$.
\item{\textit{GW bursts from close encounters}:}  Another signal from primordial black holes comes from the \ac{GW} bursts from hyperbolic encounters in dense halos~\cite{Garcia-Bellido:2017qal,Garcia-Bellido:2017knh}.  The signal frequency can lie in the frequency range of ground-based detectors for stellar-mass black holes, with a duration of order of milliseconds. 
\end{description}

\subsubsection{Phase transitions}
\label{phase transitions}

\paragraph{First-order phase transitions and gravitational-wave production}
As the universe cools down, it goes through thermal phase transitions, linked to the spontaneous breaking of symmetries. 
A first-order phase transition is characterized by a barrier 
in the effective potential between the true and false vacua that are degenerate at the critical temperature $T_\text{c}$.
During a thermal first-order phase transition, bubbles of the true vacuum, typically characterized by a nonzero vacuum expectation value of a scalar field, indicating a spontaneously broken symmetry, nucleate in the false vacuum and expand. Upon collision, they generate a quadrupole moment in the cosmic fluid, setting off primordial \acp{GW}~\cite{Kosowsky:1991ua,Kosowsky:1992rz,Kosowsky:1992vn,Caprini:2007xq,Huber:2008hg}. The subsequent evolution of the cosmic fluid  features  magnetohydrodynamical turbulence \cite{Kamionkowski:1993fg,Caprini:2006jb} and sound waves, often  the dominant component of the stochastic \ac{GW} signal \cite{Hindmarsh:2013xza,Hindmarsh:2017gnf,Ellis:2019oqb}.

The spectra generated by phase transitions are characterized by a peak frequency, which is related to the inverse duration of the phase transition, and exhibit power-law-scaling at larger and smaller frequencies, leading to a peaked power spectrum. The calculation of such a spectrum involves two steps. First, key parameters of the phase transition, such as the cosmological temperature $T_{\ast}$ at the end of the phase transition (typically below $T_\text{c}$ due to the expansion of the universe), the energy released during the transition, the speed of the bubble walls and the inverse duration of the transition are determined from the finite-temperature effective potential of a given particle-physics model \cite{Quiros:1999jp}. Second, the \ac{GW} spectrum is determined from these parameters. While this can only be done by numerical simulations, phenomenological fits to the \ac{GW} spectra from simulations exist \cite{Caprini:2015zlo,Caprini:2019egz} which are typically used to calculate the spectrum in the effective potential in the considered microscopic model. Last, a comparison to the sensitivity curve of a given \ac{GW} spectrometer allows to calculate the signal-to-noise ratio. When the signal-to-noise ratio is large, the reconstruction of the signal is expected to be accurate~\cite{Croon:2018erz, Figueroa:2018xtu, Hashino:2018wee, Flauger:2020qyi}.

Depending on the specifics of the potential and the expansion rate, strong supercooling with $T_{\ast}\ll T_\text{c}$ can occur. In this regime, a large energy difference between the stable and metastable vacua  provides a larger amount of energy released during the phase transition, often leading to a stronger \ac{GW} signal; see  \cite{Ellis:2019oqb,Lewicki:2020jiv,Ellis:2020nnr} for recent studies and \cite{Cutting:2020nla} for a numerical simulation.

\paragraph{Phase transitions in the Standard Model}
Within the Standard Model, the electroweak phase transition would only be of first order for a much lower Higgs mass than the observed one. The phase transition in \ac{QCD} at which confinement and spontaneous chiral symmetry breaking occur, is a cross-over; see \cite{Addazi:2018ctp,Addazi:2019mlo} for an alternative scenario where \acp{GW} are sourced by domain wall formation in the \ac{QCD} vacuum. Accordingly, \acp{GW} from a first-order phase transition are thus commonly seen as a characteristic of \ac{BSM} physics throughout the literature.

\subsubsection{Other sources: Superradiance and environmental effects} 
\label{section3.1.8}
\Acp{GW} could also arise from superradiantly formed boson clouds around black holes~\cite{Arvanitaki:2009fg, Arvanitaki:2010sy, Arvanitaki:2014wva, Brito:2015oca, Arvanitaki:2016qwi} and binary systems that are being affected by their environments (e.g., dark matter spikes or ultralight bosonic clouds around the black holes)~\cite{Barausse:2014tra, Barausse:2014pra, Baumann:2018vus, Berti:2019wnn, Baumann:2019eav}. 
In particular, it has been hypothesized that ultralight boson ``clouds'' could form around rotating black holes by extracting rotational energy from the black hole through superradiance, a classical wave amplification process that has been studied for decades~\cite{Press:1972zz}. 

We could plausibly identify these clouds in the stochastic, continuous and binary inspiral channels using current~\cite{Arvanitaki:2009fg, Arvanitaki:2010sy, Arvanitaki:2014wva, Brito:2015oca, Arvanitaki:2016qwi, Brito:2017wnc, Brito:2017zvb, Isi:2018pzk, Palomba:2019vxe} and future~\cite{Audley:2017drz, Sathyaprakash:2019yqt, Baibhav:2019rsa, Sedda:2019uro} \ac{GW} detectors. 
Indeed, the superradiant instability would cause the black hole to spin down and form a cloud which could emit continuous, monochromatic \acp{GW}~\cite{Brito:2015oca}. 
The spin-induced signatures could be searched for using measurements of the black holes spins from binary coalescences~\cite{Brito:2017zvb, Ng:2019jsx, Fernandez:2019qbj} and the monochromatic \ac{GW} signatures using continuous wave \cite{Sun:2019mqb} and stochastic searches \cite{Tsukada:2018mbp}. 
Moreover, if a binary inspirals near the black hole/cloud system, the cloud could alter the orbit of the binary inspiral,
possibly causing gravitational drag, mass screening, accretion, and excitation of orbital resonances~\cite{Barausse:2014tra, Barausse:2014pra, Baumann:2018vus, Hannuksela:2018izj, Berti:2019wnn, Baumann:2019eav}, which would imprint signatures of the cloud on the \acp{GW} from the inspiral. 
Therefore, measurements of the black hole properties could be used to theoretically predict the shape of the cloud. 
Through extreme-mass ratio inspiral measurements, the shape of the cloud can also be directly and simultaneously measured. 
It was therefore suggested that extreme mass ratio inspirals may allow for a unique cross-verification to test the light boson hypothesis by matching the theoretical predictions with direct measurements of the cloud~\cite{Hannuksela:2018izj}.  

However, ultralight boson clouds are not the only objects that could modify a binary inspiral. 
It has been suggested that other environmental effects, in particular so-called ``dark matter spikes'', formed as a consequence of adiabatic growth of massive black holes in galactic centers~\cite{Gondolo:1999ef}, could cause the orbits of extreme mass ratio inspirals to be altered due to mass screening, dynamical friction, and dark matter accretion by the smaller, inspiraling body~\cite{Eda:2013gg, Macedo:2013qea, Eda:2014kra, Yue:2017iwc, Barack:2018yly}. 
If these effects are embedded in the \ac{GW} signal, we could reconstruct the dark matter spike profile. 
However, let us note that it is not yet clear how much various uncertain astrophysical processes may play a role in hampering the reconstruction, or if dark matter spikes would form with sufficient density to be detected~\cite{Ullio:2001fb, Merritt:2002vj, Merritt:2003qk, Gnedin:2004cx, Bertone:2005hw, Pato:2015dua}.
Typically, other environmental effects are too small to significantly alter the inspiral~\cite{Barausse:2014pra, Barausse:2014tra}.\footnote{See, however, research on the possible effect of the accretion disk on the inspiral~\cite{Yunes:2010sm}. } 

If these dark matter spikes exist and are made of the \ac{QCD} axion, inspiraling neutron stars into the spikes would trigger radio emission whose profile could be predicted from the reconstructed dark matter spike profile, allowing for a unique opportunity to detect \ac{QCD} axions~\cite{Edwards:2019tzf}. 
Meanwhile, numerical and analytical  studies of  the  mergers of black hole/spike systems indicate that the distribution of DM around them can dramatically affect the evolution of binaries~\cite{Kavanagh:2018ggo, Bertone:2019irm}.  
Furthermore, it has been suggested that these spikes are incompatible with very light bosonic and fermionic DM and with self-annihilating DM, and thus their observation may hint at the nature of the DM particle~\cite{Hannuksela:2019vip}. 

Indeed, if either ultralight boson clouds or dark matter spikes were identified, they would have the potential to reshape our understanding of fundamental physics and dark matter. 

\subsection{Propagation}
\label{section3.3}

Before reaching the Earth's atmosphere, cosmic messengers from astrophysical sources travel through Galactic and intergalactic spaces, interacting with matter, radiation, and magnetic fields. This can lead to changes in the energy of the particles and even their nature. As a consequence, propagation effects have considerable impact in the observables measured by observatories at Earth, such as the energy spectrum, particle type, and the angular and temporal distribution of the arriving particles.  

All messengers undergo adiabatic energy losses (redshift) due to the expansion of the universe.
Gamma rays and cosmic rays can interact with low-energy photon backgrounds, most notably the \ac{CMB} with typical photon energies~$\varepsilon \sim 1$\,meV, the infra\-red/visible/ultra\-violet \ac{EBL} with~$\varepsilon \sim 1$\,eV, and the \ac{CRB} with~$\varepsilon \sim 1 \, \text{neV}$. Electrically charged particles such as cosmic rays and charged leptons are deflected by \acp{IGMF} and \acp{GMF}, so that their arrival directions do not directly correspond to the position of their sources, and their propagation takes longer than in the rectilinear case. Conversely, the other messengers considered here follow geodesics in spacetime, which are at first approximation straight lines, but whose trajectories can be affected by the presence of massive objects via gravitational lensing. Additionally, neutrinos undergo flavor oscillations, and the flavor ratio detected at the Earth has been averaged over the long distances traveled, and does not reflect the flavor ratio at the source. Thus, many searches for \ac{BSM} physics that depend on the measurement of the relative neutrino flavors at the Earth must depend on assumptions on the production mechanism at the source. The naive assumption that neutrinos are produced from the decays of pions and kaons created in proton interactions in the source gives a flavor ratio of $1:2:0$ for $\nu_e : \nu_\mu : \nu_\tau$ at the production site which, after vacuum oscillations over cosmological distances, leads to an expected $1:1:1$ ratio at the Earth. However the complex, and unknown in most cases, inner structure of astrophysical sources of neutrinos can lead to energy losses of pions and muons before they decay (in the presence of strong magnetic fields or high matter densities for example) and the flavor ratio can then deviate from the expected $1:2:0$ and be energy-dependent. Neutron-rich sources would also lead to a deviation from the $1:2:0$ ratio at origin, and therefore at Earth. 

The energy spectra, composition (which includes the masses of cosmic rays, the flavors of neutrinos and the polarization of gamma rays and \acp{GW}),
and arrival directions of cosmic messengers detected at Earth thus depend both on intrinsic properties of their sources and on phenomena they undergo during their propagation. It is therefore essential to study these effects in detail.  Several simulation codes have been developed for this purpose, for example \cite{Allard:2005ha,Hooper:2006tn,Aloisio:2008pp,Aloisio:2010he,Kalashev:2014xna,Heinze:2019jou,Domenico:2013daa,Batista:2016yrx,Aloisio:2017iyh} for~\acp{UHECR} and \cite{Batista:2016yrx, Blytt:2019xad} for gamma rays.

Some of the quantities relevant to the propagation of cosmic messengers are poorly known, for example the spectral energy density of the \ac{EBL} and \ac{CRB} at certain redshift and wavelengths, the cross sections of certain channels of photodisintegration of atomic nuclei~\cite{Batista:2015mea, AlvesBatista:2019rhs}, and the \acp{IGMF} and \acp{GMF}~\cite{AlvesBatista:2017vob, Batista:2021rgm}. This introduces non-negligible uncertainties in the reconstruction of the source properties from the observations.

\subsubsection{Adiabatic losses due to the expansion of the universe}
\label{redshift}
All particles traveling over cosmological distances undergo adiabatic energy losses due to the expansion of the universe, given by \begin{equation}
    \frac{1}{E} \frac{dE}{dt} = -H(z), 
    \label{eq:adloss}
\end{equation} 
where $H(z)$ is the Hubble rate at redshift $z$.
In the case of messengers traveling rectilinearly and reaching Earth with no other energy losses, this means that the quantity~$(1+z)E$ stays constant during the propagation, i.e.\ their energy at Earth is~$E_\text{final} = E_\text{initial}/(1+z_\text{source})$. At redshifts $z\ll1000$ radiation can be neglected and one has $H(z)=H_0\sqrt{(1+z)^{3}\Omega_{\mathrm{m}}+\Omega_{\Lambda}}$ where the measured values for today's Hubble rate $H_0$ range within an interval $[67,74]$km/s/Mpc, depending on whether they are determined from \ac{CMB}~\cite{Aghanim:2018eyx} or local measurements~\cite{Riess:2016jrr} (known as the Hubble tension, see, e.g. \cite{Freedman:2021ahq}), the matter density parameter~$\Omega_\text{m} = 0.315 \pm 0.007$, and the dark energy (cosmological constant) density parameter~$\Omega_\Lambda = 1 - \Omega_\text{m}$.

\subsubsection{Interactions with background photons}

\begin{figure}[htb]
    \centering
    \includegraphics[width=0.8\textwidth]{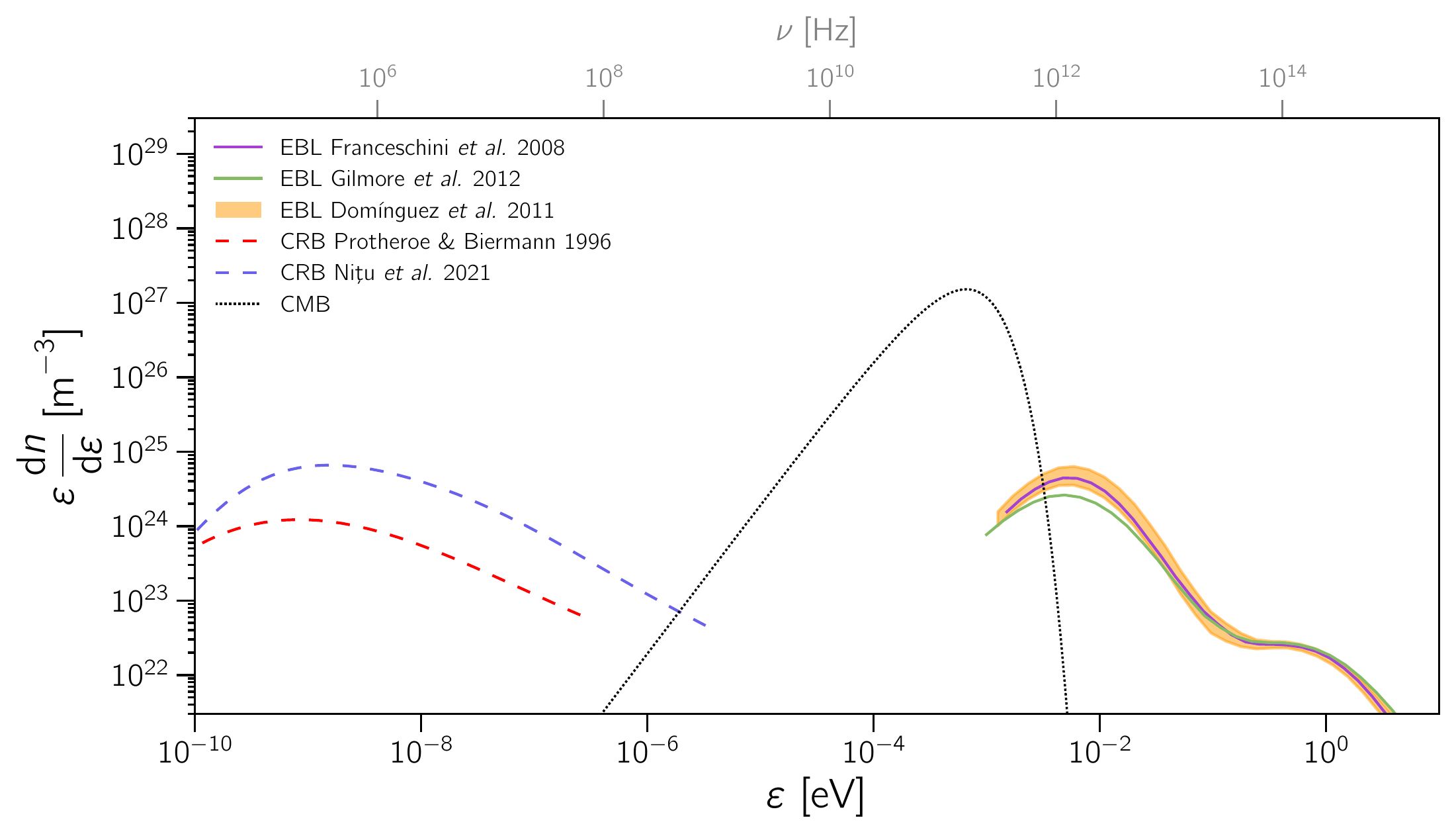}
    \caption{Compilation of the density of background photons ($n(\varepsilon)$) with different energies ($\varepsilon$) at present time for a wide range of frequencies ($\nu$, shown at the top). Here the dashed lines correspond to the \ac{CRB}, the dotted line to the \ac{CMB}, and the \ac{EBL} is represented by solid lines. Lines of different colors correspond to different models: Protheroe \& Biermann~\cite{Protheroe:1996si},  Ni\c{t}u \textit{et al.}~\cite{Nitu:2020vzn}, Franceschini \textit{et al.}~\cite{Franceschini:2008tp}, Gilmore \textit{et al.}~\cite{Gilmore:2011ks}, and the model by Domínguez \textit{et al.}~\cite{Dominguez:2010bv}, with uncertainties. }
    \label{fig:photonFields}
\end{figure}

The universe is permeated by electromagnetic radiation at a wide range of wavelengths (see \autoref{fig:photonFields}), which can affect the propagation of particles. Gamma rays are absorbed producing electron-positron pairs which, in turn, can generate high-energy gamma rays via inverse Compton scattering. \Acp{UHECR} can lose energy via Bethe-Heitler pair production, photodisintegration, and pion production. Charged leptons produced as by-products of these processes, in particular electrons and positrons, are an essential part of this paradigm because they can upscatter background photons to high energies triggering an electromagnetic cascade. Neutrinos and \acp{GW}, on the other hand, are not significantly affected by background radiation.

The dominant background radiation is the \ac{CMB} ($\varepsilon \approx 0.7$\,meV, where $\varepsilon$~is the energy at the peak of the spectrum, in the laboratory frame), and it consists of the relic black-body radiation from the Big Bang, which became free to propagate when the universe cooled down to~$T \approx 3{,}000$~K allowing nuclei and electrons to combine into neutral atoms, and has since been redshifted by a factor of~$1+z \approx 1{,}100$. 

The second most significant background light after the \ac{CMB} is the \ac{EBL}, which comprises the \ac{IR}--\ac{UV} (about 0.1 to 1000\,\textmu{m}) part of the electromagnetic spectrum emitted throughout the whole history of universe. 
It has not been completely determined nor explained yet. At near-\ac{IR}, visible and \ac{UV} wavelengths ($\lambda \lesssim 1$\,\textmu{m}), the \ac{EBL} is believed to consist mainly of starlight, with subdominant contributions from \acp{AGN}, whereas at mid- and far-\ac{IR} wavelengths, it is thought to be mostly made up of thermal radiation re-emitted by dust heated by shorter-wavelength \ac{EBL}~\cite{Cooray:2016jrk,DeAngelis:2013jna,Biteau:2015xpa}. 

\begin{figure}[h]
     \centering
     \includegraphics[width=.37\linewidth]{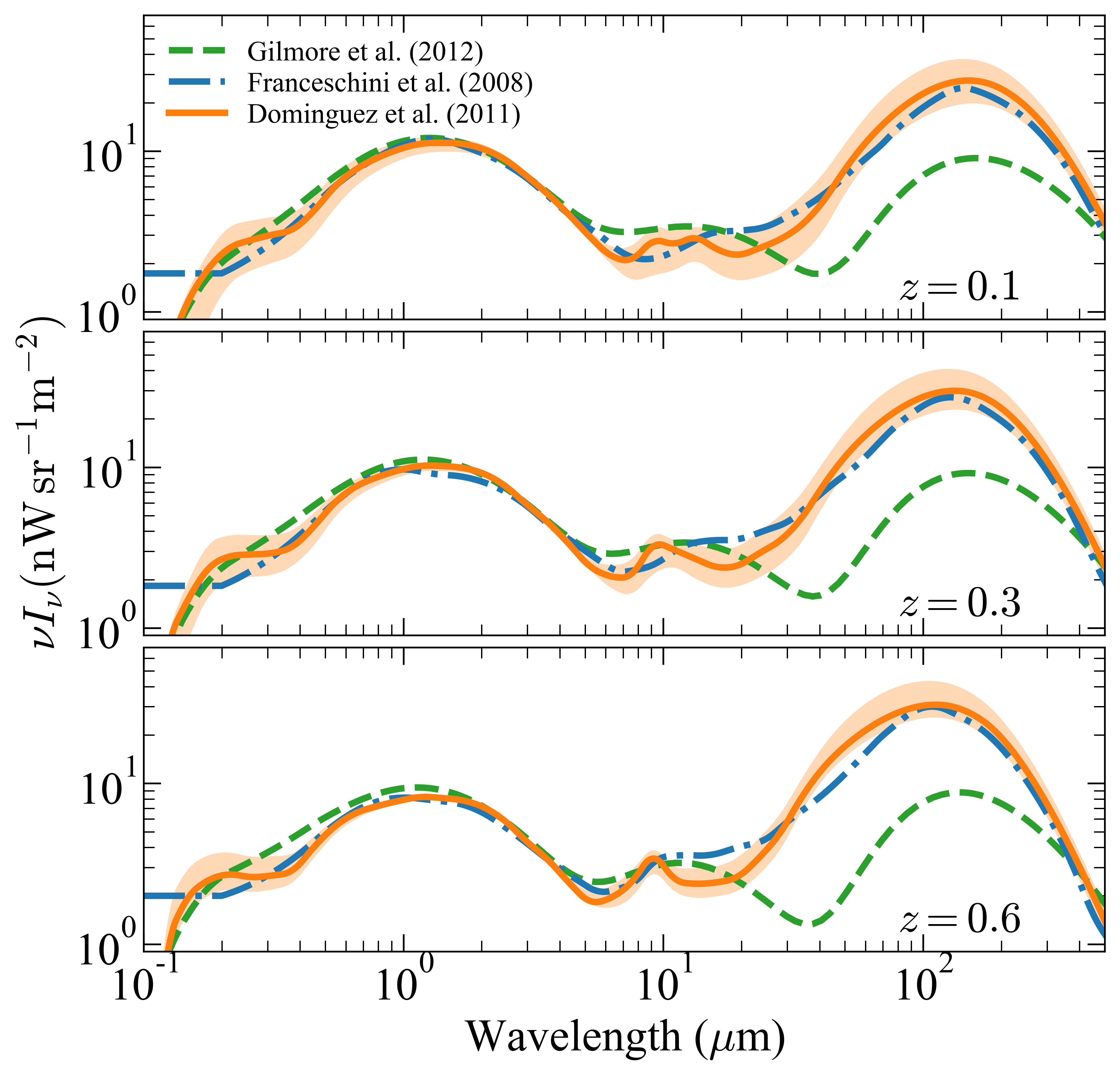}
     \includegraphics[width=.5\linewidth]{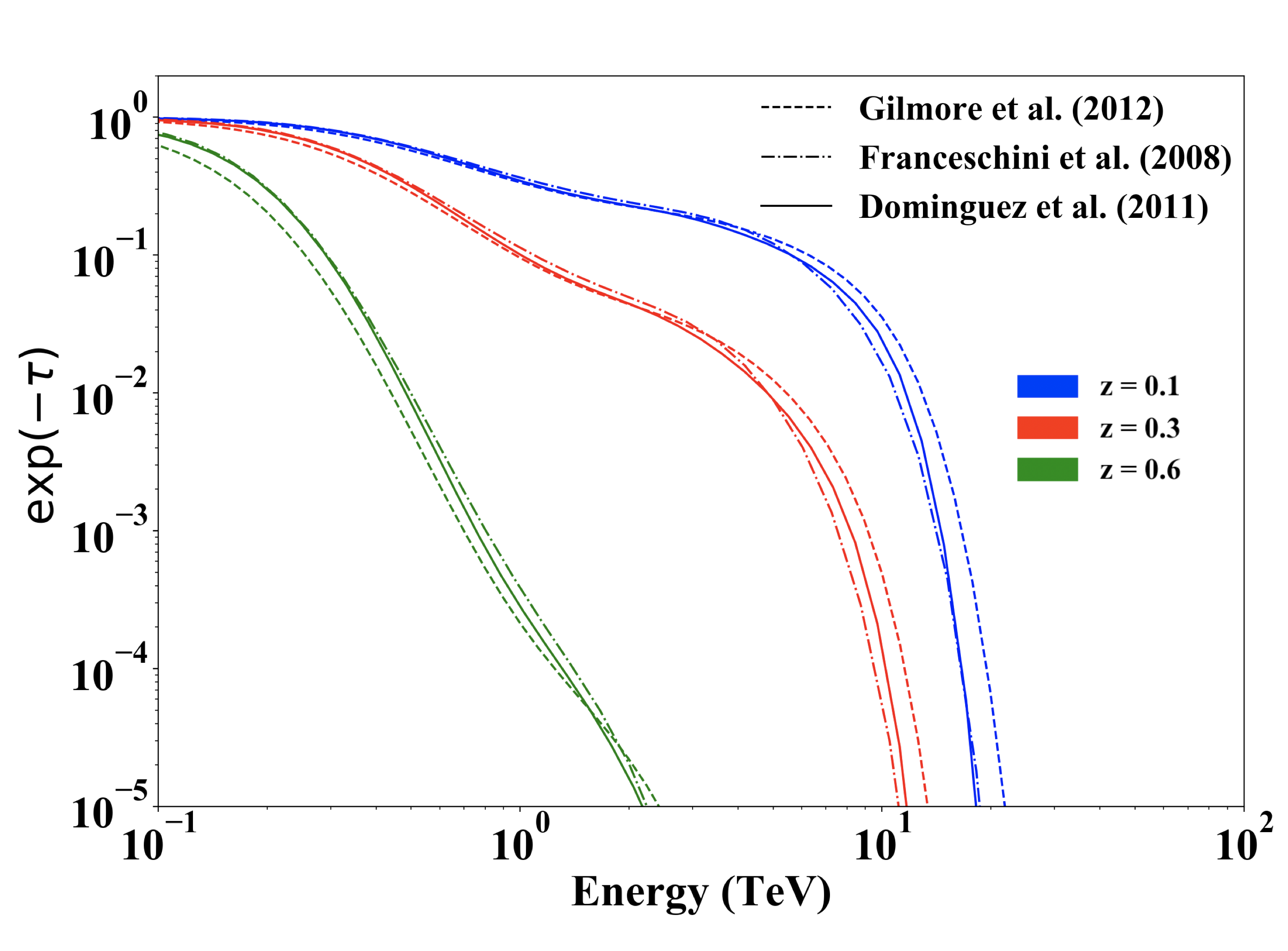}
     \caption{
     {\it Left}. Comparison of the \ac{EBL} intensity of three models~\cite{Gilmore:2011ks,Franceschini:2008tp,Dominguez:2010bv}. The orange bands represent the model uncertainties reported in~\cite{Dominguez:2010bv}. Each panel considers a different redshift as indicated. 
         {\it Right}. The absorption coefficient as a function of the observed gamma-ray energy for gamma rays coming from sources at different redshifts (indicated by different colors), obtained from three different \ac{EBL} models as indicated. 
          }
     \label{figure3.2.1-EBL}
\end{figure}

%%%%%%%%%%%%%%%%%%%%%%%%%%%%%%%%%% 
Direct measurements of the \ac{EBL} are particularly difficult and are subject to large uncertainties since its origin and magnitude vary with energy. At optical wavelengths, the intensity of the \ac{EBL} suffers mainly from the zodiacal light that can be several orders of magnitude higher than the \ac{EBL} itself. However, an agreement has recently emerged between some \ac{EBL} models that can be broadly classified by their approach. The left panel of Figure~\ref{figure3.2.1-EBL} compares three of them between $0.1$ and $500$\,\textmu{m}. In the figure, the two most prominent galaxy emission features are seen; from left to right, the stellar photospheric peak is at about $1$\,\textmu{m}, while the dust re-radiation peak occurs at around $100$\,\textmu{m}. 
The model in~\cite{Gilmore:2011ks} is a semi-analytical model of galaxy formation and evolution, which considers starlight re-emitted by dust at the mid and far-\ac{IR}. It starts from first principles and given initial conditions, after which
its implications are extrapolated to the present time. This and other models such as~\cite{Primack:1998wn} are known as {\it forward evolution} types.
In contrast, the {\it backward evolution} models extrapolate backward in time the present stage of a given population of galaxies. For example, using gamma-ray attenuation the model presented in~\cite{Franceschini:2008tp} 
studies known sources of galaxies and \acp{AGN} as the only background source of light, and evolves luminosity functions to $z = 1.4$. Complementary, there are the empirical models or {\it observed evolution} types. 
The model in~\cite{Dominguez:2010bv} is based on a sample of about 6000 galaxies in a range of redshift less than one, whose extrapolations extend to redshift 4.
The uncertainties in the model can be seen as the orange bands in the Figure~\ref{figure3.2.1-EBL}.
In this very general classification, the {\it inferred evolution} types can also be mentioned, where the evolution of galaxy properties is inferred in some range, for example~\cite{Finke:2009xi, Kneiske:2002wi}. It can be expected that the models that rely directly on observations, as~\cite{Dominguez:2010bv},
{\it Dom\'inguez et al.\ (2011)}
will improve substantially with the advent of new data and optimized observation tools and techniques. All these \ac{EBL} models assume a $\Lambda$CDM cosmology.

%%%%%%%%%%%%%%%%%%%%%%%%%%%%%%%%%%%%%%%%%%%%%%%%%%%%%%%%%%%%%%%%%%%%

Assuming standard Lorentz invariance, an interaction between a a cosmic messenger with mass~$m$ ($0$~for a gamma ray), energy~$E$, and speed~$\beta$ (in $c=1$~units) with a background photon with energy~$\varepsilon$ at an angle~$\theta$ 
is possible if the
squared center-of-mass energy of the system $s=m^2 + 2E\varepsilon(1 - \beta\cos\theta)$
is above the kinematic threshold $s_{\min}$.
Here, we are interested in the ultrarelativistic limit, $\beta \to 1$ (exact for gamma rays, and to an excellent approximation for cosmic rays and secondary electrons).
For a given~$E$ and~$\varepsilon$, the maximum possible~$s$ is obtained in head-on collisions ($\theta = \pi$), for which~\(
    s_{\max}(E,\varepsilon) = m^2 + 2 E\varepsilon(1+\beta) 
\),
meaning that for a given background energy~$\varepsilon$ a process is only possible if the photon energy is~\(
    E \ge E_{\min}(\varepsilon) = \frac{s_{\min} - m^2}{2\varepsilon(1+\beta)} 
\), and vice versa for a given~$E$ only if~\(
    \varepsilon \ge \varepsilon_{\min}(E) = \frac{s_{\min} - m^2}{2E(1+\beta)} 
\).
The interaction rate of a process can be computed by integrating its cross section~$\sigma(s)$ over the distribution of background photon energies~$\frac{d n}{d\varepsilon}(\varepsilon,z)$ (number of background photons per unit volume and unit energy) and angles, assumed isotropic, which, after converting the integration over~$d\theta$ to one over~$ds$, works out to
\begin{equation}
    \lambda^{-1}(E,z) = \frac{1}{8\beta E^2}
    \int_{\varepsilon_{\min}(E)}^{+\infty}
    \frac{1}{\varepsilon^2}\frac{d n(\varepsilon,z)}{d\varepsilon}
    \left.\int_{s_{\min}}^{s_{\max}(E,\varepsilon)} (s-m^2) \sigma(s) d s\right.
    d\varepsilon. \label{eq:mfpIntegral}
\end{equation}  
(In practice, the integration over~$d\varepsilon$ can be limited to an energy~$\varepsilon_{\max}$ above which the contribution of the background photon field becomes negligible.)
Since this is an immediate consequence of the kinematics of special
relativity, any modification of the kinematics will modify it and affect the expected interaction rates.

%%%%%%%%%%%%%%%%%%%%%%%%%%%%%%%%%%%%%%%%%%%%%%%%%%%%%%%%%%%%%%%%%%%%
\paragraph{Interactions of electrons and photons}

\begin{description}
\item{\textit{Pair production}\label{pairprod-photon}:}

Pair production ($\gamma_\text{HE} + \gamma_\text{bg} \to e^+ + e^-$) through the scattering of a cosmic gam\-ma ray of energy~$E$ off a background photon of energy~$\varepsilon$ is an allowed process in special relativity, with a threshold~
\begin{equation}
s_{\min} = 4m_e^2
\quad\to\quad
E_{\min}(\varepsilon) = m_e^2/\varepsilon,
\qquad
\varepsilon_{\min}(E) = m_e^2/E \label{pp-thres-sr}
\end{equation} (see, e.g., Ref.~\cite{Carmona:2020whi}) and the Breit--Wheeler cross section~$\sigma_\text{PP}(s)$.
Modifications of this kinematics can affect the expected transparency of the universe~\cite{Carmona:2020whi}.  We will explore this possibility in Sec.~\ref{transparency}, especially in relation with the transparency anomaly reported in several works~\cite{DeAngelis:2013jna}.

For gamma rays with energy between $\sim$~10\,GeV and 100\,TeV, the \ac{EBL} is the dominant background. The \ac{CMB} overtakes the \ac{EBL} for energies between 100\,TeV and $10\,\mathrm{EeV}$. For higher energies the \ac{CRB} dominates. As a consequence of the pair-production process ($\gamma\gamma \to e^+e^-$), incoming high-energy photons are absorbed and their spectrum is attenuated, such that for a source at redshift $z_\text{s}$, the flux is absorbed by a factor corresponding to the convolution of the source intrinsic spectrum with the corresponding background, as follows:
\begin{equation}
    \Phi_\text{obs}(E_{0}) = \Phi_\text{int}(E_{0}(1+z_{s})) \times \mathrm{e}^{-\tau(E_{0},z_\text{s})}.
    \label{eq:obs_spectrum}
\end{equation}
Here $\tau(E_{0},z_\text{s})$ is the optical depth which depends on the observed energy  $E_{0} = E / (1+z)$ of the high-energy gamma ray and the source redshift $z_{\mathrm{s}}$:
\begin{equation}
\tau\left(E_{0}, z_{\mathrm{s}}\right)=\int_{0}^{z_\text{s}}  \lambda^{-1}(E, z) \frac{d l}{d z}  \, d z,
\label{eq:optical_depth}
\end{equation}
where $\lambda^{-1}(E, z)$ is given in Eq.~(\ref{eq:mfpIntegral}).  
The integration over $dz$ accounts for the distance covered by gamma rays. It includes the assumed cosmology and is performed along the line of sight. Gamma-ray absorption increases with the source redshift (see Figure~\ref{figure3.2.1-EBL} right). The redshift for which optical depth reaches unity defines the gamma-ray horizon for a given gamma-ray energy. Beyond the gamma-ray horizon, the universe becomes progressively opaque for high-energy gamma rays.

%%%%%%%%%%%%%%%%%%%%%%%%%%%%%%%%%%
\item{\textit{Inverse Compton scattering:}}
The scattering of charged leptons (in particular electrons and positrons) by background photons can produce high-energy gamma rays: $e^\pm + \gamma \rightarrow \gamma + e^\pm$. The \ac{CMB} is the dominant background photon field for electron energies of up to 1\,EeV, above which the photons from the \ac{CRB} start to dominate the interaction probabilities. Inverse Compton scattering is one of the most important processes at high energies because electrons created, for example, via pair production, can then produce more high-energy gamma rays and initiate a cascade process in the intergalactic space. For energies below $\sim$~100\,TeV the typical distance traveled by electrons is of the order of tens of kpc, such that the observational signatures of this process are manifested through the secondary gamma rays they produce.
For inverse Compton scattering, 
the mean free path can be computed from Eq.~\eqref{eq:mfpIntegral}
with the cross section~$\sigma_\text{IC}$
and the kinematic threshold~$s_\text{min}=m_e^2$.

The higher-order counterpart to inverse Compton scattering is triplet pair production ($e^\pm + \gamma \rightarrow e^\pm + e^+ + e^-$), which becomes comparable to the former for energies larger than~$\sim~1\,\mathrm{EeV}$. In this case, the \ac{CMB} dominates over the other backgrounds.
\end{description}

%%%%%%%%%%%%%%%%%%%%%%%%%%%%%%%%%%%%%%%%%%%%%%%%%%%%%%%%%%%%%%%%%%%%
\paragraph{Interactions of nuclei}\label{sec:UHECRprop}

In this section we describe the interactions of \acp{UHECR} with background photons. We neglect the nucleus recoil in the interactions, and we separate the discussion of the processes changing the Lorentz factor of the particle from the ones changing the particle type, as done in \cite{Aloisio:2008pp,Aloisio:2010he}. Therefore if the energy of a nucleus of mass~$A$ is written as~$E=\Gamma Am_{N}$, where $\Gamma$~is the Lorentz factor and $m_{N}$~is the mass of the nucleon, the energy losses can be written as
\begin{equation}
\frac{1}{E} \frac{dE}{dt} = \frac{1}{A} \frac{dA}{dt} + \frac{1}{\Gamma} \frac{d\Gamma}{dt} .
\label{eq:eneloss}
\end{equation}

\begin{figure}[htb]
    \centering
    \includegraphics[width=0.72\columnwidth]{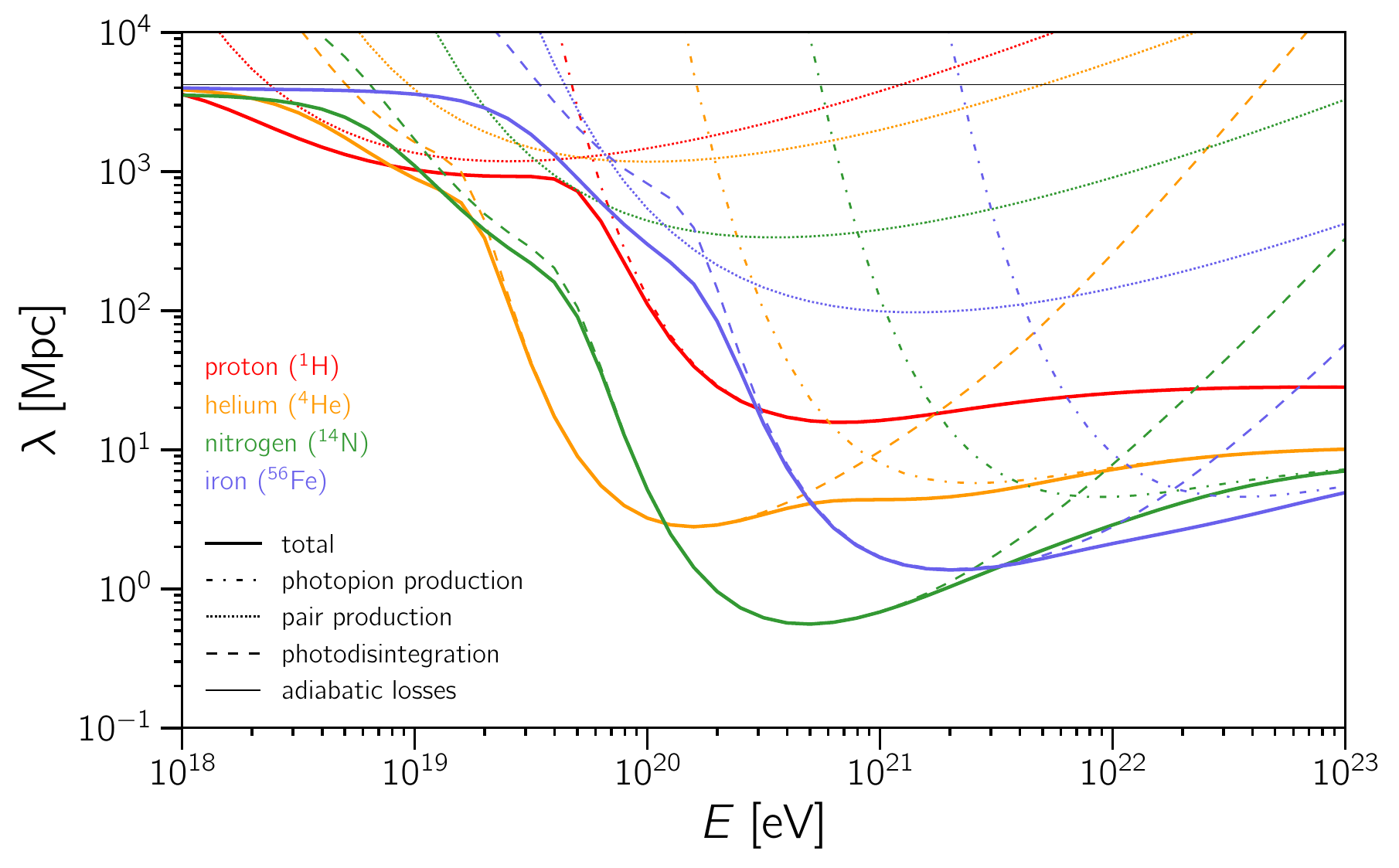}
    \caption{Energy loss length ($\lambda$) as a function of the \ac{UHECR} energy ($E$), assuming the EBL model from~\cite{Gilmore:2011ks} and the photodisintegration model from \cite{Rachen:1996zeh} (helium), \cite{Khan:2004nd} (nitrogen and iron).}
    \label{fig:uhecr_mfp}
\end{figure}

The mean free path~$\lambda$ for a cosmic ray of mass~$m = Am_N$ interacting with a background photon at redshift~$z$ can be computed from eq.~\eqref{eq:mfpIntegral}, which is often written in terms of the photon energy in the nucleus rest frame $\varepsilon' = (1-\beta\cos\theta)\Gamma\varepsilon = (s-m^2)/2m$ rather than of~$s$; the integral~$\int_{s_{\min}}^{s_{\max}(E,\varepsilon)} (s-m^2) \sigma(s) \,d s$ then becomes $4 m^2 \int_{\varepsilon'_{\min}}^{\varepsilon'_{\max}(E,\varepsilon)} \varepsilon' \sigma(\varepsilon') \,d\varepsilon'$, where~$\varepsilon'_{\min}= \left.(s_{\min}-m^2)\middle/2m\right.$ and~$\varepsilon'_{\max}(E,\varepsilon) = (1+\beta)\Gamma\varepsilon$.
Depending on the $\varepsilon'$ range, several types of interactions between nuclei and photons are possible, one consequence of which is an upper limit to the energy with which \acp{UHECR} from distant extragalactic sources can reach us, the \ac{GZK} limit, which is a possible cause of the observed cutoff in their energy spectrum (Sec.~\ref{sec:uhecr}).  These processes include:

\begin{description}
\item{\textit{Bethe-Heitler pair production:}}
When the energy of the photon in the nucleus rest frame is $\varepsilon' \gtrsim 1$\,MeV, 
i.e.\ when the energy of the nucleus in the laboratory frame is $E \gtrsim 0.5 A (\mathrm{meV}/\varepsilon)$\,EeV,
pairs of electrons and positrons can be created \cite{Blumenthal:1970nn}: \(
    ^{A}_{Z}X + \gamma \rightarrow {^{A}_{Z}X} + e^{+} + e^{-}
\). This process has a short mean free path ($\sim 1$\,Mpc) but each interaction results in a very small energy loss by the nucleus ($\lesssim 0.1\%$), thus it is usually treated as continuous. 
Pair production on CMB photons is the dominant energy loss mechanism for protons with $2.5\,\mathrm{EeV} \lesssim E \lesssim 60\,\mathrm{EeV}$, with an energy loss length (mean free path divided by inelasticity) of the order of 1\,Gpc.  For other nuclei, the energy loss length is $Z^2/A$~times that of protons. 
\item{\textit{Nuclear photodisintegration:}}
When the photon energy in the nucleus rest frame is $\varepsilon' \gtrsim 8$\,MeV ($E \gtrsim 4A (\mathrm{meV}/\varepsilon)$\,EeV, the precise value depending on the nuclide),
the nucleus can be stripped of one or more nucleons or occasionally larger fragments, e.g.\ \begin{align*}
{^A_Z X} + \gamma &\to {^{A-1}_{\phantom{0+}Z}X} + n, &
{^A_Z X} + \gamma &\to {^{A-1}_{Z-1}X'} + p, &
{^A_Z X} + \gamma &\to {^{A-2}_{Z-2}X''} + {^4_2\mathrm{He}}, &
&\text{etc.}
\end{align*}
All the fragments will inherit the Lorentz factor of the parent nucleus, hence the inelasticity is $1/A$ for one-nucleon ejection, $4/A$ for alpha-particle ejection, etc\@.  This is the dominant energy loss for nuclei at most energies, mainly on EBL photons for nucleus energies below a few EeV per nucleon and on CMB photons above.  The secondary neutrons will undergo beta decay~$n \rightarrow p + e^- + \bar\nu_e$ within~$9(E/\mathrm{EeV})$\,kpc in average, so all secondary nucleons can be treated as protons beyond Galactic distance scales.  The giant dipole resonance dominates the cross sections for $\varepsilon'<30$\,MeV, and envisages the emission of one or two nucleons. At larger energies, the quasi-deuteron effect takes over, where the photon interacts with a nucleon pair, with the consequent emission of several nucleons \cite{Puget:1976nz}.
The branching ratios of the various photodisintegration channels are difficult to measure, especially those involving the ejection of a charged fragment; in particular, those of alpha-particle ejection have only been measured for a very few nuclides.  Certain phenomenological models are available, but they sometimes overestimate the little available data by an order of magnitude or neglect such channels altogether (see Ref.~\cite{Batista:2015mea} and references therein).
\item{\textit{Pion production:}}
At larger energies $\varepsilon' \gtrsim 145$\,MeV  ($E \gtrsim 72A (\mathrm{meV}/\varepsilon)$\,EeV), the photon can interact with a nucleon (whether free or bound in a nucleus), typically via the $\Delta$ resonance, producing pions with $\approx 20\%$ energy of the nucleon ($E/A$): 
\begin{align*}
	p + \gamma &\to p + \pi^0, &	p + \gamma &\to n + \pi^+, &
	n + \gamma &\to n + \pi^0, &\text{or\quad}n + \gamma &\to p + \pi^-.
\end{align*}
(In the case of a bound nucleon, it is thereby ejected from its nucleus.) Secondary particles are then generated, such as neutrinos~\cite{Kotera:2010yn, Romero-Wolf:2017xqe, AlvesBatista:2018zui, Heinze:2019jou}, from the decay of charged pions, and photons, from the decay of neutral pions. Pion production on CMB photons is the dominant energy loss for protons with $E \gtrsim 60\,\mathrm{EeV}$.
At even higher $\varepsilon'$, the production of multiple pions, and/or heavier hadrons such as kaons, becomes possible.
\end{description}
Figure~\ref{fig:uhecr_mfp} shows a comparison of the energy loss lengths  for these different interaction processes, for four types of nuclei: hydrogen ($^{1}$H), helium ($^{4}$He), nitrogen ($^{14}$N), and iron ($^{56}$Fe).

\subsubsection{Effects of magnetic fields}
\label{magnetic}
Charged particles like cosmic rays and the charged leptons produced in electromagnetic cascades can be deflected by \acp{IGMF} and \acp{GMF}. The former comprises fields in clusters ($\sim 1$\,\textmu{G}), filaments ($\sim 1$--$100\,\text{nG}$), and cosmic voids, whose strength is mostly unknown, ranging from  $\mathcal{O}(10^{-6}\,\mathrm{nG})$ to $\mathcal{O}(1\,\mathrm{nG})$ \cite{Durrer:2013pga,Vachaspati:2020blt}. The coherence length of these fields, i.e.\ the distance scale over which they can be approximated as constant, is also poorly constrained, especially in the voids. Theoretically motivated values lie between $\sim 10^{-6}\,\text{Mpc}$ and $\sim 10^3\,\text{Mpc}$~\cite{Durrer:2013pga}, and there are no strong observational constraints. For example, the detection of a flare of the blazar TXS0506+056 in neutrinos and gamma rays suggest a range $\sim 0.1$--$100\,\text{Mpc}$~\cite{AlvesBatista:2020oio}.

The \ac{GMF} is known to have both a regular component roughly following the spiral arm structure of the Galaxy and a turbulent component (with coherence length $\mathcal{O}(100\,\mathrm{pc})$), both of the order of a few \textmu{G}, but there are large uncertainties on the details.
The main difficulty is that we cannot measure the \ac{GMF} as a function of position in 3D, but only line-of-sight integrals weighed by electron densities which are themselves poorly known (e.g.\ the Faraday rotation, proportional to the line integral of the radial \ac{GMF} times the total electron density, or the polarized synchrotron emission, proportional to the line integral of the transverse \ac{GMF} squared times the relativistic electron density), and the measurements themselves are contaminated by large backgrounds (see Ref.~\cite{galaxies7010017} for a recent review of these issues).
Several models of the \ac{GMF} are available \cite{Sun:2010sm,Jansson:2012rt,Jaffe:2013yi}, but none of them is fully consistent with all available data \cite{Adam:2016bgn}.

\bigskip

Two regimes of cosmic-ray propagation can be identified depending on the Larmor radius~$r_\text{L}$, defined as
\begin{equation}
    r_\text{L} = \frac{R}{B_\perp}
    \approx 1.081 \frac{E/Z}{\mathrm{EeV}} \left(\frac{B_\perp}{\mathrm{nG}}\right)^{-1} \mathrm{Mpc}
    = 1.081 \frac{E/Z}{\mathrm{EeV}} \left(\frac{B_\perp}{\text{\textmu{G}}}\right)^{-1} \mathrm{kpc} \, ,
\end{equation}
where the magnetic rigidity~$R$ of a particle is its momentum divided by its electric charge ($R=E/Z$~for fully ionized nuclei in $c=e=1$~units) and $B_\perp$ is the magnetic field perpendicular to the trajectory of the particle (i.e., in the case of particles hitting the Earth, perpendicular to the line of sight).  If the Larmor radius is much less than the coherence length~$L_B$, the propagation is diffusive; otherwise the magnetic deflections can be considered a perturbation on top of rectilinear propagation (quasi-ballistic propagation).
The typical angular deflections over a distance~$D$, in this case, are
\begin{equation}
    \Delta\theta_\text{CR} \simeq 
    \begin{cases}
    2 \arcsin \dfrac{D}{r_\text{L}} \approx
    \dfrac{D}{r_\text{L}}\,\mathrm{rad} \approx 53^\circ Z
    \left(\dfrac{E}{\mathrm{EeV}}\right)^{-1}
    \dfrac{B_\perp}{\text{\textmu{G}}} \dfrac{D}{\mathrm{kpc}}\,, & D \ll L_B\,, \\[9pt]
    53^\circ Z \left(\dfrac{E}{\mathrm{EeV}}\right)^{-1}
    \dfrac{B_\perp}{\mathrm{nG}} \dfrac{\sqrt{L_B D}}{\mathrm{Mpc}}\,, & D \gg L_B\,,
    \end{cases}
\end{equation}
and the corresponding time delays are
\begin{equation}
    \Delta t_\text{CR} = 
    \begin{cases}
    \dfrac{D}{c} \dfrac{\Delta\theta^2}{24} + \mathcal{O}(\Delta\theta^4)
    \approx 116\, Z^2 \left(\dfrac{E}{\mathrm{EeV}}\right)^{-2}
    \left(\dfrac{B_\perp}{\text{\textmu{G}}}\right)^2
    \left(\dfrac{D}{\mathrm{kpc}}\right)^3\,\mathrm{yr}\,, & D \ll L_B\,, \\[9pt]
    1.2\times10^5\,Z^2 \left(\dfrac{E}{\mathrm{EeV}}\right)^{-2}
    \left(\dfrac{B_\perp}{\mathrm{nG}}\right)^2
    \dfrac{D}{\mathrm{Mpc}}
    \left(\dfrac{L_B}{\mathrm{Mpc}}\right)^2\,\mathrm{yr}\,, & D \gg L_B\,.
    \end{cases}
\end{equation}
For the turbulent \ac{GMF}, the transition between the diffusive and the quasi-ballistic regime is expected to occur at rigidities $E/Z \sim 10^{17}$--$10^{18}$\,eV, hence cosmic rays are generally assumed to have Galactic origin below such energies. 
Note that realistic values of the time delay are much longer than the typical duration of \ac{UHECR} experiments (a few years to a few decades), meaning it is not feasible to use charged cosmic rays to study transient extragalactic phenomena.

Concretely, using recent estimates, deflections of cosmic rays by the regular \ac{GMF} are expected to range from $\sim 15$~to $\sim 40 \times \tfrac{10\,\mathrm{EeV}}{E/Z}$~degrees depending on the direction \cite{Smida:2015kga,Unger:2019xct}; root-mean-square de\-flec\-tions by the turbulent \ac{GMF} are a few times smaller, ranging from $\lesssim 3.5$ (most likely $\sim 0.5$)${}\times \tfrac{10\,\mathrm{EeV}}{E/Z}$~degrees at high Galactic latitudes to $\lesssim 27$ (most likely $\sim 5$)${}\times \tfrac{10\,\mathrm{EeV}}{E/Z}$~degrees at low Galactic lat\-i\-tudes \cite{Pshirkov:2013wka}. 
De\-flec\-tions due to IGMFs depend on the distribution of these fields in the cosmic web, but even in more extreme scenarios with highly magnetised voids, cosmic rays from sources closer than $\sim 50$\,Mpc would be deflected, on average, by $\lesssim 75 \times \tfrac{10\,\mathrm{EeV}}{E/Z}$~degrees~\cite{AlvesBatista:2017vob}.
Less extreme scenarios suggest that extragalactic deflections are comparable to those due to the \ac{GMF}~\cite{Hackstein:2017pex}.

One might expect that by causing \acp{UHECR} to travel longer distances, \acp{IGMF} would alter their spectrum and composition; but the propagation theorem \cite{Aloisio:2004jda} states that in the case of a ``continuous'' distribution of sources (i.e.\ average distance between sources much less than all other relevant length scales, such as the Larmor radius, the \ac{UHECR} energy loss length, and the age of the sources times the speed of light), the spectrum at Earth is not affected by the magnetic fields. 
Conversely, in the case of a discrete distribution of sources with separations comparable to or larger than the Larmor radius, the effect of \acp{IGMF} can be modeled as a low-energy exponential cutoff \cite{Mollerach:2013dza}, although \acp{IGMF} may be too weak for this effect to significantly impact observations~\cite{Batista:2014xza}.  
It is important to stress that it is not clear if the conditions required for the validity of the propagation theorem hold, so caution should be taken when invoking this theorem to justify the omission of magnetic-field effects.

\begin{figure}
    \centering
    \includegraphics[width=0.45\textwidth,trim=0 0 254 0, clip]{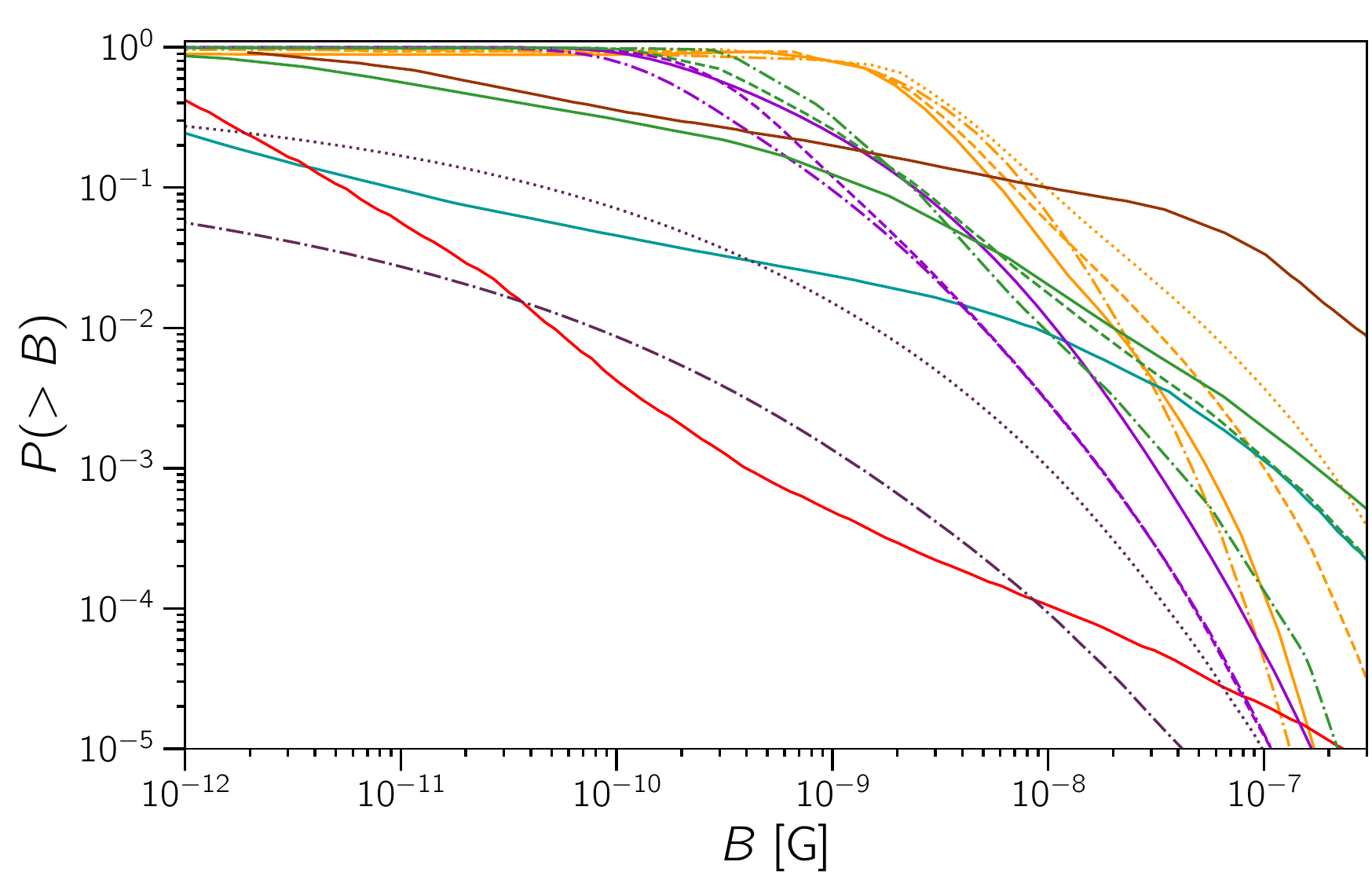}
    \hspace{0.05\textwidth}
    \includegraphics[width=0.45\textwidth]{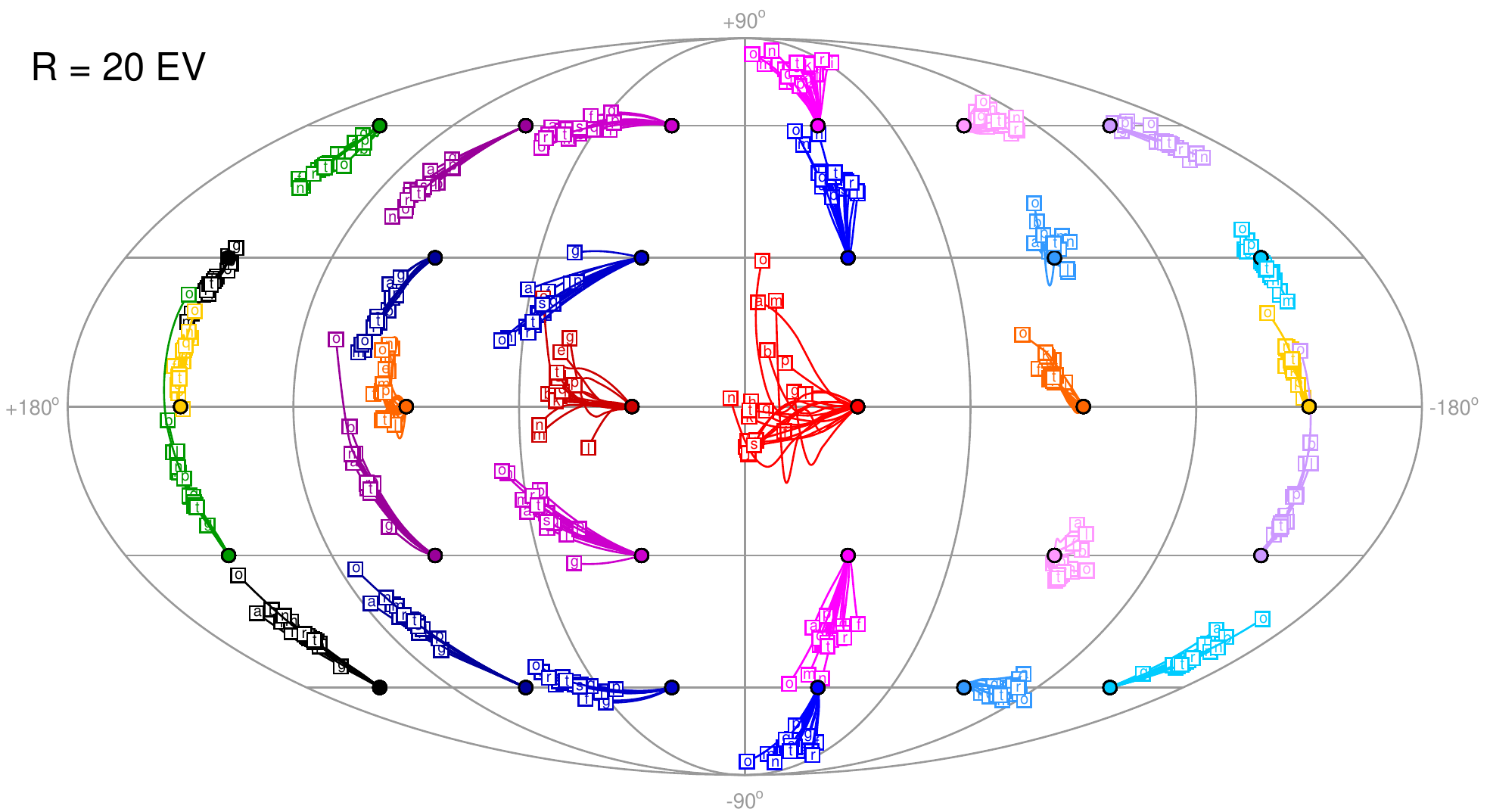}
    \caption{Left: the ``filling factor'' of \acp{IGMF} (i.e.\ the volume fraction of the universe in which the magnetic field strength is greater than the specified amount) as predicted by various models, from Ref.~\cite{AlvesBatista:2019tlv}.
    Right: deflections of cosmic rays with~$E=20Z$\,EeV (squares are source positions and circles are arrival directions at Earth) by the regular \ac{GMF} as predicted by various models, from Ref.~\cite{Unger:2019xct}.}
    \label{fig:section5.1-magnetic}
\end{figure}

Our poor knowledge of \acp{IGMF} and \acp{GMF} is exemplified in \autoref{fig:section5.1-magnetic}, which shows the major differences existing among the various available models.

\bigskip 

Magnetic fields are not directly important for the propagation of gamma rays, but they can affect the charged component of the electromagnetic cascades they induce, namely the electron-positron pairs generated via high-energy gamma-ray interactions with background photons. The pairs are deflected by an angle proportional to the magnetic field and they travel for a certain distance comparable to the mean free path for inverse Compton scattering $\lambda_\text{IC}$. The local deflection angle~$\delta$ of the electrons with respect to their original direction depends on the coherence length~$L_B$ of the magnetic field in comparison with the mean free path. In the quasi-rectilinear regime ($L_B \gg \lambda_\text{IC}$), this angle is $\delta \simeq L_B / r_\text{L}$. If the propagation is diffusive ($L_B \ll \lambda_\text{IC}$), then $\delta \simeq (\lambda_\text{IC} L_B)^{1/2} / r_\text{L}$. 

It follows from this underlying picture that the secondary gamma rays detected at Earth can be separated from the primary gamma rays by a given angle $\Delta \theta_\gamma$ even if they were originally emitted in the same direction. For gamma rays emitted with energy $E_\gamma$ between $\sim$~1\,GeV and 100\,TeV, $\Delta \theta_\gamma$ is roughly~\cite{Neronov:2009gh},
\begin{equation} 
    \Delta \theta_\gamma \simeq
    \begin{cases}
    0.07^\circ (1+z)^{-\frac{1}{2}} \left( \dfrac{\tau_\theta}{10} \right)^{-1} \left( \dfrac{E_{\gamma}}{0.1 \, \text{TeV}} \right)^{-\frac{3}{4}} \left( \dfrac{B}{10^{-14} \, {\rm G}} \right) \left( \dfrac{L_{B}}{1 \, {\rm kpc}} \right)^{\frac{1}{2}} & L_{B} \ll \lambda_{\rm IC}\,, \\
    0.5^\circ (1+z)^{-2} \left( \dfrac{\tau_\theta}{10} \right)^{-1} \left( \dfrac{E_{\gamma}}{0.1 \, {\rm TeV}} \right)^{-1} \left( \dfrac{B}{10^{-14} \, {\rm G}} \right) & L_{B} \gg \lambda_{\rm IC}\,,
    \end{cases}
    \label{eq:gammaHalo}
    \end{equation}
where $\tau_\theta$ is equal to the angular diameter distance to the source of primary gamma rays divided by the mean free path for pair production, i.e., a proxy for the optical depth \eqref{eq:optical_depth}. In practice, this deflection will lead to the formation of a ``halo'' around point-like astrophysical sources of gamma rays (see e.g.~\cite{Aharonian:1993vz}). More complex morphologies such as spiral-like structures can arise depending on the angle of observation of the object and properties of the magnetic field such as its helicity~\cite{AlvesBatista:2016urk, Duplessis:2017rde, Batista:2021rgm}.

If all gamma rays are emitted by an astrophysical object simultaneously, then the magnetic deflection of the charged component would incur delays ($\Delta t_\gamma$) in the arrival times of the secondary photons~\cite{Plaga:1995ins}, approximately given by~\cite{Neronov:2009gh}:
\begin{equation}
    \Delta t_\gamma \simeq
    \begin{cases}
    7 \times 10^{5} \, {\rm s} \, (1 - \tau_{\theta}^{-1}) (1 + z)^{-5} \kappa \left( \dfrac{E_{\gamma}}{0.1\,{\rm TeV}} \right)^{-\frac{5}{2}} \left( \dfrac{B}{10^{-18}\,{\rm G}} \right)^{2} & L_{B} \gg \lambda_{\rm IC}\,,\\
    1 \times 10^{4} \, {\rm s} \, (1 - \tau_{\theta}^{-1}) (1 + z)^{-2} \kappa \left( \dfrac{E_{\gamma}}{0.1\,{\rm TeV}} \right)^{-2} \left( \dfrac{B}{10^{-18}\,{\rm G}} \right)^{2} \left( \dfrac{L_{B}}{1\,{\rm kpc}} \right) & L_{B} \ll \lambda_{\rm IC}\,.
    \end{cases}
    \label{eq:gammaDelay}
\end{equation}
The parameter $\kappa$ is of order unit and accounts for EBL model uncertainties.

An immediate consequence of Eqs.~(\ref{eq:gammaHalo}) and~(\ref{eq:gammaDelay}) is that, in the presence of \acp{IGMF}, the main gamma-ray observables will change. First, the angular broadening might spread the total flux from a source over a larger angular window, decreasing the measured point-like flux. Second, in the case of limited-duration transient events, part of the flux might only reach Earth millions of years later. Therefore, if \acp{IGMF} are stronger than $B \gtrsim 10^{-16}$~G, as the results~\cite{Neronov:1900zz, Biteau:2018tmv} indicate, they need to be taken into account in the modelling, given their impact in the modelling of the propagation and their degeneracy with respect to the source properties~\cite{Dolag:2010ni, Saveliev:2020ynu}.

Electromagnetic cascades are well understood, although there are many uncertainties related to the distribution of photon fields. Nevertheless, the role played by \acp{IGMF} on the cascades has been recently questioned in a number of works~\cite{Broderick:2011av, Schlickeiser:2012hn}, which claim that plasma instabilities arising due to the back-reaction of the intergalactic medium on the electrons of the cascade could effectively halt the production of high-energy secondary gamma rays, cooling down the electrons before they can up-scatter background photons. Nevertheless, while this effect might actually play a substantial role, there are indications that this process does not completely quench the development of the cascades~\cite{AlvesBatista:2019ipr, Perry:2021rgv}. For a more detailed discussion, see Ref.~\cite{Batista:2021rgm}.

\subsubsection{Gravitational lensing}

\newcommand{\srcs}{QSO\,B0218+357}
\newcommand{\srcp}{PKS\,1830-211}

Light rays are deflected when propagating through a gravitational field. Long suspected before \ac{GR}, it was only after Einstein’s final formulation of his theory that the effect was described quantitatively. 
According to the Einstein’s theory of gravity, the deflection angle does not depend on the photon energy, so we assume that gravitational lensing is achromatic. 
Measurements of light propagation at the highest energies of the electromagnetic spectrum, i.e.\ \ac{VHE} gamma rays, in comparison with the propagation at the lower energies, are of utmost importance for testing \ac{GR}.

The simplest approximation for calculations of gravitational lensing is by a point mass lens. It works very well for many objects, such as stars and planets, but also in cases of radiation from quasars that is lensed on large scales by galaxies with regular distribution of mass which is concentrated in the centers of spiral and elliptical galaxies. In case of strong lensing, multiple images are observed, dependent on the geometry of the event. A special case of a ring-shaped image (``Einstein ring") is observed for the lensing alignments when the source, lens and the observer lie on the same line. If the multiple images cannot be resolved, amplification (or magnification) of the source is observed, and such case of strong lensing is called microlensing.
Gravitational lensing of electromagnetic radiation passing close to centers of massive galaxies is manifested in several ways:
i) different lensed images (created as light passes different paths through the lens) arrive with certain time delays, allowing us to measure the mass of the lens, as well as to probe the Einstein’s theory of gravity by testing the gravitational lensing achromacity, and possible effects of quantum gravity. ii) microlensing by single stars and planets in the lens galaxy can help to constrain the sizes of emission regions of electromagnetic radiation at different energies.  
Measurements of the time delays in different passbands, for instance \srcs\ with radio \cite{Biggs:2018svq} and \acs{HE} gamma rays \cite{Cheung:2014dma} are not precise enough to constrain possible \ac{QG} induced energy dependence. In addition, the measured time delay could depend on the position of the emitting region (in the \ac{AGN} core or in the jets), which needs not be the same at all wavelengths \cite{Barnacka:2015yyo}. 
Instead, one could compare the time profiles of a flare and its lensed counterpart with detectors such as \acp{IACT}, which cover a range of more than 2 order of magnitude in energy~\cite{Glicenstein:2019rzj}. The shape of the delayed flare would be affected by the \ac{QG} energy-dependent travel time.  
The comparison of the 2 pulses could also be affected by the finite size of the emission region, but not by the time profile of the emission or by possible microlensing induced changes in the amplitude ratio of the pulses.
The main drawback of the method is the small travel time difference between the arms of gravitational lenses which is of the order of tens of days. This is to be compared with the travel time of photons from AGNs or GRBs which are of the order of 1\,Gyr.   
Requirement of co-linearity of source-lens-observer system makes the strongly gravitationally lensed active galaxies rare. Thus, they are observable mostly at large cosmological distances ($z_\text{s}\gtrsim1$), allowing only brightest of those sources to be studied in \acs{HE}.
In addition, in the \acs{VHE} range, the large distance of those sources causes strong absorption in \ac{EBL}, which further hampers detection possibilities with \acp{IACT}.\par
%%%%%%%%%%%%%%%%%%%%%%%%%%%%%%%%%%%%%%%%%%%%%%%%%%%%%%%%%%%%%%%%%%%%%%%5
At GeV energies, two strongly gravitationally lensed systems were detected: \srcp\ \cite{Barnacka:2010fz,Abdo:2014bka} and \srcs\ \cite{Cheung:2014dma}. 
The former  
is a blazar located at $z_s=2.507$. It was discovered as a single source in the Parkes catalog, but later radio observations by the Very Large Array (VLA) and Australian Telescope Compact Array (ATCA) clearly revealed two sources, one in the northeast and one in the southwest, separated by $0.98''$  and connected by an Einstein ring~\cite{1988MNRAS.231..229P,1991Natur.352..132J}. 
The images cannot be resolved by \textit{Fermi}-\ac{LAT}. The emission from the two images can be distinguished by measuring a time delay from variable gamma-ray light curves. The autocorrelation of the light curve gives the time delay between two GeV components of  $27.1 \pm 0.6$\,days \cite{Barnacka:2010fz}, in agreement with radio measurements~\cite{Lovell:1998ka,1995ApJ...444..561V,2001ASPC..237..155W}. 
However, in a later independent analysis~\cite{Abdo:2014bka} of these two large gamma-ray flares of \srcp\, such a delayed activity was not confirmed. The authors set a lower limit of about 6 on the gamma-ray flux ratio between the two lensed images.
\srcs\ was discovered by NRAO 140-ft telescope in its strong source survey~\cite{1972AJ.....77..797P}. 
Radio imaging revealed it to be a gravitationally lensed blazar with the smallest separation double-image known ($335''$), and an Einstein ring with a similar angular diameter~\cite{1992AJ....104.1320O,1993MNRAS.261..435P}.
The lens galaxy is at redshift $z=0.6847$~\cite{1993MNRAS.263L..32B},
and the blazar was later securely measured at $z_s=0.944\pm0.002$~\cite{Cohen:2002ej}.
\srcs\ is the only strongly gravitationally lensed source detected in the \acs{VHE} band so far\cite{Ahnen:2016vlr}.
Detection by the MAGIC telescopes was achieved during a short \acs{HE} flaring activity observed by \textit{Fermi}-\ac{LAT} in 2014. 
Unfortunately observations could be only performed during the trailing image of the source, however the time of the flare agreed with the delay observed at lower energies. 
As reported in~\cite{Abdalla:2019pfk}, in August 2014, following a flare alert by the \textit{Fermi}-\ac{LAT} collaboration, the \ac{H.E.S.S.} array observed \srcp\ with the aim of detecting the gamma-ray flare delayed by 20--27 days from the alert flare, as expected from observations at other wavelengths. 
No photon excess or significant signal was detected, but an upper limit  on  the flux above 67\,GeV was computed and compared to the extrapolation of the \textit{Fermi}-\ac{LAT} flare spectrum.

To fully exploit the gravitational lensing of \ac{AGN} objects, it is essential to detect the \acs{VHE} gamma-ray emission from leading and trailing images of the same flare.

%\newpage
\section{Receiving the message}
\label{section4}

The Earth's atmosphere is opaque to gamma rays and \acp{UHECR}, rendering direct ground-based observations not possible. 
Low-energy gamma rays and cosmic rays can be observed via satellite-borne detectors;  the fluxes of higher-energy ones are too low for that, but they are detected by arrays of ground-based detectors via the particle cascades they produce in the atmosphere, known as \acfp{EAS}. 
Neutrinos, on the other hand, easily penetrate our atmosphere, the Earth, and ourselves, but for that precise reason, they are notoriously difficult to detect, and require detectors of immense sizes (neutrinos of very large energies, above tens of TeV, have non-negligible interaction with the Earth, but their flux is very low).
Regardless of the differences between gamma rays, \acp{UHECR}, and neutrinos, ground-based detectors are based on the same principle: a primary particle entering a medium interacts with an atomic nucleus in air, water, or ice. The interaction triggers a formation of a cascade of particles. Further down the line, charged particles from the cascade emit Cherenkov radiation recorded by the detectors, and the nitrogen in the air emits \ac{UV} fluorescence light.  Similar processes occur in satellite-borne detectors of gamma rays. 
In contrast to the first three messengers, which are detected as individual particles, \acp{GW} are oscillations of the spacetime curvature. Therefore, they are detected with antennas which measure their amplitude and phase. \acp{GW} interact very weakly with matter, thus propagate nearly uncontaminated by absorption and scattering from the source to the detector. 

\subsection{Particle showers}
\label{section4.1}
\label{section4.1.1}

Particle showers are cascades of secondary particles initiated by a primary particle (cosmic ray, gamma ray, or neutrino) in a certain medium.
Satellite-borne gamma-ray detectors use this phenomenon as one of the detection techniques. However, detectors of \acp{UHECR}, \ac{VHE} gamma rays, and neutrinos use naturally occurring media: air, water, and ice. We will explain the process on the example of air showers, keeping in mind that the underlying principle is the same for other media as well. Subtle differences result mostly due to different densities and optical properties.

A high-energy photon or electron/positron colliding with a nucleus~$N$ in the atmosphere  ($\approx 78\%$ nitrogen and $\approx 21\%$ oxygen) undergoes Bethe--Heitler pair production~$\gamma + N \to N + e^{+} + e^{-}$ or Bremsstrahlung~$e^{\pm} + N \to N + e^{\pm} + \gamma$, respectively, and the outgoing $e^\pm,\gamma$ in turn interact with other nuclei, and so on, generating an electromagnetic shower.  Energetic hadrons, such as \acp{UHECR}, produce hadronic showers via repeated hadronic interactions, $\text{hadron} + N \to N' + \text{hadrons}$, which include electromagnetic subshowers initiated by decay products of unstable hadrons (mainly $\pi^0 \to 2\gamma$) produced in previous interactions.  Muons just lose energy by ionization $\mu^{\pm} + N \to N^+ + e^-$ at a rate of~$\approx 2\,\mathrm{MeV/g\,cm^2}$, so they produce a track which can be detected via Cherenkov and/or fluorescence light but they still generally reach the surface with most of their initial energy and can penetrate deep into the ground, being a background for underground neutrino detectors. Neutrinos generally pass through the Earth unaffected, but can occasionally interact with electrons in matter via a neutral-current interaction (resulting in a high-energy electron, which then initiates an electromagnetic cascade) or a charged-current interaction (resulting in a high-energy charged lepton of the same flavor as the neutrino; the electron, muon or tau lepton then initiates a cascade, track or ``double bang'' respectively). 

In contrast to the showers produced in particle calorimeters, air showers develop in the atmosphere, whose density~$\rho \lesssim 10^{-3}\,{\rm g\,cm^{-3}}$, and extend over length scales of kilometers, hence the name \acf{EAS}. It is customary to measure the distance traveled by the shower as an atmospheric depth $X$ in ${\rm g\,cm^{-2}}$, giving the amount of air traversed by the shower from the top of the atmosphere in the direction of the \textit{shower axis}, i.e.~parallel to the direction of the primary particle initiating the shower. The total vertical atmospheric depth at sea level is $X_\text{v}\approx 1030\,{\rm g\,cm^{-2}}$, increasing with zenith angle $\theta$ approximately as $X_\text{v}/\cos\theta$, with $\theta$ measured from the vertical ($\theta=0^\circ$) to ground. 

A toy model to describe electromagnetic showers is due to Heitler \cite{Heitler:1936jqw}, while for hadronic showers a model was developed by Matthews \cite{Matthews:2005sd}. Despite their many simplifications, these models reproduce many important features of air showers that are also obtained with detailed Monte Carlo simulations and observed experimentally. 
The models also serve to shed light on the relation between the main shower observables and the nature of the primaries producing them, as well as the features of hadronic and electromagnetic interactions at UHE. 

\subsubsection{Electromagnetic showers}
The first interaction of a primary \ac{VHE} gamma ray in the atmosphere is mainly  
the electron-positron pair production in the Coulomb field of an atomic nucleus in the air --- so-called Bethe--Heitler process \cite{Bethe:1934za}. At high energies the Bethe--Heitler cross-section \cite{Bethe:1934za} is independent of the photon energy, 
\begin{equation}
 \sigma_\text{BH} = \frac{28Z^2\alpha^3}{9m_e^2}\left( \ln\frac{183}{Z^{1/3}}-\frac{1}{42}\right) ,
\end{equation}
where $\alpha$ is the fine structure constant and $Z$ is the atomic number of the nucleus. This is the standard result based on special relativity. In several \ac{QG} models with broken Lorentz symmetry the Bethe--Heitler cross-section may be reduced at high energies \cite{Vankov:2002gt, Rubtsov:2012kb}, as discussed in Sec.~\ref{sec:QEDLIV}.

The products of the first interaction, electron and positron, are deflected in the Coulomb field of another atomic nucleus, emitting photons through the Bremsstrahlung process \cite{Bethe:1934za, Tsai:1973py}. The resulting secondary photons, in turn, again produce electron-positron pairs through the Bethe--Heitler process. The cycle repeats, and at each step the number of particles (photons, electrons, positrons) increases exponentially while the mean energy of particles decreases. The cascade dies out when the mean particle energy falls below a certain critical energy $E_\text{c} \simeq 80\,\mathrm{MeV}$. Below this threshold, the energy loss rate on the process of ionization of air molecules starts to dominate over the rate of Bremsstrahlung.
At this stage (called the shower maximum) the shower contains the maximal number of particles. 

The mean free path $X_{0,\text{BH}}$ or radiation length $X_{0,\text{Br}}$ of a primary or secondary particle in the shower is inversely proportional to the corresponding cross-section, and is $X_{0,\text{BH}} = 47\,\mathrm{g\,cm}^{-2}$ for Bethe--Heitler process, and $X_{0,\text{Br}} = 37\,\mathrm{g\,cm}^{-2}$ for Bremsstrahlung \cite{Tsai:1973py} 
(compared to the vertical depth of the atmosphere at sea level of $\approx 1030\,\mathrm{g\,cm^{-2}}$). 
Hence, the shower usually starts and develops in high layers of atmosphere, at an altitude $\sim\,20$--$30$\,km \ac{asl}.

Some basic features of an electromagnetic shower can be obtained analytically in a simplified Heitler model \cite{1954qtr..book.....H}. The assumptions are as follows: i)~the mean free paths for Bethe--Heitler and Bremsstrahlung processes are set  equal, $X_{0,\text{Br}}=X_0$; ii)~in the final states of each reaction, the energy is distributed equally among the products. The Heitler model shows that the number of particles grows with distance along the shower axis, $N(x) = 2^{x/X_0}$, while the energy of each particle decreases with the distance, $E(x) = E_0\times 2^{-x/X_0}$, where $E_0$ is the energy of primary particle. At the shower maximum distance $x=X_{\max}$ the number of particles $N_{\max}=2^{X_{\max}/X_0}=E_0/E_\text{c}$ is proportional to the energy of the primary particle. Thus, the number of particles at the maximum can be used to reconstruct the primary energy. The second prediction of the Heitler model is  the logarithmic dependence of the shower maximum on primary photon energy,
\begin{equation}\label{eq:X_EM_max}
X_{\max} = X_0 \log_2 \frac{E_0}{E_\text{c}}.
\end{equation}
Although the Heitler model is based on rather rough assumptions (i and ii), its predictions are quite good and can be improved by more complete calculation including numerical simulations. Note that in \ac{QG} models the above assumptions may be violated, leading to a somewhat different shower development. 

Let us also note some important features of electromagnetic showers induced by \ac{EHE} gamma rays. First, for primary photon energies above $10^{18}$\,eV scattering in the Coulomb field of several nuclei becomes important. The destructive interference between several scattering centers somewhat suppresses both the Bethe--Heitler and Bremsstrahlung processes (LPM effect) \cite{Landau:1953um, Migdal:1956tc}. Second, for primary photon energies more than $\sim 10^{19.5}$\,eV the dominant channel for the first interaction in the Earth's neighbourhood is no longer the Bethe--Heitler process but the electron-positron pair production in the geomagnetic field. This process occurs at altitudes of several hundred kilometers above the Earth, outside of the atmosphere. The resulting electrons and positrons emit synchrotron radiation in the geomagnetic field, giving rise to a small electromagnetic cascade called \textit{preshower},  which gives a unique signature for the subsequent \ac{EAS} \cite{Risse:2007sd,Homola:2003ru}. 

As already mentioned, electromagnetic showers can also result from the decay of neutral pions, thus constituting subshowers of larger \acp{EAS}. 

%-------------------------------------------------
\subsubsection{Hadronic showers}
\label{section4.1.2}

When \iac{UHECR} proton or nucleus interacts with a nucleus of air, a large number of secondary particles are produced, with the multiplicity $N_{\rm mult}$ defined as the number of particles per collision, reaching values of tens to hundreds at \ac{UHE}. On average the collision takes place after the particle traverses one interaction length $\lambda_\text{i}(E_0)$, decreasing with the primary energy $E_0$ as the cross-section increases.  On average $\sim 30-40\%$ of the energy of the primary \ac{UHECR} is carried by a leading baryon or nucleus contributing to further developing the shower. The remaining energy is employed in the creation of ultra-relativistic secondary particles, most of them 
charged ${\pi^+, \pi^-}$ and neutral pions $\pi^0$, with a smaller number of heavier mesons such as charged and neutral kaons, $\rho$-mesons, etc\@. This implies that the energy of the primary particle rapidly degrades into a large number of secondaries that can subsequently interact or decay depending on the nature and energy of the particle, hence further contributing to shower development. 

\begin{figure}[!ht]
\includegraphics[width=\textwidth]{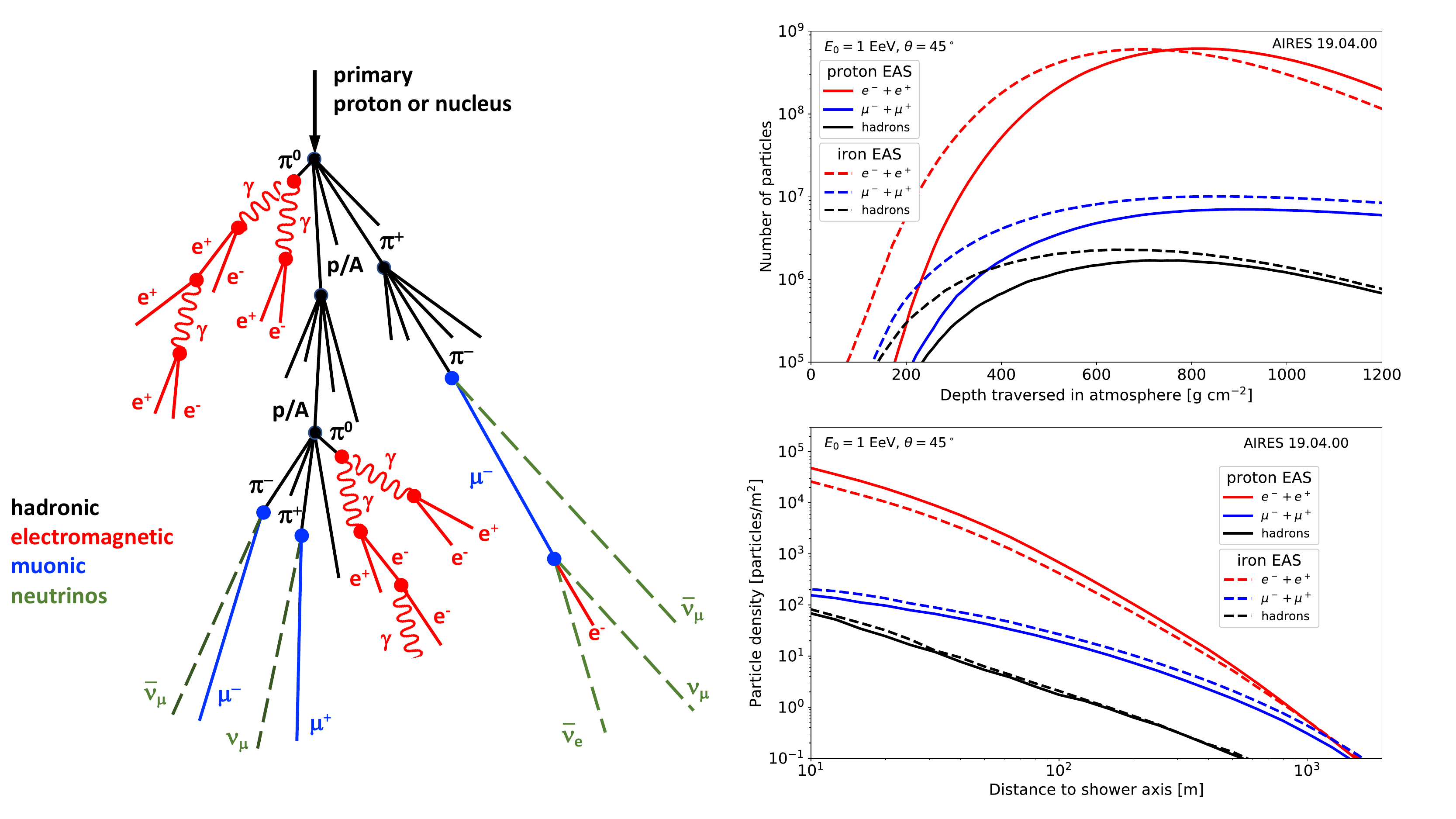}
\caption{Left panel: Sketch of an \ac{EAS} induced by \iac{UHECR} proton (p) or nucleus (A). Depicted are the hadronic (black), electromagnetic (red), muonic (blue) and neutrino (green) components. Right panel top: Profile along the direction of the shower axis of the number of particles in \acp{EAS}. Right panel bottom: Particle densities at ground as a function of the radial distance to the shower core measured in the shower plane perpendicular to shower axis. In both panels electrons and positrons (red), muons (blue), hadrons (black) were obtained with AIRES 19.04.00 Monte Carlo simulations of \acp{EAS} induced by a primary proton (solid lines) and an iron nucleus (dashed lines) of primary energy $E_0=1$\,EeV and arriving at angle $\theta=45^\circ$ with respect to the vertical to the ground.
}
\label{fig:EAS}
\end{figure}

Charged pions decay through the weak interaction and have a lifetime $\tau_{\pi^\pm}=2.6 \times 10^{-8}\,{\rm s}$. On average they travel a distance $d_{\pi^\pm}=\Gamma \rho c\tau_{\pi^\pm} = 780\rho\times\Gamma\,{\rm g \,cm^{-2}}$ before decaying, where $\Gamma=\left.E_{\pi^\pm}\middle/m_{\pi^\pm}\right.$ is the Lorentz factor.  
On the other hand, the interaction length of pions is $\lambda_{\pi}\simeq 120\,{\rm g\,cm^{-2}}$, also decreasing with energy. Once produced, if the $\pi^{\pm}$
live long enough they can encounter an air nuclei and interact before decaying. This happens if $d_{\pi^{\pm}} \gtrsim \lambda_\pi$, otherwise they typically decay. 
In general terms, high-energy $\pi^\pm$ with $E_{\pi^{\pm}}\gtrsim 10$\,GeV mostly interact because $\Gamma$ is large, while lower energy $\pi^\pm$ decay. At high altitudes, where the atmosphere is less dense, charged pions of higher energy are more likely to decay than at low altitudes. 
The stochastic competition between interaction and decay of charged mesons determines the details of the development of hadronic showers. Interacting charged pions along with the leading baryon and other hadrons, form a penetrating core of shower particles constituting the {\it hadronic component} of the cascade (see Fig.\,\ref{fig:EAS}). Charged pions decaying into muons, $\pi^+ \rightarrow \mu^+ + \nu_\mu$, $\pi^- \rightarrow \mu^- + \bar\nu_\mu$, represent the main contribution to the {\it muonic component} of the shower, being also responsible for the production of atmospheric neutrinos (Fig.\,\ref{fig:EAS}). 

On the other hand neutral pions, continuously produced in the hadronic core of the shower, have a lifetime $\tau_{\pi^0} = 8.4 \times 10^{-17}$\,s characteristic of electromagnetic interactions, much smaller than that of $\pi^\pm$. As a result most $\pi^0$ do not travel long enough distances to interact with an air nucleus unless their energy is well above $10^{18}\,\mathrm{eV}$. Instead, $\pi^0$ typically decay almost immediately into two photons $\pi^0\rightarrow \gamma\gamma$, initiating electromagnetic subshowers as described above. Neutral pions are mainly responsible for the {\it electromagnetic component}, that constitutes the bulk of the particles in an \ac{EAS}. 

The three distinct and intertwined shower components, hadronic, electromagnetic and muonic are sketched in Fig.\,\ref{fig:EAS}.

The decay of charged pions and muons within the showers generate a strong flux of atmospheric neutrinos. 
The essential features of the atmospheric neutrino flux were described from the convolution of the cosmic-ray flux and the neutrino yields, based on the hadronic interaction model~\cite{Zatsepin:1962}. Their spectrum is a steeply-falling power-law ($\sim E^{-3.7}$). Study of atmospheric neutrinos is interesting in its own, but also because at energies above a few tens of TeV they represent a major background in searches of astrophysical neutrinos.

From the point of view of \ac{VHE} gamma-ray observations, a very important task is to distinguish photon-induced showers from hadronic ones. The main difference being in their spatial structures. Photon showers are more homogeneous, thinner and axially symmetric about the primary gamma-ray direction. Hadronic showers, as we have seen, can contain several distinctive subshowers, and have a much higher muon content. 

\paragraph{Simple model of shower development} The basic features of a hadronic shower can be obtained in the Matthews model \cite{Matthews:2005sd}, where it is assumed that after the interaction of a UHE proton with energy $E_0$, a number $N_{\rm mult}$ of pions is produced, all with the same energy~$E_0/N_{\rm mult}$, and $1/3$~of them being $\pi^0$ and $2/3$~being $\pi^{\pm}$. The effect of the leading baryon in the collision is neglected. All neutral pions are assumed to decay immediately feeding the electromagnetic component of the shower, which, after the first generation of pions, carries an energy $E_0/3$, while the energy carried by the hadronic core of the shower (i.e.\ the charged pions) is $2E_0/3$. All charged pions are assumed to interact if their energy is above~$E_{\pi}^{\rm dec}\simeq 20\,{\rm GeV}$, below which $\pi^{\pm}$ are assumed to decay feeding the muonic component of the shower. In each collision of a $\pi^{\pm}$, it is again assumed that only pions are produced, $1/3$~of them being neutral, decaying immediately and starting new electromagnetic subshowers. The process goes on, and after $k$ generations the energy carried by the hadronic component of the shower is~\(
    E_{\rm had} \simeq (2/3)^k E_0
\),
while that of the electromagnetic component is~\(
    E_{\rm EM} = E_0 - E_{\rm had} \simeq \left( 1 - (2/3)^k \right)E_0 
\).
This implies that after only $k=6$~generations, more than $90\%$ of the primary energy goes into the electromagnetic component of the shower, reflecting its predominance over the hadronic and muonic components as observed experimentally. After $k$~generations, the energy per pion is $E_\pi=E_0/(N_{\rm mult})^k$. The production of pions stops when $E_\pi=E_{\pi}^{\rm dec}$ and all charged pions decay producing muons. 
The maximum number of generations in the shower $k_{\rm max}$ is then given by the number of generations needed for the energy per pion $E_\pi$ to reach $E_{\pi}^{\rm dec}$,
\begin{equation}
E_\pi=\frac{E_0}{(N_{\rm mult})^{k_{\rm max}}} = E_{\pi}^{\rm dec} 
\Rightarrow 
k_{\rm max} = \frac{\ln(E_0/E_{\pi}^{\rm dec})}{\ln N_{\rm mult}}.
\end{equation}
Since one muon per decaying charged pion is produced, an estimate of the number of muons in the shower $N_\mu$ is given by:
\begin{equation}
N^p_\mu = N_{{\pi}^{\pm}} = 
\left(\frac{2}{3}\,N_{\rm mult}\right)^{k_{\rm max}} 
\Rightarrow
N^p_\mu=\left(\frac{E_0}{E_{\pi}^{\rm dec}}\right)^\beta
~{\rm with}~
\beta=1+\frac{\ln(2/3)}{\ln{N_{\rm mult}}}.
\label{eq:mu}
\end{equation}
Typical values of $\beta$ range from 0.85 to 0.95, reflecting that the number of muons is almost proportional to the primary energy $E_0$. The average energy per muon is roughly $E_\pi^{\rm dec} /2 \simeq 10\,{\rm GeV}$. Eq.~\eqref{eq:mu} also reflects the dependence of the scaling of $N_\mu^p$ with energy on a property of hadronic interactions such as the total multiplicity. Since energetic muons only lose energy through ionization and barely interact or decay, they travel through the atmosphere almost unattenuated, typically reaching ground level, except for those produced in showers with very large zenith angles $\theta$. 

Exploiting the Matthews model, we can obtain an estimate of the average depth $\langle X^p_{\rm max}\rangle$ at which the number of particles in a proton-initiated shower is maximum. Most of the particles in an \ac{EAS} are $e^-$, $e^+$ and $\gamma$ produced in the electromagnetic subshowers initiated by photons produced in the decay of neutral pions. Also, the largest subshowers in terms of number of particles are those initiated by the most energetic $\pi^0$s of the first generation. Taking these two facts into account, an approximate estimate of $\langle X^p_{\rm max}\rangle$ is given by the depth of maximum $X^\gamma_{\rm max}$ of electromagnetic subshowers initiated by the photons generated in the decay of $\pi^0$s produced in the first generation and having energy $E_\gamma=E_0/2N_{\rm mult}$. Using Eq.~(\ref{eq:X_EM_max}) for $X^\gamma_{\rm max}$:
\begin{equation}
\langle X^p_{\rm max} \rangle \simeq \lambda_\text{i} + X^\gamma_{\rm max}\!\left(\frac{E_0}{2N_{\rm mult}}\right) \simeq \lambda^p_\text{i} + X_0 \log_2\frac{E_0}{2 N_{\rm mult}E_\text{c}},
\label{eq:Xmax_p}
\end{equation}
where $\lambda^p_\text{i}$ is the mean free path of the proton. 
Eq.\,(\ref{eq:Xmax_p}) reflects the dependence of 
$X^p_{\rm max}$, an observable accessible in many experiments, on the properties of
the first UHE interaction: namely, on one hand the cross-section~$\sigma_{p\text{--air}}$   through $\lambda_\text{i}=\langle A \rangle m_p/\sigma_{p\text{--air}}$ where $\langle A \rangle\simeq14.5$~is the average mass number of air and $m_p$~is the proton mass, and on the other hand the total multiplicity~$N_{\rm mult}$ with a weaker dependence \eqref{eq:mu}.

%-------------------------------------------------
If the primary cosmic rays are heavier nuclei the main observables, namely the depth of maximum development~$\langle X^A_{\rm max}\rangle$ and the number of muons~$N^A_\mu$, can be easily obtained on the basis of the so-called {\it superposition model} \cite{Glauber:1970jm} considering that the shower produced by a cosmic-ray nucleus of mass number~$A$ and energy~$E_0$ is equivalent to a superposition of $A$ showers produced by $A$~individual nucleons of energy~$E_0/A$ each. Following this model:
\begin{align}
& \langle X^A_{\rm max}(E_0) \rangle \simeq \langle X^p_{\rm max} (E_0/A) \rangle = \lambda^A_\text{i} + X_0 \log_2 \frac{E_0}{2N_{\rm mult}E_\text{c}} - X_0 \log_2 A \\
&\Rightarrow \langle X^A_{\rm max}(E_0) \rangle - \langle X^p_{\rm max}(E_0) \rangle \simeq X_0 \log_2 A,
\end{align}
making apparent the dependence of $X_{\rm max}$ on the mass $A$ of the primary particle initiating the shower (see Fig.\,\ref{fig:EAS}).
As for the number of muons in a nucleus-initiated shower, using the superposition model and Eq.\,(\ref{eq:mu}) it is given by
\begin{equation}
N_\mu^A (E_0) = A\times N_\mu^p\!\left(\frac{E_0}{A}\right) =  
A\times\left(\frac{E_0}{A E_{\pi}^{\rm dec}}\right)^\beta = 
A^{1-\beta}\times N_\mu^p(E_0).
\end{equation}
This implies that for a given primary energy the number of muons is larger in showers induced by heavier nuclei than in showers induced by protons. For instance an iron-induced shower has $\sim 1.5$ times as many muons than a proton shower of the same energy, making this observable a valuable discriminator of primary mass (see Fig.\,\ref{fig:EAS}). 

On the other hand the number of electrons and positrons in showers induced by protons or heavier primaries is similar, 
\begin{equation}
N_e^A(E_0)\simeq A\times N_e^p\!\left(\frac{E_0}{A}\right) \simeq A\times \frac{E_0}{A E_\text{c}} = N_e^p(E_0),
\end{equation}
implying that the electromagnetic size is a good estimator of shower energy regardless of the composition of the \ac{UHECR} flux. 

%-------------------------------------------------
\paragraph{Monte Carlo simulations} Besides this simplified but yet powerful model, a more detailed treatment of shower development accounting for the stochastic nature of the hadronic, weak and electromagnetic interaction and transport processes involved requires the use of sophisticated Monte Carlo simulations of shower development in the atmosphere. While the electromagnetic and weak interactions are rather well understood, \ac{QCD} cannot be applied to calculate from first principles the features of hadronic interactions such as the $p/A$--air cross section, multiplicity and energy and momentum distribution of secondaries; instead, one has to rely on hadronic interaction models, based on theoretical ideas and empirical parameterizations and extrapolations tuned to describe collider and accelerator data.  The hadronic interactions in the first few generations of an \ac{EAS} occur with very high center-of-mass energies where we lack knowledge of their properties: the first interaction of a cosmic ray with energy~$E$ has~$\sqrt{s} \approx 43\sqrt{E/\mathrm{EeV}}$\,TeV, which for~$E \gtrsim 10^{17}$\,eV is greater than achieved at the LHC\@.  Subsequent interactions are mostly pion-initiated rather than nucleon-initiated, and there are few direct laboratory measurements of the former.  Moreover, most secondaries in the \acp{EAS} are moving in the forward direction, almost parallel to the direction of the primary \ac{UHECR}, where typically collider experiments do not register particles because of the presence of the beam itself, but where most of the energy of the \acp{EAS} flows.

Widely used Monte Carlo codes simulating in great detail \acp{EAS} initiated by UHE photons, nucleons, or nuclei (and even neutrinos) are the AIRES \cite{AIRES:2019} and CORSIKA \cite{Heck:1998vt} programs, which incorporate different models to describe hadronic interactions at high energy such as \textsc{EPOS}~\cite{Pierog:2019opp}, \textsc{QGSJet}~\cite{Ostapchenko:2019few} and \textsc{Sibyll}~\cite{Engel:2019dsg}.  There are sizable differences among these models, severely limiting our ability to estimate the mass numbers of primary \acp{UHECR}: for example, the $\left< X_{\max} \right>$ predictions by the latest versions of \textsc{Sibyll} and \textsc{QGSJet} differ by~$26\,\mathrm{g}/\mathrm{cm}^2$, corresponding to a factor of~$3$ in the mass number estimated from a given observed~$X_{\max}$ (the latest \textsc{EPOS} giving intermediate predictions).  Furthermore, all of these models predict average numbers of muons in showers significantly in deficit of the observations \cite{Cazon:2020zhx}; the observed shower-to-shower fluctuations in muon number are in agreement with the predictions \cite{PierreAuger:2021qsd}, showing that the mismatch in the averages cannot be due to only a major effect affecting the extreme-energy interactions at the top of the showers, but must be due to a small effect accumulating throughout the shower development including in lower-energy interactions near the bottom.

In Fig.~\ref{fig:EAS} the longitudinal profile of electrons and positrons, muons and hadrons is shown for showers induced by protons and iron nuclei simulated with AIRES. As can be seen in the figure, the depth at which the number of electrons + positrons is maximum, namely $X_{\rm max}$, is smaller in  iron-induced showers than in proton-induced \acp{EAS}, while the number of muons $N_\mu$ is larger, in agreement with the simple models described above. 

Another important property of the showers is the lateral distribution of the density of particles at ground level. This is a decreasing function of the radial distance from the shower core measured in the shower plane orthogonal to the shower axis, as shown in Fig.~\ref{fig:EAS}. The lateral distribution of the electromagnetic component of an \ac{EAS}, which gets constantly regenerated by the hadronic core and by muon decays, is mainly due to multiple Coulomb scattering. The muonic component of hadronic showers, which is typically produced high in the atmosphere, has a flatter lateral profile inheriting the lateral spread of the parent charged pions. At UHE energies the signals can be measured up to a few km from the shower core. 

Monte Carlo simulations also allow to model fluctuations between properties of showers of the same energy. In particular, fluctuations of $X_{\rm max}$, generally quantified by the standard deviation of its distribution $\sigma(X_{\rm max})$, are quite different for different primary masses. The spread in $X_{\rm max}$ in proton-induced \acp{EAS} arises mostly from the fluctuations in the depth at which the first interaction occurs, with contributions from the fluctuations in the subsequent shower development, reflecting the dependence of $\sigma(X_{\rm max})$ on hadronic interactions at \ac{UHE} and \ac{UHECR} composition. For showers initiated by heavier nuclei the fluctuations are typically smaller because these can be considered as a superposition of $A$ showers of energies $E_0/A$ whose fluctuations get averaged especially at large values of $A$. In addition to the fluctuations in the depth at which the first interaction occurs, the width of the distribution reflects the spread of nuclear masses at the \ac{UHECR} sources and the modifications that occur during their propagation to the Earth \cite{PierreAuger:2013xim}. 

In general any physics process capable of modifying the interaction or decaying lengths of particles with respect to the values predicted by the Standard Model, is prone to have sizable effects on \acp{EAS} development (Sec.~\ref{hadronic}), mainly on $X_{\rm max}$ but also on the muon size of the shower. For instance the kinematics of $\pi^0$ decay can be affected by \ac{LIV}, forbidding $\pi^0$ decay above some energy threshold which instead interact, affecting $X_{\rm max}$ \cite{Lang:2020geh} and the number of muons produced. Monte Carlo simulations are also a powerful tool to test the effect that \ac{BSM} physics may have on shower observables.

\subsubsection{Cherenkov and fluorescence radiation from showers}
\label{sec:cherenkov-and-fluorescence}
The most efficient method for observing gamma rays of astrophysical origin in energies exceeding hundreds of GeV
is to observe \acp{EAS} by ground-based detectors (see Sec.~\ref{section4.2.1}). The particles composing the shower can be detected directly (by water Cherenkov detectors) or indirectly, through the Cherenkov radiation emitted by the charged particles from the shower in the atmosphere (observed by \acp{IACT}).

The particles inside the air shower have velocities larger  than the speed of light in the air. This is a condition for Cherenkov radiation: a charged particle, traveling faster than the speed of light in a dielectric medium, emits Cherenkov radiation in a narrow cone with angle $\theta_\text{C}$ with respect to its trajectory. The angle $\theta_\text{C}$ depends on the refractive index of media $n$ and the charged particle velocity~$\beta$ (vacuum speed of light is set to unity),
\begin{equation}
\cos \theta_\text{C} = \frac{1}{\beta n}.
\end{equation}
Since $n \simeq 1.00029$ for air (at sea level), the angle $\theta_\text{C} \sim 1^\circ$ for an ultrarelativistic charged particle. Thus, the shower core is accompanied by a cone composed of a large number of Cherenkov photons, which can be detected by air Cherenkov detectors. The observed spectrum of Cherenkov photons peaks at near-UV regime, $\lambda \sim 330$\,nm \cite{McCann:2012gef}. The most of the observed Cherenkov photons comes from the shower maximum.  Similarly, neutrino telescopes (Sec.~\ref{section4.2.2}) detect the Cherenkov light from tracks and cascades in water and ice; in this case, $n \approx 1.3 \to \theta_\text{C} \approx 40^\circ$.

For \acp{EAS} with energy above~$10^{17}$\,eV, such as those initiated by \acp{UHECR}, the dominant emission is the \ac{UV} fluorescence light (300\,nm--430\,nm wavelengths) emitted near-isotropically by the atmospheric N$_2$ molecules excited by the charged particles in the shower.  The number of fluorescence photons emitted is proportional to the calorimetric energy, i.e.\ the energy deposited in the atmosphere due to electromagnetic energy losses by the charged particles, which is $\sim 90\%$ of the total energy of the shower; the rest is the energy dumped into the ground by muons and neutrinos reaching the surface, known as ``invisible energy''.  See Sec.~\ref{uhecr-detectors} for more details about \ac{UHECR} energy estimation.

\subsection{Detection of gravitational waves}
\subsubsection{Single sources}
\newcommand{\Mc}{\mathcal{M}_\text{c}}

\Ac{GW} detectors are nearly omni-directional antennas that register \acp{GW} coherently, measuring both their amplitude and their phase \cite{2011LRR....14....5P}. Single sources of \acp{GW} include compact-object binary systems, signals from asymmetric dynamical processes for individual compact objects, and short burst-like signals. The latter are detected by looking for excess power in the detector outputs. The main detection method for the former two is based on matched filtering, where the data is cross-correlated with theoretical models; other methods also exist but are less sensitive~\cite{LIGOScientific:2019hgc}. For brevity, we focus here on binary systems.

For a binary system with component masses $m_1$ and $m_2$, the dimensionless time-domain \acp{GW} polarizations are to lowest-order given by
\begin{align}
    h_+ &= -A \frac{1+\cos^2\iota}{2} \cos\left(\int 2\pi f_\text{GW} \,dt\right);&
    h_\times &= -A    \cos  \iota  \sin\left(\int 2\pi f_\text{GW} \,dt\right),
\end{align}
where $\iota$~is the inclination angle between the orbital angular momentum and the line of sight. The amplitude is, to lowest order, given by 
\begin{equation}
        A = \frac{ 4 (G\Mc)^{5/3} (\pi f_\text{GW})^{2/3 }}{d_\text{L} c^4},
\end{equation} 
where $d_\text{L}$ is the luminosity distance and \begin{equation}
    \Mc = (1 + z) \left(m_1 m_2\right)^{3/5} \left(m_1 + m_2\right)^{-1/5}
    \label{eq:chirpmass}
\end{equation}
is the \textit{redshifted chirp mass} of the binary and $f_\text{GW}(t)$ is the frequency of the emitted \acp{GW}. 
To lowest order, the frequency increases with time according to 
\begin{equation}
    f_\text{GW}(t) \approx \frac{1}{8\pi} \left[\frac{125}{(G\Mc/c^3)^5 (t_\text{c} - t)^3}\right]^{1/8},
\end{equation}
where $t_\text{c}$ is the time of coalescence. This illustrates the characteristic `chirp' signals from binary inspirals associated with an increase in amplitude and frequency over time. It also shows that the phase evolution depends on the intrinsic parameters of the binary (here: the chirp mass), while the amplitude also involves the relation to the observer (such as the luminosity distance). Such chirp signals have different properties than noise, and can be extracted by matched filtering. Higher order relativistic corrections to the phase and amplitude depend on different source parameters, including the mass ratio, spins of the objects, various coefficients characteristic of their interior structure, and properties of the dynamics such as eccentricity. Although not shown in detail here, the higher order effects in the waveform are also impacted by several of the \ac{QG} phenomena discussed in the previous sections, including modified strong-field dynamics, corrections to black hole horizons, new gravity-matter couplings in binaries involving neutron stars, the presence of \ac{BSM} fields either as exotic compact objects or in the surrounding environments of black holes or neutron stars, and propagation effects. Several of these effects enter through generic phenomena such as spin-and tidally induced deformations, or modified quasi-normal-mode spectra. Other features of the waveform are theory-dependent.

It is often useful to consider the properties of the frequency-domain Fourier amplitude of the time-domain signal. Its root-mean-square value, averaged over inclinations and polarizations, is to lowest order
\begin{equation}
    \bigl|\tilde{h}(f)\bigr| = \sqrt{\frac{4\pi}{3}} \frac{c}{r} \left(G\Mc/c^{3}\right)^{5/6} (\pi f)^{-7/6}.
\end{equation}
Signal amplitudes are conveniently compared
with detector noise, by first converting all spectra into \textit{characteristic strain}, which is
    $h_\text{c}(f) = 2f\bigl|\tilde{h}(f)\bigr|$ for the signal and 
    $h_\text{n}(f)= \sqrt{f S_\text{n}(f)}$ for the detector noise, 
where $S_\text{n}(f) $ is the power spectral density of latter.

A useful benchmark for what kinds of binary systems are promising \acp{GW} sources and how many events one can expect is obtained by considering that binary systems that emit \acp{GW} at a frequency of $f_\text{GW}$ will coalesce within a timescale of 
\begin{equation}
    \tau_\text{merge} = 1.4 \times 10^{3}\,\mathrm{yr}
        \left(\frac{\Mc}{2M_\odot}\right)^{-5/3}
        \left(\frac{f_\text{GW}}{0.01\,\mathrm{Hz}}\right)^{-8/3}.
\end{equation}
There are many different types of binary systems whose merger time is less than a Hubble time and are thus promising \acp{GW} sources for current and planned detectors described in the next subsection. For black holes, the merger frequency is inversely proportional to the total mass, which sets an upper limit on the masses of black holes observable with each instrument. Roughly speaking, current detectors are mainly sensitive to stellar-mass binaries, while signals involving supermassive black holes are important for future space-based detectors.
The center of every galaxy likely hosts a super massive black hole with mass 
in the range of $10^6$--$10^9 M_\odot$ \cite{1997CQGra..14.1411R}.  When galaxies collide, friction between the super massive black holes and the stars and gas of the merged galaxy may cause the super massive black holes to spiral into a common nucleus and get close enough for their orbit to decay due to \ac{GW} emission within 
a Hubble time, in some systems \cite{2003ApJ...582L..15K}. For a wide range of super massive black hole masses, the frequency of \acp{GW} produced during the inspiral, merger and ringdown of a binary black hole merger is within the sensitivity range of \ac{LISA} (which will be able to observe any merger that happens in its frequency band anywhere in the universe). 

Sources with mass ratio $m_1/m_2$ (with $m_1>m_2$) in the range $\sim 10^{-3}$--$10^{-6}$ are called extreme mass ratio inspirals, whereas possible sources with mass ratio in the range $\sim 10^{-2}$--$10^{-3}$ are termed intermediate mass
ratio inspirals. The signals from such extreme or intermediate mass inspirals will contain an enormous number of $\sim$ $m_1/m_2$ cycles of waveform, which will enable extremely high precision study. For instance, the emitted \acp{GW} \cite{1995PhRvD..52.5707R,2002PhRvD..66d4002G,2002PhRvD..66f4005G,2004PhRvD..69h2005B,2006PhRvD..73f4037G,2007PhRvD..75b4005B} encode a very
clean map of the spacetime geometry of the massive black hole \cite{1997PhRvD..56.1845R,2001CQGra..18.4067H,2005CQGra..22S.605G,2006CQGra..23.4167G,2007PhRvD..75d2003B}.

Beyond the inspiral epoch, a binary black hole system will undergo a nonlinear merger into a final remnant black hole, which is initially highly perturbed, and radiates away the perturbations through its characteristic quasi-normal modes. This is called the ringdown epoch. For a pair of super massive black holes, one can estimate the effective amplitude in the ring-down phase as %\begin{equation}
%    h_\text{eff} \simeq 3 \times 10^{-17}
%    \left(\frac{\nu}{0.25}\right)
%    \left(\frac{M}{2 \times 10^6 M_\odot}\right)
%    \left(\frac{r}{6.5\,\mathrm{Gpc}}\right)^{-1},
%\end{equation}
\begin{equation}
    h_\text{eff} \simeq 3 \times 10^{-17}
    \left(\frac{\nu}{0.25}\right)
    \left(\frac{m_1 + m_2}{2 \times 10^6 M_\odot}\right)
    \left(\frac{r}{6.5\,\mathrm{Gpc}}\right)^{-1},
\end{equation}
where $\nu = m_1 m_2/(m_1 + m_2)^2 \in (0, 0.25]$ is the \textit{symmetric mass ratio}. A pair of $10^6 M_\odot$ super massive black holes would thus produce a signal with very large signal-to-noise ratio in comparison to the \ac{LISA} background noise. For a black hole, the no-hair theorem predicts a specific relationship between different quasi-normal modes, which depend only on its mass and spin. Accurately measuring multiple modes and looking for potential echoes associated with horizon-scale modifications~\cite{Abedi:2020ujo,Cardoso:2019rvt}, as will be possible for high signal-to-noise ratios, will enable stringent tests of the no-hair theorem, and constraints on horizon-scale physics. Such tests, albeit with large statistical errors, are already being performed with current detectors for the stellar-mass black hole remnants.

Mergers of compact object binary systems can also be used as standard sirens to estimate the Hubble constant \cite{Schutz:1986gp,Holz:2005df,Dalal:2006qt,Nissanke:2009kt,Abbott:2017xzu}. This can be exploited to test a feature shared by many \ac{QG}, namely a non-perturbative effect which underlines the way the dimension of spacetime changes with the probed scale. This dimensional flow influences the \ac{GW} luminosity distance, the time dependence of the effective Planck mass, and the instrumental strain noise of interferometers. Investigating the consequences of \ac{QG} dimensional flow for the luminosity distance scaling of \ac{GW}
s in the frequency ranges of LIGO and \ac{LISA}, it was shown that the quantum geometries of \ac{GFT}, spin foams and \ac{LQG} can give rise to observable signals in the \ac{GW} spin-2 sector~\cite{Calcagni:2019ngc,Calcagni:2019kzo}. 

Quantum black holes are expected to have a discrete energy
spectrum, and to behave in some respects like excited
atoms \cite{1974NCimL..11..467B,1995PhLB..360....7B,1986JETPL..44...63M}. In \cite{2020arXiv200713761A} the authors find that black hole area discretization could impart observable imprints to the \ac{GW}  signal from binary black hole merger, affecting the absorption properties of the black holes during inspiral and their late-time relaxation after merger. The rotation of binary black holes seems to improve the prospects to probe such quantum effects.

The most common detection methods for binary signals are based on matched filtering, searching over large parameter spaces of potential, pre-calculated signals. For burst searches, morphology-independent methods are used. For source localization, simultaneous observations using several detectors are required, so that the
source can be triangulated in the sky (the time differences in the signal arrival times
at the various detectors are used for this). For very long-lived sources, however, one can take advantage of the motion of a single detector around the Sun. With a baseline of 2\,AU, the angular resolution can be better than $1''$ for a source emitting at 1\,kHz, for ground-based detectors or of order 1\,rad for a space-based detector, like \ac{LISA} (observing at frequencies several orders of magnitude lower). For very strong \ac{LISA} sources, with signal-to-noise ratios in the range of  $\sim 10^3$--$10^4$, the angular resolution can be arcminutes.

The O1, O2, and O3a science runs of the LIGO-Virgo detector network resulted in dozens of clear binary black hole merger detections, as well as a few events involving one or two neutron stars.  Independent analyses of the publicly released LIGO-Virgo data reported a few additional significant binary black hole event candidates \cite{Zackay:2019tzo,Venumadhav:2019tad,Venumadhav:2019lyq}.  From these observations, a binary black hole merger event rate of 9.7--101\,${\rm Gpc}^{-3}\,{\rm year}^{-1}$ has been inferred~\cite{LIGOScientific:2018mvr}. From the \ac{BNS} merger event GW170817 alone, the \ac{BNS} event rate has been estimated as 110--3840\,${\rm Gpc}^{-3}\,{\rm year}^{-1}$ \cite{LIGOScientific:2018mvr}. The prospects of observing \ac{GW} transients in the next planned scientific runs of the LIGO-Virgo-KAGRA network (O4 and O5) are discussed in \cite{KAGRA:2013rdx}.

\subsubsection{Stochastic GW background \label{sec:stochGWsect}}
The superposition of countless discrete sources, such as white dwarf binaries are expected to produce a \ac{GW} foreground (detectable with \ac{LISA}), whereas fundamental processes, such as the Big Bang, may have produced \iac{SGWB}, which would have been redshifted to low frequencies.
Therefore, the \ac{SGWB} is produced by two main classes of sources that are unresolved in both the time and angular domain: the astrophysical component generated by galactic and extra-galactic sources, and the cosmological component induced by phenomena that occurred in the early universe.

The energy density of a stochastic field of \acp{GW} per unit logarithm of frequency is \cite{Allen:1996vm,Thorne:1987af} $d \rho_\text{GW}/{d \ln f}=4 \pi^{2} f^{3} S_\text{GW}(f)$, where $S_\text{GW}(f)$ is  a statistical mean
square amplitude per unit frequency. The energy density as a fraction of the closure or critical
cosmological density, given by the Hubble constant $H_0$ as $\rho_\text{c} = 3H_0^2/8\pi$. The resulting ratio is
called $\Omega_\text{GW}(f)$, which is the energy density per logarithmic interval of frequency normalized over the critical energy density:
\begin{equation}
\Omega_\text{GW}(f)=\frac{10 \pi^{2}}{3 H_{0}^{2}} f^{3} S_\text{GW}(f).
\end{equation}

The characterization of the \ac{SGWB} sources requires to disentangle the \ac{SGWB} components. A natural way to do so is by fitting the frequency profiles $\Omega_{\rm GW}(f)$ of each class of \ac{SGWB} sources. Indeed, the frequency dependence of $\Omega_{\rm GW}(f)$ tend to differ from source to source. A further method is to exploit the different level of anisotropy provided by the different sources. 
In the long term, both approaches are considered promising and we discuss them in detail in later on.

\Iac{SGWB} signal can be split into the sum of an isotropic and an anisotropic component. In this distinction,  the angular resolution of a given detector has to be taken into account. Current detectors are for instance sensitive to the anisotropy level of the galactic sources (aligned to the galactic disk), but not to the one of the large structure formations. Detectable anisotropies appear as periodic modulation of the overall signal when the detector is rotating in the background reference frame. Such rotation does not instead affect the isotropic component of the signal, and $\Omega_{\rm GW}(f)$  tends to differ from a given class of sources to an other. For instance, for detectors not sensitive to the anisotropic distribution of the extra-galactic binaries, the overall unresolvable extra-galactic events are expected to be detected as \iac{SGWB} with energy spectrum $\Omega_{\rm GW}(f)\propto f^{2/3}$, whereas the \ac{SGWB} energy spectra from the early universe have radically different predictions for $\Omega_{\rm GW}(f)$.

There are several \ac{SGWB} searches in place in the current \ac{GW} detectors. The LIGO and Virgo Collaborations have performed searches with the power-law template  $\Omega_{\rm GW}(f)\propto f^{\alpha}$, with $\alpha \in [-8,8]$, and found no evidence of \iac{SGWB} signal in the Hz frequency range~\cite{LIGOScientific:2019vic}. On the contrary, the NANO\-Grav Collaboration has performed a similar search in the nHz frequency range,  unveiling \iac{SGWB} at a certain level of confidence~\cite{Arzoumanian:2020vkk}. Analyses with more data confirming or disproving this excess are expected in around a year from now.

The power-law template is an excellent approximation for some astrophysical \acp{SGWB} but by far not ideal for other expected signals. Searches adopting more complex frequency shape exist. Guessing right the template enhances the experimental sensitivity to a given signal (see ref.~\cite{Abbott:2017mem} for a detailed example). On the other hand, nature may surprise us with an unexpected \ac{SGWB} signal. For this reason, template-free searches giving hints on the frequency shape of the overall \ac{SGWB} signal, are worthy~\cite{Karnesis:2019mph, Caprini:2019pxz, Pieroni:2020rob, Flauger:2020qyi}. Till now, no clear evidence of \iac{SGWB} incompatible with a power-law \ac{SGWB} has been detected.

Apart from the isotropic component, the astrophysical and cosmological \acp{SGWB} are characterized by anisotropies, generated both at the moment of production and during the propagation. Such anisotropies contain information about the angular distribution of the sources and can be used as a tracer of astrophysical or cosmological structure.
They can also be used to test the particle physics content of the universe~\cite{DallArmi:2020dar}. 

The literature proposes several theoretical formalisms for calculating the energy density of the \ac{SGWB} and its anisotropies~\cite{Contaldi:2016koz, Cusin:2017fwz, Jenkins:2018nty,Bartolo:2019oiq, Bartolo:2019yeu,Bertacca:2019fnt}, and techniques to detect them at current and future detectors are nowadays very sophisticated~\cite{Allen:1996gp, Giampieri:1997ie, Cornish:2001hg, Ungarelli:2001xu, Seto:2004np, Kudoh:2004he, Ballmer:2005uw, Taruya:2005yf, Taruya:2006kqa, Hotinli:2019tpc, Mingarelli:2013dsa,Renzini:2019vmt, Contaldi:2020rht, Mentasti:2020yyd}. 
Recently, some of them have been tuned to probe the anisotropies induced by cosmic strings~\cite{Jenkins:2018nty} and compact binary mergers~\cite{Jenkins:2018uac, Payne:2020pmc}. The overall outcome is that the characterization of the anisotropic component has enormous potential as a novel probe of the large-scale structures of the universe, and of late-universe cosmology~\cite{Jenkins:2018kxc}, although several subtleties must be properly taken into account to achieve the correct data interpretation~\cite{Bartolo:2018evs, Bartolo:2018rku, Bertacca:2019fnt, Jenkins:2019nks, Margalit:2020sxp}.

The LIGO and Virgo Collaborations have produced upper limits on the measurements of the \ac{SGWB} anisotropy in the 20--500\,Hz band for different frequency spectra (namely for sources of astrophysical or cosmological origin), and for point as well as extended source distributions on the sky~\cite{LIGOScientific:2019gaw}. These results probe different types of anisotropy by means of complementary techniques, namely spherical harmonic decomposition of the \ac{GW} power on the sky, a broad-band radiometer analysis, and a directed narrow-band radiometer at the frequency spectrum for  astrophysically interesting directions.  Similar techniques for measuring \ac{SGWB} anisotropy in the 1\,mHz band using \ac{LISA} are being developed~\cite{Adams:2013qma}.

Beside the frequency-shape and anisotropy tests, there have been explored other tools potentially useful to disentangle and characterize the \ac{SGWB} components. Among these, the chirality can have an important role. To our knowledge, there are no astrophysical sources producing more of a given circular \ac{GW} polarization than the other. The observation of chirality in the \ac{SGWB} would then be a clear indication of detection of  a cosmological component~\cite{Lue:1998mq,Jackiw:2003pm, Alexander:2004wk,Contaldi:2008yz,Takahashi:2009wc,Satoh:2010ep,Sorbo:2011rz, Barnaby:2011qe,Maleknejad:2011jw, Shiraishi:2013kxa,Creminelli:2014wna,Maleknejad:2016qjz,Dimastrogiovanni:2016fuu, Domcke:2016bkh, Bartolo:2017szm,Domcke:2017fix,Domcke:2018rvv,McDonough:2018xzh,Qiao:2019hkz,Mirzagholi:2020irt,Watanabe:2020ctz, Bartolo:2020gsh,Bordin:2020eui, Almeida:2020kaq}. 
 For an isotropic \ac{GW} background, single planar detectors are unable to detect chirality~\cite{Seto:2007tn, Seto:2008sr, Smith:2016jqs}. A planar interferometer indeed responds in the same way to a left-handed \ac{GW} arriving perpendicular to the plane of the detector or to a right-handed \ac{GW} of the same amplitude coming from the opposite direction. Such a degeneracy is however broken if the \ac{SGWB} is anisotropic, in which case a single planar detector becomes sensitive to chirality~\cite{Domcke:2019zls}. For a recent estimate of the sensitivity to circular polarization of \ac{LISA} alone or together with Taiji, see e.g.~Refs.~\cite{Domcke:2019zls,Orlando:2020oko}. 
 
Another crucial observable to distinguish a cosmological \ac{SGWB} from other \ac{SGWB} signals  is the measurement of non-Gaussianity. In fact  it is plausible that the \ac{SGWB} produced by incoherent astrophysical sources is Gaussian, due to the central limit theorem. So, a detection of non-Gaussianity would be a signal of large scale coherency, likely causes by the presence of a cosmological component in the signal~\cite{Bartolo:2019oiq, Bartolo:2019yeu}. 
 
Finally, discovering extra-polarizations in the \ac{SGWB} (besides the ``plus" and ``cross" polarization predicted by \ac{GR}) would be a milestone in the field. Many alternative theories of gravity predict the existence of vector and/or scalar modes in addition to standard tensor polarizations. The detection of these extra polarization modes would represent a clear violation of \ac{GR}, while a non-detection would help to experimentally constrain extended theories of gravity. Till now, no evidence has been found~\cite{Abbott:2018utx}.

\subsection{Currently operating and planned detectors}
\label{section4.2}

\subsubsection{Gamma-ray detectors}
\label{section4.2.1}\label{section4.2.1.1}\label{iact} \label{water-cherenkov}
Due to the opaqueness of the Earth's atmosphere to gamma rays, direct ground-based observations are not possible. Instead, the observations are performed using either satellite-borne detectors, or ground-based observations of \acp{EAS}. Space-based and ground-based gamma-ray telescopes offer complementary capabilities for the study of \ac{LIV}. 
Measurements with the satellite-borne instruments have the big advantage to operate all the time, and cover the whole sky, allowing to gather statistics on multiple events (\acp{GRB}, \ac{AGN} flares, etc.). 
They probe a large range of distances because the universe is almost transparent to MeV and GeV photons. 
Their effective areas, however, are not very large, and given that the gamma-ray flux generally decreases as a power of energy, satellite-borne detectors will be less suited for detection of gamma rays with energies $\gtrsim100\,\mathrm{TeV}$. Here is where ground-based instruments, \acp{IACT} and water Cherenkov arrays, with their much larger effective areas take over. Similarly to space-based detectors, water Cherenkov arrays have duty cycles close to 100\%, however, being fixed to a particular location, they can cover about $2/3$ of the sky. 
Sensitivities of gamma-ray experiments in different energies is shown in Figure~\ref{figure3.4.3_DiffSensCTA}.

\begin{description}
\item{\textit{Satellite-borne gamma-ray telescopes}:} 
utilize various techniques for gamma-ray detection, depending on the target energy bands. Namely, more energetic soft gamma rays are typically measured using coded-mask technique. It consists of two type of elements, one made of very thick and heavy material with high atomic number, and the other are either empty or made of very light and thin material. The incoming gamma-ray radiation is stopped by the thick elements, but is unaffected by the thin ones. Any source will cast the shadow of the coded mask on the imaging detection plane, placed a few meters below the mask. The pattern of the mask is organized in a special way, so that from the detected shadow the position and shape of the sources in the \ac{FOV} can be reconstructed in a unique way. For example, the Imager on Board the  INTEGRAL Satellite (IBIS, \cite{Ubertini:2003})  uses tungsten coded-aperture mask.  IBIS provides fine imaging (angular resolution $12'$ below 500\,keV), source identification and spectral sensitivity to both continuum and broad lines between 15\,keV and $10\,\mathrm{MeV}$. Future missions like \textit{SVOM} \cite{2016arXiv161006892W} or \textit{THESEUS} \cite{Amati:2018},  will have an improved sensitivity   over a much broader energy range, from few keV up to~$20\,\,\mathrm{MeV}$. 
To measure photons at even higher, GeV, energies one need to use different technique, like the one used by the \textit{Fermi} Gamma-ray Space Telescope. It is an international and multi-agency space observatory which  carries two instruments: the Gamma-ray Burst Monitor \cite{2009ApJ...702..791M} and the \acf{LAT} \cite{Atwood:2009}. 
The Gamma-ray Burst Monitor flight hardware comprises of 14 scintillation detectors, sensitive to X-rays and gamma rays with energies between~8\,keV and~$40\,\mathrm{MeV}$. 
It can detect \acp{GRB} in that energy range across the whole of the sky not occluded by the Earth.
The \ac{LAT} is an imaging gamma-ray detector which detects photons with energies from about~20\,MeV to~$300\,\mathrm{GeV}$. It is a pair-conversion instrument, where incoming photons are converted to electron-positron pairs. The \ac{LAT} has \iac{FOV} of about 20\% of the sky. 
On average, it observes the entire sky every 3 hours. 

Some of the current high energy missions are also able to measure polarimetry at hard X-rays and soft gamma rays. Measuring the polarization of hard X-rays and soft gamma rays  is most easily done by Compton scattering.

\item{\textit{Imaging atmospheric Cherenkov telescopes}:} 
are instruments optimized for detecting \acs{VHE} gamma rays. They observe Cherenkov radiation within \ac{EAS} induced by gamma or cosmic rays entering the atmosphere, thus making atmosphere a part of detector. This allows for effective areas much larger than satellite borne detectors, enabling observations in energy band from few tens of GeV to $\sim100\,\mathrm{TeV}$. In order to record dim Cherenkov radiation, cameras of \acp{IACT} are composed of photomultipliers. Characteristics of primary particles (type, energy, direction) are reconstructed from the shower images. Sensitivity of \acp{IACT} is usually defined as the weakest flux a source needs to exhibit in order to be detected with $5\sigma$ significance in a certain observation time. Comparison of different instruments can be seen in Figure~\ref{figure3.4.3_DiffSensCTA}. It should be noted that energy thresholds for observations with \acp{IACT} are not sharp, rather they are peaks of the distributions of excess. Therefore, spectra can be reconstructed at somewhat lower energies. 
The most important existing \ac{IACT} observatories are \ac{H.E.S.S.}, \ac{MAGIC}, and \ac{VERITAS}.
\ac{H.E.S.S.} 
is an array of five \acp{IACT} located on the Khomas Highland plateau of Namibia \cite{Giavitto:2017ghe}. 
Its original four-telescope array (each telescope with a 12\,m diameter reflector) started science operations beginning of 2004. In 2012 a fifth telescope with a 28\,m diameter (currently the largest \ac{IACT} in the world) reflector was commissioned. 
\ac{MAGIC}
is a system of two \acp{IACT}, 
located in El Roque de los Muchachos Observatory in the Canary Island of La Palma, Spain \cite{Aleksic:2014poa, Aleksic:2014lkm}. 
The construction of \ac{MAGIC}-I was completed in 2003, and it became fully operational in 2004.
The second telescope, \ac{MAGIC}-II, was commissioned in 2009. 
Both telescopes have 17\,m diameter reflector dishes, and usually operate as a stereoscopic mode. 
\ac{MAGIC} is located in the same place as the Northern site of the next generation gamma-ray observatory CTA, which is currently being built. 
\ac{VERITAS}
is located at the historical Fred Lawrence Whipple Observatory in southern Arizona, USA\@.  
It stands at the site where the Whipple collaboration pioneered the \ac{IACT}, with the famous Whipple 10\,m telescope. The Whipple 10\,m was in operation from 1968 until it's decommissioning in 2013, most notably detecting the Crab Nebula as the first ever TeV gamma-ray source in 1989 \cite{1989ApJ...342..379W}.
\ac{VERITAS}, designed to follow up on this achievement saw first light in 2004 \cite{2005Ap&SS.297..275C,Weekes:2001pd}, but it wasn't until 2007 when 4-telescope observations with the whole array were realized. 
Over the years, it went through several upgrades and improvements to reach its present performance level \cite{Park:2015ysa}.

The \acf{CTA} 
will operate in the energy range extending from~20\,GeV to~$300\,\mathrm{TeV}$. It will consist of two arrays of \acp{IACT} in Northern and Southern Hemispheres. While the primary goal of the Northern array will be study of extragalactic objects at the lowest possible energies, the Southern one will cover the full energy range and concentrate on Galactic sources. 
Building on the technology of current generation ground-based gamma-ray detectors (\ac{MAGIC}, \ac{H.E.S.S.}, and \ac{VERITAS}), \ac{CTA} (see~\cite{CTAConsortium:2018tzg}) will be about ten times more sensitive and have unprecedented accuracy in the detection of high-energy gamma rays. 
It will consist of more than 100 telescopes of different sizes (large size telescopes (LSTs) \cite{Mazin:2017wdc}, medium size telescopes (MSTs) \cite{Puhlhofer:2016ubq}, and small size telescopes (SSTs) \cite{Montaruli:2015xya}), located in the Northern and Southern Hemispheres. This will allow to significantly boost detection area, and hence photon rate, providing access to the faint sources and shortest timescale phenomena. 
The differential sensitivity of \ac{CTA} and comparison to other existing gamma-ray instruments is shown in Figure~\ref{figure3.4.3_DiffSensCTA}.
\begin{figure}
    \centering
    \includegraphics[width=0.9\textwidth]{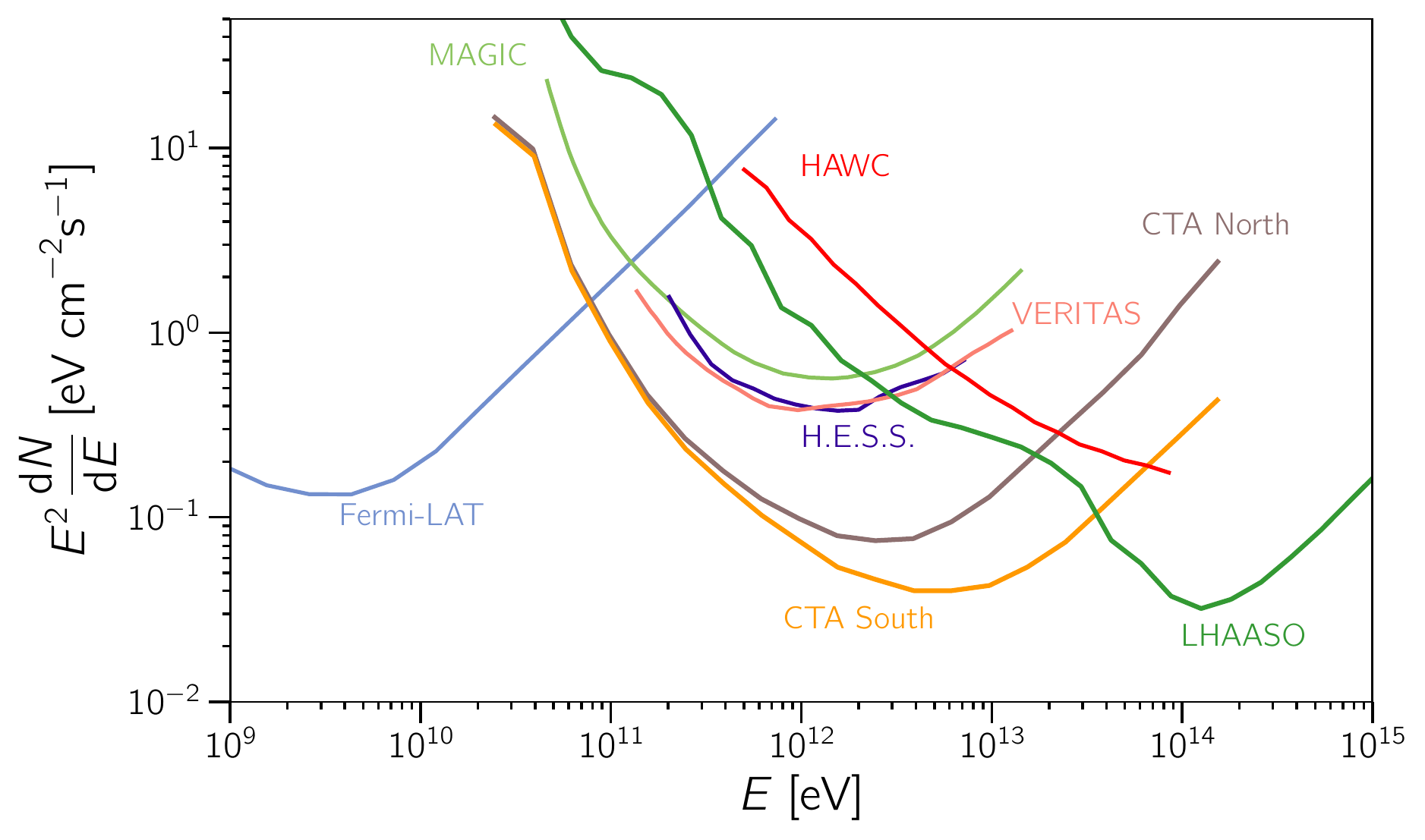}
    \caption{The differential sensitivity of several gamma-ray detectors operating in the GeV--PeV energy range. Sensitivities \ac{CTA}~\cite{CTAConsortium:2018tzg}, \ac{H.E.S.S.} \cite{Holler:2015tca}, \ac{MAGIC} \cite{Aleksic:2014lkm}, and \ac{VERITAS} \cite{VERITASperformance} are for 50~hours of observation. The flux sensitivities for \ac{HAWC}\label{acro:HAWC}~\cite{Abeysekara:2017mjj} and \ac{LHAASO}\label{acro:LHAASO}~\cite{Bai:2019khm} correspond to approximately 3000~hours and 1~years, respectively. The Fermi-LAT band includes sources at various positions in the sky, for the \texttt{P8R3\_SOURCE\_V2} instrument response function~\cite{Fermi-LATperformance}.}
    \label{figure3.4.3_DiffSensCTA}
\end{figure}

The searches for \ac{QG} effects performed with observations of gamma rays mostly rely on observations of strong sources. 
For such sources the systematic uncertainties connected with observations with \ac{IACT} instruments often dominate the statistical uncertainties stemming from the event numbers. 
The effect of systematic uncertainties of the measurements performed with \acp{IACT} is often split into three kinds: uncertainty on the flux normalization, uncertainty on the energy scale, uncertainty on the spectral shape (photon index), apart from others less relevant for this case as uncertainties on absolute pointing and angular resolution. 
Since the intrinsic flux of cosmic sources is normally not known, the pure uncertainty on the flux normalization (due to e.g. non perfectly known trigger efficiency, or imperfections of Monte Carlo simulations) is often not so relevant. 
An important exception is, however, the case a of combination of measurements from multiple \ac{IACT} instruments, particularly those of different  sizes, sensitive to different energy ranges. Each  will then show independent systematic uncertainties and the combination may translate into undesired artificial spectral features. Furthermore, comparison of sources, observed by different instruments, and their spectral combination is often dominated by uncertainties  on the absolute flux scale. 

The systematic uncertainty on the spectral shape (due to e.g.\ instrumental non-linearities, or spectral unfolding methods) can be important in methods that rely on finding features in gamma-ray spectra. 
It is also the type of systematic uncertainty that is the most difficult to estimate in particular due to its dependence at other observation and source parameters (e.g.\ whether the flux of the source is measured at low/medium/high energies, whether the spectral shape is a simple power-law or more complicated, and whether the source is weak, i.e.\ background-dominated). Finally, atmospheric disturbances, if uncorrected, can modify the reconstructed spectral  shape  of the source.

In most of the cases the dominant type of \ac{IACT} systematic uncertainty is the one on the energy scale of the instrument (due to e.g.\ non-perfectly known atmospheric transmission and calibration of the light yield of the telescopes). 
In the methods that relay the estimated energies of gamma-ray events to the \ac{QG} energy scale, or search energy-dependent emission features, such systematic uncertainty would directly propagate to the uncertainty of the final result. 
Moreover, uncertainties on the energy scale also cause an independent uncertainty on the flux normalization, which, as mentioned above, is important in cases of combination of data from multiple experiments. 
It should be noted that since most gamma-ray sources have photon indices softer than $-2$, the resulting effect on flux normalization is large.
As an example, for a source with an index $-2.6$, the corresponding flux uncertainty is $ _{-20\%}^{+30\%}$ for an uncertainty on the energy scale of $\pm 15\%$.
For very steep sources the uncertainty can be much larger than this, even close to a factor of two (see e.g.~\cite{Ahnen:2016vlr}).
The systematic uncertainties on the energy scale of the currently operating major \ac{IACT} installations are about 15\% \cite{Aleksic:2014lkm} (see also \cite{Aharonian:2006pe,Abeysekara:2015pjl}). 
In the case of featureless spectrum like the one of the  Crab Pulsar, such systematics add between 5\% and 29\% uncertainty to the derived limits on the \ac{QG} scale~\cite{HESS:2011aa,Ahnen:2017wec,Acciari:2020kpi}, depending on the assumptions on spectral behavior of the source and the leading order of the energy dependence of the \ac{LIV}. 

Requirements for \ac{CTA} include a maximally allowed systematic uncertainty on the energy scale of 10\%, dominated by contributions from the atmosphere~\cite{Gaug:2014rja}. That ``atmospheric'' uncertainty budget can then be more or less equally distributed between contributions from air shower simulation codes and simplifications made in the simulations for speed reasons, residual uncertainties in the atmospheric density profile and consequently the profile of the refractive index of air~\cite{Munar-Adrover:2019cst}, light absorbing molecules (particularly ozone) and additional extinction of Cherenkov light due to aerosols and optically thin clouds~\cite{Garrido:2013rja,Dorota:2020mqx}. Major efforts are devoted to reduce the last contribution~\cite{Gaug:2017nci}, where a continuous monitoring of the extinction profile across the large \ac{FOV} of CTA is envisaged through a combination of Raman \ac{LIDAR} data~\cite{Gaug:2019qij,Ballester:2019eax} and wide-field stellar photometry~\cite{Ebr:2018oeg}. Besides atmospheric contributions, the optical throughput calibration of \acp{IACT} themselves is critical, but can be relatively well controlled (to better than a few percent uncertainty) with a careful analysis of ring images captured from the Cherenkov light of local muons~\cite{Gaug:2019hca}. Finally, telescope cross-calibration can be carried out efficiently using selected stereo shower images, reconstructed with a same impact distance between two telescopes~\cite{Mitchell:2015xca}.  

\item{\textit{Water Cherenkov experiments}:}
are detectors of charged particles, which usually cover a vast surface to maximize the probability that such particles interact with a medium. They do not image the Cherenkov light from extensive air showers as \acp{IACT}; instead, they sample the atmospheric shower of secondary particles at ground level using highly purified water as a media in an optically isolated container. In each of them, there is one or more photomultiplier tubes to detect the Cherenkov light produced by charged particles reaching the detector. Water Cherenkov experiments have wide \ac{FOV}, about $90^{\circ}$, and close to 100\% duty cycle, compared to a typical \ac{FOV} of \acp{IACT} of a few degrees, and a duty cycle of 10 -- 20\%. The relatively low cost of detecting \acs{VHE} gamma rays with extensive detectors is one of the most notable advantages when compared to other techniques. 

The \acf{HAWC} 
observatory
is a wide-\ac{FOV} array 
located near Sierra Negra volcano, in Puebla, M\'{e}xico, at $4,100$\,m \ac{asl}. The full \ac{HAWC} array covers $22,000$\,m$^2$, and it is sensitive to primary cosmic and gamma rays in the energy range between~100\,GeV and~$100\,\mathrm{TeV}$. It has been operating since 2015 at 95\% duty cycle, as the most sensitive currently-operating gamma-ray observatory in the world above~$10\,\mathrm{TeV}$. 

The Tibet AS$\gamma$ Experiment 
is an air-shower array at an altitude of 4,300\,m \ac{asl}, located in Yangbajing, Tibet, China. With 397 plastic scintillation detectors of 0.5\,m$^2$, it covers an area of 65,700\,m$^2$. Tibet started to operate in 1990 to observe cosmic and gamma rays in the TeV energy range. In 2014, it improved its sensitivity above 10 TeV with a muon detector array made of 64 underground water Cherenkov detector, covering a total area about 3,400\,m$^2$ and installed 2.3\,m under the previous detector array, which makes it possible to suppress $99.92\%$ of cosmic-ray background events above 100\,TeV~\cite{Amenomori:2008wa,Amenomori:2019rjd}. 

The \acf{LHAASO} observes cosmic and gamma rays at 4,410\,m \ac{asl} in Daocheng, Garzê, Tibetan Autonomous Prefecture in Sichuan, China~\cite{Bai:2019khm}. It started the first stage operations in April 2019 and is scheduled to be completely operational by 2021~\cite{David:2019nat}. It is expected to be sensitive to gamma rays in the energy range between~$10^{11}$ and~$10^{15}\,\mathrm{eV}$, and cosmic rays in the energy between~$10^{12}$ and~$10^{18}\,\mathrm{eV}$. 
The \ac{LHAASO}-WCDA will potentially detect hundreds of thousands of photons above 100\,GeV from \ac{GRB}s. Therefore, using \ac{GRB}s observed with \textit{Fermi}-\ac{LAT}, reference scenarios were made to obtain preliminary sensitivity limits on \ac{LIV} studying energy-depending time delay from hypothetical long \ac{GRB}s.
In fact, \ac{LHAASO} has recently reported~\cite{Cao:LHAASO} indications of the existence of PeVatrons (cosmic accelerators to the PeV scale) in our galaxy, thanks to the observation of the highest-energy photons ever detected, up to 1.4\,PeV, what has already implications for Lorentz invariance violation studies~\cite{Li:2021tcw,Chen:2021hen,Satunin:2021vfx}.

The Southern Wide field-of-View Gamma-ray Observatory (SWGO) 
is a next-generation ground-based type experiment in the \acs{VHE} band, proposed to be constructed in the Southern Hemisphere.
It is foreseen to be made of two arrays. The densely populated inner part will consist of 4000 detection units covering an area of 80,000\,m$^2$ (about 4 times the \ac{HAWC} effective area), which will be surrounded by the outer array with 1000 detection units housing an area of 221,000 m$^2$. The inner part will conduct the low-energy performance of the SWGO proposal, while the outer with the extended effective area will increase the high-energy range. 
SWGO will be the first facility of the kind operating in the south, hence effectively complementing current and future instruments at the global multi-messenger effort to understand extreme astrophysical phenomena. Worth to mention that in the scientific proposals of SWGO, there is a confirmed interest in testing \ac{LIV} phenomenology, for which the access to south astrophysical sources and the extended and complementary sensitivity to the northern experiments as \ac{HAWC}, will open a new frontier of opportunities to such endeavor~\cite{Abreu:2019ahw, Albert:2019afb}.

\end{description}

\subsubsection{Neutrino telescopes} 
\label{section4.2.2}

The design of neutrino telescopes follow the idea proposed for the first time in the 1960s~\cite{Markov:1960vja}. The weak interaction with a detector target and the low flux of cosmic neutrinos determine the size of neutrino telescopes to be in the order of a km$^3$, which represents a technological challenge in many ways. The principle is to detect the Cherenkov light emitted by the products of the interaction of a high-energy neutrino inside or in the vicinity of the detector. The design of the existing and planned neutrino telescopes is basically the same: a large array of optical sensors with ns time resolution deployed in a very large volume (100\,kt to Gt) of a natural transparent medium, like water or ice. They are installed as deep as possible to attenuate the flux of muons produced by the interaction of cosmic rays in the atmosphere, which otherwise can mimic the signal of charged-current $\nu_\mu$ interactions. The known location of the optical modules, the relative time they detect light and the amount of photoelectrons detected are used to reconstruct the characteristics of the events: flavour, energy and direction of the neutrino. 

There are currently several neutrino telescopes in operation or in construction phase at different locations: IceCube, at the South Pole~\cite{Aartsen:2016nxy}, which with 1\,km$^3$ instrumented volume is currently the largest neutrino telescope in operation; ANTARES, off the coast of Toulon (France)~\cite{2011NIMPA.656...11A}; and Baikal, in Lake Baikal (Russia)~\cite{Aynutdinov:2009zza}. All these projects have plans for upgrades that increase the detection volume and sensitivity to high-energy neutrinos: IceCube-Gen2~\cite{Aartsen:2014njl}, KM3NeT~\cite{Adrian-Martinez:2016fdl} and Baikal-GVD~\cite{Avrorin:2021vpz}, respectively. Even if the underlying technology is similar in all these projects, photon scattering and absorption in the detector medium are key variables to take into account in the design of a neutrino telescope. Deep sea and lake water typically contain less suspended impurities and present less scattering than ice. On the other hand, water is less transparent than deep ice. Scattering and absorption in the detector medium are directly related to the pointing and energy resolution achievable in a neutrino telescope. Note that, given the steeply decreasing flux of astrophysical neutrinos with energy, the detection of extremely high energetic neutrinos ($E \gtrsim 10^{16}$~eV) would require impractically huge detection volumes, even for open geometries like the projects mentioned above. The attenuation of light in water or ice makes it economically unrealistic to instrument the required volume with optical modules. At such energies another technique is being explored: the detection of the radio pulses that are emitted by electromagnetic showers produced in neutrino interactions, the so called Askaryan effect~\cite{Askaryan:1961pfb}. This technique works only in ice, water being opaque to radio waves, and there are currently several prototypes and ongoing R\&D efforts. See~\cite{Huege:2017khw}for a review.  

Neutrino telescopes are all-flavor detectors. Events are usually classified as ``tracks'' or ``cascades'', or a mix of the two. Tracks are the signature of the muons from charged current muon-neutrino interactions while cascades arise from electron- or tau-neutrino charged current interactions or neutral current interactions of any flavor. The distinct characteristics of these events in a neutrino telescope are due to the different penetrating capabilities of muons, electrons and tau leptons in matter. Muons of $\mathcal{O}(100)$\,GeV or higher can travel several kilometers in ice or water, while electrons quickly loose energy through scattering in the medium, and tau leptons quickly decay.  Due to the typical sparse instrumentation of neutrino telescopes (from 10s to 100s of meters between optical modules), the burst of Cherenkov radiation from the secondary particles (``cascades'') emitted by electrons or in the decay of the tau appears as a point source of light in the detector, as opposed to the long track of a muon. Due to the transparency of water or ice the interactions of the neutrinos can occur well outside the detector and still produce a detectable signal. The trigger effective volume of a neutrino telescope is therefore much larger than its instrumented volume. It is important, however, to keep in mind that the real measure of the efficiency of a neutrino telescope to a given signal is the effective volume of the detector calculated for that specific signal. In this sense, neutrino telescopes can not be considered to have a unique effective volume, but it depends on the analysis performed, and it can be very different for different searches. It is important to take this into account when using public data from neutrino telescope collaborations to probe theoretical models.

Tracks and cascades have complementary characteristics. Tracks provide pointing with a good angular resolution (depending on energy and down to $\sim$0.1$^\circ$ above TeV energies) thanks to the long lever arm, but rather poor energy resolution since the neutrino interaction can occur outside the detector and only part of the track is seen. They benefit from the highest effective area but suffer from a huge background of atmospheric muons, several orders of magnitude larger than the atmospheric neutrino flux, not to mention the astrophysical neutrino flux. Cascades have poorer angular resolution due to the roughly spherical topology of the events, but excellent energy resolution if the interaction occurs inside the instrumented volume since all the initial neutrino energy is then deposited within the detector. The atmospheric muon background can be in principle eliminated by either considering only events entering the detector from below, i.e.\ using the Earth as a filter, or by using the outer layers of the detector as a veto region for tracks entering from outside, considering only events with their interaction vertex inside the fiducial volume. In the first approach, only half the sky is visible at any time, while the second approach opens the whole sky to observation, but at the expense of reducing the energy reach of the detector (since, for example, only starting or fully contained tracks are considered).

Flavor identification is crucial in neutrino telescopes to be able to measure \ac{QG} effects. These  would typically appear as deviations of the expected flavor ratio from standard neutrino oscillations both in the atmospheric neutrino flux and in the astrophysical neutrino flux, in the latter case under certain assumptions about the flavor composition at the  source~\cite{Aartsen:2017ibm}. The distinction between atmospheric and astrophysical neutrinos is in reality a distinction of energy and pathlengths, and the sensitivity to \ac{QG} effects reflects this separation. Atmospheric neutrinos travel at most one Earth diameter, but they are copious (IceCube detects about $10^5$ per year), while astrophysical neutrinos are still few, but tiny effects on the flavor oscillations due to \ac{QG} effects in their propagation can be amplified by the large path lengths involved. Besides, many effects in the neutrino flavor mixing due to \ac{QG} effects scale as (length $\times$ energy), instead of (length/energy) as in standard oscillations. Astrophysical neutrinos present therefore an advantage for this kind of searches. On the other hand, the copious flux of atmospheric neutrinos can be useful to test scenarios that predict annual variations or directional asymmetries in the detected flux due to rotational invariance violations, as discussed in Sec.~\ref{section7.5}. In general, neutrino telescopes have proven to be very versatile detectors able to address physics topics well beyond their original target of study, the high-energy neutrino universe~\cite{PerezdelosHeros:2020tan}.

\subsubsection{UHECR detectors} \label{uhecr-detectors}
Modern \ac{UHECR} experiments are hybrid detectors, consisting of an array of \ac{SD} stations surrounded by \ac{FD} telescopes overlooking the \ac{SD} area.
\Acp{SD} are sensitive to the secondary particles of the \ac{EAS} (muons, electrons, positrons and photons) as they reach the surface of the Earth, providing information on their arrival times and lateral distribution, whereas \acp{FD} measure the UV fluorescence emitted by charged particles in the shower (Sec.~\ref{sec:cherenkov-and-fluorescence}).  By measuring the rate of fluorescence emission as a function of atmospheric slant depth~$X$, the detector measures the longitudinal development profile of the air shower. The integral of this profile gives the calorimetric energy, to which an estimate of the ``invisible energy''dumped into the ground by muons and neutrinos is added to estimate the total energy of the shower.
Whereas \acp{FD} provide a nearly calorimetric measurement of the shower energy, they can only operate in clear, moonless nights, corresponding to a duty cycle of about 15\%.  Hybrid events, i.e.\ those simultaneously detected by both \acp{FD} and \acp{SD}, are used to calibrate measurements performed by \acp{SD} with a duty cycle of about 100\%.
Moreover, hybrid detection enhances the reconstruction capability with respect to the individual \ac{FD} and \ac{SD} reconstruction. 

The largest \ac{UHECR} array in the world is the Pierre Auger Observatory \cite{ThePierreAuger:2015rma}, located in Malarg\"ue, Mendoza Province, Argentina  ($35.2^\circ$~S, $69.2^\circ$~W, 1400\,m \ac{asl}).
Its main array for detecting the highest-energy cosmic rays (full detection efficiency above 3\,EeV) consists of 1600 water-Cherenkov \ac{SD} stations deployed in a isometric triangular grid of 1500\,m spacing covering an area of 3000\,km$^{2}$ overlooked by four \ac{FD} sites on its perimeter with six telescopes each. It has been taking data since 1~January 2004 and is sensitive to \acp{EAS} with zenith angles up to~$80\degree$.
In order to extend the sensitivity to lower energies, there are two denser regions within the boundary of the 1500\,m array, one extending over 28\,km$^2$ within which the stations are spaced by 750\,m and one covering 2\,km$^2$ with 433\,m spacing, overlooked by three extra \ac{FD} telescopes with higher elevation.
Each \ac{SD} station consists of a polyethylene tank with a diameter of~3.6\,m and a height of~1.5\,m containing 12,000 liters of ultra-pure water, surmounted by three photomultiplier tubes detecting the Cherenkov light emitted when relativistic secondary particles from an air shower propagate in water.
In data analyses, all detector effects are corrected in a way as independent of simulations as possible: the zenith-angle dependence of atmospheric attenuation via constant intensity cuts \cite{Hersil:1961zz}, the ``invisible energy'' via \ac{SD} measurements of the muon flux at the ground \cite{Aab:2019cwj}, and the energy resolution and bias from the distributions of the ratios between \ac{SD} and \ac{FD} energy estimates.

The Telescope Array experiment is the largest \ac{UHECR} experiment in the Northern Hemisphere~\cite{Kawai:2008zza}.  It is located in a semi-desert area in Millard County, Utah, USA ($39.3^\circ$~N, $112.9^\circ$~W, 1400\,m \ac{asl}).
Its \ac{SD} array consists of 507~plastic scintillator detectors in a 1.2\,km-spacing square grid covering an area of \(\approx 700\,\text{km}^2\), and the three \ac{FD} stations at the periphery of the \ac{SD} array contain 38~fluorescence telescopes in total.  Events can be observed by either the \ac{SD} or \ac{FD} separately, or in coincidence.  The ``main'' array has been fully operational since March 2008, and is sensitive to showers with~\(E \gtrsim 1\,\text{EeV}\) and zenith angles up to~$45\degree$ (and up to~$55\degree$ above~$10^{18.8}$\,eV); the Telescope Array Low-energy Extension \cite{Thomson:2011gke}, consisting of 10 extra \ac{FD} telescopes looking at higher elevations (operating since 2013) and 103 extra \ac{SD} stations with closer spacing (deployed in 2018), extends the energy range down to a few PeV, allowing the whole energy range from a few PeV to hundreds of EeV to be covered by one experiment.  The energy of \ac{FD} events is estimated calorimetrically plus a $7$--$9\%$~correction for ``invisible energy'' from \textsc{QGSJet}-II.03 proton simulations, and that of \ac{SD} events by comparing the particle density at 800\,m from the shower core to that of \textsc{QGSJet}-II.03 proton simulations and dividing the result by~1.27 to bring the \ac{SD} and \ac{FD} estimates of hybrid events into agreement.

The Pierre Auger Observatory is currently undergoing an upgrade known as AugerPrime \cite{Aab:2016vlz} to improve the mass composition in the whole energy range and particularly for energies above 40\,EeV, where the intrinsic duty cycle of the \ac{FD} and the scarce accuracy of the \ac{SD}-based composition measurements become important.  This upgrade includes improvements of existing technology as well as new detectors. New \ac{SD} electronics and an additional small photomultiplier tube will be implemented to increase the sampling rate and to extend the dynamic range. Moreover each \ac{SD} station will be equipped by a 3.8\,m$^2$ plastic scintillator detector on top, as well as a radio antenna to detect air showers in the frequency range from 30 to 80\,MHz \cite{Pont:2021pwd}. 
The sampling of the shower particles by different detectors, which
have different responses to the muonic and electromagnetic components, will enable disentangling these two components providing an estimate of both, mass and energy of the shower, on an event-by-event basis, also providing data with
a duty cycle of almost $100\%$. The deployment of AugerPrime is at the moment in progress, and will allow to extend the operation of the Auger Observatory until 2030.
The Telescope Array is expanding its operations with the TA\(\times\)4 project~\cite{Abbasi:2021hO}, to cover an area that four times as wide, reaching a combined coverage of \(3000\,\text{km}^2\).  The expanded \ac{SD} array of TA\(\times\)4 is overlooked by two new \ac{FD} stations.  This setup was designed to study cosmic rays with the focus on energies above \(57\,\text{EeV}\), where, owing to the quadrupled covered area, it is expected to collect about four times as many events as the original array.  At the time of writing, the two new \ac{FD} stations and 257 out of 500 new \ac{SD} stations have been deployed and taking data.
Other projects scheduled for the 2020s include the radio array GRAND~\cite{GRAND:2018iaj} and the space-based \ac{FD} missions EUSO~\cite{Ricci:2016bpj} and POEMMA~\cite{Olinto:2019mjh}, and there are ongoing discussions about a Global Cosmic Ray Observatory (GCOS)
in the farther future.

\subsubsection{Gravitational wave detectors} \label{GW-detectors}
\paragraph*{Ground based detectors}
The current network of ground-based \ac{GW} detectors comprises Advanced \ac{LIGO}~\cite{TheLIGOScientific:2014jea}, with sites in Hanford, Washington and Livingston, Louisiana, Advanced Virgo~\cite{TheVirgo:2014hva} (named for the Virgo constellation) near Pisa in Italy,  KAGRA~\cite{Akutsu:2018axf} (KAmioka GRavitational-wave Antenna) in Japan, and a 600 meter-scale detector GEO600 in Germany~\cite{Luck:2010rt}. Multiple such detectors are important to verify the astrophysical origin of a signal, determining its sky localization which is important for multimessenger counterparts and cosmology, and for tests of \ac{GR}. 
The interferometer detectors have armlengths of three to four kilometers and are sensitive to \acp{GW} with frequencies of $\sim$ 20–-2000\,Hz, owing to their kilometer-scale arms and the intricate technology implemented to reduce, for instance, seismic noise at low frequencies, thermal noise at intermediate frequencies and laser shot noise at high frequencies. Mitigating noises from various sources is an extremely challenging task, because the \ac{GW} signals even from the strongest sources have only a tiny amplitude on earth, of order $\sim 10^{-21}$, requiring the detectors to sense changes in the armlength much smaller than the size of an atomic nucleus.

 The
era of \ac{GW} astronomy was initiated with the historic detection of the first binary black hole merger event GW150914 \cite{Abbott:2016blz}. The first detection of the \ac{BNS} merger event, GW170817 \cite{TheLIGOScientific:2017qsa}, was
accompanied by detections across the electromagnetic spectrum \cite{GBM:2017lvd}. Binary black hole mergers may also be accompanied by electromagnetic emission,  if there is surrounding material \cite{Schnittman:2013qxa}. Observational searches of transient signals were reported in \cite{Abbott:2016ezn,Abbott:2017vtc,Abbott:2017gyy,Abbott:2017oio,TheLIGOScientific:2017qsa,Abbott:2016cjt,LIGOScientific:2018mvr,Abbott:2018oah,TheLIGOScientific:2016qqj,TheLIGOScientific:2016uux,Abbott:2016ymx,TheLIGOScientific:2016pea}. 

The detector facilities were already funded in the 1990s. They underwent an initial stage that lasted several years. This was intended mainly as a proof of concept to facilitate the development of the highly sophisticated technology and experience required for the advanced, second generation of detectors, which were planned to achieve sufficient sensitivity to make the first discoveries. The advanced detectors have completed three observing runs O1-O3~\cite{Abbott:2020qfu}:
\begin{enumerate}
\item The first observing run (O1) involved only the \ac{LIGO} detectors and officially lasted from 18 September 2015 to 12 January 2016. Data of good quality was also collected a week before and after these dates. The detectors achieved a 80 Mpc range for \ac{BNS} detections, and measured three binary black hole mergers. 
\item O2 lasted from 30 November 2016 to 25 August 2017, and the typical sensitivity was 80--100\,Mpc for \acp{BNS}. The Virgo detector joined on 1 August 2017 with a 30\,Mpc range. The three detector network in operation significantly improved the sky localization of \ac{GW} events~\cite{LIGOScientific:2018mvr}. During this run, seven binary black hole mergers and one \ac{BNS} inspiral were observed. 
\item O3 started on April 1, 2019 and ended on March 27, 2020 due to the COVID-19 pandemic, with a commissioning break from October 1 to November 1, 2019. Various changes to the hardware led to a significant increase in sensitivity and more than tripled the number of \ac{GW} candidate detections. A further change compared to O1-O2 was that alerts about candidate events were made public immediately, for instance on the GCN network. In the first half of O3 before the commissioning break, 39 events were detected~\cite{Abbott:2020niy}. The KAGRA detector came online on February 25, 2020. The data analysis of the second half of O3 has not yet been published, only a special events paper on neutron star-black hole discoveries.
\end{enumerate}
The observations to date are summarized in the O1-O2 first \ac{GW} transient catalog GWTC-1~\cite{LIGOScientific:2018mvr} and in the O3a catalog GWTC-2~\cite{Abbott:2020niy}. Altogether, they include 11$+$39 events. The next observing run is planned to start in 2022 with a duration of one year and the detectors operating close to their design sensitivity. There are plans for further detector upgrades and a fifth observing run in the mid-2020s, including also a detector in India~\cite{Abbott:2020qfu}.  

The \ac{GW} community is currently very active in planning for new facilities and third-generation ground-based detectors. The main concepts are known as the Einstein Telescope in Europe~\cite{Punturo:2010zz,Maggiore:2019uih}, and the Cosmic Explorer in the US~\cite{Reitze:2019iox}, both of which target an order of magnitude better sensitivity than the current detectors and a significantly wider sensitive frequency bandwidth. In Australia, current ideas focus on smaller scale facilities that could work in unison with the current detector network such as a dedicated high-frequency detector~\cite{Ackley:2020atn}. The Einstein Telescope vision was recently selected to the 2021 ESFRI roadmap, which paves the way for concrete funding investments, and a technology pathfinder is under construction in Maastricht. The third-generation detectors will observe hundreds of thousands of source per year, and take the next step for precision physics with \ac{GW} sources in that frequency range. 

The current \ac{GW} measurements are impacted not only by statistical but also by systematic uncertainties. On the experimental side, these are due to the calibration of the detectors, as described in detail in~\cite{Cahillane:2017vkb,Sun:2020wke}. Calibration is necessary to convert from the electronic detector output to the difference in armlength, which is a direct measure of the dimensionless strain \ac{GW} amplitude, the quantity of interest for extracting the physics and astrophysics of the source. Detector calibration requires thoroughly determining the response of the detector, for example, by inducing  well-defined small changes to the armlength by using the lasers to vary the photon pressure on the mirrors. Most of the time, this is performed only at specific frequencies to avoid contaminating measurements, and is less often applied across the entire sensitive frequency range. In addition to such quantitative measurements, computer modeling is used to further characterize the detector behavior. This enables accounting for numerous other effects in the feedback control system and the opto-mechanical response of the detectors. The overall calibration uncertainty is usually determined in two stages, with a first low-latency estimate superseded by a later refined estimate obtained after an offline characterization of the errors based on additional measurements on individual detector components. Currently, the typical calibration uncertainties are less than $10\%$ in the wave amplitude and 10 degrees in phase over the calibrated range in frequencies, but vary over time, as explained in ~\cite{Cahillane:2017vkb,Sun:2020wke}. 

A second source of systematic uncertainties, especially for \acp{GW} from binary systems, are the theoretical model waveforms used to extract the fundamental information from the data. Not only the detections but also the measurements rely on cross-correlating the data with theoretical model waveforms, which facilitate the highly nonlinear connection between the source and the asymptotic \acp{GW}. Using MCMC sampling over millions of waveforms compared with the data over a range in parameter space, the Bayesian data analysis methods determine the posterior probability distributions of the source parameters. Shortcomings in the models, such as lack of sufficient accuracy, parameter space coverage, missing physical effects, and the like, may thus bias our interpretation of the signals. To quantify the impact of the theoretical uncertainties on recent detections, the data analysis is performed using different available waveform models that combine analytical and numerical relativity information, or are surrogates to numerical relativity data. So far, the waveform systematics have been subdominant compared to the statistical errors as far as could be quantified in this way but are starting to become noticeable in some cases~\cite{Abbott:2020niy}. 

\paragraph*{Space based detectors}
The most advanced planned and approved space-based detector is \ac{LISA}~\cite{Audley:2017drz}, which after the successful \ac{LISA}-Pathfinder mission,
which tested part of the needed technology, in particular the drag-free
control \cite{Armano:2018kix}, is now under development under the leadership of ESA with
important contribution from NASA\@. 
It aims to measure \ac{GW}
s directly by using laser interferometry.
The \ac{LISA} concept has a constellation of three spacecraft arranged in an
equilateral triangle with sides 2.5 million km long, flying along an
Earth-like heliocentric orbit. The distance between the satellites will be precisely monitored to detect a passing \ac{GW}.

The scope of \ac{LISA} is to detect and study low-frequency \ac{GW}
from about 0.1 mHz to 1 Hz, and thus to complement ground-based gravitational
observatories. \ac{LISA} opens new possibilities
for astrophysical studies by
 allowing, for instance, to detect
supermassive black holes (typically of $10^6 - 10^7~M_{\odot}$)
merging at cosmological distances.
Mergers of a supermassive black hole with another compact object (such as another black hole or a neutron star) produce a very clean \ac{GW} 
signal which \ac{LISA} will be able
to measure with high precision. Alternative gravity theories influence the dynamics of such mergers
and hence \ac{LISA} is expected either to directly see the imprints of
certain alternative theories or to put severe constraints on them.
Another class of objects, which will be observed by \ac{LISA}, 
are ultra-compact binaries, in particular of white dwarfs in
our Galaxy. They are important sources of \ac{GW}s in the 
mHz  frequency  range.
Moreover, it will be possible to detect or put strong constraints
on the primordial \ac{GW} background, which is just, as the
\ac{CMB}, a leftover from the Big Bang.

The observation of long-wavelength \acp{GW} with \ac{LISA} is particularly promising as a probe of fundamental physics \cite{Barausse:2020rsu}. Anomalies related to violations of the Equivalence Principle or Lorentz invariance could produce modifications in the dispersion relation of matter or horizon-scale modifications in black hole physics due to \ac{QG} effects. Observations of the dispersion relation of \acp{GW} could constrain a large class of modified theories, which include massive gravity models that attempt to explain the late-time acceleration of the universe, as well as other Lorentz-violating theories, whose renormalizability makes them attractive candidates for \ac{QG}.

\subsection{Multiwavelength and multimessenger  astronomy}
\label{section3.1.9}
The most powerful tool for studying astrophysical objects, is the multi-wavelength (now also multi-messenger) characterisation of the sources obtained by simultaneous observations with different telescopes across the whole electromagnetic spectrum. 
The study of  the flux of photons, from the radio to the \acs{VHE} gamma-ray band, is a decisive tool for the understanding of the emission scenario of \acsp{AGN}, pulsars and \acsp{GRB}. The lightcurves, that represent the evolution of the flux of photons over time, give insights into the origin of the observed emission. The \acsp{SED} define a typical spectral shape (a two-bumped structure)  that is the fingerprint of each object and is directly connected to the source and mechanism of acceleration of the most energetic photons~\cite{Ghisellini:2017ico}. The first bump of the \acs{SED} is located in the optical-UV part of the electromagnetic spectrum, while the second bump appears at higher energies of the photons, in the \acs{HE} or \acs{VHE} gamma-ray band. While the former bump is universally explained as produced by synchrotron radiation, the latter can have different interpretations and this is why modeling the entire \acs{SED}, especially with gamma-ray data, is of extreme importance in this field. For this reason, recently the astrophysics community is pushing towards the creation of new facilities which can detect \acs{VHE} gamma rays with better sensitivity than the currently operating telescopes~\cite{CTAConsortium:2018tzg}, and at the same time, to refine the observational strategies in order to collect simultaneous data from very different types of telescopes and experiments.  In an event of a flare from an astrophysical object, the analysis of simultaneous data can then provide a scenario for the emission mechanism. The \acsp{SED} can be interpreted by different models, that can be leptonic, hadronic, or hybrid, depending on how they explain the high-energy part of the \acs{SED}. 
In 2013 the important discovery of IceCube Collaboration of an extraterrestrial neutrino~\cite{Aartsen:2013jdh} opened a new era in astrophysics: if neutrinos can be detected from a blazar, then their detection could be simultaneous to photons, and a \acs{SED} could be built which includes other than photons, another particle, or messenger, a neutrino. This possibility became a reality in 2018, when the first multi-messenger study was performed during the flare of the blazar TXS~0506+056 ~\cite{IceCube:2018dnn,Ahnen:2018mvi}: in that occasion a neutrino was detected simultaneously to photons across the entire electromagnetic spectrum. 
Other messengers are \acp{GW}, for the first time directly detected  the first observation of a binary black hole merger~\cite{Abbott:2016blz}. In 2017 another important multi-messenger observation was performed, when a \ac{GW} generated by a \ac{BNS} inspiral was associated with a \acs{GRB}~\cite{TheLIGOScientific:2017qsa}, later reclassified as a kilonova. 
The detection of \acp{AGN} in the \acs{VHE} gamma-ray band~\cite{Acciari:2019dxz,Acciari:2019dbx,Arakawa:2019cfc}, roused the expectations in the astrophysical community of a nearby connection between multi-wavelength observations and \acp{GW}~\cite{Seglar-Arroyo:2019yso}. Gathering simultaneous data from all the available messengers would improve sensibly our knowledge of astrophysical objects and emission mechanisms. 

%\newpage
\section{Phenomenology of quantum gravity}
\label{section5}

In the preceding chapters we have outlined the theory of \ac{QG} scenarios, we have described and summarized the various cosmic messengers that one can use as probes, including how they are produced in their sources and how they propagate to the observer, and we have discussed how those messengers are detected. In the present chapter we finally describe the various effects \ac{QG} models can have on these messengers and how they can be, and have been, searched for with existing and future  detectors. We include the constraints that have been derived on the parameters of the \ac{QG} model, typically the \ac{QG} scale, based on the non-observation of such effects. We separate the possible phenomena into subsections on time delays of messengers relative to some baseline, on birefringence, i.e. dependence of propagation speed on polarisation and possibly direction, on modification of interaction rates and thresholds for interactions, on decoherence of oscillations between different particle states, on the role of \ac{CPT} symmetry, and on further quantum gravity signatures in \acp{GW}. We provide the most prominent results of experimental tests as we go along. For a comprehensive and up-to-date census of experimental studies and results, we refer the reader to the QG-MM Catalogue~\cite{QGMMCatalogue}.

\subsection{Time delays}
\label{section7.1}

Among the effects that have favored the birth and the development of \ac{QG}  phenomenology, the possibility of in-vacuo dispersion in particle propagation has played undoubtedly a prominent role~\cite{AmelinoCamelia:2008qg,Mattingly:2005re,AmelinoCamelia:1997gz,Schaefer:1998zg,Gambini:1998it,Alfaro:1999wd,AmelinoCamelia:1999pm,Ellis:2002in,Myers:2003fd,Ellis:2005wr,RodriguezMartinez:2006ee,Jacob:2006gn,Jacob:2008bw,AmelinoCamelia:2009pg,Abdo:2009zza,Ackermann:2009aa,Aharonian:2008kz,Albert:2007qk,Bolmont:2010np,HESS:2011aa,Ellis:2011ek,Vasileiou:2013vra,Xu:2016zxi,Amelino-Camelia:2016ohi}.
It is indeed presently well understood that the large cosmological distances, generally characterizing the propagation of particles in astrophysical observations, can provide a source of amplification of the tiny possible Planckian effect, such that it could be within the reach of present experimental sensitivity.
In the frame of \ac{QG}, the emergence of in-vacuo dispersion is related mostly to the hypothesis of a Planck-scale \ac{MDR} characterizing particles propagation. 
One can expect the Planck-scale modification either violating the relativistic symmetries (\ac{LIV}, \cite{AmelinoCamelia:1999pm,Ellis:2002in,Ellis:2005wr,RodriguezMartinez:2006ee,Jacob:2008bw}) or deforming them (\ac{DSR}, \cite{AmelinoCamelia:2000mn,Magueijo:2002am,KowalskiGlikman:2001gp,KowalskiGlikman:2004qa,AmelinoCamelia:2010pd,AmelinoCamelia:2012it,Rosati:2015pga}; see also Secs.~\ref{section2.2.1},~\ref{section2.2.2} and~\ref{section2.2.4}).

The key to link in-vacuo dispersion with observations is the analysis of time delays in the arrival of ultra-high energy particles from a distant source~\cite{AmelinoCamelia:1997gz}, assuming they are emitted simultaneously. 
While in flat (Minkowski) spacetime, it is relatively straightforward to associate a time delay formula to a \ac{MDR}, in order to comply with the cosmological distances of the relevant astrophysical sources (redshift $\gtrsim 1$), one has to take into account the contribution of spacetime curvature due to cosmological expansion.
In the \ac{LIV}  scenario, after several stages~\cite{Ellis:2002in,Ellis:2005wr,RodriguezMartinez:2006ee}, a consensus has been reached for an empirical formula for \ac{QG}  induced time delays including the contribution of redshift~\cite{Jacob:2008bw}. The time delay formula proposed in~\cite{Jacob:2008bw} has been so far the only one used for testing \ac{QG} time delays against observations.
For the \ac{DSR} scenarios, due to the much greater complexity demanded by the (deformed) relativistic framework (see also Secs.~\ref{section2.2.2.1},~\ref{section2.2.2.2} and ~\ref{section2.2.2.4}), a formulation of in-vacuo dispersion time delays including the contribution of curvature/expansion has been obtained only recently~\cite{Rosati:2015pga}, showing that the same empirical formula of~\cite{Jacob:2008bw} is obtained naturally also in the \ac{DSR} scenario, but depending on the specific of the deformed symmetries also some alternatives may arise (see below). 
The \ac{DSR}  scenario is expected to require a quantum description of spacetime, in most studies formalized as non-commutative spacetime (see also Sec.~\ref{section2.2.2.3}) and since the operative understanding of spacetime quantization has still not reached full maturity, the phenomenology at present still relies on some effective classical spacetime coordinates and classical translation transformation parameters.
While most proposed versions of the effective classical coordinates and transformation parameters all lead to the same result for the time delay formulas (see e.g.~\cite{AmelinoCamelia:2012it} and~\cite{AmelinoCamelia:2011nt}), there are some of these effective schemes that give different time delay formulas and in some cases~\cite{Carmona:2017oit,Carmona:2018xwm,Carmona:2019oph,Relancio:2020mpa} do not lead to any time delay.

The aim of this section is to discuss the search for traces of \ac{QG}  in-vacuo dispersion in the propagation of messengers (photons, neutrinos, \acp{GW}) over astrophysical scales through a comparison of their times of arrival and their spectral observations in association with astrophysical sources. In contrast, charged cosmic rays cannot be used for such studies because they are deflected by magnetic fields, so they incur time delays even assuming standard physics.

\Acp{GRB} have particularly interesting features for \ac{LIV} searches. Being among the most energetic explosions in the universe, the extremely transient \acp{GRB} can be seen on cosmological distances (redshifts up to $z \sim 8$),  emitting in different energy bands (from keV to GeV in the prompt phase), up to very-high-energy photons (0.2–3\,TeV \cite{Acciari:2019dxz,HESS:2021dbz}). Furthermore, \acp{GRB} are believed to be truly multimessenger objects, even if there is still no significant association with a neutrino signal \cite{Aartsen:2016qcr, Aartsen:2017wea}. GW170817 was at first associated with GRB 170817A \cite{Monitor:2017mdv}, but then this burst was reclassified as a kilonova \cite{Smartt:2017fuw}.

Due to the limited amount of multimessenger observations, the works on testing in-vacuo dispersion focus mostly on the electromagnetic observations of \acp{GRB}. Other possible probes are \acp{AGN}  which have lower redshift and broader time structure but higher energy photons, and pulsars, with better defined transient structure, but visible on much lower distances. Refer to Sec.~\ref{section3.2} for a more detailed description of astrophysical sources. 

An important limitation comes from the fact that the observed time delay is a combination between the intrinsic time delay due to the emission mechanism of the source, the \ac{QG} time delay and other three (at least) possible terms (due to non-zero rest mass of the messenger, dispersion by the line-of-sight free electron content, and Shapiro effect \cite{Minazzoli:2019ugi} due to the propagation in gravitational potentials). Thus, in order to evaluate the \ac{QG} time delay, it is needed to make assumptions on the source, on the contributions of the different time delay mechanisms and on the underlying cosmology. While the three additional effects can be assumed negligible for high-energy photons, neutrinos and \acp{GW} \cite{Pan:2020zbl} and the effect of different cosmologies can be evaluated numerically \cite{Biesiada:2007zzb, Biesiada:2009zz}, the problem of the intrinsic time delay is more complicated, since it requires precise knowledge of the physics of the source. The study of emission and acceleration models in order to constrain intrinsic delays in astrophysical sources is a recent effort (see e.g. \cite{Perennes:2019sjx} in the case of blazar flares) and a lot of progress is still needed in that field. 

Methods to alleviate this problem have been developed \cite{AmelinoCamelia:1997gz, Ellis:2007qt, Bolmont:2006kk, Biesiada:2009zz, Pan:2015cqa, Wei:2015hwd} and are based on the idea that the \ac{LIV} time delay will depend on the redshift, while the intrinsic time delay will not, thus allowing for statistical sampling on different redshift bins and using linear regression analysis with the slope corresponding to the \ac{QG} scale related to the \ac{LIV} effect, and the intercept representing the possible intrinsic time delay. In \cite{Zhang:2014wpb}, the authors used different intercepts for different groups of \acp{GRB}, making the strong assumption that the \acp{GRB} falling on the same line have the same intrinsic time delay.

Another problem is that \acp{GRB} are divided into two groups, short (sGRB) and long (lGRB), with different progenitors (affecting the intrinsic time delays), different time-series and selection bias with respect to $z$ (lGRBs are seen on higher $z$) \cite{Scargle:2006kr}. There are different theories on the jet-producing mechanism in \acp{GRB}~\cite{Beloborodov:2009be, Meszaros:2011hr, Duran:2010et, Bosnjak:2011pt} and Ref.~\cite{Chang:2012gq} shows that even the variation of the distance from which the jet is launched has an effect on the estimated \ac{QG} energy. This can be aggravated by high energy and low energy photons being emitted from different regions \cite{Kumar:2009ps} or by including stochastic effects \cite{Ellis:2005wr}. Since lGRBs usually have a spectral lag (a delay between high and low energy photons related to their spectral evolution), while sGRBs do not have it, another method to avoid the intrinsic time delay problem is to take only sGRB and use their duration as a conservative upper limit for the \ac{QG} time delay \cite{Chang:2015qpa, Bernardini:2017tzu}. Finally, in Ref.~\cite{Wei:2016exb} a single \ac{GRB} (GRB 160625B) is used, but with a unique positive-to-negative lag transition, which serves as an anti-correlation between the intrinsic time delay and the \ac{QG} time delay. 

The results so far do not show strong evidence of in-vacuo dispersion and different methods used in the characterization of the timing and the gaps in our understanding of the sources, lead to contradictory conclusions about the absence or presence of hints of observable effects of \ac{QG} in time delays. The most recent first order effective \ac{QG} energy scale varies between  
$E_{\text{QG},1} \ge 2.23\times10^{14} \,\text{GeV}$ \cite{Du:2020uev} and $E_{\text{QG},1} \ge 0.58\times10^{19}\,\text{GeV}$ from GRB~190114C, while the result obtained with GRB~090510 is $E_{\text{QG},1}>9.3\times10^{19}\,\text{GeV}$ \cite{Vasileiou:2013vra}. Assuming a physical association between a PeV neutrino and the blazar PKS~B1424-418, a limit of $E_{\text{QG},1} > 1.09 \times 10^{17}\,\text{GeV}$ was obtained in \cite{Wang:2016lne}.
Other analysis propose to reconsider the bounds adopting a statistical approach combining data from multiple sources. This type of analysis have been applied to both \ac{GRB}  photons~\cite{Zhang:2014wpb,Xu:2016zxi,Xu:2016zsa} and \ac{GRB}-neutrinos~\cite{Amelino-Camelia:2016fuh,Amelino-Camelia:2016wpo,Amelino-Camelia:2016ohi,Huang:2018ham} (and the combination of the two~\cite{Amelino-Camelia:2016ohi}). The results suggest a significant correlation between time delays and particle energies, with a quantum gravity scale $E_\text{QG}$ roughly compatible with the order of Planck-scale ($E_{\text{QG},1}\sim 5\times10^{17} \,\text{GeV}$). However, on the photon side they must deal with the tightest (and solid) bound provided by GRB~090510~\cite{Vasileiou:2013vra}, as well as with the assumptions on the intrinsic time lag, and on the neutrino side with the ambiguities in determining the associated sources. As already pointed out, (see also Sec.~\ref{section7.1.4.4}), no significant association between a GRB and a neutrino signal has been reported so far.

Finally, a possible origin of different dispersion relations can be a modified theory of gravity introducing a massive graviton and/or extra scalar fields \cite{Bettoni:2016mij, Yang:2018ioy, Mirshekari:2011yq,Mottola:2016mpl}. While such theories have been severely limited by the \ac{GW} observations \cite{Monitor:2017mdv}, some specific classes are still considered viable \cite{Ezquiaga:2018btd}. It has been estimated that a modified gravity will have a significant effect on the derived bounds of \ac{QG}  \cite{Biesiada:2007zzb,Biesiada:2009zz}. 

\subsubsection{Formulae for time delay}
\label{section7.1.2}

Since the energies of particles relevant for the analysis of \ac{QG}  in-vacuo dispersion are ultra-relativistic, we can limit ourselves to discuss the massless case.
In flat spacetime, the velocity of a massless particle with (spatial) momentum $p$ can be derived from its dispersion relation $E(p)$ (in unmodified special relativity $E=|p|$) through the relation (group velocity) $v=dE/dp$.
If the previous relation still holds (this is straightforward for the \ac{LIV}  case, and it is also possible to prove in the \ac{DSR}  case, once the properties of translation generators and relative locality are carefully taken into account~\cite{AmelinoCamelia:2011cv,Mignemi:2019yzn}), then, \iac{MDR} relation of the type $E = p \left(1 + \xi \frac{n+1}{2} \frac{E^n}{E_\text{QG}^n}\right)$ yields an energy dependent particle velocity
\begin{align}\label{velocity}
v \approx  1 + \xi  \frac{n+1}{2}\frac{E^n}{E_\text{QG}^n}\,,
\end{align}
where $\xi = \pm 1$ is a parameter which is either $+1$ for superluminal or $-1$ for subluminal propagation, and $n$ labels the leading order of the Lorentz invariance deformation or violation.

This result allows one to consider the following thought experiment: two massless particles, whose energies differ by an amount $\Delta E$, are simultaneously emitted from a given source at coordinate distance $R$ from the detector. Due to their different velocities they arrive at the detector with a time difference
\begin{equation}
\Delta t_\text{QG}\sim \xi R\frac{\Delta E}{E_\text{QG}}.
\end{equation}
Therefore, the distance $R$ behaves as a natural amplifier of this effect, enhancing the contribution due to the ratio $\Delta E/E_\text{QG}$, and ultra-relativistic particles emitted simultaneously with low energy photons at large distances from Earth are suitable candidates to observe such a time delay \cite{AmelinoCamelia:2008qg}.

In order for the sensitivity to the Planckian effect to be within the reach of observations, one has to consider sources at high redshift ($z\gtrsim 1$) cosmological distances. 
This requires to generalize Eq.~(\ref{velocity}) to include the contribution of spacetime curvature/expansion.
More precisely, several features have to be taken into account,
\begin{enumerate}
    \item curved spacetime effects (redshift and expansion of the universe) in the \ac{MDR}, in the context of cosmology assuming a homogeneous and isotropic universe;
    \item a careful characterization of the observer who detects the time of arrival of the massless particles, in particular the observer measurements of time intervals and measurements of energy;
    \item lack of precise knowledge of the physics of the source, and in particular of the mechanism of emission;
    \item  other possible features that could affect the characterization of time delays, like interaction of the massless particles with the intergalactic medium.
\end{enumerate}

A first attempt of deriving an empirical formula that could be used in these phenomenological investigations was made by Ellis \textit{et al.} \cite{Ellis:2002in} in the context of \ac{LIV}. After several stages~\cite{Ellis:2005wr,RodriguezMartinez:2006ee}, this result was later amended in~\cite{Jacob:2008bw, Bolmont:2006kk}.
Since this formula is the only one that is used for the phenomenological analyses, and it has proved some robustness also in respect to various conceptual arguments~\cite{RodriguezMartinez:2006ee,Rosati:2015pga} (see below), we discuss its derivation in more detail.

Let us fix the notation referring to the unmodified Lorentz invariant background. In a FLRW universe, where the universe expansion is parametrized by the scale factor $a(t)$, with $t$ the co-moving time coordinate, the dispersion relation for massless particles can be described as
\begin{align}\label{eq:DRFLRW}
    E = \frac{c p}{a(t)}\,,
\end{align}
where $c$ is the speed of light and $p$ is the conserved co-moving momentum. The redshift $z$ is given by $a(t) = \frac{1}{1+z}$, with at present time $z=0$ and $a=1$. 

While it is not obvious how to describe leading order Planck-scale corrections to~(\ref{eq:DRFLRW}), a natural ansatz, pursued in~\cite{Jacob:2008bw} (see also~\cite{RodriguezMartinez:2006ee}), is to modify~(\ref{eq:DRFLRW}) by powers of the ``redshifted" comoving momentum $p/a(t)$, so that one assumes a dispersion relation of the type 
\begin{equation}\label{DispPiran}
E = \frac{cp}{a(t)}\sqrt{1- \xi \left( \dfrac{cp}{a(t) E_\text{QG}}\right)^n} ,
\end{equation}
and obtain, to first non-vanishing order in $\frac{1}{E_\text{QG}}$, the velocity $v=dE/dp \sim c a^{-1}(t)\left[1 - \xi \frac{1+n}{2}  \left( \tfrac{c p}{a(t) E_\text{QG}}\right)^n\right]$, leading to the time delay formula ($\xi=\pm 1$ for subluminal resp.~superluminal propagation)
\begin{equation}\label{jacob-piran}
    \Delta t_\text{QG}=\xi \frac{1+n}{2H_0}  \left(\frac{ E_0}{E_\text{QG}}\right)^n\int_0^z\frac{(1+z')^n\, dz'}{\sqrt{\Omega_\text{m}(1+z')^3+\Omega_{\Lambda}}},
\end{equation}
where $E_0$ is the redshifted energy of a high energetic massless particle measured at present, $H_0$ is the Hubble parameter at present, $\Omega_\text{m}$ the pressureless matter density, and $\Omega_{\Lambda}$ the dark energy density of the universe.
    
Here $\Delta t$ is the time interval measured by an observer at a fixed position, i.e. the $t$-coordinate interval between the time of arrival of a low energetic photon and high energetic photon. Effects coming from the spatial curvature parameter $k$ of the universe have been neglected, since for the derivation $k=0$ was assumed.

\paragraph{Proposals for alternative and extended formulae}
\label{section7.1.2.1}
While the formula proposed in~\cite{Jacob:2008bw} starts from a minimal ``natural" modification of the FLRW dispersion relation, some extensions of \eqref{jacob-piran} have been suggested in the literature.

A first possible generalization arises noticing, as suggested in~\cite{RodriguezMartinez:2006ee}, that there is some arbitrariness in the choice of the dependence on the scale factor in~(\ref{DispPiran}).
Considering just linear (leading order) modifications of (\ref{eq:DRFLRW}), i.e. setting $n=1$ in the above equations, in~\cite{Rosati:2015pga} a more general ansatz for the \ac{LIV}  modified FLRW dispersion relation was assumed,
\begin{align}
    E \simeq \frac{c p}{a}\left[1 -\frac{cp}{E_\text{QG}}\left(\lambda' a(t) +\lambda'' + \frac{\lambda}{a(t)} + \frac{\lambda'''}{a^2(t)}\right)\right],
\end{align}
with $\lambda',\lambda'',\lambda''',\lambda$ dimensionless parameters, and such that for $\lambda'=\lambda''=\lambda'''=0$, $\lambda=\xi$, one obtains (\ref{DispPiran}). The corresponding expression of the time delay (for $n=1$) then reads
\begin{equation}\label{Giacomo1}
    \Delta t_\text{QG} \simeq \frac{E_0}{H_0E_\text{QG}}\int_0^z \frac{dz'}{\sqrt{\Omega_\text{m}(1+z')^3+\Omega_{\Lambda}}}\left[\lambda(1+z')+\frac{\lambda'}{1+z'}+\lambda''+\lambda'''(1+z')^2\right].
\end{equation}
It is worth mentioning that for a particular choice of the parameters, e.g. assuming $\lambda=-\lambda'$ and $\lambda''=\lambda'''=0$, the \ac{LIV}  modification is triggered by the spacetime curvature/expansion (i.e. it vanishes as $a(t)\rightarrow 1$). The phenomenology of such a so called ``curvature induced" quantum gravity effect has been recently analyzed in \cite{Amelino-Camelia:2020bvx} using data from \acp{GRB} photons (see also below).

Besides that, also in~\cite{Rosati:2015pga}, the time delay was calculated when the translation and Lorentz symmetries are not violated, but deformed, in a \ac{DSR}  scenario. 
The analysis for the \ac{DSR}  case turns out to be much more complex, requiring to take carefully into account the effects of relative locality~\cite{AmelinoCamelia:2010qv} (see Secs.~\ref{section2.2.2.1},~\ref{section2.2.2.2} and~\ref{section2.2.2.4}). 
The starting point of the analysis in~\cite{Rosati:2015pga} is a deformation of the de Sitter on-shell relation (see also~\cite{AmelinoCamelia:2012it})
\begin{equation}
m^2 \simeq E^2 - p^2 -2 H N p + (\alpha E^3 + \beta E p^2)/E_\text{QG},
\end{equation}
with $H$ the constant (de Sitter) expansion rate and $N$ the boost/charge generator (in 1+1 dimension).
The relative locality setting requires to describe carefully the action of the time and space translation symmetry generators $E$ and $p$ corresponding to energy and momentum, and the role of the observers measuring the times and positions of the photons.
This is obtained in~\cite{Rosati:2015pga} through a suitable procedure of foliation of FLRW spacetime in terms of (deformed) de Sitter slices, so that, finally, the following time delay formula is obtained
\begin{equation}\label{Giacomo2}
\Delta t_\text{QG} \simeq \frac{E_0}{E_\text{QG}}\left[\beta\int_0^z \frac{dz'(1+z')}{H(z')}+\alpha\int_0^z\frac{dz'}{(1+z')H(z')} \left(1+z'-H(z')\int_0^{z'}\frac{d\bar{z}'}{H(\bar{z}')} \right)^2\right],
\end{equation}
where $H(z)=H_0\sqrt{\Omega_\text{m}(1+z)^3+\Omega_{\Lambda}}$ is the Hubble parameter and $\alpha$, $\beta$ and $\gamma$ are dimensionless parameters that govern the deformation of the Lorentz and translation symmetries. For $\alpha=0$ and $\beta=\xi$, one has again~(\ref{jacob-piran}). It was pointed out in~\cite{Rosati:2015pga} that the formula~(\ref{jacob-piran}) proposed in~\cite{Jacob:2008bw} has the special role, both for \ac{LIV} and \ac{DSR} , of complying with translational invariance in the de Sitter limit (see~\cite{Rosati:2015pga} and~\cite{AmelinoCamelia:2012it}).

In \cite{Pfeifer:2018pty} general perturbations of the dispersion relation \eqref{eq:DRFLRW}, parametrized by a free perturbation function $h$ have been considered. Starting from the general dispersion relation $E = c p/a(t) + \xi h(t,E_0/a,p)$, the time delay for photons of different red shifted particle energies $E_{01}$ and $E_{02}$ is, to first order in the perturbation function $h$,
\begin{align}\label{Christian}
    \Delta t_\text{QG} = t_2 - t_1 = \frac{\xi}{H_0}\int_0^z \frac{f(z', E_{02}) - f(z', E_{01})}{\sqrt{\Omega_\text{m} (1+z')^3 + \Omega_k (1+z')^2 + \Omega_\Lambda}} dz',
\end{align}
where the photon of energy $E_{01}$ arrives at time $t_1$ and the photon of energy $E_{02}$ arrives at time $t_2$. The function $f(z,E_0)$ is given by
\begin{align}
    f(z,E_0) = \frac{1}{2 E_0^2 (1+z^2)^2} \left( h- E_0\frac{\partial}{\partial E_0}h - p \frac{\partial}{\partial p}h\right)\,.
\end{align}
Depending on the choice of the perturbation function $h(t(z),E_0 (1+z),p)$ this formula reproduces the above equations \eqref{jacob-piran}, \eqref{Giacomo1} and \eqref{Giacomo2}.

Another possible extension of the time delay formula~\eqref{jacob-piran}, comes from the observation that, in some string/brane-inspired microscopic models of quantum gravity, the \ac{QG}-induced space-time ``foamy" effects might be associated with interaction of photons with effective ``D-particles defects''~\cite{Ellis:1999uh,Ellis:2004ay,Ellis:2009vq,Li:2009tt}.
This would result, due to the line density of D-particle defects at a given redshift $z$, on a redshift dependent effective \ac{QG}  scale
\begin{equation}
E_\text{QG} \rightarrow E_\text{QG}(z).
\end{equation}
While such a dependence on the redshift depends on the details of the models, a natural expectation is that it scales linearly with the expansion scale factor $a(t)$. In that case the phenomenological time delay formula would coincide with one of the above formulae for a specific parametrization (for instance with~(\ref{Giacomo1}) for only $\lambda'''\neq 0$).

\paragraph{Time delays from other models}
\label{section7.1.2.2}

\par\noindent
In the framework of \ac{GUP}, see Sec.~\ref{section2.2.4}, the speed of a particle is modified via the standard dispersion relation which is GUP-modified. In addition, in a gravitational field, the speed of  a particle is  modified due to the curvature via the dispersion relation. Combining both frameworks, the speed of gravitons emitted by  a \ac{GW} event is modified due to gravitational and quantum effects. In the context of a gravitational theory beyond \ac{GR}, the speed of photons can also be modified accordingly. 
Therefore, using data from the \ac{GW} detectors (to “hear” the gravitons) and the Zwicky Transient Facility (to “see” the photons emitted during the specific \ac{GW} event) as well as employing a value for the \ac{GUP} parameter given by other experiments/observations, one can compute the difference in the speeds of gravitons, i.e., $\tilde{v}_{g}$, and photons, i.e., $\tilde{v}_{\gamma}$, coming from the same \ac{GW} event \cite{Das:2021lrb}. 
In particular, for the case of a \ac{GUP} with a quadratic term in momentum \cite{Kempf:1994su}, the difference in the speeds of gravitons and photons reads
\begin{equation}
\Delta \tilde{v}_{g}=
\left| \tilde{v}_{\gamma}-\tilde{v}_g \right|
=
\frac{3\,\beta}{c}
\left|(E_{g}^{2}-E_{\gamma}^{2})\right| \left(1+\frac{2GM}{r\,c^2}\right),
\label{diff_photon_1}
\end{equation}
where  $\beta=\beta_{0}\ell_\text{P}^{2}/\hbar^2= \beta_0/M_\text{P}^{2}c^2$  is  the dimensionful \ac{GUP} parameter, while $\beta_{0}$ is the corresponding dimensionless \ac{GUP} parameter. For the case of a \ac{GUP} with a linear and quadratic term in momentum  \cite{Das:2008kaa, Das:2010sj,Das:2009hs,Ali:2009zq, Ali:2010yn}, the difference in the speeds of the gravitons and photons reads
\begin{equation}
\Delta \tilde{v}_{g}=
\left| \tilde{v}_{\gamma}-\tilde{v}_g \right| =
\,2\alpha
\left|(E_{g}-E_{\gamma})\right|
\left(1+\frac{GM}{r\,c^2}\right),
\label{diff_photon_5}
\end{equation}
where  $\alpha= \alpha_{0}\ell_\text{P}/
\hbar = \alpha_{0}/M_\text{P}c$
is the dimensionful \ac{GUP} parameter, while $\alpha_{0}$ is the corresponding dimensionless \ac{GUP} parameter.
Finally, assuming that $\Delta t = r / \Delta \tilde{v}_{g}$, with $r$ to be the distance between our detectors and the \ac{GW} event,  we can compute the time delay $\Delta t$ between  the signals. \\

\par\noindent

Besides \ac{QG}, it is possible to modify the dispersion relation also trough some theories of modified gravity. In this case one may expect \ac{LIV}  due to a generalized dispersion relation of the form:
\[E^2=p^2c^2+m_g^2c^4+Ap^\alpha c^\alpha,\] 
where $\alpha$ and $A$ are parameters depending on the modified theory of gravity and $m_g$ is the graviton mass. One can derive \[v_g^2/c^2=1-m_g^2 c^4/E^2-A E^{\alpha-2}(v_g/c)^\alpha,\] which for  $AE^{\alpha-2} \ll 1$ simplifies to:   \[v_g/c=1+\frac{(\alpha-1)}{2} A E^{\alpha-2}.\] 
For $\alpha >1$ and $A<0$, the \ac{GW}
 travels more slowly than  the  speed  of  light. For the simplest case, $A=0$ (or $\alpha=0$), corresponding to massive graviton, from the multimessenger observation of GW170817/GRB170817 , one obtains the limit of $m_g \le 1.3\times 10^{-19}\,\mathrm{eV}/c$ \cite{Monitor:2017mdv}. Other possible values (omitting the value of $A$) are: $\alpha = 3$ corresponding to \ac{DSR}, $\alpha=4$ -- Ho\v{r}ava--Lifshitz theory, extra-dimensional theories and non-commutative geometries, multifractional  spacetime  theory for $\alpha=2-3$ etc.
~\cite{Mirshekari:2011yq, Yunes:2016jcc,Ezquiaga:2018btd}. From the LIGO measurements \cite{Monitor:2017mdv} putting a strict bound of $|c_g-c|\le 10^{-15}$ follows that the only surviving theories of modified gravity are either these starting with $c_g=c$ and applying a conformal transformation, or those that start with $c_g\neq c$ and compensate the difference with a disformal transformation \cite{Ezquiaga:2018btd}.

\subsubsection{Towards comparison with the experiment}
\label{section7.1.2.3}

In \cite{Ellis:2005wr} (see also the analyses in~\cite{Zhang:2014wpb,Xu:2016zxi}) it has been suggested that the above time delay formulas can be extended in the following way
\begin{align}
    \Delta t_{\textrm{obs}} = \Delta t_\text{QG} + (1+z) \Delta t_{\textrm{int}}
\end{align}
to include time delays due to a non simultaneous emission of the photons of different energies, as mentioned in Sec.~\ref{section7.1.2}. For the comparison with experiment one introduces $K_n:= \frac{E_\text{QG}^n}{\xi}\Delta t_\text{QG}$ for $n$-power expansion of the \ac{MDR}. Then, if there is an energy-dependence of light speed, it will manifest as a linear relation between $\Delta t_\text{obs}/(1+z)$ and $K_n$. This means that photons with the same $t_\text{int}/(1+z)$ will fall on the same line on a $K_n-t_\text{int}/(1+z)$ plot and from this slope one can find $\xi/ E^n_\text{QG}$. Here $\Delta t_{\textrm{int}}$ is often called ``intercept" \cite{Pan:2020zbl}. 
This method is referred to as the Figure of merit formula \cite{Abdo:2009zza, Ackermann:2009aa}.

It is important to remember that finding the time delay is based on the assumption that the observed time delay is a sum between the \ac{QG}  time delay and the intrinsic time delay. In general, however, there are more terms contributing to the observed time delay $\Delta t_\text{obs}= \Delta t_\text{int} + \Delta t_\text{QG} + \Delta t_\text{spec}+ \Delta t_\text{DM}+\Delta t_\text{gra}$. Here $\Delta t_\text{int}$ is the delay due to a delay of the emission of high and low energy photons at the source, $\Delta t_\text{spec}$ --  is caused by  special-relativistic effects for non-zero rest mass photons, $\Delta t_\text{DM}$ -- by dispersion of photons by the free electrons in the line of sight, $\Delta t_\text{gra}$ is the relative Shapiro time delay due to the difference in the arrival times of two particles moving in a potential well (if Einstein’s equivalence principle is violated). Some of these contributions are directional (for example  $\Delta t_\text{DM}$), other should not depend on the direction  (like $\Delta t_\text{int}$). Studies on the effect from these contributions can be found in \cite{Gao:2015lca, Wei:2015hwd} and it is generally considered that they should be negligible for the high energy photons emitted from GRBs, AGNs or pulsars.

\subsubsection{Analysis of energy and time profiles from gamma-ray data}
\label{section7.1.3}

As discussed earlier in Sec.~\ref{section3}, the two classes of effects which are commonly used in order to look for a possible signature of \ac{QG} with gamma-ray experiments are energy-dependent time delays and anomalous spectral features at the highest energies, for example due to gamma-ray decay into $e^+ e^-$ pairs in vacuum or to \iac{LIV}-induced decreased absorption by the \ac{EBL}. 

The analyses of energy and time profiles require a large amount of good quality data to be collected in the first place. In particular for the search for 
\ac{QG} induced effects, and in order to provide stringent constraints on the corresponding models, the study of the two classes of effects mentioned above implies the detectors to have a good sensitivity and energy resolution. A good sensitivity on a wide energy range not only allows the detection of more sources with a broad distribution of distances, but also the observation of the high variability needed to measure time lags. Satellites such as \textit{Fermi}-LAT constantly monitor a large fraction of the sky in the search for these events. For ground-based detectors such as \ac{IACT}, which have a limited \ac{FOV}, and in order to maximize the number of transient and random flares recorded, it is essential they can quickly move to a new target when receiving alerts issued by other instruments.

Analysis procedures have the same goal for satellites or ground-based experiments: to identify the incoming particles and reconstruct their energy and direction. Selection cuts are carefully chosen from Monte Carlo numerical simulations to minimize the amount of misidentified events, to optimize background event rejection and to maximize energy and angular resolutions. In addition, on all present-day experiments, time-stamping is provided for each event by GPS receivers which guarantee a sub-microsecond accuracy. The end result of the overall analysis procedure is simply a list of events which can be used as the main ingredient for time delay or anomalous spectral feature searches. To be complete, it is necessary to stress that some of the methods listed below also require to take into account detector response functions such as energy resolution and effective area.

\paragraph{Basic time profile analyses}
\label{section7.1.3.1}

The goal of time profile analyses in the context of searches for
\ac{QG} induced time delay effect, is mainly to look for energy-dependent time delays. This type of search has been performed in many different ways, either starting directly from individual photon properties or from energy and time distributions obtained from a larger data set.

\begin{figure}
    \centering
    \includegraphics[width=3.5in]{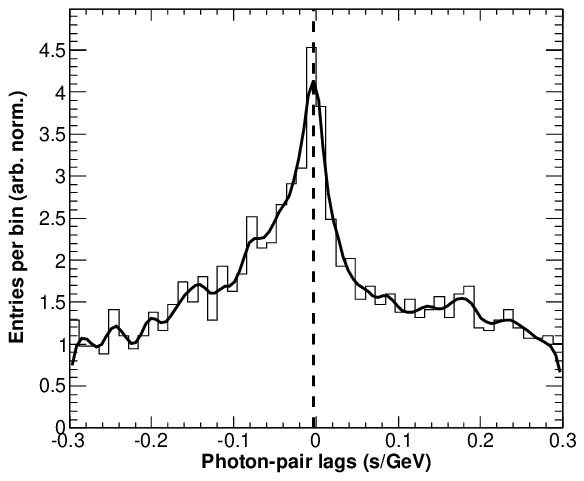}
    \includegraphics[width=3.5in]{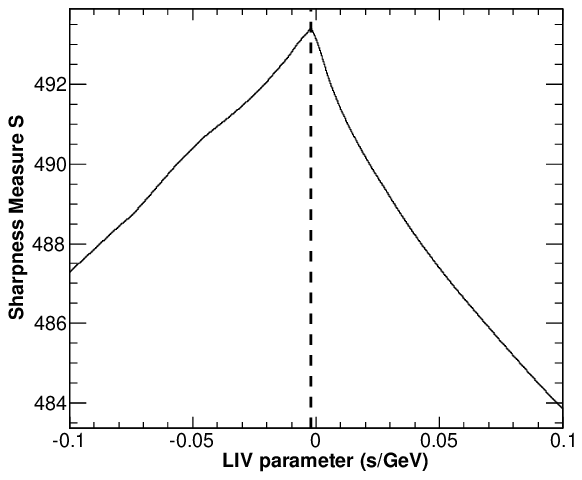}
    \caption{Application of PairView and the sharpness-maximization method on \textit{Fermi} data collected for GRB\,090510, in the case of a linear \ac{LIV} effect. Left: distribution of spectral lags $l_{i,j}$. A kernel density estimate (black curve) is used to locate the maxima of the distribution. Right: sharpness as measured by the sharpness-maximization method algorithm. The maximum of the curve corresponds to the best estimate for the \ac{LIV} parameter. Reprinted with permission from~\cite{Vasileiou:2013vra}.}
    \label{figure7.1.3_PV_SMM}
\end{figure}

A simple method consists in comparing directly the time when the highest energy photon in the sample is detected with the time when the detector is triggered \cite{Ackermann:2009aa,Abdo:2009zza}. This is done assuming this particular photon is really emitted by the source, and cannot be emitted before the main emission starts. Another technique to be applied directly to a photon list was used in \cite{Vasileiou:2013vra} to analyze \textit{Fermi} data. The procedure, called PairView, consists in constructing the distribution of spectral lags defined as 
\begin{equation}
\label{eq:7.1.3-PV}
    l_{i,j} = \frac{t_i - t_j}{E_i^n - E_j^n}
\end{equation}
for all pairs of photons $(i,j)$ in the studied sample with no duplicate ($i > j$), where $n$ is the order of the \ac{LIV} effect. The lag can then be estimated directly from the maximum of the distribution (Figure~\ref{figure7.1.3_PV_SMM}, left).

The next class of analysis techniques uses distributions of the detection times or light curves as an input. The light curves are usually produced in two or more energy bands and they are then compared with each other, either directly, or using sophisticated signal processing algorithms. The simplest method which has been used consists in a direct comparison of the position of the main maximum of a light curve in different energy ranges \cite{Boggs:2003kxa,Kaaret:1999ve}. As far as signal processing methods are concerned, \acp{CCF} or wavelet transforms were used. The \ac{CCF} is a standard method to measure the time shift between two time series. The lag can be extracted simply from the position of the maximum of the \ac{CCF} (see~e.g.~\cite{Li:2004kp}). When using wavelet transforms, several steps are required: the light curves are first de-noised using a discrete wavelet transform and then the extrema are located using a continuous wavelet transform. The difficulty here comes from the fact the extrema have to be associated in pairs (one in the low-energy light curve, the other in the high-energy one). The lags are then computed for each pair and averaged to get the overall lag for a particular source \cite{Ellis:2002in,Bolmont:2006kk}. The last class of methods to be discussed in this section uses the fact that any kind of dispersion tends to decrease the sharpness of peaks in the light curves. Several methods were developed to use this particular feature. Assuming the kind of dispersive effect at play, the opposite effect is applied to the data in order to maximize sharpness.  The sharpness-maximization method \cite{Vasileiou:2013vra}, dispersion cancellation, \cite{Scargle:2006kr}, and energy cost function, \cite{Albert:2007qk}, all use this principle to extract the amount of dispersion present in the data. In the case of the sharpness-maximization method, each photon (energy $E_i$ and time $t_i$) is first shifted to a new time $t'_i = t_i - E_i^n \times \tau_n$, where $\tau_n$ is the \ac{LIV} parameter to be estimated. Then the sharpness of the corresponding light curve is computed as
\begin{equation}
    \label{eq:7.1.3-SMM}
    \mathcal{S}(\tau_n) = \sum_{i=1}^{N-\rho} \ln\left(\frac{\rho}{t'_{i+\rho}-t'_i}\right),
\end{equation}
where $\rho$ is a parameter of the method and $N$ the number of photons in the sample. It allows to probe the light curve with different levels of details and has to be carefully chosen, usually using a large number of simulated data sets. The best estimate of $\tau_n$ is obtained maximizing function $\mathcal{S}$ (Figure~\ref{figure7.1.3_PV_SMM}, right).

In the case of the ECF method~\cite{Albert:2007qk, Abdo:2009zza, Ackermann:2009aa}, the idea is to restore the lost power of an EM pulse due to its propagating trough a linearly-dispersive medium. It aims to find a transformation restoring the original power of the signal, with $M_{\text{QG},n}$ corresponding to the maximum restored power. In practice, it finds the time interval in which the burst is most active with Kolmogorov--Smirnov statistic and then it looks for all the photons shifted by such $\delta t (E)$ and sums their energies. By repeating this procedure for different $\delta t$, one obtains the maximal energies as a function of $\delta t$, with the position of the maximum corresponding to the lag that best recovers the signal. 

The so called Band function proposed in  \cite{Band:1997yj, Aharonian:2008kz, Abdo:2009zza, Ackermann:2009aa}  is based on studying the \ac{CCF} and the \ac{ACF} of an observed transient and the temporal correlation of two time series ($v_1,v_2$) separated by a lag $\tau$. For an observation of duration $T$ consisting of $N$~data points divided between time bins of length $\Delta t= T/N$, one defines the \ac{ACF} and the \ac{CCF} as:
\begin{align*}    
    v_i &= d_i - b_i, &
    \sigma_v^{\prime \, 2} &= \frac{1}{N} \sum_{i=1}^N (v_i^2 - d_i),
\end{align*}\begin{align*}
    \operatorname{ACF}(\tau=k\Delta t; v) &=
    \frac{1}{N\sigma_v^{\prime \,2}}
    \sum\nolimits_{i=\max(1,1-k)}^{\min(N,N-k)} v_i v_{i+k}, \\    
    \operatorname{CCF}(\tau=k\Delta t; v_1, v_2) &=     
    \frac{1}{N\sqrt{\sigma_{v1}^\prime\sigma_{v2}^\prime}}
    \sum\nolimits_{i=\max(1,1-k)}^{\min(N,N-k)} v_{1i} v_{2(i+k)},
\end{align*}
where $d_i$ is the observed signal from the $i$-th bin, $b_i$ is the background rate and thus $v_i$ is the background-substracted signal.  The noise is assumed to be Poissonian, and $ \sigma_v^{\prime \, 2} $ is the variance with removed noise. The lag here is $\tau=k\Delta t$. 

To connect them with the physical properties of the \ac{GRB}, one approximate them empirically with the Band function -- a smoothly joined power law with high-energy exponential cutoff and a high-energy power law -- from which one may derive the needed time lag (and other quantities such as the peak energy) \cite{Yu:2016epf}. The simplest application is to consider two signals for the high and low energy of the type $v_\text{h}=\exp(-t/t_\text{h}), v_\text{s}=\exp(-t/t_\text{s})$ for which $t_\text{s}>t_\text{h}$ for hard-to-soft evolution. Then:
\begin{align*}
    \operatorname{ACF}(\tau,v_\text{h}) &= \exp\left(-\left|\tau\right|\middle/t_\text{h}\right),\\
    \operatorname{CCF}(\tau,v_\text{h},v_\text{s}) &= \begin{cases}
        2\frac{\sqrt{t_\text{h} t_\text{s}}}{t_\text{h}+t_\text{s}}\exp\left(-\tau\middle/t_\text{s}\right), &\tau>0,\\
        2\frac{\sqrt{t_\text{h} t_\text{s}}}{t_\text{h}+t_\text{s}}\exp\left(\tau\middle/t_\text{h}\right), &\tau<0.
    \end{cases}
\end{align*} Fitting such theoretical \ac{ACF} and \ac{CCF} to the observational one allows to describe the spectral evolution of the burst. Other possible approximations for the signal can be used, depending on its shape and other properties.  

In order to find the peak energies and the spectral indexes, the Band function can be defined as:
\begin{equation}
    f_\text{Band}(E)= A\times\begin{cases}
        \left(\frac{E}{100\,\mathrm{keV}}\right)^{\alpha}
        \exp\left(-\frac{(\alpha+2)E}{E_\text{p}}\right) & E<E_\text{c},\\
        \left(\frac{E}{100\,\mathrm {keV}}\right)^{\beta}
        \exp\left(\frac{(\beta-\alpha)E_\text{c}}{100\,\mathrm {keV}}\right) & E \ge E_\text{c},
    \end{cases}
\end{equation}
where
$E_\text{c}=\frac{\alpha-\beta}{\alpha+2}E_\text{p}$, \ 
$\alpha$~is the low-energy power-law photon index,
$\beta$~is the high-energy power-law photon index,
$E_\text{p}$ is the peak energy,
$E_\text{c}$ is the characteristic energy,
and $A$~is a normalization factor. This is one possible empirical approximation; others also are used, such as the smoothly broken power law, the cutoff power law, etc.~\cite{Yu:2016epf}.  

Since the \ac{ACF} and \ac{CCF} average over the entire burst, they assume that quantities separated by a given time-lag are related the same way throughout the burst. An improvement on this method can be found in \cite{Aharonian:2008kz} with the modified \ac{CCF}, which is a \ac{CCF} applied to oversampled light curves. Its main benefit is that it can resolve time delays below the duration of the fluxbins \cite{Aharonian:2008kz}.

\paragraph{Advanced likelihood techniques for time and energy profile analyses}
\label{section7.1.3.2}

\begin{figure}
    \centering
    \includegraphics[width=3.5in]{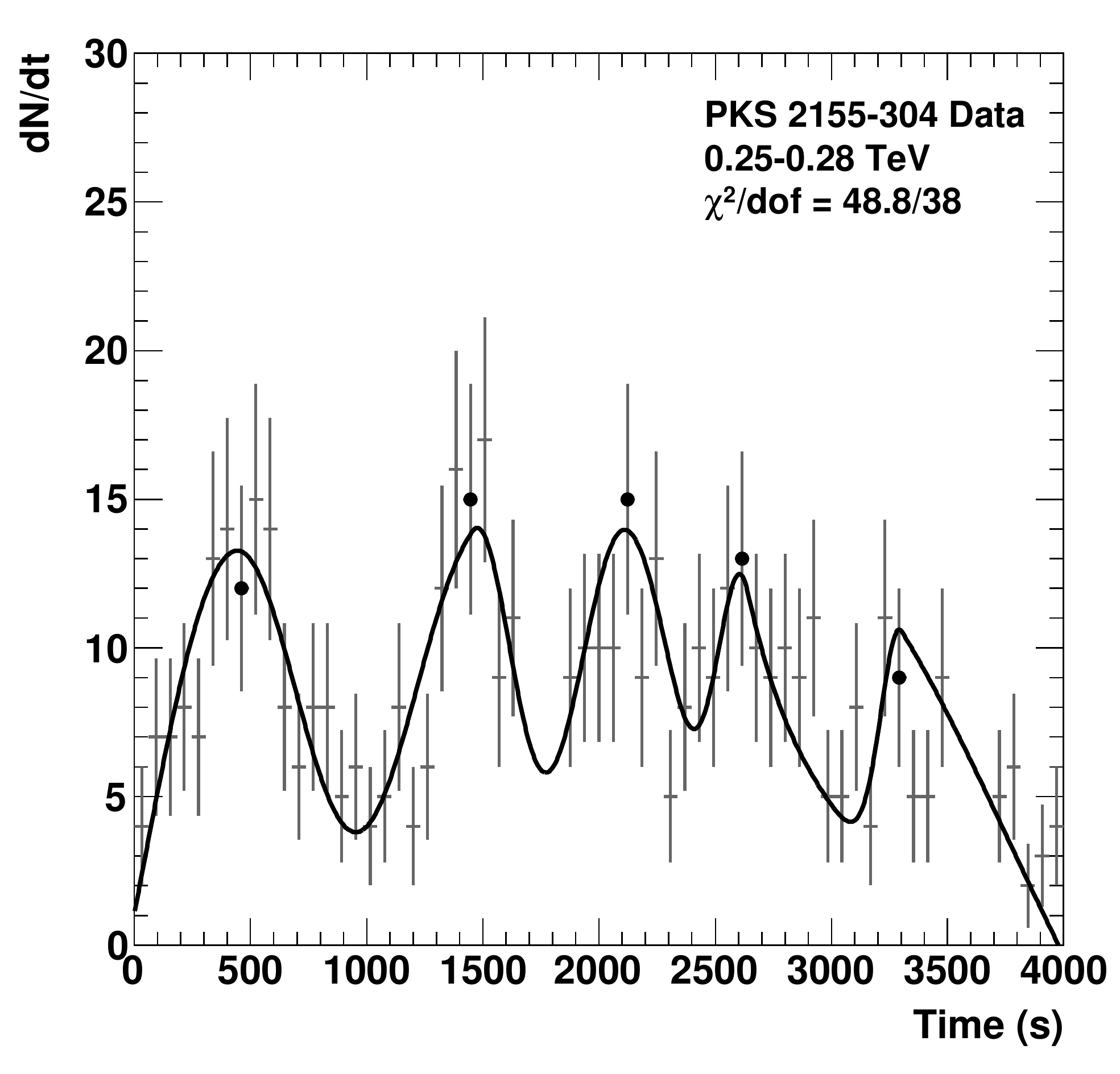}
    \includegraphics[width=3.5in]{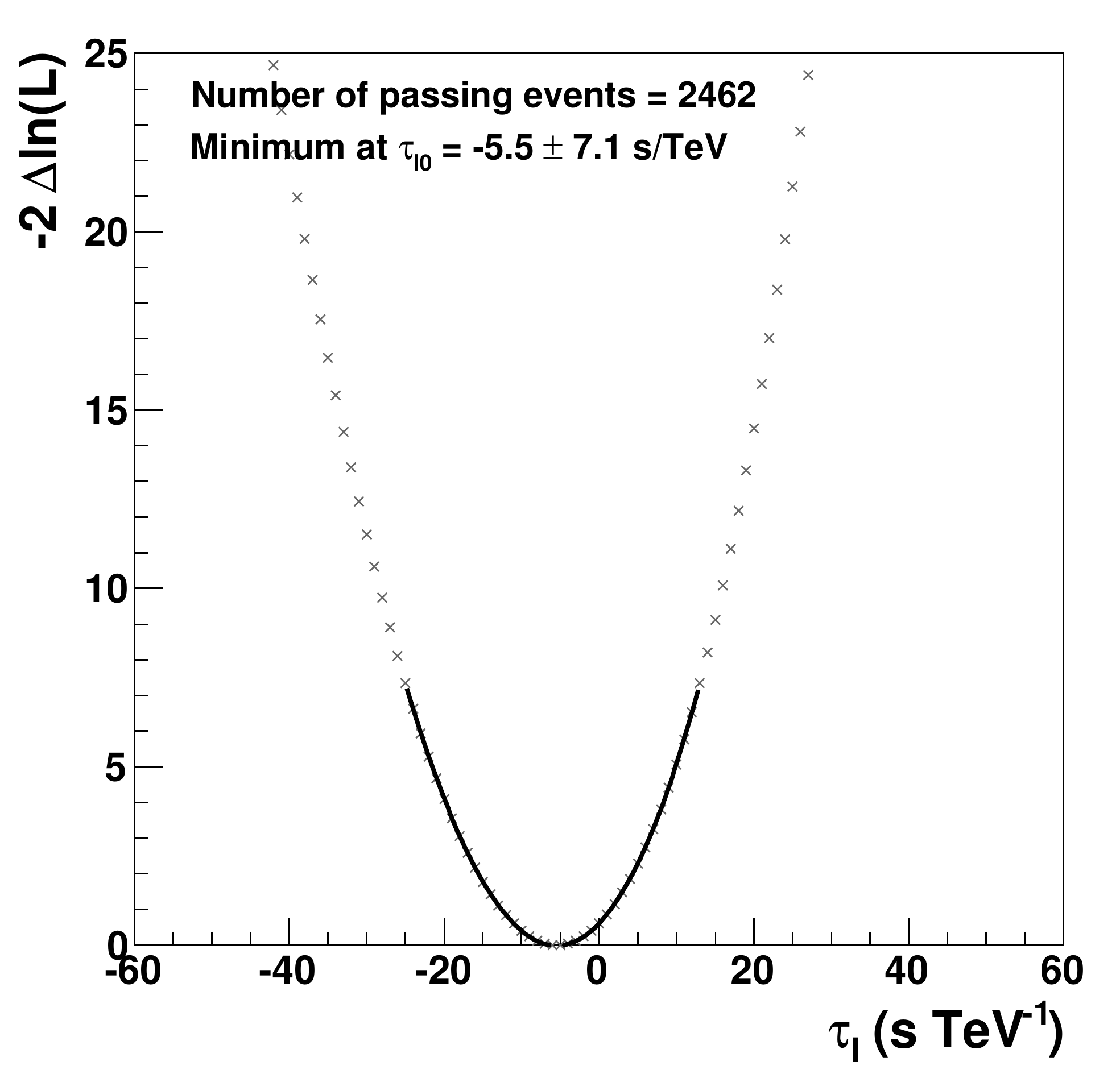}
    \caption{Application of a likelihood method on \ac{H.E.S.S.} data collected for the flare of PKS 2155-305 in 2006. Left: light curve in the range 0.25-0.28~TeV, with a 61~s bin width. The parameterization (black curve) is used as an input for the calculation of the probability density function. Right: likelihood curve, here shown as $-2\Delta\ln \mathcal{L}$ in the case $n = 1$. The minimum corresponds to the best estimate for the \ac{LIV} parameter~$\tau_{n=1}$. Reprinted with permission from~\cite{HESS:2011aa}.}
    \label{figure7.1.3_LIKE}
\end{figure}

Likelihood techniques are commonly used in many areas of physics to evaluate to what extent a set of data agrees with a particular theoretical model. In the context of \ac{QG} phenomenology with gamma rays, they have been used extensively to look for energy-dependent time delays as well as non-standard spectral features at the highest energies.

Concerning the search for anomalous phenomena in \acs{HE} gamma-ray spectra of distant sources, the way the likelihood is used is straightforward and consists in fitting the energy distribution with a model taking into account \ac{QG} induced effects. The fit provides estimated values, or limits on model parameters.

For energy-dependent time delay searches, a specific likelihood technique was developed to take into account both energy and time distributions~\cite{Martinez:2008ki}. The likelihood to be maximized is computed as the product:
\begin{equation}
    \label{eq:7.1.3-like}
    \mathcal{L} = \prod_{i} P(E_i,t_i|\tau_n),
\end{equation}
over all events, where $P$~is the probability density function of observing one event at energy~$E_i$ and time~$t_i$, given a \ac{QG} \ac{LIV} parameter~$\tau_n$. The probability density function is obtained from a propagation model, including \ac{QG} \ac{LIV} effects, and parameterizations of measured energy and time distributions:
\begin{equation}
    \label{eq:7.1.3-like-pdf}
    P(E_i,t_i|\tau_n) = N \int_0^\infty D(E_i, E_S)\,  \Lambda(E_S)\, F(t_i - \tau_n E^n_S)\, dE_S,
\end{equation}
where $D$~combines instrument response functions for energy resolution and effective area, $\Lambda$~is the energy distribution, $F$~is a parameterization of the light curve at emission, i.e.\ with no \ac{QG} induced delay, and $N$~is a normalization factor (Figure~\ref{figure7.1.3_LIKE}).
While the energy distribution is obtained from the full data set, the parameterization of the light curve is taken below a fixed separation energy, where  \ac{QG} induced effects are assumed to be small. In order to minimize bias, photons used to obtain this parameterization are usually excluded from the data sample when computing the likelihood. Therefore, the separation energy has to be chosen as a compromise to maximize the number of photons used in the likelihood fit, while keeping high enough statistics for the low energy parameterization to be as accurate as possible.

The likelihood procedure as quickly described here has become a standard for time lag searches since it has a very good sensitivity. It was applied to \ac{AGN} flares (see e.g. \cite{HESS:2011aa}), \acp{GRB} \cite{Vasileiou:2013vra,Acciari:2020kpi} and pulsar data~\cite{Ahnen:2017wec}. 

A very interesting feature of the likelihood analysis is the fact that it provides an easy way to combine several data sets, even if these were collected by different experiments on different sources, under various observing conditions, and ultimately with different systematic effects. The combined log-likelihood is simply obtained by summing log-likelihoods obtained for individual datasets. A common working group, formed from members of the H.E.S.S., MAGIC and VERITAS collaborations, has implemented this technique \cite{Bolmont:2022yad}, and is now preparing to make use of it to obtain the first combined limit from observations of GRB, AGN and pulsars collected by these three experiments.

\paragraph{Additional remarks}
\label{section7.1.3.3}

All the methods briefly described in this section have their advantages and drawbacks. As already mentioned, likelihood techniques have a very good sensitivity but their implementation can be complex and often require some simplifications. Also, the model used has to be as accurate as possible. In addition, algorithms can require long computation times, especially when all systematics are included as nuisance parameters. On the contrary, basic techniques for time profile analysis are easy to implement and quicker to execute, but are less sensitive. Since the different methods use different ways to probe the data, it is generally a good practice to compare their results (see e.g. \cite{Vasileiou:2013vra}).

To end this section, let us stress also that since no significant \ac{QG}-induced effect has been detected so far, either from time delay or \ac{EBL} absorption analyses, the limits published in the literature rely on a careful assessment of statistical and systematic errors. For a more detailed comparison between different methods and results, in particular when used on \acp{IACT} data, we refer the reader to \cite{Terzic:2021rlx}.

\subsubsection{Experimental bounds}
\label{section7.1.4}

The methods introduced in Sec.~\ref{section7.1.3} allow to measure a time delay with a statistical error and most of the time with systematic errors. To find the energy scale from the time delay, one uses the formulas given in Sec.~\ref{section7.1.2} with the  Figure of merit formula \cite{Abdo:2009zza, Ackermann:2009aa}. The principal experimental bounds derived from time delay measurements are summarized here for each type of source separately: \acp{AGN}, pulsars, and \acp{GRB}.  For an exhaustive list of bounds the reader is referred to~\cite{QGMMCatalogue}.

\paragraph{Tests with AGNs}
\label{section7.1.4.1}

\Acp{AGN} are galaxies mainly characterized by accretion of matter onto supermassive black holes at their center (see Sec.~\ref{section3.1.1} for more details). This results in emission of jets of highly relativistic plasma producing VHE gamma rays likely through inverse Compton scattering. For experimental purposes, only \acp{AGN} with their jet pointing towards the Earth (blazars and flat spectrum radio quasars) can be actually exploited. Among these variable sources, some are slowly evolving while another flaring type has strongly variating gamma-ray flux on period from minutes to months. The energy spectra of \acp{AGN} has an exponential cut-off in the TeV regime, likely due to absorption of VHE gamma rays by pair-production. The interest of \ac{AGN} flares is that they can be observed with large statistics of VHE photons up to a few tens of TeV, contrary to \acp{GRB} for which typical energies are $\sim\mathcal{O}(100)\,\text{GeV}$. On the other hand, \ac{EBL} gives rise to extinction of VHE photons, thus limiting the efficiency of these sources to redshift $z \lesssim 1$. 

The best limits for the linear deviation are obtained from PKS 2155-304 data from H.E.S.S.\ \cite{HESS:2011aa}, a blazar at $z=0.116$. The corresponding data sample with threshold $\sim120\,\text{GeV}$ has a large signal-to-background ratio. The emission time distribution is fitted by a superposition of five Gaussian spikes (left panel of Fig.~\ref{figure7.1.3_LIKE}) while the energy spectrum is consistent with a power law in the range~$0.25\text{--}4.0\,\mathrm{TeV}$. From the calibration studies, it is found that the main uncertainty is related to the event selections and to the parameterizations for the emission time distribution. The obtained 95\% CL limit is $E_\text{QG,1}>2.1\ 10^{18}\,\text{GeV}$ for the subluminal case with closely comparable value for the superluminal case.
The best limits for the quadratic deviations are obtained from one flare of high flux from Mrk~501 observed by the H.E.S.S.\ telescopes in 2014, involving gamma rays at multi-TeV, fast flux variability and energy spectrum up to 20\,TeV~\cite{HESS:2019rhe}. Mrk~501 is a blazar at $z=0.034$. The usual Jacob-Piran formula for the relative energy-dependent time delay is confronted to the data, using the Maximum Likelihood Method. Since the data have a large signal-to-background  ratio,  the contribution of the background is neglected. Events whose energy is in the range 1.3--3.25\,TeV are used to fit the emission time distribution at the source while the events with energy above 3.25\,TeV serve to calculate the likelihood and obtain the best estimate for the time delay. Statistical as well as systematic uncertainties are taken into account. The obtained 95\% CL limits are $E_\text{QG,2}>8.5 (7.3)\times 10^{10}\,\text{GeV}$ in the subluminal (superluminal) case.

\paragraph{Tests with pulsars}
\label{section7.1.4.2}

Pulsars (see Sec.~\ref{section3.1.5}) have the following advantages for \ac{LIV}  searches: they are not random events thus allowing for crosschecks with different instruments and observations for a extended periods of time, which increase the statistics and improves the limits. Moreover, the origin of the delays, intrinsic or induced by \ac{LIV}, can in principle be determined since intrinsic lags should be constant when normalized by the rotation period. In addition, pulsars observed are generally nearby galactic sources, so that the time delay do not depend on a particular lag-redshift relation. 

The best limits from Crab pulsar data for linear and quadratic departures (both subluminal and superluminal) from dispersion relations have been recently reported by MAGIC collaboration in \cite{MAGIC:2017vah}. The Crab pulsar, located at $2.0\pm0.5$ kpc, has a rotational period of 33.7 ms, slowly increasing, is a well-studied pulsar due to its brightness and large spectrum coverage. Its emission involves a main pulse (P1), an inter-pulse (P2) and a bridge region between P1 and P2. The energy spectrum of both pulses can be described by power-laws, extending to $\sim1.5$ TeV for P2. \\
In the analysis, only events close to P2 above 400 GeV have been considered among all recorded events. Two methods have been used, leading to comparable bounds although more sensitive for the second method. The first method is a direct comparison of peak positions of a pulse (the differences in mean fitted pulse positions at different energies), combined with the standard formula of average phase delay between photons of different energies. One considered the high energy band 600-1200 GeV and 2 lower bands (400-600 GeV and below 400 GeV which is likely free of change in the emission process). The second method is based on a maximum likelihood method first proposed in \cite{Martinez:2008ki}, applied to the events with energy above~$400\,\mathrm{GeV}$. The systematic uncertainties have been included in the bounds and the profile likelihood have been calibrated for energies above~$400\,\mathrm{GeV}$.\\
The best 95\% CL limits obtained from the Crab pulsar data analysis are $E_\text{QG,1}> 5.5 (4.5)\ 10^{17}\,\text{GeV}$ for subluminal (superluminal) linear departures and $E_\text{QG,2}> 5.9 (5.3)\ 10^{10}\,\text{GeV}$ for subluminal (superluminal) quadratic departures. In the quadratic case, the bounds are slightly weaker by a factor $\sim2$ than the bounds from GRB~090510 published in \cite{Vasileiou:2013vra}, while in the linear case the obtained limits are two order of magnitude below the best bounds from GRB~090510. Pulsars then appear also as appropriate sources to investigate quadratic departures from dispersion relations. 

\paragraph{Tests with GRBs}
\label{section7.1.4.3}

\Acp{GRB} are prime sources in the search for \ac{QG} effects due to their extreme brightness, high redshift and high energy emission (see Sec.~\ref{section3.1.2} for more details). Before considering test with GRBs a word of caution should be added. At present intrinsic temporal dependencies of the spectrum are not well understood and it is difficult to isolate those from LIV effects. A particularly interesting result was that of \cite{Ellis:2005wr} using 35 \acp{GRB} with known redshifts and the wavelet transform technique to search for genuine features. The obtained bound was $E_{\text{QG},1} \ge 0.9\times10^{16}\,\text{GeV}$ \ ($\ge 1.6 \times 10^{16}\,\text{GeV}$ with stochastic time-lags) and in it, the intrinsic time delay and the \ac{QG} time delay are found with statistical sampling on different redshift bins for all \acp{GRB}. 

A bound is obtained \cite{Ackermann:2009aa} by analyzing a 31\,GeV photon from GRB\,090510.  Depending on the choice of the starting (ending) time, one obtains different \ac{LIV} bounds. In the sub-luminal case, the most conservative estimate comes from assuming that the GeV photon has been emitted after the onset of the initial soft gamma-ray spike (the \ac{GRB}  trigger) and it implies $>1.45 \times 10^{19}\,\text{GeV}$. If instead, one assumes that the GeV photon comes from the main soft gamma-ray emission epoch, the bound is $>4.17\times 10^{19}\,\text{GeV}$, while if it comes from the onset of the 100\,MeV (or 1\,GeV) photons, then one gets $>6.24$ ($>12$) $\times 10^{19}\,\text{GeV}$. For the super-luminal case (negative time delay), estimates give $> 1.24 \times 10^{21}\,\text{GeV}$ if the high-energy photon was emitted before the end of the low-energy spike, or $>1.63 \times 10^{19}\,\text{GeV}$ if one takes the 750 MeV photon observed during the first soft gamma-ray spike. Final bound is obtained from the spectral lag procedure in the 30\,MeV--30\,GeV range giving $>1.49 \times 10^{19}\,\text{GeV}$.

Another estimate from GRB\,090510 using 3 different methods (PairView, sharpness-maximization method and maximum likelihood) gives $E_{\text{QG},1}>9.27 \times 10^{19}\,\text{GeV}$ and $E_{\text{QG},2}>1.3\times10^{11}\,\text{GeV}$ \cite{Vasileiou:2013vra}. Using GRB\,169625B, which is the only burst so far with a well-defined transition from positive lags to negative ones, the bounds $E_{\text{QG},1} \ge 0.5 \times10^{16}\,\text{GeV}$ and $E_{\text{QG},2} \ge 1.4 \times10^{7}\,\text{GeV}$ are obtained \cite{Wei:2016exb}. A sample of 56 \acp{GRB} was analyzed by fitting their spectral lag  with known redshifts and by using a maximization of the likelihood function and $\chi^2$ statistic, leading to  $E_\text{QG} \ge  2.0 \times 10^{14}\,\text{GeV}$ \cite{Wei:2017qfz}. Also, no redshift dependence for $\Delta t_\text{QG}$ is found. 

More recently, results based on the maximum likelihood analysis of GRB\,190114C and the detection of more than 700 photons above $0.3\,\text{TeV}$ and up to $\sim2\,\text{TeV}$ \cite{Acciari:2020kpi} found the bounds  $E_\text{QG,1}>0.58\times10^{19}\,\text{GeV}$ ($E_\text{QG,1}>0.55\times10^{19}\,\text{GeV}$ for the superluminal) for lineal corrections and $E_\text{QG,2}>6.3\times10^{10}\,\text{GeV}$ ($E_\text{QG,2}>5.6\times10^{10}\,\text{GeV}$ for the superluminal) for quadratic corrections. The 2 TeV \ac{GRB}  is currently the most energetic observed event but the limits obtained from it in the linear case are below those from GRB 090510 due to the lower redshift (the redshift of GRB\,090510 is~$z=0.9$, vs $z=0.4225$ for GRB\,190114C) and also due to the limitations in the duration of the observation.   

 Ref. \cite{Du:2020uev}, using a global fit of the spectral lag data with Bayesian analysis on GRB 190114C, found $E_\text{QG,1} \ge 2.23\times10^{14} \,\text{GeV}$ and $E_\text{QG,1} \ge 0.87 \times10^{6} \,\text{GeV}$ (by seeing an evidence of a transition from positive lags to negative lags well measured at $\sim 0.7\,\text{MeV}$). 

In \cite{Pan:2020zbl} GRB 169625B along with 23 other time delay \acp{GRB}  measured by \cite{Ellis:2005wr} were analyzed with a Gaussian process to reconstruct $H(z)$, utilizing a more complicated relation between the intrinsic and the
\ac{QG}  time delay, to produce the following bounds: $E_{\text{QG},1} \ge 0.3 \times 10^{15}\,\text{GeV}$ and $E_{\text{QG},2} \ge 0.6 \times 10^9\,\text{ GeV}$. Their results showed a strong correlation between the parameters describing intrinsic and propagation time-lags pointing to a need of a better model of the intrinsic lags. Another analysis based on data from multiple GRBs is given in \cite{Agrawal:2021cim}.

A method based on a statistical analysis of multiple \ac{GRB}  photons was proposed in~\cite{Zhang:2014wpb}, looking for regularity patterns between energy dependent time delays and some fitted values of the intrinsic time lag in the observations. The same analysis was repeated in~\cite{Xu:2016zxi}, but, differently from~\cite{Zhang:2014wpb}, where the trigger time of the Fermi Gamma-ray Burst Monitor was taken as the reference time for the time delays, in~\cite{Xu:2016zxi} it was taken the peak time of the first main pulse of \ac{GRB}  low energy photons. It was found that 8 out of the 13 \ac{GRB}  photons relevant for the analysis manifested a strong regularity along a line with $\Delta t_{int} \sim -10$\,s, yielding an estimated $E_\text{QG} \sim 3.6 \times 10^{17} \,\text{GeV}$. 
A following analysis~\cite{Xu:2016zsa} considered a further \ac{GRB}  photon from GRB\,160509A, finding that it falls on the same line of those previous 8 \ac{GRB}  photons considered, supporting the conclusion of~\cite{Xu:2016zxi}.

\paragraph{Tests with multi-messenger data:  gravitational waves,  photons and neutrinos}
\label{section7.1.4.4}

Multi-messenger astrophysics is still in its infancy. As this review is being written, only two confirmed multi-messenger detections have been reported. The first one associates the \ac{GW} signal GW170817 with the gamma-ray signal from a kilonova. The second event, the one associated with blazar TXS~0506+056, was detected in both photon and neutrino sectors. For the other results mentioned in this section, the associations between multi-messenger signals were assumed.

The \ac{GW} event GW170817 detected by the LIGO and Virgo detectors \cite{TheLIGOScientific:2017qsa} was first associated with GRB~170817A detected by the \textit{Fermi}-GBM with a very high probability: the probability of these two signals occurring by chance at nearby locations and in a short period of time is $5\times10^{-8}$ \cite{LIGOScientific:2017zic}. In the same reference, the delay between the GW and the photon signals was measured to be $1.74\pm0.04$ seconds. This delay was then used to provide limits on SME parameters. GRB~170817A was eventually found to be extremely under-luminous and was reclassified as a kilonova (see e.g.~\cite{Smartt:2017fuw}).

For completeness, let us mention another work \cite{Ellis:2016rrr} in which GW150914 signal is compared with a low-significance signal from \textit{Fermi}-GBM. The limit derived in this work is weak, with $E_{\text{QG},1} > 100\ \text{keV}$.

The object TXS~0506+056 is a blazar located at redshift $z = 0.34$. The muon neutrino signal was recorded on 22 September 2017 by IceCube \cite{IceCube:2018dnn}, while a significant (6.2~$\sigma$) detection, corresponding to a gamma-ray flare of the source, was reported by MAGIC after a 13 hour-long observation period starting on 28 September. This event was analyzed searching for the first time a \ac{LIV} evidence from both photon and neutrino signals \cite{Ellis:2018ogq}. Based on a time lag of $\sim$10 days between the neutrino and the flaring episode in the VHE gamma-ray range, the limits obtained are $E_{\text{QG},1} > 3\times 10^{16}\ \text{GeV}$ and $E_{\text{QG},2} > 4\times 10^{4}\ \text{GeV}$. This delay of $\sim$10 days illustrates one of the main difficulty when searching for \ac{LIV}-induced delays between photon and neutrino signals. Indeed, in the presence of \ac{LIV} at the Planck scale, the expected time delay between a neutrino with energy $O(10-100)$ TeV and low energy photons would be of the order of several days for redshifts $z\sim 1-2$. As a result, establishing the association between the neutrino and \ac{GRB} signals is even more challenging with \ac{LIV} than in a no-\ac{LIV} scenario.

Another limit obtained with the blazar PKS~B1424-418 ($z=1.522$) and a PeV neutrino gives $E_{\text{QG},1} >1.09 \times 10^{17}\ \text{GeV}$ and $E_{\text{QG},2} > 7.3 \times 10^{11}\ \text{GeV}$ \cite{Wang:2016lne}. However, it is important to stress that in this reference, the association between the neutrino and gamma-ray signals is assumed.
Until recently, it was believed that \acp{GRB} were natural high energy neutrino emitters. However, as of now, no significant detection of a neutrino signal associated with a \ac{GRB} was ever reported (see e.g. \cite{IceCube:2017amx}). Due to their high variabilities and large distances, \acp{GRB} are nonetheless extremely promising candidates to perform time-of-flight \ac{LIV} analyses using both neutrino and gamma-ray data when a significant association is ascertained in the future. In the remaining part of this section, we report on several analyses which were performed looking for time delays between gamma-rays and yet speculative neutrinos from \acp{GRB}.
Such a study of the correlation of time delays between IceCube TeV neutrinos and \ac{GRB} photons was proposed for the first time in~\cite{Amelino-Camelia:2016fuh}. In~\cite{Amelino-Camelia:2016wpo}, the analysis was extended considering possible effects related to dual-gravity lensing and discussing in greater detail the role of the background signal.
In~\cite{Amelino-Camelia:2016ohi}, the analysis for the neutrinos was combined with a similar one for the \ac{GRB}-photons, drawing from the results of~\cite{Xu:2016zxi,Xu:2016zsa}. A significant correlation was obtained for both kinds of signals, suggesting a quantum gravity scale $E_\text{\text{QG},1} \sim 3-6 \times 10^{17}\ \text{GeV}$. 
In~\cite{Huang:2018ham}, four additional IceCube events were considered\,\footnote{The reader's attention should be drawn to the fact that in this reference, the authors claim that IceCube collaboration suggested a probable association between neutrinos and \acp{GRB}, quoting references \cite{Aartsen:2016qcr} and \cite{IceCube:2017amx}. However, on the contrary, no association is reported in these papers.}, leading to $E_\text{\text{QG},1} \sim (6.5\pm0.4)\times 10^{17}\ \text{GeV}$.
Such a correlation, even if reported in several articles by different teams, has nonetheless to be taken cautiously. In particular, it has to be stressed again that as of now, neutrino--\ac{GRB} associations are only assumed. In addition, the number of data points is still too low to draw any conclusions, especially considering the unknown aspects of the mechanisms of emission at the sources.

\subsubsection{Additional features}
\label{section7.1.5}

The aforementioned time delays~\eqref{jacob-piran} refer to individual photons of a given energy, viewed as particles.
That description is suitable for systematic effects that concern the comparison of times of arrival between single (isolated) high energy photons and some reference time fixed by the detection of the low energy part of the signal associated to a given source (\acp{GRB} or AGNs etc.).
This kind of analysis is particularly powerful since typically only a few (rarely more than one) very high energy photons (e.g. with energy at the source $\gtrsim 40\ \text{GeV}$) are associated to a rather wide low energy signal, so that the tightest constraints on in-vacuo dispersion are the ones on systematic effects (see e.g.~\cite{Ackermann:2009aa}).
It is however interesting to consider possible non-systematic \ac{QG}  contributions that may affect the shape and width of the signal distribution itself, even at relatively lower energy ranges.
In such a case, the time delays \eqref{jacob-piran} refer to the position of a peak of a photon pulse. However, once there are \ac{QG}-induced \acp{MDR}, the \emph{shape} of the photon distribution will be affected, in the sense that the width of the path will exhibit non trivial, frequency(energy)-dependent in general,  modifications in its spread as the time evolves~\cite{Ellis:1999sd}, compared to the Lorentz-invariant case. If we assume a generic dispersion relation $\omega = \omega (k)$, a standard analysis using Fourier transforms shows that at time $t$ a Gaussian wave packet, which is a typical example of a photon pulse, sufficient for our purposes here, will have the form:
\be
   \left|f(x,t)\right|^2 = \frac{A^2}{\sqrt{1 + \frac{\alpha ^2 t^2}{(\Delta x_0)^4}}}  
\exp\Biggl(-\frac{(x - c_\text{g} t)^2 }{2(\Delta x_0)^2
\bigl(1 + \tfrac{\alpha ^2 t^2}{(\Delta x_0)^4}\bigr)}\Biggr),
\label{timegauss}
\ee
where $\alpha \equiv \frac{1}{2} \left(d^2\omega /d^2 k\right)$, and $c_\text{g} \equiv d\omega/dk$ is the group velocity, which is the velocity with which the peak of the distribution moves in time, and $\Delta x_0$ indicates the spread of the pulse at the emission time ($t=0$) at the source, which is an intrinsic source effect. 

We see immediately in Eq.~(\ref{timegauss}) that the quadratic term  $\alpha$ in the dispersion relation does not affect the motion of the peak, but only the spread of the Gaussian wave packet:
\be
 | \Delta x | = 
\Delta x_0\, \sqrt{1 + \frac{\alpha ^2 t^2}{(\Delta x_0)^4}},
\label{spread}
\ee
which thus increases with time. The quadratic term $\alpha$ also affects the amplitude of the wave packet: the latter decreases together with the increase in the spread (\ref{spread}), in such a way that the integral of $|f(x,t)|^2$ is constant. In \ac{QG}  models entailing  photon refractive indices that scale linearly with the photon energy (frequency $\omega$), the quantity $\alpha$ in \eqref{spread} is constant, proportional to the inverse power of the \ac{QG}  energy scale. This leads to a spread \eqref{spread} in the width of the photon wave packet independent of the energy of the photon to leading order in the \ac{QG}  scale.

For generic \ac{QG}  dispersive models, one can use statistical estimators, based on the
irregularity, kurtosis, and skewness of the photon sources, e.g.\ \acp{GRB}~\cite{Ellis:2018lca} that are relatively bright, to constrain possible \ac{QG}  dispersion effects during propagation (scaling linearly or quadratically in the photon energy) in a rather robust way, and also to disentangle source from propagation effects.
For instance, in the case of relatively bright \acp{GRB}, 
in the 100 MeV to multi-GeV energy
band, detected by the \textit{Fermi}-LAT detector~\cite{Atwood:2009ez}, such a detailed statistical analysis implies~\cite{Ellis:2018lca} that the energy scale
characterizing a linear energy dependence of the refractive index should exceed a few $10^{17}\,\mathrm{GeV}$.

In \ac{QG} , however, one may encounter additional effects, associated with stochastic fluctuations of the light cone.  
In the context of the (string theory) D-particle foam models~\cite{Ellis:1999uh,Ellis:2004ay,Ellis:2009vq,Li:2009tt}, such effects would result in 
stochastic fluctuations in the velocity of light 
of order~\cite{Ellis:1999uh} 
$
\delta c \sim E/E_\text{QG}^{\rm stoch} c^2, 
\label{stochc}
$
where $E_\text{QG}^{\rm stoch}$ is the scale associated to stochastic effects, that in principle can differ from the scale $E_\text{QG}$ associated to systematic effects\,\footnote{Notice moreover that in general the ``effective'' scale of the \ac{QG}  effects relevant for the analyses of this section could differ from the fundamental \ac{QG}  scale (usually taken to be around the Planck scale), especially in \ac{QG}  theories like string theory, where, depending on the details of compactification, the effective scale can be some combination of the string scale and the fundamental \ac{QG}  scale. For economy, in this section we refer in general to $E_\text{QG}$ as the effective scale for the relevant effects.}.
Such
an effect would
motivate the following parametrization of any possible
stochastic spread in photon arrival times from a distance $L$ (here we absorb, for notational brevity, the universe expansion effects in the definition  of $L$):
\be
\left(\delta \Delta t \right) = \frac{L E}{cE_\text{QG}^{\rm stoch}}\,.
\label{stochc2}
\ee
We emphasize that,
in contrast to the variation \eqref{DispPiran} 
in the refractive index -
which refers to photons of different energy -
the fluctuation (\ref{stochc2}) 
characterizes the statistical spread in the velocities
of photons {\it of the same energy}. 
We note that, in the case of \ac{QG}  models entailing
linear-in-energy refractive indices, 
for astrophysical sources at cosmological distances
with redshifts $z \simeq 1$, and with an initial $\Delta x_0
$ of a few km, one finds that the correction (\ref{spread})
is very suppressed for the relevant \ac{QG}  scales of order  $E^{(1)}_\text{QG} \gtrsim 10^{17}~{\rm GeV}$. 
Therefore, in such cases the dominant broadening effect would be the stochastic quantum-gravitational effect on the refractive index \eqref{stochc2}.

\subsection{Birefringence}
\label{section7.2}

Similarly to the time delays discussed in the previous section, birefringence is associated to anomalous in-vacuo dispersion of astrophysical messengers. 
Historically, birefringence has been mostly studied in the electromagnetic sector, but lately also \ac{GW} have been shown to be potentially affected. In this section we will cover both kinds of birefringence, respectively in Secs.~\ref{section7.2.2} and \ref{section7.2.3}.

In the electromagnetic sector, birefringence produces a rotation of the polarization plane of photons, which accumulates over cosmological distances to become potentially observable.  Such rotation can be  generated by several extensions of the standard electromagnetic theory, if they predict different speed of propagation for opposite photon polarization states (see Sec. \ref{section7.2.2} for details). Focusing on \ac{QG}-motivated modifications of electromagnetism, which produce effects that are suppressed by the Planck mass scale, the first proposal for photon birefringence was put forward in the context of (semiclassical) \ac{LQG} \cite{Gambini:1998it}. A few years later, it was shown that birefringence is also predicted by mass-dimension five operators in the \ac{EFT} extension of the \ac{QED} sector of the Standard Model \cite{Myers:2003fd}. Both these derivations lead to the same kind of birefringence effect, encoded by eq.\eqref{eq:theta} in the following section.

Regarding gravitational birefringence, it is caused by differences in propagation speed between the  $h_{+}$ and $h_{\times}$ eigenstates of a \ac{GW}, which is potentially observable at the detectors without the need for an electromagnetic counterpart. For mass dimension  5 coefficients and higher, it may yield delays between such eigenstates  causing either scrambling or splitting of the signal (shorter/longer delays between the eigenstates than the duration of the signal); changes in the amplitude between the left versus right polarized models; and also polarization-dependent lensing deflection allowing for multiple images. These effects are of particular relevance for testing \ac{QG} phenomenology in regions with matter sources where \ac{QG}-modifications to \ac{GR} predictions are non-negligible. This is the case, for instance, of theories with screening mechanisms, such as in scalar-tensor theories due to the existence of additional scalar degrees of freedom \cite{Langlois:2017dyl}. The latter modifies the background geometry around a matter-filled region and triggers the mixing of the polarization eigenstates in a way dependent on the particular screening mechanism chosen. Birefrigence can also test \ac{LIV} theories \cite{Mewes:2019dhj} (see Sec.~\ref{section7.2.3}), while parity-violating modifications may induce birefringent effects in circular polarizations for the amplitude \cite{Alexander:2004wk} and/or time delay \cite{Conroy:2019ibo,Zhao:2019xmm} of the \ac{GW} with observable cosmological imprints (see Sec.~\ref{section7.5} for other effects of \ac{CPT} violations). For instance, in Chern-Simons models  amplitude-birefringent effects at the onset of inflation affect the scalar-to-tensor ratio in the \ac{CMB} \cite{Alexander:2004wk}, and can also be tested with the \ac{LIGO}--Virgo Collaboration detections \cite{Okounkova:2021xjv} (bounding the Chern-Simons length scale by $l_0 \lesssim 1.0 \times 10^3$\,km) and with LISA \cite{Alexander:2007kv,Yunes:2010yf}.
Gravitational birefringence can also be induced by PT breaking terms coupled to gravitation. A well-known example is provided by the gravitational Chern-Simons term, also present in models including axions. This term  yields a very rich phenomenology, related to fundamental physics and dark matter, see e.g. \cite{Alexander:2004wk,Chu:2020iil}.

\subsubsection{Electromagnetic birefringence} 
\label{section7.2.2}
    \paragraph{Main formulas} 
    
    As was mentioned in the introduction, in the \ac{SME} when considering the \ac{QED} sector, mass dimension 5 \ac{CPT} breaking operators induce slight different speeds of propagation for opposite photon ``helicities''\footnote{In the \ac{LQG} model mentioned in the introduction, the dispersion relation of the two photon helicities is written as $\omega_\pm=\sqrt{k^2\mp4\chi l_\text{P}k^3}\sim \left|k\right|\left(1 \mp 2\chi l_\text{P}\left|k\right|\right)$, with $\chi$ a constant  \cite{Gambini:1998it}. This is equivalent to \eqref{eq:helicity} upon the identification $2\chi l_\text{P}=\left.\xi\middle/M_\text{P}\right.$, so that in the following we will only make reference to the parameterization of \eqref{eq:helicity}.}, 
    \begin{equation}
        \omega=k\pm \xi k^2/M_{\rm P} \label{eq:helicity}
    \end{equation}
    implying that the polarization vector of a linearly polarized plane wave with energy $k$ rotates, during the wave propagation over a distance $d$, by an angle 
\begin{equation} 
\theta(d) = \frac{\omega_{+}(k)-\omega_{-}(k)}{2}d \simeq \xi \frac{k^2 d}{M_{\rm P}}\;.
\label{eq:theta}
\end{equation} 
Clearly such angle increases quadratically with the energy and linearly with the distance, so that highly energetic astrophysical sources are optimal probes of this effect.

When considering astrophysical sources one needs to account for the effects of energy redshift. For a source at redshift $z$ with measured energy $k$, the polarization rotation angle $\theta$ is given by \cite{Gubitosi:2009eu, Galaverni:2014gca}
\begin{equation}
 \alpha(k,z)=\frac{\xi}{M_\text{P}}k^2\int_0^z (1+\bar z)H(\bar z)^{-1}d\bar z\,,   \label{eq:thetaredshift}
\end{equation}
where $H(\bar z)$ encodes the universe expansion history (see Sec.~\ref{redshift}).

Constraints on vacuum birefringence are set using two different techniques: the first one aims at measuring directly the polarization rotation angle \eqref{eq:thetaredshift}, by comparing the expected direction of the linear polarization of a source with the measured one; 
the second one relies on the fact that the energy dependence of the  polarization rotation could potentially disrupt the amount of linear polarization present in a some polarized light beam traveling over long distances. If some net amount of polarization is measured in the band, say, $k_{1} < E < k_{2}$, an order-of-magnitude constraint arises from the fact that the angle of polarization rotation (\ref{eq:theta}) cannot differ by more than $\pi/2$ over this band, as in this case the detected polarization would fluctuate sufficiently for the net signal polarization to be suppressed \cite{Gleiser:2001rm, Jacobson:2003bn}.   From (\ref{eq:theta}), assuming $\Delta\theta \leq \pi/2$ implies
\begin{equation} 
\xi\lesssim\frac{\pi\,M_{\rm P}}{(k_2^2-k_1^2)d(z)}\;.
\label{eq:decrease_pol}
\end{equation} 
This constraint merely relies on the detection of a polarized signal and does not make use of the observed
polarization degree, however, one can obtain  a more refined limit by calculating the maximum
observable polarization degree, given some maximum intrinsic value. 

\paragraph{Distinguishing Quantum Gravity effects from standard effects} 
    
  The polarization rotation effect  discussed above might be caused also by different physics besides \ac{QG} (e.g. Faraday rotation in presence of a magnetic field, or the presence of axion fields \cite{Guendelman:1991se}).
  
  The main criterion that allows us to distinguish the cause of a possible birefringence effect is its energy dependence. As seen in equation \eqref{eq:theta}, the expected energy dependence for a \ac{QG}-induced effect is quadratic. The expected energy dependence for other possible sources of birefringence is different, ranging from an inverse square law to energy-independence to linear dependence, and is discussed in \cite{Gubitosi:2012rg, Galaverni:2014gca}.
  
     Of course, while in principle a measurement of the polarization rotation angle performed at a given energy can be translated into a constraint on the parameter $\xi$ of equation \eqref{eq:theta}, this measurement does not allow us to distinguish \ac{QG}-induced birefringence from other kinds of birefringence, because it does not allow us to test for the energy dependence. For this reason a strategy to determine the quantum-gravitational origin of the effect  is to compare data on the rotation of the polarization vector from different sources emitting radiation at different energies \cite{Gubitosi:2012rg, Galaverni:2014gca}. This kind of analysis assumes that the effect should be the same for all sources and isotropic. Isotropy can in fact be violated by more general birefringence effects with quantum-gravitational origin, see \cite{Gubitosi:2010dj}, but constraints on this kind of effect are much less studied at the moment and rely mainly on studies of the \ac{CMB} \cite{Gubitosi:2011ue, Gruppuso:2020kfy}.
     
     Recent analyses of \ac{GRB} polarization overcome these difficulties, since they rely on the comparison of the polarization direction within different energy bands of the same source, see below.

\paragraph{Existing bounds}

    In the following we list the main sources used to measure directly the polarization rotation angle, grouped by energy of the source and distance. We also report the constraints in Table \ref{table:birefringence}.  We distinguish the observations made within a single energy band, and those that compare several energy bands, see the discussion above. In some of the works reported below, only the constraints in terms of the rotation angle $\theta$ are provided and not in terms of $\xi$. In order to translate the constraints on $\theta$ into constraints on $\xi$ one can use the relation \eqref{eq:thetaredshift}. 
    
    \begin{itemize}
        \item \ac{UV} emission of radio galaxies \cite{Alighieri:2010eu}, energy at detection $E\sim 2.5 $ eV, distance $2\lesssim z\lesssim4$, obtaining $\theta=-0.8^{\circ}\pm2.2^{\circ}$ assuming isotropy of the effect;
        \item Radio sources \cite{Whittaker:2017hnz}, energy $E\sim 10^{-5} $ eV, $z < 1$, obtaining $\theta=-2.03^{\circ}\pm 0.75^{\circ}$ assuming isotropy of the effect;
        \item Crab Nebula (comparison between the neutron star rotation axis - measured by the HST and Chandra satellites - and the gamma-ray polarization direction - observed by INTEGRAL) \cite{Maccione:2008tq}, energy $E\sim 10^2 $ keV, distance $z\sim 10^{-7}$, obtaining $|\xi|<9\cdot10^{-10}$ at $3\sigma$ c.l.;
        \item \ac{GRB}s (method based on the comparison of the polarization angle of different energy bands of the same \ac{GRB}, so this measurement assumes a given energy dependence of the effect and for this reason is capable to determine the quantum-gravitational origin of the effect) \cite{Gotz:2013dwa}, energy bands are in the range $250-350$ keV and $350-800$ keV and the redshift is $z=1.33$, $\xi<3.4\cdot10^{-16}$.

     \end{itemize}
        
Besides the above-listed constraints which rely on just one kind of sources, Ref. \cite{Galaverni:2014gca} provides a constraint cast by combining measurements of the polarization rotation angle from different kinds of sources at different redshifts and energies (this constraint assumes isotropy), obtaining $\xi=(1.2\pm 14.1)\cdot 10^{-11}$ if \ac{GRB} data are excluded and $\xi=(0.0\pm 8.6)\cdot 10^{-17}$ including \ac{GRB} data.

    \begin{center}
 \begin{tabular}{||c| c | c| c||}   \hline\label{table:birefringence}
 Source & $E$ & $z$ & constraint\\ [0.5ex] 
 \hline\hline
  Radio sources & $10^{-5} $ eV & $z < 1$  &   $\theta=-2.03^{\circ}\pm 0.75^{\circ}$ \cite{Whittaker:2017hnz} \\
 \hline
Distant radio galaxies & $2.5 $ eV & $2\lesssim z\lesssim4$ &  $\theta=-0.8^{\circ}\pm2.2^{\circ}$ \cite{Alighieri:2010eu} \\ 
 \hline
 Crab Nebula & $10^2 $ keV & $z\sim 10^{-7}$ & $|\xi|<9\cdot 10^{-10} $ \cite{Maccione:2008tq} \\
 \hline
 GRBs & $250-350$ keV and $350-800$ keV & $z=1.33$ & $\xi<3.4\cdot10^{-16}$  \cite{Gotz:2013dwa}  \\
 \hline

\end{tabular}
\end{center}
    
    As mentioned above, upper bounds on birefringence can be also cast by observing that if the polarization rotation angle were larger than $\pi/2$ over a certain energy band, then polarization would be erased in that band, see discussion above and equation \eqref{eq:decrease_pol}. In the following we list constraints cast using this technique
    
    \begin{itemize}
    \item \ac{GRB}s: $|\xi|<\mathcal{O}(10^{-15})$ \cite{Toma:2012xa}
    \item \ac{UV} emission of radio galaxies $|\xi|<5\cdot 10^{-4}$ \cite{Gleiser:2001rm}
    \end{itemize}

    Another way to constrain birefringence is based on the analysis of the polarization of the \ac{CMB}. These constraints lie beyond the scope of this review, so we just list a few references \cite{Gubitosi:2009eu, Aghanim:2016fhp,Bianchini:2020osu,Namikawa:2020ffr,Minami:2020odp}. Regarding these studies, it is interesting to note that in reference \cite{Minami:2020odp} the authors claim to have found a birefringence angle of $0.35\pm 0.14$, excluding a zero value at $99.2\%$ c.l. Notice that \ac{CMB} data also allow us to perform a study of the energy dependence of the effect, even though just a few studies exploit this possibility \cite{Gubitosi:2012rg, Galaverni:2014gca}.

\subsubsection{Gravitational birefringence} \label{section7.2.3}

\paragraph{Birefringence constraints from multimessenger signals} 

The space of viable modified gravity theories was already heavily constrained by the observation of $\left| c_\text{GW}/c-1 \right| \lesssim 5 \times 10^{-16}$, as follows from the \ac{BNS} merger event GW170817 \cite{TheLIGOScientific:2017qsa} and its EM counterpart GRB170817 \cite{Monitor:2017mdv}, which ruled out many \ac{QG}-motivated theories predicting different in-vacuo propagation speeds between messengers \cite{Ezquiaga:2017ekz}. The consideration of birefringence effects via time delays may allow to further constraint the space parameter of specific theories. In \ac{QG}-motivated theories, in general this time delay will be a combination of the modified effective metric induced by the theory in which all eigenstates propagate, and the pure birefringence effects causing each eigenstate to propagate differently. The time delay between two \ac{GW} eigenstates $I,J$ with respective velocities $c_I,c_J$ as a matter-filled region (the lens)  is crossed in the direction of propagation of the signal, $u$, can be estimated, under some approximation, as \cite{Ezquiaga:2020dao}
\begin{equation}
\Delta t_{IJ}=\int du\left(\frac{1}{c_I}-\frac{1}{c_J}\right)+\frac{(1+z_L)}{2c}\frac{D_L D_{LS}}{D_S}\left( \vert \vec{\hat{\alpha}}_I \vert^2 -\vert \vec{\hat{\alpha}}_J \vert^2 \right)
\end{equation}
where the first term is the Shapiro delay while in the second (geometrical time delay) term $D_L,D_{LS},D_S$ are the angular diameter distances to the lens, source, and between the lens and the source, respectively, $z_L$ the redshift of the lens, and $\vec{\hat{\alpha}}_I\approx -\tfrac{1}{2}\int du \vec{\nabla}_{\perp} c_I^2(\vec{x},\hat{k})$. The exact value of the birefringence-induced time delay will thus depend on the \ac{QG} modifications to the background geometry, allowing to probe specific \ac{QG}-motivated theories. This is the case of the quartic Horndeski one, as it can be seen in Fig.~16 of \cite{Ezquiaga:2020dao}, where testable delay times from birefringence are estimated to be in the range $\Delta t_{10}>1$\,s for \ac{GW}-electromagnetic wave delay, and $\Delta t_{21}>1$\,ms for delays between polarizations, allowing to test the space parameter of this theory beyond the constraint of GW170817 + GRB170817. Additionally,  time delay constraints from multimessenger observations allow to constraint the energy scale for any theory of gravity with the lowest-order parity-violating terms (which excludes the Chern-Simons model, where this effect does not exist) to the level $M_\text{PV} \geq \mathcal{O}(10^4\,\mathrm{eV})$ \cite{Zhao:2019szi}.

\paragraph{Birefringence in the Standard-Model Extension}

In the \ac{SME}, operators of mass dimension  $d \geq 5$ induce gravitational birefringence due to a polarization-dependent dispersion of the \ac{GW} during its propagation \cite{Kostelecky:2016kfm}. 
The dispersion breaks the rotational symmetry of Lorentz invariance, contrarily to the polarization-dependent dispersion phenomena occurring at lower mass dimension that break the boost symmetry. The \ac{GW}  4-momentum is modified by the dispersion as: 
\begin{equation}
    \omega= 1  -  \varsigma^0  \pm \sqrt{(\varsigma^1)^2 + (\varsigma^2)^2 + (\varsigma^3)^ 2}\,,
    \label{smeomega}
\end{equation}
where birefringence is due to the $\varsigma^{1,2,3}$ terms that can be expressed as a function of the \ac{GW} spherical harmonics and \ac{SME} operators of mass dimensions 5 and 6, $k_{(V)jm}^{(d=5)}$, $k_{(E)jm}^{(d=6)}$ and $k_{(B)jm}^{(d=6)}$. For mass dimension 5 coefficients, the birefringence  time delay reads as follows: 
\begin{equation}
    \Delta t \simeq  2 \omega^{d-4}  \int_{0}^{z} \frac{(1+z')^{d-4}}{H_{z'}}  dz' \sum_{jm} Y_{jm}(\boldsymbol{\hat{\textbf{n}}}) k_{(V)jm}^{(d=5)}\,.
    \label{smedt}
\end{equation}
In \cite{Mewes:2019dhj} the non-observation of a double peak in the LIGO interferometers with the GW150914 event is exploited to constrain a linear combination of  mass dimensions 5 and 6 \ac{LIV} operators. They obtain the first constraints on gravitational \ac{LIV} $d=5$  operators and competitive bounds with laboratory experiments for $d=6$. 
Beyond the delay that may be observed at the peak, the modification of the \ac{GW} signal due to spacetime birefringence is derived in \cite{Mewes:2019dhj}, and sensitivity studies carried in \cite{ONeal-Ault:2021uwu} show that single events observable at advanced \ac{LIGO} and Virgo sensitivities can place constraints on the order of $\mathcal{O}\left(10^{-13}\right)$ on mass dimension 5 coefficients. 
Using the 11 samples of the catalog of \ac{GW} detections during the first \ac{LIGO}--Virgo Collaboration observational runs described on Sec.~\ref{GW-detectors},  \cite{Shao:2020shv} constrains individually the same coefficients. Bounds are consequently improved with factors  $10^2$ to $10^5$, with orders $\left[\mathcal{O}(k_{(V)jm}^{(d=5)})\right]  \lesssim 10^{-15}$ and $\left[\mathcal{O}(k_{(E,B)jm}^{(d=6)})\right]  \lesssim 10^{-11}$, respectively. \\

\paragraph{Distinguishing quantum from classical effects: the case of the gravitational spin Hall} 

   The gravitational spin Hall effect is a purely classical effect which emerges as a next-to-leading-order correction to the geodesic approximation for the propagation of wave packet solutions to Maxwell's equations or to the equations of linearized gravity of a fixed spacetime background. The effect gives polarization-dependent trajectories for circularly polarized waves and leads to the splitting linearly polarized waves. The magnitude of the effect is inversely proportional to the frequency of the wave, depends on the curvature of spacetime, and has only recently been understood to full extent for Maxwell's equations \cite{oancea2020gravitational} and for linearized gravity \cite{andersson2021propagation}. A similar treatment for the Dirac equation can be found in \cite{she_thesis}, and a general overview can be found in \cite{oancea2019overview}. In \cite{frolov2012scattering}, questions of spin Hall effect induced time delay have been considered in black hole spacetimes, albeit with an incomplete set of correction terms. Given the quite recent developments of the theoretical foundations, observational confirmation or observational bounds do not exist yet. 
   
\subsection{Modified interactions and threshold effects}
\label{section5.3}

Planck-scale modifications of elementary particles' dispersion relations, emerging in several approaches to quantum gravity (see Sec.~\ref{section2}), modify the kinematics of particle interactions. 

The kinds of modifications one might expect for a given process depend on two main factors:
\begin{itemize}
    \item whether the \ac{MDR} is universal or different for different particles;
    \item whether the \ac{MDR} is accompanied by a modified energy-momentum conservation law.
\end{itemize}

Concerning the first point, the possible non-universality of the \acp{MDR} introduces a high level of complexity in phenomenological analyses. In fact, predictions for a given process, say $a+b\rightarrow c+d$ where two particles $a$ and $b$ interact to produce particles $c$ and $d$, depend on the dispersion relation that is assigned to each of the four particles, with a large number of different possible combinations. Even assuming, as a simplifying example, that for each particle  $i$ ($i={a,b,c,d}$) just the leading order corrections to dispersion relations should be taken into account,
\begin{equation}
    E_i^2=m_i^2+|\vec p_i|^2 \left(1+ \eta_i \left|\vec p_i\right|/E_{\text{P}}\right)\,,
    \label{dispersionrel}
\end{equation}
still the values of the correction coefficients $\eta_i$ could all be different, leading to a variety of different predictions for the same process.

Concerning the second point, there are fundamentally two options for the energy-momentum conservation law, depending on whether the theory is breaking Lorentz symmetries (this is the case of \ac{LIV} theories, discussed in Sec.~\ref{section2.2.1}, where the conservation laws are unmodified with respect to \ac{SR}) or it is just deforming them (this is the case of \ac{DSR} theories, discussed in Sec.~\ref{section2.2.2}, where conservation laws are modified in order to be invariant under the same deformed symmetries that leave the dispersion relation invariant). An in-depth discussion of the differences between the two options can be found in Sec.~\ref{section2.3.2}.  
Here we briefly mention that the two options  produce different predictions for particle interactions, even if one starts from the same dispersion relation. Most notably, in the case of \ac{DSR} one cannot have an appearance of thresholds that would normally be absent in \ac{SR}, because the relativity principle is still valid and these kinds of thresholds would introduce a preferred frame. In contrast, such thresholds do appear in \ac{LIV} models.
This is the reason why the phenomenology related with thresholds has been focused on the \ac{LIV} scenario.

Besides the two kinematical issues just discussed, one has to consider the dynamical features of the theory within which the deformed dispersion relations (and possibly deformed conservation laws) are embedded in considering constraints on interaction processes. This is relevant, for example, in studying the onset of a threshold above which a process that is usually forbidden in \ac{SR} becomes allowed in the modified theory. The non-observation of such a process above the threshold energy does not completely rule out the kinematical modifications, because there could be dynamical reasons why the process does not happen.
These considerations explain the popularity of a framework to study \ac{LIV} in the \ac{SM}, known as (nonminimal) \ac{SME} \cite{Colladay:1998fq, Kostelecky:2009zp}, which includes the \ac{LIV} terms in the \ac{SM} Lagrangian (see Sec.~\ref{section2.2.1}).

In the following, we review the main processes that are relevant to constrain Planck-scale-modified kinematics in particle interactions, grouped by relevant \ac{SM} sectors. We discuss the constraints obtained in the \ac{LIV} scenario, and in the last section, we discuss the cases where the \ac{DSR} scenario can be applied.

\subsubsection{Sectors} 
\label{section5.3.1}

Here we classify the different \ac{LIV} phenomenology regarding its sector: \ac{QED}, with processes involving electrons, photons, and neutrinos; hadrons, focused on \ac{UHECR} and air shower thresholds modifications; and gravity, describing the consequences of having \iac{MDR} for gravitons. 

In the following, we consider the following dispersion relations for the $i$-th particle:
\begin{equation}
    E_i^2 = m_i^2 + p_i^2~\left( 1 + \sum_{n=0,1,2} \eta_n^i \left(\frac{p_i} {E^{(n)}_{\text{LIV},i}}\right)^n\right).
 %   m_i^2 + p_i^2\,\left( 1 + \sum_{n=0,1,2}\eta_i \left( \frac{p_i}{E^{(n)}_{\text{LIV},i}}\right)^n\right).
    \label{eq:mdr}
\end{equation}
Here the mass scale~$E^{(n)}_{\text{LIV},i}$ determines the scale of \ac{LIV} for the $i$-th particle type for the corresponding power~$n$.
In the case~$n=0$, the parameters~$\eta^i_0$ are small values determining the maximal velocity of the corresponding particle. 
In the cases~$n=1$ and~$n=2$, the parameters~$\eta^i_n$ are just~$\pm 1$ since the magnitude of \ac{LIV} is determined by the scale~$E^{(n)}_{\text{LIV},i}$. The positive sign  refers to the so-called superluminal type of \ac{LIV} while the negative sign refers to subluminal type.

\paragraph{QED sector} \label{sec:QEDLIV}
In the following, we denote the \ac{LIV} parameters for photons~$\eta^\gamma_n$ as~$\xi_n$ and those for electrons and positrons~$\eta^e_n$ as~$\eta_n$.
In some \ac{CPT}-odd models the signs of~$\xi_1$ depend on the photon polarization and the value of~$\eta_1$ depends on the electron helicity state \cite{Jacobson:2005bg}. Note that only part  of \ac{LIV} effects, referred to kinematics, can be described in terms of \acp{MDR}. The full picture including dynamic \ac{LIV} effects came from the Lagrangian for a concrete model.
 
For example, the term
\begin{equation}
-\frac{1}{4}(k_F)_{\mu\nu\rho\sigma}F^{\mu\nu}F^{\rho\sigma}\label{eq:QED_gamma}
\end{equation}
with the real and dimensionless tensor $(k_F)_{\mu\nu\rho\sigma}$ that can be taken as double-traceless leads to \ac{CPT}-even \ac{LIV}. 10~of the independent components
of $(k_F)_{\mu\nu\rho\sigma}$ lead to birefringence, 8~lead to direction-dependent effects, one can be absorbed into the photon field normalization and only the
component
\begin{align}
(k_F)_{\mu\nu\rho\sigma}&=\frac{1}{2}\left(\eta_{\mu\rho}\tilde\kappa_{\nu\sigma}-\eta_{\mu\sigma}\tilde\kappa_{\nu\rho}+\eta_{\nu\sigma}\tilde\kappa_{\mu\rho}-
\eta_{\nu\rho}\tilde\kappa_{\mu\sigma}\right),
&\text{with}\quad\tilde\kappa_{\mu\nu}&=\frac{\kappa}{2}\left[{\rm diag}(3,1,1,1)\right]_{\mu\nu}\label{eq:QED_gamma2}
\end{align}
leads to isotropic nonbirefringent \ac{LIV}, characterised by the real dimensionless parameter $\kappa$.
Apart from this example, which in particular has been considered in the context of \acp{UHECR}, for compactness, we do not provide the concrete Lagrangians here explicitly but refer to the dynamic effects as well.

In the following paragraphs we discuss some relevant \ac{QED} processes, appearing or modified in the case of \ac{LIV}, and the corresponding bounds on \ac{LIV} parameters in \ac{QED} sector.
 
\begin{description}

\item{\it Synchrotron radiation:} Synchrotron radiation is a classical process of electromagnetic radiation by a charged particle moving not parallel to  magnetic field. Most of the radiation from the charged particle is emitted around a critical frequency $\omega_c = \frac{3}{2}eB\gamma^3(E)/E$, where $B$ is the magnetic field, $E$ is the particle's energy, and $\gamma(E) \equiv \left( 1 - v^2(E)\right)^{-1/2}$ is particle's Lorentz factor; $v$ is its group velocity. In the case of \ac{LIV}, the formula for the critical frequency keep the same form but the group velocity $v$ and so the Lorentz factor $\gamma(E)$ get modified \cite{Jacobson:2002ye}. 
The best constraints on \ac{LIV} mass scale for electrons \cite{Jacobson:2002ye,Maccione:2007yc, Liberati:2012jf} came from the observation of the Crab Nebula photon spectrum in $100$\,MeV energy range which is assumed as a result of synchrotron radiation by electrons with energy up to $1$\,PeV \cite{2010A&A...523A...2M} inside the Crab Nebula. The constraints on the \ac{LIV} mass scale are two-sided, $E^{(1)}_{\text{LIV},e} < 10^{24}$\,GeV for the case~$n=1$ \cite{Maccione:2007yc}, and $E^{(2)}_{\text{LIV},e} < 2\times 10^{16}$\,GeV for~$n=2$ \cite{Liberati:2012jf}, see also Table~\ref{sec5.4:Table1}. The synchrotron constraints refers only to \ac{LIV} in electron sector since the outgoing photons are sufficiently less energetic than the initial electrons.

\item{\it Vacuum Cherenkov radiation  $e^{\pm}\rightarrow e^{\pm}\gamma$ and related processes:}

The process of photon emission by a moving electron in a vacuum is called vacuum Cherenkov radiation, in analogy with the standard Cherenkov radiation in media. Vacuum Cherenkov radiation is forbidden in \ac{SR} and occurs if the electron velocity exceeds the velocity of the emitted photon (superluminal dispersion relation for electrons).
Thus, it happens when the electron energy exceeds a threshold \cite{Coleman:1997xq, Jacobson:2002hd}.
 
The simplest case of vacuum Cherenkov radiation is related to the emission of soft photons so that the Lorentz breaking is negligible in the photon sector.
For $n=0,1,2$ the  vacuum Cherenkov threshold energy is given by \cite{Jacobson:2002hd}
\begin{equation}
 E^{\text{thr}}_{\rm VC} = \left(\frac{m_e^2 \left( E^{(n)}_{\text{LIV},e}\right)^n}{(n+1)\eta_n}\right)^{1/(n+2)}.
\label{eq:VC_th}
\end{equation}
Just above threshold this process is also extremely efficient, with a time scale of order $\tau_{\rm VC} \sim 10^{-9}$\,s \cite{Jacobson:2005bg}.
The presence of at least $1$\,PeV long living electrons inside the Crab Nebula \cite{2010A&A...523A...2M} allows to cast the bound $E^{(1)}_{\text{LIV},e} > 10^{22}$\,GeV,  $E^{(2)}_{\text{LIV},e} > 10^{14}\,\mathrm{GeV}$. 

If the parameters $\eta_n$ 
have different values for two helicity states, the decay process with changing helicity may occur
\footnote{For ultrarelativistic electrons the helicity states almost coincide with electron or positron states.}
\cite{Jacobson:2005bg,Maccione:2007yc}.
This process is usually called helicity decay. Contrary to vacuum Cherenkov radiation, helicity decay is allowed for any energy. However, there is an effective threshold for this process: at smaller energies than the effective threshold helicity decay is suppressed; for larger energies the rate  of helicity decay and vacuum Cherenkov radiation are of the same order \cite{Jacobson:2005bg}. The similar effect appears in other subsectors of \ac{SME} not reduced to the modifications of dispersion relations \cite{Borisov:2020cub}.
For some parameters ranges the process of pair emission $e^- \to e^- e^- e^+$ may be also effective  \cite{Jacobson:2005bg}.

\item{\it Photon decay $\gamma \to e^+e^-$:} One of the simplest \ac{LIV} reactions in \ac{QED} is the process of vacuum photon decay to an electron-positron pair $\gamma \to e^+ e^-$. This process is kinematically forbidden in \ac{SR}. In the case of \ac{LIV} (superluminal photon dispersion relation) the photon decay become allowed  if the photon energy exceeds a certain threshold, determined by the energy-momentum conservation \cite{Coleman:1998ti,Jacobson:2002hd, Mattingly:2002ba,Stecker:2001vb}. 
 Assuming that \ac{LIV} in the electron sector is suppressed compared to the photon sector, the threshold energy reads \cite{Jacobson:2002hd} 
 \begin{equation}
     E^\text{thr}_\gamma = \left( 4m_e^2 \bigl(E_{\text{LIV}}^{(n)}\bigr)^n \middle/ \xi_n \right)^{1/(n+2)}.
 \end{equation}
 If \ac{LIV} terms in photon and electron sector are assumed of the same order, the threshold condition is a bit more complicated, see \cite{Galaverni:2008yj}. 
 Above the threshold the rate of the photon decay is in fact very fast \cite{Jacobson:2005bg, Rubtsov:2012kb,Martinez-Huerta:2016azo}. Thus, the observation of an astrophysical photon with energy $E_\gamma$ means that this energy is less than the threshold one, $E_\gamma < E_\gamma^\text{thr}$. Inverting this condition, one obtains the constraint, which in case of suppressed \ac{LIV} for electrons reads   \[E^{(n)}_\text{LIV} > \left( \frac{E_\gamma^{n+2} }{4m_e^2} \right)^{1/n},\] where~$n=1,2$ and~$\xi_n=+1$. 
   The observation of the most energetic gamma rays leads to the constraints presented in Table~\ref{sec5.4:Table1}. Gamma rays with energy larger than $200$\,TeV have been detected~\cite{Amenomori:2019rjd,Albert:2020yty}, and very recently up to 1.4\,PeV~\cite{Cao:LHAASO}. Shortly, for $E_\gamma=1.5$\,PeV exactly, one may expect  $E^{(1)}_\text{LIV} < 3.1\times 10^{24}$~\,GeV for the case~$n=1$, and $E^{(2)}_\text{LIV} < 2.2\times 10^{15}$\,GeV for~$n=2$. Remarkably, these constraints are an order of magnitude weaker than the synchrotron constraints on \ac{LIV} in electron sector for~$n=0$ and~$2$, but on the same order of magnitude for~$n=1$. Therefore, it is really acceptable to neglect any \ac{LIV} in electrons considering the photon decay and related processes, including photons with energy $\sim 100$\,TeV or less. 
 
 One can expect a fraction of photons with energies larger than $10^{19}$\,eV in cosmic rays. Considering photon decays for these energies, one should also take into account \ac{LIV} in electrons, the perspective constraints are set on a combination between photon and electron \ac{LIV} parameters \cite{Galaverni:2007tq} (see also the discussion in Sec.\ref{hadronic}).
 
\item{\it Photon splitting $ \gamma \to 3 \gamma$:}

The process of photon splitting to three photons is forbidden in \ac{SR} but allowed in the \ac{LIV} case for any energy, when the photon dispersion relation is superluminal and $n \geq 1$. The width of the splitting process strongly depends on the photon energy, leading to observable effects when the energy is sufficiently high. In this sense, one has an effective threshold due to \ac{LIV} which induces a  cutoff in the high energy gamma-ray spectrum: the photon flux for a given energy decreases by a factor $P = \exp\left(-L_\text{s}/\langle L_{\gamma_\to 3\gamma}\rangle \right)$, where $L_\text{s}$ is the distance to the source and $\langle L_{\gamma_\to 3\gamma}\rangle$ is a photon mean free path related to the splitting. Very stringent bounds on the scale of \ac{LIV} \cite{Gelmini:2005gy,Astapov:2019xmt, Satunin:2019gsl,Albert:2019nnn} can be obtained from the non-observation of such cutoff in photon spectra which continue beyond $100$\,TeV, as we present them in Table~\ref{sec5.4:Table1}.

\item{\it Pair production on the EBL $\gamma \gamma \to e^+e^-$:}
\label{transparency}
Electron-positron pair creation by $\gamma\gamma$ interaction is a kinematically allowed process in \ac{SR}. It is responsible for attenuating the gamma-ray flux's energy propagating through cosmological distances by the interaction with the background light, such as the \ac{EBL} discussed in Sec.~\ref{pairprod-photon}. Modifications in the photon dispersion relation due to the \ac{LIV} lead to changes in the energy threshold for this process \eqref{pp-thres-sr}. Now, the modified pair-production threshold can be written as~\cite{Galaverni:2007tq,Stecker:2009hj,Scully:2008jp, Biteau_Williams:2015}:
\begin{equation}\label{eq:modified_energy_threshold}
\varepsilon_{\text{thr}}= \frac{m_{\mathrm{e}}^{2}}{E_{\gamma}} \pm \frac{1}{4}\left(\frac{E_\gamma}{E_\text{LIV}^{(n)}}\right)^{n}E_\gamma.
\end{equation}
Both superluminal and subluminal types are allowed, leading to two different outcomes~\cite{Martinez-Huerta:2020cut}. In the superluminal type, the energy threshold is lowered, compared to LI scenario, leading to a stronger absorption and consequently to a steeper source spectrum. On the other hand, in the subluminal type the energy threshold is increased and domain of the target \ac{EBL} photons for pair creation is reduced. In this case, gamma rays at higher energies are less absorbed making the spectrum less attenuated. Until now, stringent constraints on the scale of \ac{LIV} using this channel have been obtained only in the subluminal type, with the most prominent ones presented in \autoref{sec5.4:Table1}. 

\item{\it Pair production in Coulomb field of nuclei (Bethe-Heitler)  $\gamma N \to N e^+e^-$:}

This process, allowed in \ac{SR}, is also responsible for energy losses of extragalactic EeV cosmic rays, see Sec.~\ref{section3.3}. In case of subluminal photon dispersion relation the cross-section of the process decreases compared to \ac{SR} \cite{Vankov:2002gt, Rubtsov:2012kb}. Thus, the photon-initiated air showers begin deeper than in the standard case. If the shower starts significantly deeper, it cannot be recognized as a photon shower by the observations.  The predicted effect is similar to those of photon splitting: the observable photon flux from a source is suppressed by an energy-dependent factor $P$ \cite{Rubtsov:2016bea}. 
The absence of such effect in the observations of Crab Nebula photon spectrum leads to the strong constraint on \ac{LIV} of subluminal type in the photon sector \cite{Rubtsov:2016bea, Satunin:2019gsl}, see \autoref{sec5.4:Table1}. Note that although the suppression of Bethe--Heitler process seems to be a general feature whenever the photon dispersion relation is subluminal,  the quantative calculation has been made only for the case $n=2$ \cite{Rubtsov:2012kb}. The calculation related to $n=1$ have not been provided yet and may be an interesting task. The Bethe--Heitler process applied to the showers initiated by UHE photons is discussed in Sec.~\ref{hadronic}.

\end{description}

\paragraph{Neutrinos}

 The propagation of cosmic neutrinos may be affected by processes which are forbidden in \ac{SR} and allowed in \iac{LIV} scenario. Neutrino pair production $(\nu  \to \nu e^+e^-)$ is an example of a reaction which is due to \ac{LIV} and has a threshold depending on the energy scale of \ac{LIV} and on the electron mass. This reaction is dominant over, e.g., production of a pair $\mu^+\mu^-$, which has a much higher threshold because of the large muon mass with respect to the electron mass. 

Another process which is comparable to the neutrino pair production is neutrino splitting $ \nu \to \nu \nu \bar{\nu}$. In this case, there is no threshold if one assumes zero mass for neutrinos, but \ac{LIV} defines an energy scale with acts as an effective threshold, since the process has a suppression below that energy scale, which can be seen to be of the order of the threshold scale for pair production for \ac{LIV} scales around the Planck mass.

Neutrino decays according to the previous processes entail a mechanism of energy loss for neutrinos besides the one given by the expansion of the universe, together with a change in the neutrino population of the neutrino flux. It is important then to understand how the spectrum of cosmic neutrinos gets modified by these non-conventional \ac{LIV} effects.

Neutrino pair production proceeds through neutral or charged channels, mediated by a $Z^0$ boson, or a $W^{\pm}$, respectively. The first process was carefully computed in Ref.~\cite{Carmona:2012tp}, which obtained the corresponding decay rate in different \ac{LIV} scenarios. This result was then used both in Monte Carlo simulations~\cite{Stecker:2014oxa} and in an analytical model~\cite{Carmona:2019xxp} to obtain the new features in the \ac{LIV}-modified spectrum. The simulations of Ref.~\cite{Stecker:2014oxa} contained both pair production and neutrino splitting processes, neglecting charge current interactions (which only affect to electron neutrinos and are only important 1/3 of the time because of neutrino oscillations) and approximating the decay rate in the case of neutrino splitting, which is driven only by a neutral current interaction, as three times the result given in Ref.~\cite{Carmona:2012tp} for the pair production process. In contrast, the analytical model of Ref.~\cite{Carmona:2019xxp} only included the pair-production process.

The previous numerical and analytical works both found a cutoff in the neutrino spectrum around the value of the above mentioned threshold energy scales for the case of a quadratic correction (proportional to $\Lambda^{-2}$, where $\Lambda$ is a high-energy scale) of the neutrino dispersion relation as the most characteristic feature of the modified spectrum.

Choosing $\Lambda$ in such a way that the cutoff is of the order of $10\,$PeV (which corresponds to a value of $\Lambda$ around two orders of magnitude below the Planck scale), the feature is compatible with the IceCube data for the detected cosmic neutrinos of the highest energies~\cite{Aartsen:2020aqd}. While the initial absence of detected neutrinos around the Glashow resonance at $6.3\,$PeV (corresponding to a resonant formation of a $W^-$ boson in the interaction of a high-energy electron antineutrino with an electron)~\cite{Glashow:1960zz} could, in view of the previous results, support \iac{LIV} scenario, the IceCube collaboration has very recently reported~\cite{IceCube:2021rpz} the detection of a particle shower consistent with being created at the Glashow resonance. This means that the previous \ac{LIV} studies should better be interpreted as estimates of the sensitivity of present and future experiments detecting cosmic neutrinos to \ac{LIV} corrections to standard physics, which would then be able to put constraints on the high-energy scale controlling this departure from special relativity at or near the Planck scale.

Neutrino pair production as well as neutrino splitting can also be due to a difference in the maximal velocity of propagation of neutrinos and charged leptons or among different neutrinos, which is another possible correction due to \ac{LIV}. For a recent detailed analysis of this possibility, see Refs.~\cite{Somogyi:2019yis,Jentschura:2020nfe}. 

On the other hand, in the context of the \ac{SME}, $SU(2)_L$ gauge invariance may imply tight limits on \ac{LIV}  in the neutrino sector from the strong constraints in the charged-lepton sector; see the discussion in Ref.~\cite{Crivellin:2020oov}.

Lorentz invariance can also be tested through its effect on the atmospheric neutrino oscillation pattern. Anomalous flavor-changing effects in the zenith angle and energy distributions can be looked for in the rates of atmospheric muon neutrinos. This has been tested by large water Cherenkov neutrino detectors like IceCube \cite{Aartsen:2017ibm}. 

Moreover, already existing limits on neutrino \ac{NSI} obtained looking also for anomalies in the atmospheric neutrino pattern \cite{Aartsen:2018jd}, \cite{Thakore:2020} can be transformed into limits in Lorentz invariance \cite{diaz:2015}. As pointed out in \cite{diaz:2015}, even if we can directly relate the parameters that control these non-standard effects, the physics behind them is completely different, as follows from the fact that \ac{NSI} effects require propagation in matter, which is not the case for \ac{CPT} violation in this context. When doing the transformation of \ac{NSI} experimental limits into \ac{CPT} limits, underlying assumptions like those on the matter composition should be taken into account.

Further improvements on these results can be obtained with neutrino beam projects as DUNE \cite{Barenboim:2018ctx} or P2O \cite{Akindinov:2019}, thanks to a better sensitivity to neutrino oscillation parameters.

\paragraph{Hadronic sector}\label{hadronic}
The description of the kinematical effects of \ac{LIV} in the hadronic sector are given by the \ac{MDR} for the protons, pions,  
and photons~\eqref{eq:mdr}. As mentioned above, $\eta^n_i$ determines a maximal velocity for $n=0$ and represents signs $\pm1$ for $n=1$ and $n=2$. Due to their energy, \acp{UHECR}, and their sub-products in air showers, are the most used messengers to test  \ac{LIV} scenarios. The constraints to \ac{LIV} are given by the possible modifications on the thresholds of processes. 
\begin{description}

\item {\it Proton (nuclei) vacuum Cherenkov radiation $ p(N) \to p(N) \gamma$:} This is a process which is forbidden in \ac{SR}, but becomes allowed above a given energy threshold in \ac{LIV} scenarios. Since protons and nuclei above the energy threshold lose their energy rather quickly, an observation of a cosmic UHE particle from a distant source can be used to put constraints on the energy threshold and consequently on \ac{LIV}. Using this ansatz, a
stringent bound on isotropic, nonbirefringent subluminal \ac{LIV} in the photon sector was obtained~\cite{Klinkhamer:2007ak,Klinkhamer:2008ss}, corresponding to $\kappa\lesssim6\times10^{-20}$ in eqs.~(\ref{eq:QED_gamma}) and~(\ref{eq:QED_gamma2}).

\item { \it GZK processes, e.g.\ $p \gamma_\text{CMB} \to p \pi_0$:} 
     The expected suppression of the \ac{UHECR} flux at the highest energies due to interactions with background photons (Sect.~\ref{sec:UHECRprop}) can be used in different ways to constrain Lorentz breaking dispersion relations up to order~$n=2$.
     \Iac{MDR} can lead to a shift of the standard thresholds.  Depending on the modification, the attenuation length for photo-pion production in the case of \ac{UHECR} protons and for photo-disintegration in the case of \ac{UHECR} nuclei can be significantly larger than in the standard propagation~\cite{Aloisio:2000cm}.
     As a consequence, \acp{UHECR} could travel farther than they would under Lorentz invariance~\cite{Stecker:2009hj,Boncioli:2015cqa,Boncioli:2017nec,Lang:2020geh,Torri:2020fao}.  
     This effect can be tested taking advantage of the measurements of the suppression of the flux~\cite{Aab:2020gxe,Aab:2020rhr,Ivanov:2020rqn}. However, the outcome of studies interpreting \ac{UHECR} results in terms of astrophysical scenarios shows that the maximum \ac{UHECR} energy at the sources could be smaller than was assumed in some earlier works, thus limiting also the capability of testing \ac{LIV} scenarios.  The  measured  energy  spectrum  and $X_{\max}$~distributions, fitted together for the first time in
     \cite{Boncioli:2017nec,Lang:2020geh,PierreAuger:2021tog} with the aim of constraining \ac{LIV} parameters, provide limits at the 5$\sigma$ C.L.~\cite{PierreAuger:2021tog}, although the systematic uncertainties severely complicate the interpretation of results.
     \item {\it GZK photons:} An alternative route is to use the secondary photons, so-called \ac{GZK} photons, which are produced by the decay of the $\pi^0 \to 2\gamma$~\cite{Galaverni:2007tq,Maccione:2008iw}. For some \ac{LIV} in pion and \ac{QED} sectors this process is kinematically forbidden \cite{Coleman:1998ti}, the other type of \ac{LIV} reproduces the standard pion decay, producing \ac{GZK} photons. These photons are mainly absorbed by pair production onto the \ac{CMB} and \ac{CRB}. Thus, the fraction of UHE photons in \acp{UHECR} is theoretically predicted to be less than 1\% at $10^{19}$~eV \cite{Gelmini:2005wu}. The most recent limits on the presence of photons in the \ac{UHECR} spectrum are imposed by the Pierre Auger Observatory \cite{Rautenberg:2020yof} and the Telescope Array \cite{Kuznetsov:2020jgr}. 
     Very tight constraints can be then cast thanks to the strong dependence of the photon pair production on \ac{LIV}  modifications. In particular, not only the usual threshold energy can be shifted but also an upper threshold can be introduced \cite{Jacobson:2002hd}, i.e.~the reaction can be shut down at high energies. If such an upper threshold energy is for some \ac{LIV}  parameters lower than $10^{20}$\,eV  for the \ac{GZK} secondary UHE photons, then they would be no longer attenuated by the \ac{CMB} and would be able to reach the Earth and so make up a significant fraction of the total \ac{UHECR} flux~\cite{Lang:2017wpe}. However, in this case they would  violate the above mentioned experimental bounds. This implies that such values of the \ac{LIV}  parameters could be excluded leading to severe constraints on Lorentz breaking \ac{QED} at order $n=1$ and $n=2$~\cite{Galaverni:2007tq,Maccione:2008iw,Galaverni:2008yj,Lang:2017wpe}.  Of course, once \ac{LIV}  is introduced, these extra UHE photons could also be removed via the above discussed photon decay \cite{Maccione:2008iw} or photon splitting process for superluminal photon dispersion relation. However, for photons around $10^{19}$\,eV  the analysis proposed in \cite{Jacobson:2002hd} is correct (the same it is not true at lower energies) and the splitting time scale to be negligible at $E_{\gamma} \simeq 10^{19}\,\mathrm{eV}$. In a more recent paper~\cite{Lang:2020geh}, with updated results, it was shown that in the case that the cutoff is due to the sources, the limits almost disappear. 
     
     On the other hand, UHE photons with subluminal dispersion relation initiate deeper air showers that probably cannot be experimentally recognised as photon events  \cite{Rubtsov:2013wwa}. This effect decreases the expected flux of UHE photons and may remove the tension related to the pair production on \ac{CMB}. The similar effect of the observed flux suppression for the UHE photons  with energy larger than $10^{19.5}$\,eV (subluminal dispersion relation as well) came from the suppression of preshower formation  \cite{Satunin:2013an} [see Sec.~\ref{section3.3} for the preshower formation]. 
     
    \item {\it Development of air showers:} The kinematics of the $\pi^0$ decay in the development of the atmospheric showers could be modified by \ac{LIV} effects; this has been tested in \cite{Boncioli:2015cqa}, showing that the change in the particles' decay has the effect to move the shower maximum to higher altitudes as the electromagnetic part of the shower consumes faster. The suppression of the pion decay implies the number of their interactions to be higher. The final effect could be then visible in the number of muons at ground, in particular in the fluctuations of the number of muons. This has been studied in \cite{Abreu:2021Ul}; \ac{LIV} effects in this observable are expected to be larger than the change in the shower maximum.
    \item {\it Secondary photons in air showers:} Finally, some \ac{LIV} bounds can be obtained from the analysis of secondary photons in UHE showers initiated by protons/hadrons. 
    In the first stages of the development of an extensive air shower initiated in the Earth's atmosphere by a UHE particle, it is expected that very-high energy photons are produced as secondary particles. These photons can have energies far beyond those accessible by other means, but they cannot be directly observed. However, if these photons undergo the decay process $\gamma \to e^+e^-$, the development of the electromagnetic part of the shower is accelerated. Indeed, it was found that \ac{LIV} at a level allowed by previous bounds would significantly reduce the average atmospheric depth of the shower maximum $\left<X_\text{max}\right>$~\cite{Diaz:2016dpk,Boncioli:2015cqa}. Using this ansatz and recent $\left<X_\text{max}\right>$ measurements from the Pierre Auger Observatory, a stringent bound on
isotropic, non birefringent superluminal \ac{LIV} in the photon sector
was obtained~\cite{Klinkhamer:2017puj}, corresponding to $\kappa\gtrsim-3\times10^{-19}$ in eqs.~(\ref{eq:QED_gamma}) and~(\ref{eq:QED_gamma2}).
A subsequent extension of the analysis including data on the fluctuations of $\left<X_\text{max}\right>$ led to a strengthening of the bound to
$\kappa\gtrsim-6\times10^{-21}$~\cite{Duenkel:2021gkq}.
\end{description}
 
\subsubsection{Further discussions: LIV vs DSR}
As commented in the introduction, the main ingredient of \ac{DSR} scenarios, differing from \ac{LIV}  theories, is a non-additive composition law for the momenta. This produces a cancellation of effects such that the possible phenomenology induced by \ac{LIV} corresponding to forbidden processes in \ac{SR}, is not allowed in this case. Moreover, stringent constraints coming from \ac{LIV} modifications of allowed processes in \ac{SR} (such as those obtained from the development of air showers mentioned above) do not immediately apply to \ac{DSR}, 
since in this case, and in contrast with the \ac{LIV} case, the fractional change of the threshold energy is proportional to the threshold energy divided by the high-energy scale~\cite{Carmona:2018xwm,Carmona:2020whi}. Whilst this fractional change is really small when the high-energy scale is of  the order of the Planck scale, one could contemplate the possibility of a much lower high-energy scale if one discards possible effects in time delays in a \ac{DSR} scenario. Then, this opens up a completely new phenomenology usually not contemplated in the literature. In particular, there are recent works considering this possible scenario in particle accelerator physics~\cite{Albalate:2018kcf}, in particle decays \cite{Lobo:2021yem} and in astrophysical processes, like photon interaction with \ac{EBL}~\cite{Carmona:2020whi}, allowing to get a constraint of the order of some TeV on the high-energy scale.

\subsubsection{Current experimental limits on LIV parameters} 
A list of the constrains on the \ac{LIV} parameters in the QRD sector obtained from different observations are collected in Table~\ref{sec5.4:Table1}.

\begin{table}[h!]
\begin{tabular}{cccccccc}
\hline

$e^-/\gamma$ &  \multicolumn{1}{c}{\begin{tabular}[c]{@{}c@{}}Test \\ of \\ QG\end{tabular}} &
\multicolumn{1}{c}{\begin{tabular}[c]{@{}c@{}}Sub($-$) or \\ super($+$) \\ luminal\end{tabular}} &
\multicolumn{1}{c}{\begin{tabular}[c]{@{}c@{}} \\ \\  $|\xi_0|$($|\eta_0|$)\end{tabular}} &
\multicolumn{1}{c}{\begin{tabular}[c]{@{}c@{}}  \\Limits \\  $E_\text{LIV}^{(1)}$ (eV)\end{tabular}} & 
\multicolumn{1}{c}{\begin{tabular}[c]{@{}c@{}} \\ \\  $E_\text{LIV}^{(2)}$ (eV)\end{tabular}} &
\multicolumn{1}{c}{Source} & Ref. \\ \hline \hline

$e^-$ &  Synch. & both &  $2\times 10^{-20}$ & $ 10^{33}$  & $2\times 10^{25}$  & CRAB &  \small{\cite{Maccione:2007yc,Liberati:2012jf,Rubtsov:2016bea}} \\

$e^-$ &  VC & (+) & $10^{-20}$ & $ 10^{31}$ & $10^{23}$ & CRAB &  \small{\cite{Jacobson:2002hd, Jacobson:2005bg, Stecker:2013jfa}} \\

$\gamma$ & PD & (+)  &  $7.1\times 10^{-19}$ & $1.7\times 10^{33}$   & $1.4\times 10^{24}$  &LH. J2032+4102& \cite{Chen:2021hen} \\

$\gamma$ & PD & (+)  &  $1.3\times 10^{-17}$ & $2.2\times 10^{31}$   & $8\times 10^{22}$   & MultiSrc& \cite{Albert:2019nnn} \\

$\gamma$ & PD & (+)  &  $1.8\times 10^{-17}$ & $1.4\times 10^{31}$   & $5.8\times 10^{22}$  & eHWCJ1825-134& \cite{Albert:2019nnn} \\

$\gamma$ & PD & (+)  &  $2.2\times 10^{-17}$ & $9.9\times 10^{30}$   & $4.7\times 10^{22}$  & eHWCJ1907+063& \cite{Albert:2019nnn} \\
  
 $\gamma$ &   $3\gamma$ & (+)  &  - & -  & $2.5\times 10^{25}$    &LH. J2032+4102& \cite{Chen:2021hen} \\ 
  
$\gamma$ &   $3\gamma$ & (+)  &  - & -  & $1.2\times 10^{24}$  & eHWC J1825-134 &  \cite{Albert:2019nnn} \\

$\gamma$ &  $3\gamma$ & (+)  &  - & -  & $1.0\times 10^{24}$  & eHWC J1907+063 &  \cite{Albert:2019nnn} \\

$\gamma$ &   $3\gamma$ & (+) & - & -  & $4.1\times 10^{23}$  & CRAB & \cite{Satunin:2019gsl} \\

$\gamma$ &  AS  & ($-$) & - & -  & $1.7\times 10^{22}$  & diffuse (Tibet) & \cite{Satunin:2021vfx} \\

$\gamma$ &  AS  & ($-$) & - & -  & $6.8\times 10^{21}$  &LH. J1908+0621 & \cite{Satunin:2021vfx} \\

$\gamma$ &  AS  & ($-$) & - & -  & $1.4\times 10^{21}$  & CRAB & \cite{Satunin:2019gsl} \\

$\gamma$ &   AS  & ($-$) & - & -  & $9.7\times 10^{20}$ & CRAB & \cite{Satunin:2019gsl} \\

$\gamma$ &   AS  & ($-$) & - & -  & $2.1\times 10^{20}$ & CRAB &  \cite{Rubtsov:2016bea}\\

$\gamma$ &   PP  & ($-$) & - & $1.2\times10^{29}$ & $2.4 \times 10^{21}$   & MultiSrc (6) & \cite{Lang:2019PhRvD}\\

$\gamma$ &   PP  & ($-$) & $2\times10^{-16}$ & $2.6\times10^{28}$  & $7.8 \times 10^{20}$   & Mrk~501 & \cite{Stecker:2001vb,Abdalla_HESS_Mrk501:2019ApJ}\\

$\gamma$ &   PP  & ($-$) & - & $1.9\times10^{28}$ & $3.1\times10^{20}$   & MultiSrc (32) & \cite{Biteau_Williams:2015}\\

\hline
\end{tabular}
\caption{Strong and recent astrophysical bounds to \ac{LIV} in the \ac{QED} sector using synchrotron radiation (Synch.), vacuum Cherenkov radiation (VC), photon decay (PD), photon splitting ($3\gamma$), air shower suppression (AS), and  pair production (PP) on the \ac{EBL}. 
The bounds from Ref.~\cite{Biteau_Williams:2015} were translated to the pure photon sector and $n=2$ term. 
\label{sec5.4:Table1}
}
\end{table}

\subsection{Decoherence}
\label{section7.4}

The most widely used formalism to describe fundamental decoherence is the one based on the Lindblad equation describing the evolution of open quantum systems. 
    The Lindblad  
    evolution equation~\cite{Lindblad:1975ef,Gorini:1975nb,Alicki:1987al} is a linear equation for the density matrix $\rho$ of an open system, interacting with an environment, which ensures the complete positivity of $\rho$ at any moment $t$ of the evolution, and the conservation of probability, $\operatorname{tr}\rho=1$:
    \begin{align}\label{lind}{\dot \rho} &= -i [H,\rho] + \mathcal D[\rho], &
    \mathcal D[\rho] &= \sum_j \big( \{ \rho, D_j^\dagger, D_j \} - 2 D_j\, \rho \, D_j^\dagger \big),
    \end{align}
    where the overdot denotes time derivative, and $D_j$ are the Lindblad operators, associated with the quantum interactions of the open system with its environment, assuming a well-defined separation between system and environment, which is valid for weak couplings of the system with the environment (a case of relevance to quantum gravity). Below we discuss some physical applications of \ac{QG}-induced decoherence relevant to this review.

    \subsubsection{Meson decoherence}\label{section7.4.1}
    In the case of oscillating neutral mesons, say for definiteness kaons in a $\phi$ factory, or single-state kaons, as e.g. 
    in the CPLEAR experiment at CERN~\cite{Adler:1995xy}, one may use the so-called $\sigma$-matrix basis, comprising of $\{\sigma_0 = \mathbf 1_{2\times 2}, \sigma_1, \sigma_2, \sigma_3\}$, $i=1,2,3$, where $\sigma_1,\sigma_2,\sigma_3$ are the $2\times 2$~Pauli matrices, in which the density matrix is expanded as~$\rho = \sum_{\alpha=0}^3 \rho_\alpha \sigma_\alpha$. In this 
    way the evolution \eqref{lind} can be represented as $\dot \rho_\alpha = h_{\alpha\beta}\, \rho_\beta + L^D_{\alpha\beta [\rho]}$.
    Imposing complete positivity of $\rho(t)$ during the evolution, one can parametrize the non-commutator, decoherening, term of the evolution \eqref{lind}, in the most general case, by six real constant parameters~\cite{Benatti:1999gw},
    \begin{equation}\label{LD} 
    L^D_{\alpha\beta} = -2 \begin{pmatrix}
        0 & 0 & 0 & 0 \\
        0 & a & b & c \\
        0 & b & \alpha & \beta \\
        0 & c & \beta & \gamma
    \end{pmatrix},    
    \end{equation}
    with the conditions 
    \begin{align}\label{ineq}
    a & < \alpha + \gamma, & 4b^2 &\le \gamma^2 - (a - \alpha)^2\nonumber \\
    \alpha & < a + \gamma, & 4c^2 &\le \alpha^2 - (a -\gamma)^2\nonumber \\
    \gamma & < a + \alpha, & 4\beta^2 &\le a^2 - (\alpha-\gamma)^2.
    \end{align}
    Eq.~\eqref{LD} generalizes the special cases considered in \cite{Ellis:1983jz,Ellis:1995xd} with just three real parameters. The decoherence parameters lead to in principle observable effects in neutral mesons (or other oscillating particles, such as neutrinos, to be discussed in this subsection and in Sec.~\ref{section7.4.2}, as well as in Sec.~\ref{section7.5}). 
    By purely dimensional grounds the authors in \cite{Ellis:1995xd} have studied the possibility of maximal decoherence effects in quantum gravity, where the
    coefficients above are taken to be of the order $E^2/M_{\rm QG}$, where $E$ is an average energy of a neutral meson, and $M_{\rm QG}$ the quantum gravity scale, expected to be of the order 
    of the Planck mass, although, strictly speaking, such parameter might depend on the 
    specific model considered. 
    In this case it can be shown that the sensitivity of current neutral meson experiments~\cite{Domenico:2020bbk} has reached the Planck scale sensitivity. Under similar phenomenological assumptions, in the case of neutrinos, 
    in some cases, one can even exceed the Planck scale sensitivity by several orders of magnitude (see discussion in Sec.~\ref{section7.4.2}). 
    
   A specific model is obtained by imposing energy conservation on the average and monotonic increase in the von Neumann entropy $S=-\!\operatorname{tr}\rho\ln\rho$, where $\operatorname{tr}$ is over the environmental degrees of freedom. In this case one needs to impose self-adjointness of Lindblad operators: $D_j^\dagger=D_j$, and also require that 
    these operators commute with the Hamiltonian of the open system $[D_j, H]=0$, which leads to $\mathcal D[\rho]= \sum_j \bigl[D_j, [D_j, \rho]\bigr]$. Although in general, in such systems one has the evolution of an initially pure state into mixed ones, nonetheless one may consider a specific subclass (energy-driven decoherence models~\cite{Percival:1994pe,Houston:1996ho,Adler:2000vfa} in which the purity of states is preserved, $\operatorname{tr}\rho^2=\operatorname{tr}\rho=1$.
    For these specific Lindblad systems, the evolution equation can be written in terms of state vectors $|\psi\rangle$~\cite{Gisin:1992xc} in an Ito form, employing appropriate stochastic differential random variables. In addition, we may assume $D_j=\lambda_j H$, with $\lambda_j$ c-number constants. Without loss of generality, we may assume a single environmental operator~\cite{Adler:2000vfa}:
    $    D=\lambda H,\, \lambda^2=\sum_J\lambda_j^2,$
    which leads to 
    \begin{equation}\label{dct}
    d\rho=-i[H, \rho] -\frac{\lambda^2}{8}\bigl[H, [H,\rho]\bigr] +
    \frac{\lambda}{2}\bigl[\rho, [\rho, H]\bigr]\, dW_t,
    \end{equation}
    where $t$ is the time and $dW_t$ are the stochastic Ito differentials, satisfying the white noise conditions
    $dW_t^2 = dt$, \ $dt\, dW_t=0$.
    
    In \ac{QG} 
    motivated models of decoherence, it is natural to assume $\lambda=1/M_{\rm QG}$.
    In view of the double commutator term involving the Hamiltonian in \eqref{dct}, there is a much stronger suppression of the decoherence effects in such energy-driven decoherence models, compared to generic phenomenological models of Lindblad decoherence~\cite{Mavromatos:2004sz,Adler:2000vfa}, used e.g.\ in neutral mesons~\cite{Ellis:1983jz}, given that in the former case the decoherence term is proportional to the square of the energy variance~$\Delta E^2$, \ $\gamma \equiv \left<\mathcal D[\rho]\right> = \operatorname{tr}\left(\rho \frac{\lambda^2}{8}\bigl[H, [H,\rho]\bigr]\right) = \frac{\lambda^2}{8}\left(\left<H^2\right>-\left<H\right>^2\right) 
    \sim \frac{\Delta E^2}{M_{\rm QG}}^2$, while in the latter, the decoherence terms are proportional to powers of the energy of the neutral mesons, e.g.~$\gamma \sim E^2/M_{\rm QG}$ in the minimally suppressed case of decoherence~\cite{Ellis:1995xd}, discussed above. 
    
    Another scenario in which a Lindblad-type evolution equation emerges is within the framework of Planck-scale deformation of relativistic symmetries discussed in Sec.~\ref{section2.2.2.1}. In \cite{Arzano:2014cya} it was shown that the non-trivial Hopf algebra structure of the generators of time-translations in a specific realization of the $\kappa$-Poincar\'e algebra \cite{Lukierski:1992dt,Lukierski:1991pn,Borowiec:2009vb} naturally lead to a {\it non-symmetric} Lindblad equation which is covariant under the action of the deformed Lorentz generators.  
    
    The Lindblad formalism is based on a linear evolution of the density matrix $\rho$ of %the open matter 
    an open quantum system, interacting with an environment. Quantum gravity may also entail non-linear evolutions, which go beyond the Lindblad formalism. For instance, it is known~\cite{Goldin:1994go1,Goldin:1994go,Doebner:1997ry} that, if one has a \ac{QM} system described by a wave function $\psi (t,\vec x)$, whose probability density $\psi^\star \, \psi$ satisfies a Fokker-Planck equation, so that the theory admits diffusion currents, then $\psi$ satisfies a specific non-linear complex equation (Doebner-Goldin form).
    The diffusion currents stem from unitary representations of an infinite-dimensional Lie algebra of vector fields and group of diffeomorphisms. Such non-linear 
    wave-function equations can be generalised to include mixed states, thus providing non-linear evolution equations of density matrices, generalising the Lindblad formalism~\cite{Dodonov1995}. 
    It has been argued in \cite{Mavromatos:1999ci}, that a toy model of space-time ``foam'' in the context of string/brane theory, in which the non-trivial space-time
    structures are provided by point-like brane defects, does satisfy such a non-linear equation, with the diffusive  non-linearities stemming from quantum recoil stochastic fluctuations of populations of such defects. This, then, could induce non-linearities in the temporal evolution of the matter subsystem. 
    A simple example of a non-linear generalization of Lindblad equation  is given by \cite{grabert1982nonlinear, Ottinger2010}
     \begin{equation}\label{LDNL} 
      \dot\rho=i [H, \rho]-\beta {\lambda}[D,[D,H]_\rho]-\lambda [D,[D,\rho]],
      \end{equation}
where $D^\dagger=D$ and
      \begin{equation}\label{INTP}
      [D,H]_\rho= \int_0^1 \ d s\, \rho^s\,  [D,H] \,\rho^{1-s}.
      \end{equation}
This non-linear generalization of Lindblad equation has an interesting fixed point which corresponds to Boltzmann thermal equilibrium $\rho=e^{-\beta H}$. 
In this way  Hawking radiation can  emerge as a quantum decoherence phenomenon induced by black hole formation.

\subsubsection{Neutrino decoherence}\label{section7.4.2}

Neutrinos propagate as a superposition of mass and flavor eigenstates, which in combination with the non-zero differences between the mass state masses results in the phenomenon of neutrino oscillations. Neutrinos interact only very weakly with other matter and are stable, allowing this quantum superposition effect to be maintained over macroscopic (perhaps even cosmological) distances.   

Such superposition, however, will eventually degrade if the neutrino couples to a stochastic environment, resulting in a loss of coherence and the damping of neutrino oscillations. Quantum gravity may provide just such a stochastic environment, with spacetime itself expected to be fluctuating or uncertain at tiny distance scales. Such a scenario is often referred to as \textit{quantum foam}, \textit{space-time foam} or \textit{fuzzy spacetime}. Some specific \ac{QG}  scenarios that can induce neutrino decoherence include:

\begin{description}
    \item{\it Lightcone fluctuations} If spacetime is fluctuating/uncertain in nature in \ac{QG}, the travel distance/time between two points may fluctuate, resulting in an otherwise coherent neutrino `beam' becoming out of phase~\cite{and_2007, 2103.15313}.
    \item{\it Neutrino--virtual black hole interactions} It is postulated that fluctuating spacetime may be permeated by virtual black hole pairs, the \ac{QG}  analogue of the electron-positron pairs of vacuum polarization in \ac{QED}. A neutrino encountering a virtual black hole may experience severe perturbation to its state and undergo e.g.~flavor-violating processes~\cite{Anchordoqui:2005gj}, resulting in a potentially strong decoherence effect. Specific decoherence operators for a range of scenarios are proposed in~\cite{PhysRevD.102.115003}.
\end{description}

Whilst various particle species are in principal 
prone to such effects, neutrinos are particularly sensitive probes. In addition to their oscillating behavior (allowing them to act as quantum interferometers), neutrinos are observed over vast (including cosmological) distances, allowing even weak effects to accumulate into potentially measurable signals, and at high energies (up to PeV), which allows them to partially overcome any suppression of \ac{QG}  effects at energies below the Planck scale.

The phenomenological neutrino decoherence models tested typically feature damping terms of the form $e^{ - \Gamma  L }$, where $\Gamma$ quantifies the strength of the damping effects, resulting in a coherence length $L_{\rm{coh}} = 1 / \Gamma$. In experimental searches, energy-dependence of $\Gamma$ is frequently modeled phenomenologically as~\cite{PhysRevLett.85.1166}:

\begin{equation}
\label{eqn:gamma_energy_dep}
\Gamma(E) = \Gamma(E_0) \left( \frac{E}{E_0} \right)^n\,,
\end{equation}
where $n$ is a power-law index to be defined by theory or tested by experiment. When considering Planck scale physics this can be written~\cite{Anchordoqui:2005gj,PhysRevD.102.115003}:

\begin{equation}
    \label{eqn:gamma_planck}
    \Gamma(E) = \zeta \frac{E^n}{ M_{\rm{P}}^{n-1}}\,,
\end{equation}
where $\zeta$ is a dimensionless constant. Heuristically, $\zeta \sim \mathcal{O}(1)$ for a `natural' Planck scale theory, or equivalently $L_{\rm{coh}}(M_{\rm{P}}) \sim \mathcal{O}(\ell_{\rm{P}})$~\cite{PhysRevD.102.115003}. In such a scenario, strong decoherence effects would have already been identified in terrestrial neutrino experiments if $n \leq 2$, indicating that such effects are either more strongly suppressed at low energies or $\zeta \ll 1$ in Nature~\cite{PhysRevD.102.115003}. 

In neutrino--virtual black hole interaction scenarios, $1/\Gamma$ can be be heuristically interpreted as the interaction mean free path~\cite{PhysRevD.102.115003}, relating experimental constraints on $\Gamma$ to the virtual black hole number density and interaction cross section. In lightcone fluctuation scenarios, travel distance fluctuations, $\delta L$, can be modeled as~\cite{2103.15313}:
\begin{equation}
\delta L(E, L) = \delta \ell_{\rm{P}} \left( \frac {L} {\ell_{\rm{P}}} \right)^m \left( \frac {E} {M_{\rm{P}}} \right)^n \,,
\label{eq:deltaL_planck}
\end{equation}
where $\delta \ell_{\rm{P}}$ is the distance fluctuation/uncertainty for a particle traveling $ \ell_{\rm{P}}$, and $m$ is a power-law index characterizing the accumulation of distance fluctuations over travel distance. Similar parameterizations have been employed in X-ray and gamma-ray searches for fluctuating spacetime via stochastic arrival time spread~\cite{Vasileiou2015} or image degradation~\cite{Perlman_2015}. 

Neutrino decoherence results in such scenarios when the distance fluctuations become comparable to the oscillation wavelength. However, as oscillation wavelengths are macroscopic (and increase with neutrino energy), Planck scale fluctuations must accumulate to macroscopic scales to yield significant effects. This is only achieved in the most optimistic scenarios, where either there is no energy-suppression of the effects below the Planck scale, or distance fluctuations are highly correlated along the neutrino path (a more natural scenario would be that the neutrino velocity fluctuates, rather than the travel distance). And even then, measurable effects only accumulate over cosmological baselines. \\
Experimental searches for neutrino decoherence have been conducted with a broad range of neutrino sources and detectors. In general, large baselines are preferable to allow significant effects to accumulate, favoring astrophysical or atmospheric neutrino searches. For coherent neutrino sources, the signal sought is a distance- and energy-dependent modification to the neutrino detection rate resulting from the damping of oscillations, or indeed flavor transitions outside of the normally expected oscillation region (for example at higher energies~\cite{PhysRevD.102.115003}).

The most stringent bounds on energy-suppressed ($n > 0$) scenarios result from the non-observation of such signals in atmospheric neutrinos by the IceCube neutrino observatory~\cite{Aartsen_2017}. A range of decoherence operator textures for energy-dependence scenarios with $n \in \{-2, -1, 0, +1, +2\}$ were tested in \cite{Coloma:2018idr}, producing world leading limits, and significant increases in sensitivity are expected using more recent IceCube data. Earlier pioneering limits were set using Super Kamiokande data~\cite{PhysRevLett.85.1166}. For energy-independent scenarios ($n = 0$), solar neutrinos presently yield the strongest limit~\cite{de_Holanda_2020}. For energy-enhanced ($n < 0$) scenarios, the best limits to date were obtained using MINOS (long baseline accelerator)~\cite{PhysRevD.89.053002} and KamLAND (reactor)~\cite{PhysRevD.95.113005} data, although these are generally less amenable to \ac{QG}  motivated explanations (although such cases can result from lightcone fluctuations due to the neutrino oscillation wavelength dependence). Constraints have also been derived from MINOS and T2K long baseline accelerator data~\cite{Gomes:2020muc} and from NO$\nu$A and T2K \cite{Coelho:2017zes}.

Naively, astrophysical neutrinos should offer strong sensitivity to neutrino decoherence due to the unparalleled travel distances and very high energies observed for the extragalactic astrophysical neutrino flux observed by IceCube~\cite{Anchordoqui:2005gj}. However, the diffuse neutrino flux is an incoherent source, limiting its use for neutrino decoherence searches. For example, constraints from the flavor composition of the diffuse astrophysical neutrino flux are not possible in cases where the effect of decoherence is to average oscillations, 
since this is entirely degenerate with the standard expectation for an incoherent source~\cite{PhysRevD.102.115003}. Scenarios which result in an equal population of neutrino flavors (for example those result from in some neutrino--virtual black hole interaction scenarios, 
featuring \textit{relaxation} effects) 
do differ somewhat from the standard expectation, but depending on the (as yet 
unknown) production mechanism for these neutrino the effect is small and not currently distinguishable by current generation neutrino telescopes. 

\subsection{CPT symmetry}
\label{section7.5}

\ac{CPT}  invariance is one of the cornerstones of  \ac{QFT}  and thus of high energy physics. 
It is worth recalling somewhat precisely the set of assumptions underlying the proof of the \ac{CPT}  theorem for QFT in flat spacetime carried out a long time ago by Jost \cite{Jost:1957zz} within the Wightman framework of QFT \cite{Wightman:1956zz}. 
Besides the positivity of the energy, the assumptions can be conveniently formulated as: i) the theory is Poincar\'e invariant, ii) the Poincar\'e transformations of fields yield a (finite dimensional) representation of the Lorentz group in the space of indices, iii) each field comes along with its charge conjugate, these items being supplemented by the condition of locality/microcausality, expressing the fact that Bose (Fermi) fields commute (anticommute) with each other for space-like distance. Whenever these conditions are all fulfilled{\footnote{There are several equivalent ways (up to technical precautions) to formulate these conditions, one popular one being to require Lorentz invariance, unitarity and locality.}}, the corresponding local theory is \ac{CPT}  invariant. Notice that these are only {\it{sufficient}} conditions. In particular, one can construct non-local gauge theories (e.g.\ on non-commutative Moyal spaces) with \ac{CPT}  invariance which however are not Lorentz invariant, or non-local field theories which violate \ac{CPT}  but are still Lorentz invariant (among them some preserving C leading to equal masses of particle and conjugate anti-particle) \cite{Chaichian:2011fc,Dolgov:2014xza}.\\

    It is widely expected that when quantum gravitational effects are taken into account they could lead to small but potentially measurable departures from \ac{CPT}  invariance. In this section we list some of the most relevant mechanisms that \ac{QG}  furnishes for such departures from \ac{CPT}. See also Sec.~\ref{section7.2} for birefringence effects produced by parity violations.

\subsubsection{CPT violation and Lorentz invariance violation in Standard-Model Extensions} 

A widely studied framework for modeling \ac{CPT}  violation is that of \acp{SME} \cite{Colladay:1998fq}. We refer the reader to Section \ref{section2.2.1} for a review. 

Let us recall here that the so-called Greenberg theorem \cite{Greenberg:2002uu},  states that a \ac{CPT}-odd \ac{QFT} necessarily violates Lorentz covariance. The theorem can be proved using the basic assumptions of axiomatic \ac{QFT} recalled above \cite{Streater:1989vi,Bogolyubov:1975ps,RalfLehnert:2016grl,Antonelli:2020nhn}, but the idea that \ac{CPT} violation automatically implies \ac{LIV} has been widely debated in literature \cite{Chaichian:2011fc,Tureanu_2013,Chaichian:2012hy,Duetsch:2012sd,Greenberg:2011cw}. The \ac{CPT}-odd terms in the Lagrangian of \acp{SME} are those constructed contracting an odd number of Lorentz indices and explicitly violate Lorentz invariance. These new perturbation terms have a direct impact on particle free kinematics modifying the dispersion relations and therefore they can influence the oscillation pattern.\\
        For instance the \ac{LIV} part of the effective neutrino Lagrangian density can be written modifying the Dirac matrices:
        \begin{equation}
            \Gamma^{\mu}_{AB}=\gamma^{\mu}\delta_{AB}+c^{\mu\nu}_{AB}\gamma_{\nu}+d^{\mu\nu}_{AB}\gamma_{\nu}\gamma_{5}+e^{\mu}_{AB}+if^{\mu}_{AB}\gamma_{5}+g^{\mu\nu}_{AB}\gamma_{\nu}+\frac{1}{2}h^{\mu\alpha\beta}_{AB}\sigma_{\alpha\beta}\,,
        \end{equation}
        with capital Latin letters indicating flavor indices and $\sigma_{\mu\nu}=\frac{1}{2}[\gamma_{\mu},\gamma_{\nu}]$. Even the mass matrix can be amended by the introduction of \ac{LIV} terms:
        \begin{equation}
           M_{AB}=m_{AB}+im_{5AB}\gamma_{5}+a^{\mu}_{AB}\gamma_{\mu}+b^{\mu}_{AB}\gamma_{5}\gamma_{\mu}+\frac{1}{2}G^{\mu\nu}_{AB}\sigma_{\mu\nu}\,.
        \end{equation}
        It is possible to demonstrate that objects with an even number of Lorentz indices are \ac{CPT}-even and on the contrary with an odd indices number are \ac{CPT}-odd, hence the \ac{CPT}-preserving terms are proportional to: $c$, $d$, $g$ and $G$ and the \ac{CPT}-odd ones are: $a$. $b$, $e$, $f$ and $h$.
        The neutrino Lagrangian density in \ac{SME} context can therefore be amended by additional \ac{LIV} terms \cite{Kostelecky:2013rta} and the most studied possibility is given by:
        \begin{equation}
            \label{neutrinoL}
            \mathcal{L}_\text{LIV}=-\frac{1}{2}\left(p^{\mu}_{AB}\overline{\psi}_{A}\gamma_{\mu}\psi_{LB}+q^{\mu}_{AB}\overline{\psi}_{A}\gamma_{5}\gamma_{\mu}\psi_{LB}+ir^{\mu\nu}_{AB}\overline{\psi}_{A}\gamma_{\mu}\partial_{\nu}\psi_{LB}+is^{\mu\nu}_{AB}\overline{\psi}_{A}\gamma_{5}\gamma_{\mu}\partial_{\nu}\psi_{LB}\right)+\text{h.c.}.
        \end{equation}
        The \ac{CPT}-odd contributions are parameterized by the coefficient:
        \begin{equation}
        (a_{L}^{\mu})_{AB}=\left(p^{\mu}_{AB}+q^{\mu}_{AB}\right),
        \end{equation}
        and assuming only the component $\mu=0$ is non-zero, choosing the simplest contribution form, the \ac{LIV} Hamiltonian \ac{CPT}-odd term can be written in the flavor basis as:
        \begin{equation}
        \label{HneutrinoCPT}
            H_\text{LIV}^\text{CPT-odd}=a^{0}_{AB}\,,
        \end{equation} an opportune Hermitian matrix. The \ac{CPT}-odd perturbations of this kind are therefore similar to the contribution given by the introduction of \ac{NSI} with matter for neutrinos, letting $a^{0}_{AB}=\sqrt{2}G_{F}N_{e}\epsilon_{AB}$, where $G_\text{F}$ is the Fermi coupling constant, $N_{e}$ is the electronic matter density and $\epsilon_{AB}$ is the Hermitian matrix containing the NSI coupling constants. The magnitude of the perturbation terms in Eq.~\eqref{HneutrinoCPT} can affect the standard oscillation pattern with a perturbation proportional to the baseline length $L$ and can be constrained analyzing the atmospheric neutrinos data. The off-diagonal elements are the \ac{CPT}-odd \ac{LIV} terms which can affect the neutrino flavor transition. These parameters are theoretically expected as highly suppressed and the current constraints on their magnitude are obtained from atmospheric neutrino data by Super-Kamiokande (at 95$\%$ C.L.)~\cite{Abe:2014wla}:
        \begin{align}
          \left|a^{0}_{e\mu}   \right| &< 2.5\times10^{-23}\,\mathrm{GeV},& \left|a^{0}_{e\tau}  \right| &< 5.0\times10^{-23}\,\mathrm{GeV},& \left|a^{0}_{\mu\tau}\right| &< 8.3\times10^{-24}\,\mathrm{GeV}.
        \end{align}
        Improved sensitivity to some of these parameters can be obtained by DUNE~\cite{Barenboim:2018ctx}:
        \begin{align}
          \left|a^{0}_{e\mu} \right| &< 7.0\times10^{-24}\,\mathrm{GeV},& \left|a^{0}_{e\tau}\right| &< 1.0\times10^{-23}\,\mathrm{GeV},
        \end{align}
        while the diagonal coefficients can also be constrained by the synergy of projected T2K and NOvA data~\cite{Majhi:2019tfi} 
        (at 95\% C.L.)\footnote{Note that bounds on diagonal coefficients can also be derived using the constraints on NSI from Super-Kamiokande~\cite{Barenboim:2018ctx}.}:
        \begin{align}
           a^{0}_{e e} &\in [-5.5, 3.3] \times 10^{-22}\,\mathrm{GeV},& 
           a^{0}_{\mu\mu} &\in [-1.1, 1.2] \times 10^{-22}\,\mathrm{GeV},& a^{0}_{\tau\tau} &\in  [-1.1, 0.9] \times 10^{- 22}\,\mathrm{GeV},
        \end{align}
        and by DUNE~\cite{Barenboim:2018ctx}:
        \begin{align}
           a^{0}_{ee} &\in [-2.5, -2.0]\times 10^{-22}\,\mathrm{GeV} \cup [-2.5, 3.2]\times 10^{-23}\,\mathrm{GeV}, &
           a^{0}_{\mu\mu} &\in [-3.7, 4.8] \times 10^{-23}\,\mathrm{GeV}.
        \end{align}
       \Ac{CPT} invariance implies that the coupling constants in the Lagrangian  and the masses present in the dispersion relations of particles and the related antiparticles must be the same. The \ac{CPT} invariance can be tested verifying the previous statement, that is different dispersion relations for particles and antiparticles or different coupling constants must imply \ac{CPT} violation.\\
        
One of the most promising method to look for \ac{LIV} effects associated to \ac{SME} is the sidereal variation search. Experimental data are often checked with day-night period (24 hours) to check systematic effects, such as temperature dependence. The sidereal day, or the rotation period of the Earth against the fixed stars is  slightly shorter, 23\,h 56\,min $4.1$\,s.  \Ac{LIV} shows up as the periodic effect with sidereal period, this means the system is coupled with background fields of the universe. This is equivalent with the search of classic {\ae}ther effect, i.e.\ physics may depend on directions and the rotation of the Earth with sidereal period can reveal this. Lorentz violating background fields are assumed uniform within the solar system, and the Sun-centered inertial frame is chosen to be the standard frame so that results from different experiments can be comparable~\cite{Kostelecky:2008ts}. 

Sidereal time dependent effect is one of many predict signals of Lorentz violation under the \ac{SME}. Extra terms in the Lagrangian can modify almost all physics observables, energy spectrum, decay rate, oscillation length, etc. Expected effects are small, and systems with sensitive to small effects are more interesting. These include quantum interference such as neutrino oscillations. All fermions and bosons, including composite particles (hadrons and atomic nuclei) are assumed to have different coupling to the background fields, and couplings are provided separately. Also, different couplings are assumed to be different dimensional operators~\cite{Kostelecky:2009zp,Kostelecky:2011gq,Kostelecky:2013rta}. Variety of limits are reported from different communities, and all limits make large tables updated every year~\cite{Kostelecky:2008ts}.        

\subsubsection{CPT symmetry implications in the gravitational sector} 
 
A fundamental issue about \ac{CPT} symmetry is related to its implications in the gravitational sector. In this context it can be useful to investigate the behavior of antimatter in a gravitational field. The \ac{CPT} theorem for curved spacetime has not yet been demonstrated. Indeed the \ac{CPT} theorem demonstration is based on Wightman axiomatic definition of \ac{QFT} set in a Minkowski flat spacetime and cannot be straightforwardly generalized to curved spacetime. A possible \ac{CPT} theorem generalization can be obtained axiomatically defining the \ac{QFT} in a curved space using the ordered product expansion \cite{Hollands:2014eia,Hollands:2008vx}.\\
        If one assumes the validity of the \ac{CPT} theorem even in curved spacetimes, one first notices that the Einstein equations:
        \begin{equation}
            R_{\mu\nu}-\frac{1}{2}g_{\mu\nu}R=8\pi G T_{\mu\nu}
        \end{equation}
        are \ac{CPT}-even since they are written using 2 indices tensor fields (even rank tensors are \ac{CPT}-even in flat spacetime). This means that antimatter should behave like ordinary matter and be gravitationally self-attractive, but this point is still under debate \cite{Nieto:1991xq,Chardin:1992mb,Chardin:1992ve,Hajdukovic_2011,Antonelli:2020nhn}. Many theoretical arguments suggest that \ac{CPT} operator can affect internal quantum numbers, but presumably leaves unaffected the gravity sector. The aspects of matter and antimatter gravitational interaction is widely debated \cite{Hajdukovic_2010,Noyes,BenoitLevy:2008ia}, indeed depending on the interpretation antimatter even if self-attractive can interact with ordinary matter attractively or repulsively. 
        Following an argument present in \cite{Villata:2011bx} the geodesic equations of motion are:
        \begin{equation}
            m_\text{i}\frac{d^2x^{\mu}}{d\tau^{2}}=-m_\text{g}\frac{dx^{\alpha}}{d\tau}\Gamma^{\mu}_{\alpha\beta}\frac{dx^{\beta}}{d\tau}
        \end{equation}
        where $m_\text{i}$ is the inertial mass and $m_\text{g}$ the gravitational one and the action of the \ac{CPT} operator can be resumed by the relations:
        \begin{align}
            \left(\frac{dx^{\mu}}{d\tau}\right)_\text{CPT} &= -\left(\frac{dx^{\mu}}{d\tau}\right)_\text{CPT} ;& \left(\frac{d^{2}x^{\mu}}{d\tau^{2}}\right)_\text{CPT} &= -\left(\frac{d^{2}x^{\mu}}{d\tau^{2}}\right)_\text{CPT} ;& \left(\Gamma^{\mu}_{\alpha\beta}\right)_\text{CPT} &= -\Gamma^{\mu}_{\alpha\beta},
        \end{align}
        using the fact that the Christoffel symbol is a 3-index object. The antimatter equation of motion becomes therefore in a ordinary matter generated gravitational field:
        \begin{align}
           m_\text{i}\left(\frac{d^2x^{\mu}}{d\tau^{2}}\right)_\text{CPT} &= -m_\text{g}\left(\frac{dx^{\alpha}}{d\tau}\right)_\text{CPT}\Gamma^{\mu}_{\alpha\beta}\left(\frac{dx^{\beta}}{d\tau}\right)_\text{CPT}\Rightarrow\\
           \Rightarrow -m_\text{i}\left(\frac{d^2x^{\mu}}{d\tau^{2}}\right)_\text{CPT} &= m_\text{g}\frac{dx^{\alpha}}{d\tau}\Gamma^{\mu}_{\alpha\beta}\frac{dx^{\beta}}{d\tau},
        \end{align}
       where the \ac{CPT} action is restricted to the antimatter terms of the equation, that is the affine connection (Christoffel symbol) is unaffected since generated by ordinary matter. In this framework matter and antimatter should interact repulsively. Such effect 
can be potentially tested in the ASACUSA \cite{Malbrunot_2018}, AEGIS \cite{Brusa_2017} and QUPLAS experiments \cite{2020NIMPA.95163019A}, where the behavior of antimatter in the ordinary matter generated gravitational field is investigated using interferometric techniques.      
       
\subsubsection{CPT from non-commutativity}
       
Many approaches to \ac{QG} based on non-commutative spacetimes (Sec.~\ref{section2.2.2.3}) predict some departures from ordinary \ac{CPT} invariance.
Such modifications of \ac{CPT} symmetry can occur whenever some of the basic structures characterizing the quantum space differ from their standard (i.e.~low-energy) counterparts, leading possibly to a deformed \ac{CPT} operator which would reduce in the low-energy limit to the usual \ac{CPT} operator plus a small correction responsible for departures from usual \ac{CPT} invariance. A representative example from the non-commutative world is provided by the $\kappa$-deformations of the Minkowski spacetime, for which different Hermitian conjugations (i.e.~involutions) and/or different differential calculi may be chosen to carry out the construction of a Lagrangian, resulting in different possible choices for C and/or P, hence of \ac{CPT} operation (see e.g.~\cite{Arzano:2016egk,Arzano:2019toz}). For a recent attempt for a systematic investigation of \ac{CPT} symmetry within field theories on $\kappa$-Minkowski, see \cite{Arzano:2020jro}. For modified hermitean conjugation, see \cite{Poulain:2018mcm}. A full investigation of these different possibilities is still missing although one such deformed \ac{CPT} invariance has been confronted to the muon and anti-muon lifetimes giving rise to a bound $\kappa>\mathcal{O}(10^{14})$ GeV \cite{Arzano:2019toz}.\\
It must be noted that the fate of causality for quantum spaces described by a non-commutative structure  
has not been thoroughly explored yet. Such question is of great relevance for the issue of departures from \ac{CPT} since the notion of causality and hence microcausality, as we recalled above, plays a major role in the proof of the \ac{CPT} theorem. It appears that the notion of causality in a non-commutative framework is not necessarily unique. It would be thus helpful to determine whether different modified notions of causality in non-commutative spaces give rise at low energy to different departures from the usual \ac{CPT} invariance. Taking again the $\kappa$-deformations of the Minkowski spacetime as an example, the modifications of usual causality stems from the fact that the light-cones can be actually modified. Starting from the non-commutative analog of the Pauli-Jordan function, one finds that the commutator of 2 bosonic (free) fields decays rapidly over a region of finite thickness (where space-like and time-like regions blur), instead of strictly vanishing in the space-like region \cite{Mercati:2018ruw}.

A different approach based on non-commutativity is given in terms of non-commutative fields \cite{Carmona:2002iv,Carmona:2003kh,Faruk:2017izo,Scholtz:2019aob,Gamboa:2017onj,Toghrai:2021cdm,Rasouli:2018lny}. Here, instead of 
considering  spacetime non-commutative coordinates, a modified commutator between fields is introduced. For example, consider the 
free complex  scalar field theory described in terms of two real fields $\phi_1,\phi_2$ with the following commutation relations 
\begin{equation}
[\phi_i({\bf x}),\phi_j({\bf x}')] =\epsilon_{ij}\theta\,\delta^{(3)}({\bf x}-{\bf x}'),\quad [\pi_i({\bf x}),\pi_j({\bf x}')] =\epsilon_{ij}B\,\delta^{(3)}({\bf x}-{\bf x}'),
\end{equation}
while fields and conjugate momenta remain unchanged and   the free Hamiltonian given by
\begin{equation}
{\cal H} = \frac{1}{2} \sum_{i}\int\,d^3x\,\left[ \pi_i^2+({\boldsymbol \nabla}\phi_i)^2+m^2(\phi_i)^2\right],
\end{equation}
which  can be expressed in terms of a set of creation and annihilation operators $\{a_{\bf p},a^\dag_{\bf p},b_{\bf_p},b^\dag_{\bf p}\}$
as follows
\begin{equation}
H=\int\,\frac{d^3 p}{(2\pi)^3}\left[E_-({\bf p}) \left(a_{\bf p}a^\dag_{\bf p}+\frac{1}{2}\right)+E_+({\bf p}) \left(b_{\bf p}b^\dag_{\bf p}+\frac{1}{2}\right)\right],
\end{equation}
with $E_{\pm}({\bf p})=\omega_{\bf p}^2\sqrt{1+\lambda_-^2}\pm\lambda_+^2$
and $\lambda_\pm=B/(2\omega_{\bf p})\pm(\theta\omega_{\bf p})/2$ and $\omega_{\bf p} =\sqrt{{\bf p}^2+m^2}$.

This model incorporates in this way  an explicit  particle-antiparticle asymmetry. The model also breaks  the Lorentz symmetry preserving 
rotational invariance.

This simple model shows an explicit violation of \ac{CPT} in the infrared regime at finite temperature \cite{Carmona:2004xc,Brito:2015csa}. The number of particles
at temperature $T$ in a volume $V$ in the limit $\theta T\ll B/T\ll m/T\ll1$ shows the explicit asymmetry due to the infrared scale $B$
\begin{equation}
\frac{n_-}{n_+}-1\sim \frac{B}{T},
\end{equation}
showing that this non-commutative extension brings a \ac{CPT} violation effect.

In general, matter  and antimatter densities ($n$ and $\bar{n}$, respectively) will depend on temperature and then corrections to its ratio 
can be parametrized by some  constant ($\delta$) with  dimensions of energy. For the case in which $\delta \ll T$, we expect a correction with the form $n/\bar{n} \sim1+\delta/T$, a sort of infrared correction, as the one previously obtained. 

Even though this model has no interactions and no fermions are considered, it exhibits exciting properties. For example, particles and antiparticles have zero mass difference since the only modification in the mass term is a shift $m^2\to m^2+ B^2/4$. The \ac{CPT} violation in this model occurs at the kinematical level, but for particles and antiparticles with equal masses. On the other hand, processes including particle-antiparticle creation should be sensitive to the 
energy difference in this model. For example, for the anomalous magnetic moment of the electron, even if the present model is a free field theory, we expect a deviation for this quantity with the form
\[
\delta a_e\approx\frac{\alpha}{\pi}\left(\frac{B}{m_e}\right)^n\approx 2^n\times 10^{-(3+6n)} \left(\frac{B}{\mbox{eV}}\right)^n.
\]
According to present data \cite{Aoyama:2017uqe}, $\delta a_e <10^{-12}$ and then, a suppression mechanism is necessary in order to 
cancel contributions with high values of $n$. For example, for $B\sim$ keV, $n\geq3$ should be suppressed. 

Finally, it is worth noting that  processes  preserving the number of particles minus antiparticles will not show this \ac{CPT} violation effect at 
kinematical effect.

\subsubsection{Decoherence, CPT and the $\omega$ effect}

    In  the presence of decoherence induced by \ac{QG} effects~\cite{Mavromatos:2004sz}, i.e.\ quantum fluctuations of the space-time metric at the microscopic  scale at which quantum gravity effects set, the quantum operator that generates the \ac{CPT} symmetry may not be well-defined~\cite{Wald:1980nm}. 
    This is a strong (`intrinsic') form of ``violation'' of \ac{CPT} symmetry, which should be contrasted with the situation in which the \ac{CPT} operator 
    $\widehat{CPT}$ 
    is well defined, but does not commute with the Hamiltonian of the system 
    $[ \widehat{CPT}, \widehat H ] \ne 0$, 
    which, for instance, characterizes the  \ac{SME}~\cite{Colladay:1998fq}, in the presence of Lorentz-violating backgrounds that induce \ac{CPT} violation~\cite{Greenberg:2002uu,Greenberg:2003nv}. The evolution of an initially pure quantum state to a mixed one which characterizes an open relativistic local \ac{QM} system, interacting with an environment (which in  the case of \ac{QG} is identified with the quantum fluctuating singular gravitational degrees of freedom, inaccessible by a low-energy observer (e.g.\ microscopic (Planck scale) event horizons in space-time foam situations~\cite{Wheeler:1998vs}), 
    may result in an ill-defined \ac{CPT} operator\footnote{The reader should have noticed that we insist on the matter system being local and relativistic, otherwise there is no a priory reason for \ac{CPT} violation, in the absence of an environment~\cite{Greenberg:2002uu,Greenberg:2003nv}.} as it follows from a mathematical proof which results in a {\it reductio ad absurdum}~\cite{Wald:1980nm}: one initially assumes a well-defined unitary \ac{CPT} operator, $\Theta$, acting on the density matrices of the mixed states of the open system,\footnote{The \ac{CPT} operator $\theta$ acting on state vectors is anti-unitary~\cite{Streater:1989vi}, due to the anti-unitary nature of the time-reversal operator, however when one considers the action of the \ac{CPT} operator $\Theta$ on density matrices $|x\rangle \langle y|$ then it is unitary, as it is essentially a product of $\theta^\dagger \theta $.} and then proves that this assumption is incompatible with the decoherent nature of the evolution, that is, the loss of information it entails for a low-energy observer, who can only observe degrees of freedom of the \ac{EFT} corresponding to the open matter quantum system.
    Indeed, consider the \ac{CPT} operator $\Theta$, acting on asymptotic (`in' and `out') density matrices,
    \begin{equation}\label{thetacpt} \Theta \rho_{\rm in} = {\overline \rho}_{\rm out}~, \quad \Theta^\dagger = \Theta^{-1}\,.
\end{equation} One should take into account that in decoherent quantum systems the asymptotic density matrices are related by the (linear) super-scattering operator of Hawking~\cite{Hawking:1982dj}:
\begin{equation}\label{dollar}
\rho_{\rm out} = \$ \rho_{\rm in}~, \qquad
\overline{\rho}_{\rm out} = \$\overline{\rho}_{\rm out}\,.
\end{equation}
In such a situation, the factorizability property of the
super-scattering matrix \$ in terms of the ordinary scattering S-matrix, $S=e^{-H t}$, with $H$ the Hamiltonian of the matter system, breaks down, \$ $\ne SS^\dagger $. This property is equivalent to the breakdown of the local effective Lagrangian formalism, which relies on well-defined perturbative scattering matrices.
It is important to notice that, as a result of ``loss of information'' for the asymptotic low-energy observer, the $\$ $-matrix has \emph{no inverse}~\cite{Hawking:1982dj,Wald:1980nm}.

From (\ref{thetacpt}) and (\ref{dollar}), we readily obtain:
\begin{eqnarray}
{\overline \rho}_{\rm in} = \Theta \rho_{\rm out} = \Theta \, \$ \, \rho_{\rm in} =
\Theta \, \$ \, \Theta^{-1} \Theta \rho_{\rm in} =
 \Theta \, \$ \, \Theta^\dagger {\overline \rho}_{\rm out} =
\Theta \, \$ \, \Theta^{\dagger} \, \$ \, {\overline \rho}_{\rm in}\,,
\end{eqnarray}
from which it follows that $ \Theta \$ \Theta^\dagger $ is the inverse of $\$ $, thus contradicting the above-mentioned property of $\$ $ of having no inverse. It implies that the assumption on the existence of a well-defined unitary \ac{CPT} operator $\Theta $, acting on density matrices, was \emph{false}.

This is really a breakdown of  microscopic time reversibility~\cite{Wald:1980nm}, which, being unrelated, in principle, to CP properties of the system, implies a breakdown of the entire \ac{CPT} symmetry. The result should not come as a surprise, since decoherence violates one of the important assumptions of the \ac{CPT} theorem~\cite{Streater:1989vi}, that of unitarity. 

The aforementioned ill-defined nature of the \ac{CPT}
operator in cases of quantum-gravity decoherence, leads 
to an entirely novel effect of \ac{CPT} violation,  
which is rather \emph{exclusive} to neutral-meson factories, the so-called $\omega$-{\it effect}~\cite{Bernabeu:2003ym}.
The effect is associated with appropriate modifications of the Einstein-Podolsky-Rosen (EPR)
correlators of entangled neutral meson states in a meson factory.
If the \ac{CPT} generator is ill-defined, the
properties of the antiparticle may be \emph{modified}, but {\it only} perturbatively, given that the antiparticle is a physical state which
should exist even in the case of an ill-defined \ac{CPT} operator of the effective matter open system. In such a case,
the neutral mesons $M^0$ and ${\overline M}^0$ should \emph{no
longer} be treated as \emph{indistinguishable particles}, as in the case where the generator of the combined \ac{CPT} transformations is a well-defined quantum operator, even in the cases where the symmetry is violated, e.g.~in \ac{SME} effective theories~\cite{Colladay:1998fq}. As a
consequence~\cite{Bernabeu:2003ym}, the initial entangled state in the meson factory $\left|i\right>$, after the initial-meson decay, 
which was symmetric under CP, in the case of a well-defined \ac{CPT} generator, will now acquire a component
with opposite permutation (${\cal P}$) symmetry (for definiteness, below we consider explicitly the case of a $\phi$ factory, where an initial $\phi$ meson decays into two neutral kaons, $K^0$ and its antiparticle $\overline K^0$): 
{\small \begin{eqnarray}\label{initialomega} &~&|i\rangle =   \frac{1}{\sqrt{2}}\left(|K_0({\vec
k}),{\overline K}_0(-{\vec k})\rangle
- |{\overline K}_0({\vec k}),K_0(-{\vec k})\rangle \right)
+  \frac{\omega}{2} \left(|K_0({\vec k}), {\overline K}_0(-{\vec k})\rangle + |{\overline K}_0({\vec
k}),K_0(-{\vec k})\rangle \right) \nonumber \\  &~& = {\cal N} \left[\left(|K_S({\vec
k}),K_L(-{\vec k})\rangle
- |K_L({\vec k}),K_S(-{\vec k})\rangle \right)
+  \omega \left(|K_S({\vec k}), K_S(-{\vec k})\rangle 
- |K_L({\vec
k}),K_L(-{\vec k})\rangle \right)\right] ~,\nonumber 
\end{eqnarray}} where ${\cal N}$ is an appropriate normalization factor, and $K_{S(L)}$ denote the short- (S) and long(L)-lived kaon mass eigenstates. 
The quantity $\omega$ is a complex parameter. Notice that, as a result of the $\omega$-terms, there
exist 
$K_SK_S$ or $K_LK_L$ combinations in the two-kaon state,
which entail important observable effects in the various decay channels. Due to the
$\omega$-effect,  
there is \emph{contamination} of ${\cal P}$-odd state with ${\cal P}$-even
terms, the  amount of which is controlled by the $\omega$-parameter. A time evolution of the $\omega$-terms, even in a purely unitary Hamiltonian evolution,
will lead~\cite{Bernabeu:2003ym,Bernabeu:2005pm,Alvarez:2004tj,Alvarez:2006ry,Bernabeu:2016kva} to observable differences in the final states, as compared with the \ac{CPT} conserving case, which can be tested experimentally in principle, and, in fact, would constitute, if observed, a rather ``smoking-gun'' evidence of this type of decoherence-induced \ac{CPT} Violation. 

These
effects are qualitatively similar
for kaon~\cite{Bernabeu:2003ym,Bernabeu:2005pm} and $B$-meson factories~\cite{Alvarez:2004tj,Alvarez:2006ry,Bernabeu:2016kva}, but in kaon factories there is a particularly
good channel, that of both correlated kaons decaying to $\pi^+\pi^-$.
In that channel the sensitivity of the $\omega$-effect increases because
the complex parameter $\omega$, parameterizing the relevant
EPR modifications~\cite{Bernabeu:2003ym}, appears in the particular
combination $|\omega|/|\eta_{+-}|$, with
$|\eta_{+-}| \sim 10^{-3}$. In the case of  $B$-meson factories
one should focus instead on the ``same-sign'' di-lepton
channel~\cite{Alvarez:2004tj,Alvarez:2006ry}, where high statistics occurs. Moreover, the real and imaginary parts of the $\omega$ parameter can be in principle measured experimentally, if they are sufficiently large~\cite{Bernabeu:2003ym,Bernabeu:2016kva}, and can also be disentangled~\cite{Bernabeu:2005pm} from 
the conventional \ac{CPT}-Violation parameters that exist in the \ac{SME}~\cite{Colladay:1998fq}, as well as other decoherence parameters that enter the Lindblad  evolution of the meson density matrices, e.g.\ the $\alpha,\beta,\gamma$ parameters~\cite{Ellis:1983jz}, discussed in Sec.~\ref{section7.4}.

To give the reader an idea on the appropriate observables for the potential detection/bounding
of $\omega$-like effects, we consider, for concreteness, the case of  
$\phi$-factories (for which, as already mentioned, one expects maximal sensitivity of the $\omega$-effect, as compared to other neutral meson factories). The relevant $\phi$-meson decay amplitude $A(X,Y)$, where one of the kaon products decays to
the  final state $X$ at $t_1$ and the other to the final state $Y$
at time $t_2$, with $t=0$ the time moment of the $\phi$-meson
decay, reads:
$A(X,Y) = \langle X|K_S\rangle \langle Y|K_S \rangle \, {\cal N}
\,\left( A_1  +  A_2 \right)$, 
with $A_1  = e^{-i(\lambda_L+\lambda_S)t/2}
[\eta_X  e^{-i \Delta\lambda \Delta t/2}
-\eta_Y  e^{i \Delta\lambda \Delta t/2}], \quad 
A_2  =  \omega [ e^{-i \lambda_S t} - \eta_X \eta_Y e^{-i
\lambda_L t}]$
denoting the \ac{CPT}-allowed and \ac{CPT}-violating parameters respectively,
and $\eta_X = \langle X|K_L\rangle/\langle X|K_S\rangle$ and $\eta_Y
=\langle Y|K_L\rangle/\langle Y|K_S\rangle$. In the above formulae, $t$
is the sum of the decay times $t_1, t_2$ and $\Delta t $ is their
difference (assumed positive). The ``intensity'' $I(\Delta t)$, \begin{eqnarray} \label{omegevoldet} I (\Delta t) \equiv \frac{1}{2} \int_{\Delta
t}^\infty dt\, |A(X,Y)|^2~,
\end{eqnarray}
which depends only on $\Delta t$, is the
desired \emph{observable} for a detection of the $\omega$-effect,

Theoretical estimates of the $\omega$ effect exist only in toy, but quite instructive, models of decoherening quantum gravity, in which the ``environment'' of singular gravitational degrees of freedom is provided by, say, point-like massive gravitational defects in bosonic string/brane theories~\cite{Bernabeu:2006av}, which recoil upon interactions with low-energy matter degrees of freedom. Such recoil effects are inaccessible by a low-energy observer. If one considers 
entangled meson states in such toy models, then the decay of the initial meson state into pairs of neutral mesons $M^0$, $\overline M^0$, and the propagation of the latter into the environment of populations of gravitational defects, leads, as a result of the recoil of the latter during their interactions with the $M^0, \overline M^0$, 
to an $\omega$-like effect, with magnitude of the $\omega$-parameter given by:
 \begin{equation}
|\omega|^{2}\sim \xi^2 \,\frac{\left(  m_{1}^{2} + m_{2}^{2}\right)  }%
{M_{\rm QG}^{2}}\frac{k^{2}}{(m_{1}-m_{2})^{2}}~,\label{finalomega}
\end{equation}
where $k$ is the amplitude of the (average) momentum of propagation of the neutral mesons $M^0$, and $M_{\rm QG}$ is a mass scale where quantum gravity effects set in (in the example of \cite{Bernabeu:2006av}, this is the mass $M_\text{s}/g_\text{s}$, of the gravitational stringy defects, which provide the ``quantum-gravity environment'', with $M_\text{s}$ the string scale, and $g_\text{s} < 1$ the string coupling.). 
The factor $\xi^2$ expresses an average  
momentum transfer during the collision of the neutral mesons $M^0, \overline M^0$ with populations of gravitational defects, encountered during their propagation. In the absence of a complete theory of quantum gravity, this parameter cannot be further specified. The important feature of \eqref{finalomega} is that, although, as expected, the effect has an overall suppression factor inversely proportional to the square of the quantum-gravity scale $M_{\rm QG}^2$, nevertheless there is significant enhancement due to the denominator which is also proportional to the (square of the) difference between the masses of the pertinent meson eigenstates, which is very small in nearly mass-generate situations, as is the case in the meson factories. 
 
The most sensitive bounds of the real ($\Re$) and imaginary ($\Im$) parts of the $\omega$-parameter to date come, as already mentioned, from the $\phi$-factory DA$\Phi$NE in Frascati National Laboratories (Italy), in the 
KLOE2 experiment~\cite{Domenico:2020bbk}, and are given by: \begin{align*}
    \Re(\omega) &= (+1.6 ^{+3.0}_{-2.1\,\text{stat}} \pm 0.4_\text{syst}) \times 10^{-4},& 
    \Im(\omega) &= (-1.7 ^{+3.3}_{-3.0\,\text{stat}} \pm 1.2_\text{syst}) \times 10^{-4},
\end{align*} 
which are consistent with the absence of such a \ac{CPT} violation.  For comparison, we mention that the corresponding bound in the case of $B^0\, \overline B^0$ systems in the Ba-Bar experiment is~\cite{Bernabeu:2016kva}:
\begin{align*}
    \Re(\omega_{B^0}) &= (+1.1 \pm 1.6) \times 10^{-2}, & 
    \Im(\omega_{B^0}) &= (+6.4 \pm 2.8) \times 10^{-2}.
\end{align*} 
The fact that there is an observed effect for the imaginary part of $\omega$ is attributed to the 
quality of the current data, as far as the measurement of this parameter is concerned.

\subsubsection{Neutrino oscillations and CPT}

      Neutrino oscillation experiments can be used to bound \ac{CPT} invariance in a model-independent way by analyzing neutrino and antineutrino oscillation data independently. If \ac{CPT} is conserved both analyses should give the same result for the oscillation parameters associated with neutrino and antineutrino oscillations. Using global neutrino and antineutrino oscillation data the current 3$\sigma$ bounds~\cite{Tortola:2020ncu} are given by 
\begin{align}
    \left|\Delta m_{21}^2 - \Delta\overline{m}_{21}^2\right| &< 4.7\times 10^{-5}\,\mathrm{eV}^2, \nonumber \\
    \left|\Delta m_{31}^2 - \Delta\overline{m}_{31}^2\right| &< 2.5\times 10^{-4}\,\mathrm{eV}^2, \nonumber \\
    \left|\sin^2\theta_{12} - \sin^2\overline{\theta}_{12}\right| &< 0.14,  \\
    \left|\sin^2\theta_{13} - \sin^2\overline{\theta}_{13}\right| &< 0.029, \nonumber \\
    \left|\sin^2\theta_{23} - \sin^2\overline{\theta}_{23}\right| &< 0.19.  \nonumber
    \label{eq:CPT-neutrino-bounds}
\end{align}
    Here, the quantities with (without) overline correspond to antineutrino (neutrino) parameters. Note that no data from atmospheric experiments was used to derive these bounds, since currently running experiments can not distinguish neutrinos from antineutrinos. This could be possible, however, in future experiments~\cite{Kumar:2017sdq,Ternes:2019sak}. Some of these bounds will be further improved by the next generation neutrino oscillation experiments Hyper-K and DUNE~\cite{deGouvea:2017yvn,Barenboim:2017ewj,Barenboim:2018lpo}. It was shown~\cite{Barenboim:2017ewj} that an independent analysis of neutrino and antineutrino data is crucial in the context of DUNE, since if \ac{CPT} was violated in nature, the combined analysis of neutrino and antineutrino data would lead to so-called impostor solutions, which do not reflect the real character of neutrino or antineutrino oscillations. This is not a problem in current experiments due to lack of sensitivity.

\subsection{Particle oscillations: other effects}
\label{section7.6}

     Neutrino physics is in general an ideal playground to also test other classes of low-energy phenomenological effects that could be induced by quantum gravity inspired models or by any kind of very high energy quantum modifications of the space-time structure and property. The golden channel for this kind of analyses is the detailed study of the flavor oscillation parameters and pattern, since this pattern would be changed by any modification of the dispersion relations.
    The introduction of \ac{LIV} perturbations for instance can modify the free particle dispersion relations, this effect can be 
    encoded in the simple relation
    \begin{equation}
        E^{2}-p^{2}=\widetilde{m}_{i}^{2}(E),
    \end{equation}
    where $\widetilde{m}_{i}^{2}(E)$ is a sort of effective mass term containing the \ac{LIV} corrections.  
    Such modification of the dispersion relation implies, for instance, a change in the free particle oscillation pattern for neutrinos. As stated in Eq.~\eqref{neutrinoL} in the SME framework \cite{Colladay:1998fq,Kostelecky:2013rta} this effect for CPT-even perturbations is parameterized by the coefficient $(c_{L}^{\mu\nu})_{AB}=\frac{1}{2}(r^{\mu\nu}+s^{\mu\nu})_{AB}$. Again in the oversimplified scenario where only the $\mu=\nu=0$ coefficients are non-vanishing, the corresponding effective \ac{LIV} and CPT-even Hamiltonian can be written as:
    \begin{equation}
        \mathcal{H}_\text{LIV}^\text{CPT-even}=\frac{4}{3}E(c^{0})_{AB}.
    \end{equation}
    It is possible to demonstrate that under the hypothesis that the Hamiltonian can be written as $\mathcal{H}=\mathcal{H}_{0}+\mathcal{H}_\text{LIV}$ with $\mathcal{H}_\text{LIV}$ a perturbative order term, the oscillation probability can be written as:
    \begin{equation}
        P(\nu_{A}\rightarrow\nu_{B})=P^{0}(\nu_{A}\rightarrow\nu_{B})+P^{1}(\nu_{A}\rightarrow\nu_{B})+\cdots,
    \end{equation}
    with $P^{0}(\nu_{A}\rightarrow\nu_{B})$ the standard probability and the following are perturbative terms.\\
    In the \ac{SME} framework a specific attempt to explain neutrino oscillations in the \ac{LIV} context is the PUMA model \cite{Diaz:2011ia,Diaz:2010ft}. The effective Hamiltonian associated to the oscillations is written in the form:
\begin{equation}
    \mathcal{H}_\text{LIV}=A(E)  
    \begin{pmatrix}
1 & 1 & 1\\
1 & 1 & 1 \\
1 & 1 & 1 
\end{pmatrix} + 
B(E)   \begin{pmatrix}
1 & 1 & 1\\
1 & 0 & 0 \\
1 & 0 & 0 
\end{pmatrix} +
C(E)   \begin{pmatrix}
1 & 0 & 0\\
0 & 0 & 0 \\
0 & 0 & 0 
\end{pmatrix},
\end{equation}
    where $A(E)$, $B(E)$ and $C(E)$ are energy dependent coefficients with $A(E)=\frac{\Delta m^{2}}{2E}$ and $\Delta m^{2}$ is the unique mass parameter required by this model instead of the usual two $\Delta m_\text{atm}^{2}$ and $\Delta m_\text{solar}^{2}$ required by the $3\nu$\acs{SM}. This model is invariant under the action of rotations but not under boosts.\\
    Another possibility is given by the introduction of modified free particle kinematics in the context of a CPT-even modification of special relativity  \cite{Torri:2019gud,Antonelli:2018fbv,Torri:2020dec}. This model provides a minimal extension of the particle SM in a covariant fashion. In this context the dispersion relations are written as:
    \begin{equation}
    \label{neutrinoMDR}
    E^{2}-p^{2}\left(1-f(p,E)\right)=m^{2},
    \end{equation}
    where $f$ is a degree-0 homogeneous perturbation function of the ratio $|\vec{p\/}|/E$. This particular choice guarantees that the metric structure in a Finsler geometry and the 
    modified special relativity covariance (Sec.~\ref{section2.2.2.4}) are preserved. The high-energy limit of this model is the isotropic sector of the SME \cite{Colladay:1998fq} and first formulation of VSR \cite{Coleman:1998ti}. 
    In this scenario the \ac{LIV} correction introduces an additional term to the standard phase ruling the neutrino oscillation:
    \begin{equation}
        \Delta\phi_{ij}=\left(\frac{\Delta m_{ij}^2}{2E}-\frac{\delta_{i}-\delta{j}}{2}E\right)L,
    \end{equation}
    where $\delta_{i}$ is the high energy limit of the perturbation $f_{i}$ introduced in the dispersion relation~\eqref{neutrinoMDR} of the $i$-th neutrino mass eigenstate. 
    The \ac{LIV} term introduced in the phase presents a peculiar energy-baseline dependence being proportional to the product $L\times E$. This \ac{LIV} effect, which is in accordance with other models such as \ac{SME}, provides a correction to the general oscillation pattern, that, clearly, can represent only a small perturbation to the standard mass generated oscillation, but would in any case present a characteristic experimental signature. 
    
    As for the possible experimental tests, a clear advantage is offered by the possibility of studying neutrino sources of different origins, spanning
    a wide range of values for the baselines (from a few meters of short-baseline experiments to cosmological distances) and energies (from keV neutrinos up to the hundreds of TeV of the IceCube detected neutrinos and to the even higher energies investigated by experiments like Auger). The very-high- and ultra-high-energy cosmic neutrinos offer a natural privileged opportunity of searching for this kind of energy dependent corrections, but, as discussed in~\cite{Antonelli:2020uui, Torri:2020dec, Antonelli:2020udn}, the investigation of the potentialities is in progress also for other experimental contexts in which one could take advantage also from the high statistics and the unprecedented energy resolution offered by experiments that are going to start their data taking or are planned 
    for the near future. This is the case, for instance, of the long-baseline accelerator experiment DUNE~\cite{Barenboim:2018ctx} or
    a possible study with the JUNO experiment of medium and high energy atmospheric neutrinos. Another experimental advantage
    offered by the isotropy preservation in models like the one
    considered in~\cite{Antonelli:2018fbv,Torri:2019gud,Torri:2020dec,Miramonti:2021tgh}, would be the possibility of including the full set of data in the statistical analyses. Notice that these considerations about neutrino oscillations
    can, in principle, be extended to oscillations of other particles such as mesons for instance \cite{Edwards:2019lfb}.
   
    %Also \ac{LIV} has been searched in neutrino oscillation experiments. The main motivation is their sensitivities. 
    Neutrino oscillation is a natural interferometer, and it is sensitive to small quantities otherwise impossible to measure, such as neutrino masses. Typical neutrino oscillations are tuned to measure the  oscillation maximum by making certain neutrino energy beam and locating the far detector at a certain distance to choose the baseline. By doing this, the combination of them, $\frac{\Delta m^2}{4E}L$ is tuned to be order $\pi/2$ to maximize the oscillation effect. Thus, if the new physics is bigger than $\frac{\Delta m^2}{2E}$, it can be seen as a distortion of the standard oscillation signal. For example, with atmospheric neutrino parameters (assuming 1\,GeV neutrinos), $\frac{\Delta m^2}{2E}\sim 10^{-21}$\,GeV, thus, this approach can explore dimension-three \ac{LIV} down to $10^{-21}$\,GeV or $10^{-21}$ to explore dimension-four LIV operators. This already exceeds the sensitivity of the latest Michelson--Morley type experiment~\cite{Nagel:2014aga}. Since the mass term goes smaller at the higher energy, in general we expect higher-energy neutrinos are more sensitive to new physics by this approach. For 1\,TeV atmospheric neutrinos, the sensitivity goes down to $10^{-24}\,\mathrm{GeV}$. This number can be comparable with the sensitivity of a high precision magnetometer~\cite{Brown:2010dt}. These examples show a naturally high sensitivity of neutrino oscillations to look for new physics including \ac{LIV}.
    
    \Ac{LIV} was speculated as the source of the excess observed by LSND~\cite{Katori:2006mz} and sidereal time dependence of LSND data~\cite{Auerbach:2005tq} and MiniBooNE data~\cite{AguilarArevalo:2011yi} were used to search for \ac{LIV}. Since the effect becomes larger for the higher-energy beam and longer baseline, it was more interesting to search for \ac{LIV} in MINOS using both near detector data~\cite{Adamson:2008aa,Adamson:2012hp} and far detector data~\cite{Adamson:2010rn,Rebel:2013vc}. Search continues to other accelerator experiments such as T2K~\cite{Abe:2017eot}. On the other hand, reactor neutrinos are sensitive to channels including electron anti-neutrinos and Double Chooz~\cite{Abe:2012gw,Diaz:2013iba} and Daya Bay~\cite{Adey:2018qsd} searched for \ac{LIV}. At the higher energy end, conventional atmospheric neutrinos can reach up to $\sim20$\,TeV and they are attractive source to test \ac{LIV}. Since they are not controlled beams, most of experiments only test spectrum distortion due to \ac{LIV} and not sidereal variation. Searches are performed by AMANDA~\cite{Abbasi:2009nfa}, Super-Kamiokande~\cite{Abe:2014wla}, and IceCube~\cite{Aartsen:2017ibm}. In particular, IceCube limits have reached down to $\sim 10^{-42}\,\mathrm{GeV}^{-2}$ on dimension-six LIV operators for preferred astrophysical production scenarios, which is beyond the naive expected signal region of quantum-gravity motivated dimension-six operators, of order the inverse square of the Planck mass, or $\sim 10^{-38}\,\mathrm{GeV}^{-2}$~\cite{IceCube:2021tdn}. As the long baseline limit, SNO~\cite{Aharmim:2018apr} used their solar neutrino data to look for \ac{LIV}. High-energy astrophysical neutrinos are the combination of the highest energy and the longest baseline, and the weakest signal of new physics can be detected from their flavor structure~\cite{Arguelles:2015dca,Bustamante:2015waa}. Search for \ac{LIV} from the astrophysical neutrino flavors will reach to the quantum gravity-motivated signal region ($\sim 10^{-38}\,\mathrm{GeV}^{-2}$), however, current statistics of high-energy astrophysical neutrino data from IceCube is low~\cite{Abbasi:2020jmh}, and we need to wait IceCube-Gen2~\cite{Aartsen:2020fgd} to conclude the search for \ac{LIV} by this approach.  
   
 Neutrino physics can also be used to test one of the features of the new formulation of quantum gravity presented in the guise of metastring theory \cite{Freidel:2015pka} 
    based on Born geometry \cite{Freidel:2013zga,Freidel:2014qna} which also introduces the concept of ``gravitization of the quantum'' \cite{Freidel:2016pls}. For example, a deformation of quantum theory with a non-canonical time evolution described by the Jacobi elliptic functions, as opposed to canonical circular functions, can be tested using neutrino oscillations and, in particular, the current knowledge about atmospheric neutrinos. Similarly, the zero modes of the metastring, called metaparticles \cite{Freidel:2017nhg,Freidel:2017wst,Freidel:2018apz}, even though consistent with Lorentz symmetry, are non-local, and thus allow for non-canonical statistics, such as infinite statistics, which may be probed in the context of a dynamical dark energy.

Spectral anomalies, that is, deviations from the standard neutrino oscillations picture, are not the only possible signature in the presence of CPT and Lorentz violation. In the \ac{SME} context, rotation-invariance violations lead to sidereal and annual variations. This induced periodic variation of observables with time represents a signature of \ac{LIV} and has been searched for in several neutrino experiments \cite{Adey:2018qsd,Abe:2017eot,Adamson:2012hp}. Rotation-invariance violations can also induce time-independent signatures, such as the observations of unexplained directional asymmetries in the detector, as pointed out in \cite{Kostelecky:2003cr}. Finally, \ac{LIV} can also manifest as neutrino-antineutrino oscillations \cite{Diaz:2013iba}, which are well-described in the framework of the \ac{SME}.
\Acp{MDR} have been also explored in the context of light sterile neutrino searches \cite{Doring:2018cob} but have been found not to be responsible for the anomalous experimental results reported \cite{Barenboim:2019hso}.

The growing relevance of neutrinos in particle physics and, more recently in the quest for signatures of QG, has prompted their investigation from a more rigorous QFT perspective. This analysis has revealed the shortcomings of the QM treatment by pointing out the nontrivial structure of the interacting-field vacuum as a generalized $SU(2)$ coherent state and the related problem of the unitary inequivalence between the Fock spaces for fields with definite flavor and mass \cite{Blasone:1995zc}. To clarify this issue, let us consider a toy model with two interacting Dirac neutrino fields, say electron $\nu_e(x)$ and muon $\nu_\mu(x)$ neutrinos. It is well-known that these fields are related to the corresponding free fields $\nu_i(x)$ of mass $m_i$ ($i=1,2$) by Pontecorvo mixing transformations \cite{Bilenky:1978nj}, here rewritten for each $\alpha$-component as $\nu^{\alpha}_e(x)=
G_{\theta}^{-1}(t)\, \nu^{\alpha}_1(x)\, G_{\theta}(t)$ (and similarly for $\nu^{\alpha}_\mu(x)$), where $G_{\theta}(t)=\exp\left[\theta\int
d^3x\left(\nu_1^{\dagger}(x)\nu_2(x)-
\nu_2^{\dagger}(x)\nu_1(x)\right)\right]$.
This operator provides the dynamical map between the Fock space $\mathcal{H}_{1,2}$ for the fields with definite mass and the Fock space $\mathcal{H}_{e,\mu}$ for the fields with definite flavor \cite{Blasone:1995zc}.
Symbolically, one can write $G^{-1}_{\theta}: {\mathcal H}_{1,2} \mapsto {\mathcal H}_{e,\mu}$. 
At level of vacuum, this gives $|0(\theta,t) \rangle_{e,\mu}= G^{-1}_{\theta}(t)\;
|0 \rangle_{1,2}$,
where $|0 \rangle_{1,2}$ and $|0(\theta,t) \rangle_{e,\mu}$ denote the mass and flavor vacua, respectively. 

Strictly speaking, the above map is well-defined only in the finite volume limit. Indeed, in this case $G_{\theta}(t)$ is a unitary operator which preserves the canonical (anti-)commutators. On the other hand, for systems having infinite degrees of freedom (such as quantum fields), $G_{\theta}(t)$ is no longer unitary and  $\lim_{V \rightarrow \infty}\; _{1,2}\langle0|0(\theta,t)\rangle_{e,\mu}\,=\,0,\forall t$. The orthogonality of the two vacua lies at the heart of the QFT description of mixing, as it expresses the unitary inequivalence between $\mathcal{H}_{1,2}$ and $\mathcal{H}_{e,\mu}$. Specifically, the flavor vacuum acquires the structure of a $SU(2)$ coherent state, 
becoming a condensate of particle-antiparticle pairs with density 
${}_{e,\mu}\langle 0(\theta,t)| \alpha_{{\textbf{k}},i}^{r \dagger}\, \alpha^r_{{\textbf{k}},i}
|0(\theta,t)\rangle_{e,\mu}= \sin^{2}\theta\, |V_{\textbf{k}}|^{2}$, 
where
 \begin{equation}
\left|V_{{\textbf{k}}}\right|=\left(\frac{\omega_{{\textbf{k}},1}+m_{1}}{2\omega_{{\textbf{k}},1}}\right)^{\frac{1}{2}}
\left(\frac{\omega_{{\textbf{k}},2}+m_{2}}{2\omega_{{\textbf{k}},2}}\right)^{\frac{1}{2}}
\left(\frac{|\textbf{k}|}{\omega_{{\textbf{k}},2}+m_{2}}-\frac{|\textbf{k}|}{\omega_{{\textbf{k}},1}+m_{1}}\right),
 \end{equation}
and $\alpha_{{\textbf{k}},i}^{r \dagger}$ creates a quantum of mass $m_i$, frequency $\omega_{\textbf{k},i}={(m_i^2+|\textbf{k}|^2)^{1/2}}$ and polarization $r$.

The flavor/mass inequivalence 
poses the problem of selecting the right (i.e.\ physical) representation for mixed fields. A clue to an answer may be found by looking at its phenomenological consequences. For instance, it has been shown that the complex structure of the flavor vacuum leads to nontrivial modifications of the oscillation probability. By introducing the flavor charge operators $Q_{\nu_\sigma}(t)= \int d^{3}{\textbf{x}}\,
\nu_{\sigma}^{\dagger}(x)\,\nu_{\sigma}(x),\, \sigma=e,\mu\,$, the exact formula for neutrino oscillations reads
\begin{equation}
\label{eqn:oscillfor1}
\mathcal{Q}^\textbf{k}_{\nu_e \rightarrow \nu_\mu}(t)
\equiv 
\left< \nu_{\mathbf{k},e}^{r} \right|
\mathopen{::} Q_{\nu_\mu}(t) \mathclose{::}
\left| \nu_{\mathbf{k},e}^{r} \right> 
=
\sin^{2}(2\theta)\left(
    \left|U_{\textbf{k}}\right|^{2}
    \sin^{2}\left.
        \frac{\omega_{\textbf{k},2} - \omega_{\textbf{k},1}}{2} t
    \right. + 
    \left|V_{\textbf{k}}\right|^{2}
    \sin^{2}\left.
        \frac{\omega_{\textbf{k},2} + \omega_{\textbf{k},1}}{2} t
    \right.
\right),
\end{equation}
where~$\left|U_\textbf{k}\right|^2 = 1 - \left|V_\textbf{k}\right|^2$
and~$\left|\nu_{\textbf{k},\sigma}^{r}\right> = \alpha^{r\dagger}_{\textbf{k},\sigma}(0) \left|0(\theta,0)\right>_{e,\mu}$ 
is the single-particle state for the mixed field~$\nu_\sigma(x)$ (defined, for convenience, at the time~$t=0$). Here $\mathopen{::} Q_{\nu_\sigma}(t) \mathclose{::}$ is the normal ordered charge with respect to the flavor vacuum $\left|0(\theta, t)\right>_{e,\mu}$.
As opposed to Pontecorvo formula \cite{Bilenky:1978nj}, one can see that $(i)$~the oscillation amplitudes in Eq. \eqref{eqn:oscillfor1} are momentum-dependent, $(ii)$~there is an extra oscillating term depending on the sum of the frequencies, whose effects might in principle be tested experimentally. Clearly, the QM result is recovered in the relativistic limit $\left|\textbf{k}\right|\gg {(m_1 m_2)}^{1/2}$, since $\left|V_{\textbf{k}}\right|\rightarrow0$.

Further attempts to test the implications of the flavor-mass inequivalence have been recently made in the context of particle decays \cite{Blasone:2018czm,Blasone:2019agu,Blasone:2020vtm} and Casimir effect \cite{Blasone:2018obn}, among others. Nevertheless, deviations from the standard QM results turn out to be far below the sensitivity of past experiments.  More solid evidences are supposed to be obtained by current/future tritium $\beta$-decay experiments and measurements on
the non-relativistic cosmic neutrinos by capture on tritium, for which  the effects of the inequivalence are expected to be maximal. Indeed, in \cite{Lee:2017cqf} it has been shown that the spectrum of the tritium $\beta$-decay near the end-point energy is sensitive to whether neutrinos interact as either mass or flavor
eigenstates. Similar considerations hold for the neutrino capture by tritium too. Therefore, it is reasonable to
expect that high-precision measurements from KATRIN, Project 8 and PTOLEMY experiments might provide important pieces of information in the problem at hand.

Manifestations of \ac{LIV} at low energy can also be tested in double-beta decay. \ac{LIV} effects can be induced in this process by the coupling of neutrinos to the countershaded (oscillation-free) operator in \ac{SME}, particularly to its time-like component, and could manifest as perturbations in the shape of the energy spectra of electrons \cite{Diaz:2013ywa}. In the absence of observation of such perturbations, one can constrain the values of the oscillation-free (`of') coefficient $(a_\text{of}^3)_{00}$, that controlled the size of these \ac{LIV} effects. Such investigations are currently been conducted in several double-beta decay experiments \cite{EXO-200:2016hbz,CUPID:2019kto,NEMO-3:2019gwo,Brofferio} and are based on theoretical predictions of the electron spectra and angular correlation between electrons in $2\nu\beta\beta$ decay \cite{Nitescu:2020xlr,Nitescu:2020qnh}. The current limit is $(a_\text{of}^3)_{00} ~  10^{-4}\,\text{MeV}$~\cite{NEMO-3:2019gwo}.

\subsection{Further quantum gravity signatures in gravitational waves}
\label{section5.7}

In addition to the effects categorized nicely in the previous subsections, we discuss some additional possible imprints of quantum gravity in \acp{GW}, the graviton mass and the cosmic strings, which do not fit in the previous subsections.

\subsubsection{Graviton mass}
Quantum gravitational modifications to the gravitational dynamics are often implemented by adding new terms to the Einstein--Hilbert action, in an analogous way in which extra \ac{LIV} terms are added to the \ac{SM}~\cite{Liberati:2013xla}. This leads to different phenomenological implications, such as modification of the dispersion relation of the graviton or the post-Newtonian corrections to weak field \ac{GR}~\cite{Mattingly:2005re}.

\begin{description} 
\item{\it Graviton dispersion relations} In modified theories of gravitation, the  graviton satisfies dispersion relations of the type \eqref{eq:mdr}
          \begin{equation}
E^2= p^2 c^2 + A p^\alpha c^\alpha,
 \end{equation}
where $\alpha\geq 0$ and $A$ are two independent parameters.
In Ho\v{r}ava--Lifshitz theories, $\alpha=4$  \cite{Horava:2009uw},  in scenarios as  \ac{LIV}~\cite{Liberati:2012jf} and \ac{DSR}~\cite{AmelinoCamelia:2002wr}, $\alpha=3$,  and in massive gravity theories, $\alpha=0$ but $A\neq 0$ \cite{Will:2014kxa}. All these families of theories are strongly constrained by the events GW150914,
GW151226, and GW170104 detected by the LIGO-VIRGO collaboration \cite{Abbott:2017vtc}.
In particular, they lead to very stringent  bounds on the graviton mass ($\alpha=0$, $A=m_g^2$)
\begin{equation}
       m_g\,\leq\,7.7\,\times\, 10^{-23} \, \frac{\mbox{eV}}{c^2}.
\end{equation}
The constrains on $\alpha$ and $A$ are coming  from the induced changes in group velocities of 
propagating \acp{GW} and can be summarized in Figure~\ref{figLV}.
\begin{figure}[h!]
\begin{center}
\hspace{.1cm}{\includegraphics[width=12.6cm]{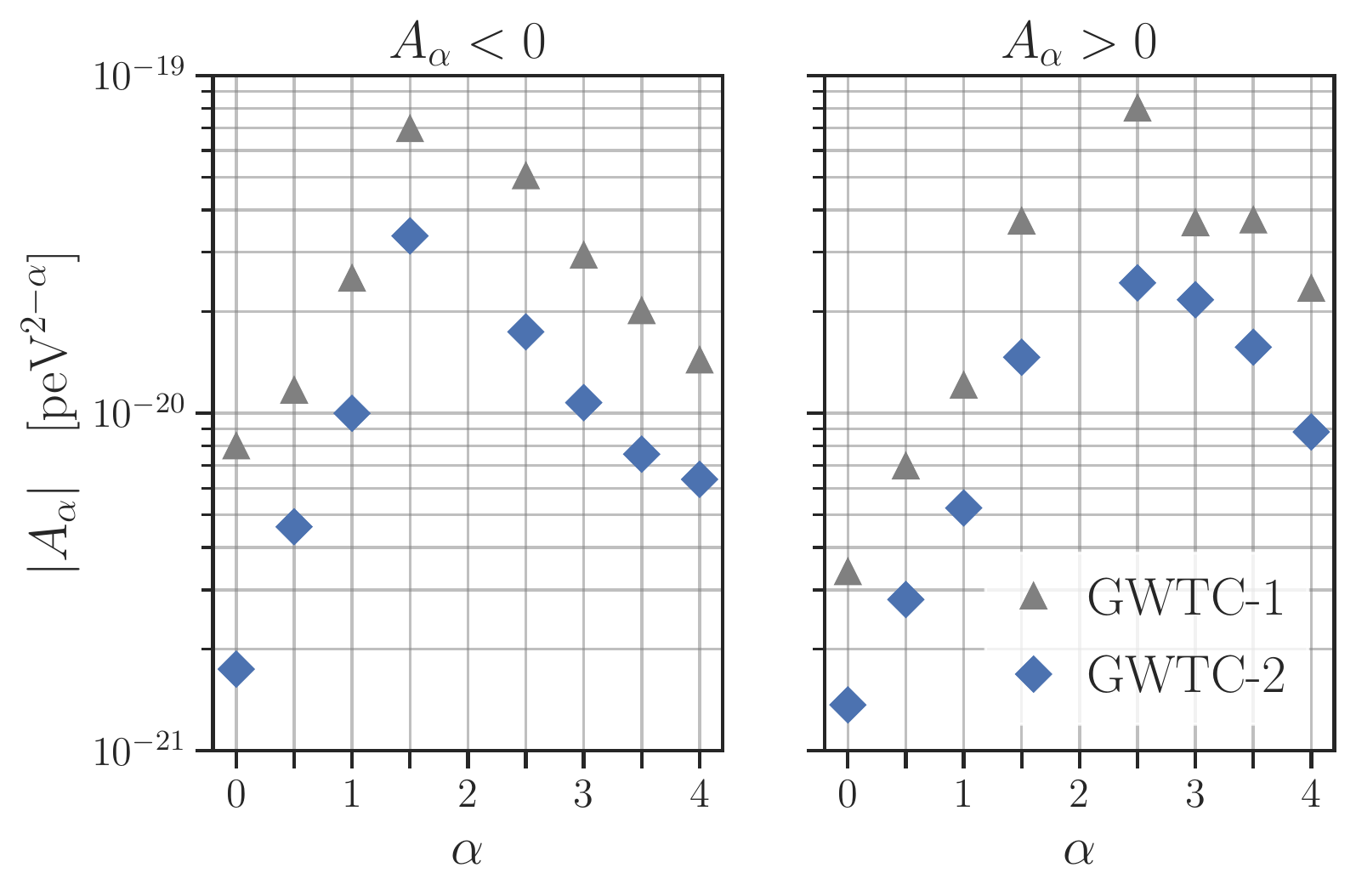}}
\end{center}
\caption{
Bounds on \acp{MDR}	
 from the second \ac{LIGO}–Virgo Gravitational-Wave Transient Catalog \cite{LIGOScientific:2020tif}. Constraints on the allowed amount
of dispersion through the 90\%-credible upper limits on $A_\alpha$ (corresponding to $A$ in the main text),
computed separately for $A_\alpha > 0$ and $A_\alpha < 0$.
Marker style distinguishes the new GWTC-2 results from the previous GWTC-1 results in \cite{LIGOScientific:2019fpa}. Figure taken from \cite{LIGOScientific:2020tif}.
\label{figLV}
}
\end{figure}
\item{ \it Cherenkov radiation}
The gravitational Cherenkov effect with radiative corrections to the graviton propagator is the analogue of the Cherenkov effect with radiative corrections to
the photon propagator. Some physical properties in extended theories of gravity have been discussed in \cite{DeLaurentis:2012fh}, where there is a clear indication that a gravitational Cherenkov radiation could be detected ranging from low energy scales up to TeV scales when gravitational degrees of freedom are considered. Furthermore, graviton masses of the order $m_g = 10-30$ eV could be detected by observing the characteristic signature of a a strong monochromatic signal in GW detectors due to relic gravitons at a frequency which falls in the range for both of space based (LISA) and earth based (\ac{LIGO} and Virgo) gravitational experiments, in the frequency interval $104\,\text{Hz} < f < 10\,\text{kHz}$.
\end{description}

\subsubsection{Spacetime dimension and discretization of area}

Many \ac{QG} theories share  a non-perturbative effect which underlines the way the dimension of spacetime changes with the probed scale. This dimensional flow influences the \ac{GW} luminosity distance, the time dependence of the effective Planck mass, and the instrumental strain noise of interferometers. 
The number of spacetime dimensions estimated from the comparison of the luminosity distance inferred from gravitational and electromagnetic radiation from GW170817 is compatible with 4 \cite{Abbott:2018lct}, a measurement sensitive to the shape of the distance posterior distributions as studied in \cite{Mastrogiovanni:2020mvm} with higher redshift events.
Investigating the consequences of \ac{QG} dimensional flow for the luminosity distance scaling of \acp{GW} in the frequency ranges of LIGO and LISA, it was shown that the quantum geometries of \ac{GFT}, spin foams and \ac{LQG} can give rise to observable signals in the \ac{GW} spin-2 sector~\cite{Calcagni:2019ngc,Calcagni:2019kzo}. 

Quantum black holes are expected to have a discrete energy
spectrum, and to behave in some respects like excited
atoms \cite{1974NCimL..11..467B,1995PhLB..360....7B,1986JETPL..44...63M}. In \cite{2020arXiv200713761A} the authors find that black hole area discretization could impart observable imprints to the \ac{GW} signal from binary black hole mergers, affecting the absorption properties of the black holes during inspiral and their late-time relaxation after merger. The rotation of black holes seems to improve the prospects to probe such quantum effects.

\subsubsection{Cosmic strings}
\label{cosmic_strings}

Cosmic strings are one-dimensional topological defects which could have been produced in a phase transition in the early universe \cite{Kibble:1976sj}. They are predicted in many high energy particle-physics models of the early universe \cite{Jeannerot:2003qv}. Cosmological scenarios based on string theory also predict in many cases the existence of a network of strings. In these models there are several objects that can play the role of a cosmic string. In particular, fundamental strings (F-strings) and the solitonic 1-dimensional Dirichlet branes (D-strings) of string theory could be cosmologically stretched until their dynamical behavior is practically
classical. These type of scenarios lead to the formation of a network of cosmic superstrings \cite{Sarangi:2002yt,Jones:2003da,Dvali:2003zj,Copeland:2003bj}. Observational constraints on cosmic superstrings could be a way to learn something about
one of the most promising avenues to quantum gravity, string theory.

At formation, the cosmological string networks are dominated by long strings whose length is larger than the Hubble size. These long strings are called infinite strings. The number and the total length of infinite strings inside the horizon increase as the size of the horizon grows. However due to their tension, they start to move relativistically eventually colliding with other string segments. In this way, the string network continuously  forms closed loops via self-intersection of a string or by the collision of two infinite strings. Once loops are formed, they oscillate with relativistic speeds and shrink by transferring their energy to GWs. This process continues until they eventually disappear. Within the Hubble horizon, while the length of infinite strings increases as the horizon extends, they can also reduce their length by forming loops. The balance between these two effects leads to a constant number of infinite strings within the horizon, a process known as the scaling law. as a result, the energy density of strings remains a fixed fraction of the overall energy density in agreement with the standard evolution of the universe.

One important feature which differentiates field theory cosmic strings from superstrings is that cosmic superstrings may have a reconnection probability $P$ significantly smaller than unity \cite{Jackson:2004zg}, while solitonic cosmic strings in field theory models always reconnect when a string collision occurs ($P\sim 1$) \cite{Shellard:1987bv,Verbiest:2011kv}. The value of $P$ can range as $10^{-3}<P<1$ in collisions between F-strings and $10^{-1}<P<1$ for D-string collisions \cite{Jackson:2004zg}. This means that superstrings may pass through each other without reconnection when they collide. As a result, cosmic superstring networks tend to produce less loops and lose energy less efficiently. However, as the network contains more infinite strings, the number of collision events increases and the network starts to produce more loops. Thus, contrary to intuition, cosmic superstring networks contain more loops than the ordinary case ($P=1$). 

Another interesting property of the cosmic superstring network is that it may contain Y-junctions. D-strings and F-strings can form bound states and produce Y-junctions where two different strings join and separate. It has been shown that the string network follows the scaling law even in the presence of Y-junctions \cite{McGraw:1997nx,Martins:2004vs,Tye:2005fn,Copeland:2005cy,Rajantie:2007hp,Avgoustidis:2007aa,Sakellariadou:2008ay}. On the other hand, the small-scale behaviour of the network is dramatically changed by Y-junctions and it may give interesting signatures in the GW signal, as discussed below. 

There is currently no observational evidence of the existence of cosmic strings. One of the most promising ways to probe cosmic strings is to search for the GW radiation \cite{Abbott:2017mem,Blanco-Pillado:2017rnf,Auclair:2019wcv}, emitted by both loops as well as infinite strings \cite{Sakellariadou:1990ne}. In addition to continuous GWs from the oscillation modes of cosmic string loops, strong GW bursts are emitted from singular points such as cusps and kinks \cite{Damour:2000wa,Damour:2001bk}. Cusps are points of the string which instantaneously acquire high Lorentz boosts. They are generically produced in loops \cite{Turok:1984cn}. Kinks on the other hand are discontinuities in the string's tangent vector produced by the reconnection of different segments of string with different directions. They exist both on infinite strings \cite{Sakellariadou:1990ne,Hindmarsh:1990xi} and loops  \cite{Olmez:2010bi} and continuously emit GWs while they propagate on a curved string until they get smoothed out by GW backreaction \cite{Quashnock:1990wv,Blanco-Pillado:2018ael,Chernoff:2018evo,Blanco-Pillado:2019nto}. 

The continuous emission of \acp{GW} from the cosmic string network and in particular the radiation emitted by loops throughout the cosmological history of the network leads to \iac{SGWB} (see Sec.~\ref{sec:stochGWsect}). 
The calculation of this \ac{SGWB} depends on several different ingredients but most importantly on the 
modelling of the string network evolution~\cite{Auclair:2019wcv,BlancoPillado:2011dq,Blanco-Pillado:2013qja,Ringeval:2005kr,Lorenz:2010sm}. The amplitude of the \ac{SGWB} strongly depends on the cosmic string tension $G\mu$. In the case of field-theoretic cosmic strings, the tension directly corresponds to the energy scale of the phase transition $\eta$ as 
\begin{equation}
G\mu\sim 10^{-6}\left(\frac{\eta}{10^{16}{\rm GeV}}\right)^2\,.
\end{equation}

Ten dimensional fundamental strings have a tension close to the Planck scale. A network of strings of that tension would be ruled out by CMB observations as well
as the non-detection of GWs in the LIGO or PTA arrays frequency bands. However, the process of compactification to a four dimensional universe allows the possibility of reducing the tension of these strings to a much lower value by placing them in specific regions of the internal manifold with a deep gravitational potential. 

Thus, observational constraints on the string tension are important for testing the underlying theory. 
Currently the most important constraints come from the low frequency band where the upper limits reported by Pulsar Timing arrays restricts the tension of the strings to be
$G\mu < O(10^{-10})$, although tighter constraints may be obtained if the reconnection probability of cosmic strings is assumed to be less than one, which, as mentioned above, appears to be the case for cosmic superstrings~\cite{Jackson:2004zg,Martins:1996jp,Avgoustidis:2005nv,Sakellariadou:2004wq}. Finally, we mention that Y-junctions in cosmic superstring networks also affect the GW spectrum~\cite{Binetruy:2010bq,Binetruy:2010cc,Matsui:2020hzi}. 

%\newpage
\section{Summary and outlook}
\label{section6}

For decades, the search for a quantum theory of gravity was exclusively a theoretical endeavor. This was mainly due to the expected realm of \ac{QG}, the Planck scale. Although some phenomenological models of \ac{QG} suggest that certain effects could manifest themselves at significantly lower energies, there is little hope that they can be tested in human-made accelerators. Cosmic messengers offer a solution to this problem. They achieve substantially higher energies and reach the Earth from enormous cosmological distances. Even though the energies of the astrophysical gamma rays, neutrinos, cosmic rays, as well as gravitational waves detected so far are still orders of magnitude below the Planck energy, large distances can serve as an amplifier to the possibly small effects of \ac{QG}. This  puts us in a position to search for and test these effects using astrophysical observations. The circumstances for experimental tests have become more favorable than ever with the dawn of multi-messenger astronomy. 

In this work we provided a comprehensive review of theoretical and phenomenological approaches to \ac{QG}. We presented up-to-date information on theoretical models and frameworks, and explored the state-of-the-art status of cosmic messenger astronomy, with the aim of discussing the most stringent bounds on the relevant \ac{QG} effects to date. Hopefully, this review has convinced even the most skeptical reader that \ac{QG} phenomenology is a viable and currently timely branch of fundamental physics research. Among the current bounds that we  listed, many of them already reach  Planck-scale sensitivity. In some cases we are even able to exclude specific Planck-scale effects with a reasonable confidence level. On the other hand, many proposed effects are still under scrutiny. Putting reliable bounds (or making a discovery) requires a thorough understanding not only of the theoretical framework of \ac{QG} and \ac{QG} phenomenology but especially of the subtleties involved in the experimental analysis and the physics of the sources and propagation of the messengers.
The goal of this review  is to equip the reader with the basic tools to embark in this endeavor: it provides an in-depth summary of the many aspects that are involved in the search for \ac{QG} signatures in astrophysical observations, ranging from the theoretical development to phenomenological modeling and to experimental techniques.

On the theoretical side, we followed both a top-down and a bottom-up approach. In the context of the former, we  presented (Sec.~\ref{section2}) brief descriptions of the basic fundamental theoretical frameworks of \ac{QG}, discussing their properties. In the context of the bottom-up approach, we gave an overview of the most relevant effective frameworks, including: doubly special relativity models, string-inspired effective field theory, higher-derivative extensions of general relativity, the standard-model extension with Lorentz invariance violation, non-commutative geometries, Finsler and Hamilton geometries, as well as models with modified uncertainty principles. The link between these effective models and the fundamental theories is yet to be established rigorously, due to the difficulties in working out definite predictions from the fundamental frameworks. For this reason, the main part of the review focused on the predictions from the effective models, and their experimental verification. We like to stress that so far there is no single figure of merit for the one or the other approach to \ac{QG}. All approaches must be simultaneously considered and investigated further. 

From a phenomenological point of view (Sec.~\ref{section5}), effective models of \ac{QG} can be classified according to the fate of Lorentz symmetries.  Besides Lorentz-invariant models, two other options are Lorentz invariance violation (for instance, of the type encountered in the so-called standard-model extension framework, or in non-critical string theory with recoiling cosmic defects) or deformed relativistic symmetries (which characterize doubly special relativity  models, relative locality  or some non-commutative geometries). One of the potential manifestations of departures from Lorentz invariance, common to some models from both approaches, are \aclp{MDR} of matter and radiation probes, which typically amount to an induced refractive index \textit{in vacuo}, with the relevant modifications increasing with (some positive power of) the energy of the probe. The experimental detection of such phenomena would constitute an important paradigm shift from our current knowledge and understanding of the fundamental interactions in nature. Whether this phenomenon is detected or not, a careful analysis is required in order to work out the implications for the different effective models, also taking into account their distinguishing features, e.g.\ concerning the laws of conservation of energy-momentum in interactions. Another important feature that could potentially characterize effective models of \ac{QG} is \acl{CPT} violation, which could exist either as a result of Lorentz invariance violation, or as a consequence of an ill-defined nature of the respective quantum generator of the \acl{CPT} symmetry in theories with space-time foamy structures. The latter entail quantum decoherence. In such cases, one may obtain ``smoking-gun'' type of effects in entangled particle-physics oscillating systems, such as neutral mesons. 

From the experimental side, we discussed the nature and origin of cosmic messengers (Sec.~\ref{section3}) and described the most important instruments and detection techniques (Sec.~\ref{section4}). We refrained from going into detailed description of specific instruments. Instead, we wanted to paint the picture from the emission in astrophysical sources until the detection. We focused especially on explaining the analysis methods used in search of effects of \ac{QG}, with a critical discussion of the results. It was our intention to point out the experimental difficulties encountered in these kinds of studies, in particular, the systematic effects which limit our sensitivity. Furthermore, we discussed the issue of degeneracy between the source-intrinsic effects, the astrophysical propagation ones, and those genuinely due to \ac{QG}. Given that our understanding of emission processes in astrophysical sources, as well as the nature of intergalactic and interstellar medium is rather limited, there is no exact way of resolving the mentioned effects. The usual approach is to exploit the expected distance and energy dependence of \ac{QG} effects in order to disentangle them from astrophysical effects. To this aim, we can perform analysis on several sources at different distances, or even on several different types of sources. These combinations should reduce the influence of the source intrinsic effects, leaving a signal that is more likely the consequence of propagation effects influenced by \ac{QG}. Nevertheless, even if an effect possibly related to \ac{QG} is detected, such as an energy-dependent modification of a messenger dispersion relation, numerous additional measurements will be necessary to understand the origin, nature, and the magnitude of this modification.

To unambiguously detect experimentally any of the effects mentioned above is a daunting task that we are only starting to be able to tackle with the latest generation of detectors. The last two decades have witnessed the development of a truly multi-messenger approach to the study of the universe. In the last twenty years large-area, stereoscopic \aclp{IACT}, large-volume neutrino telescopes, huge \acl{EAS} arrays and \acl{GW} detectors have joined the plethora of previously existing telescopes, allowing us to start peeking into the far universe with new ``eyes''. These, and future, highly sensitive instruments will surely lead to significant improvement of the bounds on current fundamental and phenomenological models of \ac{QG}, possibly reaching Planck-scale sensitivity and beyond. They might even enable us to measure some of the effects of \ac{QG} mentioned in this review. Another tantalizing possibility, always present in high-sensitivity measurements, is that we might detect effects or phenomena which we never expected. The two latter scenarios would spawn a plethora of new experiments, measurements, and theoretical interpretations. Any of the three possibilities will guide us towards a more fundamental understanding of the gravitational interaction and of the physics of the Planck-scale realm.

The title of this review, ``Quantum gravity phenomenology at the dawn of the multi-messenger era --- A review'', is inspired by that of the COST Action CA18108, within which it was prepared; indeed, multi-messenger observations have only recently become available, with only two astrophysical events having been detected with more than one type of messengers so far. We are at the very dawn of the multi-messenger era, which makes this collaboration between theorists, phenomenologists, and experimentalists in various experiments and in different messenger sectors most timely and convenient. By the time the multi-messenger observations become routine, our partnership will strive to devise the observational strategies and analysis methods optimized for testing a range of \ac{QG} models on multi-messenger data samples. Future endeavors should also include experimental efforts, such as terrestrial precision measurements and cosmological observations, which test models of quantum gravity not discussed in this review.

Victor Hess in his Nobel lecture in 1936 \cite{HESS2013351} said ``In order to make further progress, particularly in the field of cosmic rays, it will be necessary to apply all our resources and apparatus simultaneously and side-by-side; an effort which has not yet been made, or at least, only to a limited extent.'' This is exactly how we should proceed in order to ensure further progress in the search for \ac{QG} with multi-messenger astronomy. It will be necessary to combine all theoretical insights, resources and experimental efforts, something that has indeed only been done to a limited extent so far. This review article is the first step in forming such partnership. It is accompanied by a comprehensive and up-to-date census of experimental studies and results, the QG-MM Catalogue~\cite{QGMMCatalogue}, in which current observational bounds on \ac{QG} phenomenological effects are collected and made easily publicly available. A substantial progress towards the understanding of \ac{QG} and the Planck scale realm is expected to result from collaborations formed throughout this COST Action, and the activities which will follow after it ends.

%\newpage
\section*{Acronyms}
\addcontentsline{toc}{section}{Acronyms}
%%%%%%%%%%%%%%%%%%%%%%%%%%%%%%%%%%%%%%%%%%%%%%%%%%%%%%%%%%%%%%%%%%%%%%%%%%%%%%%%%%%%%%%%%%%%%%%%%%%%%%%%%%
%%%%%%%% Definitions of acronyms %%%%%
%%%%%%%%%%%%%%%%%%%%%%%%%%%%%%%%%%%%%%%%%%%%%%%%%%%%%%%%%%%%%%%%%%%%%%%%%%%%%%%%%%%%%%%%%%%%%%%%%%%%%%%%%%

\begin{multicols}{2}
\begin{acronym}
\acro{ACF}{autocorrelation function} 
\acro{AGN}{active galactic nucleus}
\acrodefplural{AGN}{active galactic nuclei} 
\acro{asl}{above mean sea level} 
\acro{BNS}{binary neutron star} 
\acro{BSM}{beyond the Standard Model} 
\acro{CCF}{cross-correlation function} 
\acro{CDT}{causal dynamical triangulation} 
\acro{CMB}{cosmic microwave background}
\acro{CP}{charge--parity}
\acro{CPT}{charge--parity--time}
\acro{CRB}{cosmic radio background}
\acro{CTA}{Cherenkov Telescope Array}
\acro{DSR}{doubly (or deformed) special relativity}
\acrodefplural{DSR}{doubly (or deformed) special relativities}
\acro{EAS}{extensive air shower}
\acro{EBL}{extragalactic background light}
\acro{EFT}{effective field theory}
\acrodefplural{EFT}{effective field theories}
\acro{EHE}{extremely high energy\acroextra{ ($E > 100\,\mathrm{PeV}$)}}
\acro{FD}{fluorescence detector}
\acro{FOV}[FoV]{field of view}
\acrodefplural{FOV}{fields of view}
\acro{GFT}{group field theory}
\acrodefplural{GFT}{group field theories}
\acro{GMF}{galactic magnetic fields} 
\acro{GRB}{gamma-ray burst}
\acro{GR}{general relativity}
\acro{GUP}{generalized uncertainty principle}
\acro{GW}{gravitational wave}
\acro{GZK}{Greisen--Zatsepin--Kuz'min}
\acro{HAWC}{High Altitude Water Cherenkov}
\acro{HDA}{hypersurface deformation algebra}
\acro{HE}{high energy\acroextra{ ($100\,\mathrm{MeV} < E < 100\,\mathrm{GeV}$)}}
\acro{H.E.S.S.}{High Energy Stereoscopic System}
\acro{IACT}{imaging atmospheric Cherenkov telescope}
\acro{IGMF}{intergalactic magnetic field} 
\acro{IR}{infrared}
\acro{LAT}{Large Area Telescope}
\acro{LE}{low energy\acroextra{ ($100\,\mathrm{keV} < E < 100\,\mathrm{MeV}$)}}
\acro{LHAASO}{Large High Altitude Air Shower Observatory}
\acro{LIDAR}{Light Detection and Ranging}
\acro{LIGO}{Laser Interferometer Gravitational-Wave Observatory}
\acro{LISA}{Laser Interferometer Space Antenna}
\acro{LIV}{Lorentz invariance violation}
\acro{LQC}{loop quantum cosmology}
\acro{LQG}{loop quantum gravity}
\acro{MAGIC}{Major Atmospheric Gamma Imaging Cherenkov}
\acro{MDR}{modified dispersion relation}
\acro{NSI}{non-standard interaction}
\acro{QCD}{quantum chromodynamics}
\acro{QED}{quantum electrodynamics}
\acro{QFT}{quantum field theory}
\acrodefplural{QFT}{quantum field theories}
\acro{QG}{quantum gravity}
\acro{QM}{quantum mechanics}
\acro{SD}{surface detector}
\acro{SED}{spectral energy distribution} 
\acro{SGWB}{stochastic gravitational-wave background}
\acro{SME}{Standard-Model Extension}
\acro{SM}{Standard Model}
\acro{SR}{special relativity}
\acro{UHE}{ultra-high energy\acroextra{ ($100\,\mathrm{TeV} < E < 100\,\mathrm{PeV}$)}}
\acro{UHECR}{ultra-high-energy cosmic ray\acroextra{ ($E > 1\,\mathrm{EeV}$)}}
\acro{UV}{ultraviolet}
\acro{VERITAS}{Very Energetic Radiation Imaging Telescope Array System}
\acro{VHE}{very high energy\acroextra{ ($100\,\mathrm{GeV} < E < 100\,\mathrm{TeV}$)}}
\end{acronym}
\end{multicols}
%%%%%%%%%%%%%%%%%%%%%%%%%%%%%%%%%%%%%%%%%%%%%%%%%%%%%%%%%%%%%%%%%%%%%%%%%%%%%%%%%%%%%%%%%%%%%%%%%%%%%%%%%%%

%\newpage
\section*{Funding Information and Acknowledgments}
\addcontentsline{toc}{section}{Funding Information and Acknowledgments}

\noindent Authors of this review have received support through the following grants: 

%\raggedright
\begin{FlushLeft}
\noindent 
Talent Scientific Research Program of College of Physics, Sichuan University, Grant No.1082204112427; the Fostering Program in Disciplines Possessing Novel Features for Natural Science of Sichuan University, Grant No. 2020SCUNL209; 1000 Talent program of Sichuan province 2021; Xunta de Galicia (Centro singular de investigación de Galicia accreditation 2019-2022); European Union ERDF, ``María de Maeztu'' Units of Excellence program (MDM-2016-0692); Red Temática Nacional de Astropartículas (RED2018-102661-T); ``la Caixa'' Foundation (ID 100010434); European Union's Horizon~2020 research and innovation program under the Marie Skłodowska-Curie grant agreement No.~847648, fellowship code LCF/BQ/PI21/11830030, and grant agreement No.~754510; Serbian Ministry of Education, Science and Technological Development Contract No. (451-03-9/2021-14/200124); FSR Incoming Postdoctoral Fellowship Ministry of Education, Science and Technological Development, Serbia (451-03-9/2021-14/200124); University of Rijeka grant (uniri-prirod-18-48); Croatian Science Foundation (HRZZ) project number IP-2016-06-9782; Villum Fonden (29405); DGA-FSE  [2020-E21-17R]; European Regional Development Fund through the Center of Excellence (TK133) ``The Dark Side of the Universe"; European Regional Development Fund (ESIF/ERDF) and the Czech Ministry of Education, Youth and Sports (MŠMT) [Project CoGraDS-CZ.02.1.01/0.0/0.0/15 003/0000437]; Blavatnik \& Rothchild grants; Basque Government grant (IT-979-16); Basque Foundation for Science (IKERBASQUE); ESA Prodex grants C4000120711, 4000132310; FNRS (Belgian Fund for Research); DGAPA-PAPIIT-UNAM [TA100122]; UNLP (X909); DICYT 042131GR (J.G.); Hungarian National Research Development and Innovation Office NKFIH (Grant No. 123996); FQXi; Swiss National Science Foundation (181461, 199307); 
Nederlandse Organisatie voor Wetenschappelijk Onderzoek - NWO  (grant numbers 680-91-119, 15MV71, and sectorplan);
Japan Society for the Promotion of Science KAKENHI Grant (20H01899, 20H05853 and JP21F21789); Estonian Research Council grants (PRG356) ``Gauge Gravity", and MOBTT5; Julian Schwinger Foundation; Generalitat Valenciana Excellence Grant [PROMETEO-II/2017/033] and [PROMETEO/2018/165]; Istituto Nazionale di Fisica Nucleare (INFN) Iniziativa Specifica TEONGRAV, Iniziativa Specifica QGSKY, Iniziativa Specifica QUAGRAP and Iniziativa Specifica GeoSymQFT; European ITN project HIDDeN (H2020-MSCA-ITN-2019//860881-HIDDeN); Swedish Research Council (2016-05996); European Research Council (668679); Advanced ERC grant TReX; Research grant ``The Dark Universe: A Synergic Multimessenger Approach'', No. [2017X7X85K] under the program PRIN 2017 funded by the Ministero dell’Istruzione, Universit`a e della Ricerca (MIUR); MIUR through the ``Dipartimenti di eccellenza'' project Science of the Universe; Research Council of University of Guilan; I.I.S.N. project 4.4501.18; Romanian Ministry of Research Innovation and Digitalization through the projects No. PN19-030102-INCDFM and CNCS - UEFISCDI No. PN-III-P4-ID-PCE-2020-2374; U.S. Department of Energy (DE-SC0020262, Task C); State project “Science” by the Ministry of Science and Higher Education of the Russian Federation (075-15-2020-778);
German Academic Scholarship Foundation; 
Deutsche Forschungsgemeinschaft - DFG - through grant numbers (408049454, 420243324, 425333893, 445990517) and under Germany’s Excellence Strategy (EXC 2121 “Quantum Universe” – 390833306, EXC 2123 ``QuantumFrontiers'' - 390837967);
Bundesministerium für Bildung und Forschung - BMBF (grants 05 A20GU2 and 05 A20PX1); 
Centro de Excelencia “Severo Ochoa” (SEV-2016-0588); CERCA program of the Generalitat de Catalunya; AGAUR, Generalitat de Catalunya (2017-SGR-1469, 2017-SGR-929);
ICCUB (CEX2019-000918-M);
National Science Centre (2019/33/B/ST2/00050) and \linebreak[4] (2017/27/B/ST2/01902); National Council for Scientific and Technological Development - CNPq [306414/2020-1];  Dicyt-USACH (041931MF); Bulgarian NSF grant KP-06-N 38/11; RCN ROMFORSK grant project.~no.~302640; Comunidad de Madrid, Spain, ``Atracción de Talento Investigador'' programme, grants No. 2018-T1/TIC-10431 and 2019-T1/TIC-13177; Comunidad de Madrid, Spain (S2018/NMT-4291 “TEC2SPACE-CM“, “Desarrollo y explotación de nuevas tecnologías para instrumentación espacial en la Comunidad de Madrid”); Science and Technology Facilities Council (STFC) UK, grants ST/T000759/1, ST/P000258/1, ST/T000732/1, ST/V005596/1; Fundação para a Ciência e a Tecnologia, Portugal, projects UIDB/00618/2020, UIDB/00777/2020, UIDP/00777/2020, CERN/FIS-PAR/0004/2019, PTDC/FIS-PAR/29436/2017, PTDC/FIS-PAR/31938/2017,  PTDC/FIS-OUT/29048/2017, and grant SFRH/BD/137127/2018; UID/MAT/00212/2020; FPU18/04571; 
Centre National de la Recherche Scientifique (CNRS), LabEx UnivEarthS (ANR-10-LABX-0023 and ANR18-IDEX-0001);
Junta de Andaluc\'{i}a [ref. A-FQM-053-UGR18], Spain; NSERC (RGPIN-2021-03644); National Science Centre Poland Sonata Bis Grant No. DEC-2017/26/E/ST2/00763 and Project No. 2019/33/B/ST2/00050; Natural Sciences and Engineering Research Council;  DGIID-DGA (2015-E24/2); Spanish Research State Agency and Ministerio de Ciencia e Innovación MCIN/AEI/10.13039/501100011033: PID2019-104114RB-C32, PID2019-105544GB-I00, PID2019-105614GB-C21, PID2019-106515GB-I00, PID2019-106802GB-I00, PID2019-107394GB-I00, PID2019-107844GB-C21, PID2019-107847RB-C41, PID2019-108485GB-I00, PID2020-113334GB-I00, PID2020-113701GB-I00, PID2020-113775GB-I00, PID2020-115845GB-I00, PID2020-118159GB-C41, PID2020-118159GA-C42, PRE2019-089024; \linebreak[4]
MCIN/AEI/10.13039/501100011033/FEDER: PGC2018-095328-B-I00, PGC2018-094856-B-I00, PGC2018-096663-B-C41, PGC2018-096663-B-C44, PGC2018-094626-B-C21, PGC2018-101858-B-I00, FPA2017-84543-P, FPA2016-76005-C2-1-P, PID2020-115845GB-I00/AEI/10.13039/501100011033; Ministerio de Universidades' BG20/00228; Ministerio de Econom\'ia, Industria y Competitividad (PID2019-105544GB-I00).
\end{FlushLeft}

S. Mukherjee, D. Minic, A. Platania and M. Schiffer acknowledge support by Perimeter Institute for Theoretical Physics. Research at Perimeter Institute is supported in part by the Government of Canada through the Department of Innovation, Science and Economic Development and by the Province of Ontario through the Ministry of Colleges and Universities. N. Stergioulas gratefully acknowledges the Italian Istituto Nazionale di Fisica Nucleare (INFN), the French Centre National de la Recherche Scientifique (CNRS) and the Netherlands Organization for Scientific Research, for the construction and operation of the Virgo detector and the creation and support of the EGO consortium. 
E. Guendelman thanks the Fundamental Questions Institute (FQXi) for finantial support.\\
%J.\ Alvarez-Muniz, R.\ Alves~Batista, D.\ Boncioli, A.\ di~Matteo, K.-H.\ Kampert, L.\ Miramonti, M.\ Niechciol, M.\ Risse, F.\ Sa\-la\-mi\-da, G.\ Sigl and C.\ Trimarelli acknowledge their membership in the Pierre Auger Collaboration.

The authors would like to acknowledge networking support by the COST Action CA18108.

\bibliographystyle{utphys} 
\bibliography{CA18108Bib}
\addcontentsline{toc}{section}{References}

\end{document}